# Addressing energy density functionals in the language of path-integrals II: Comparative study of functional renormalization group techniques applied to the (0+0)-D $O(N)$-symmetric $\varphi^4$-theory

Kilian Fraboulet[1,2,3] and Jean-Paul Ebran[1,2]

[1] CEA, DAM, DIF, 91297 Arpajon, France
[2] Université Paris-Saclay, CEA, Laboratoire Matière en Conditions Extrêmes, 91680, Bruyères-le-Châtel, France
[3] Institut für Theoretische Physik and Center for Quantum Science, Universität Tübingen, Auf der Morgenstelle 14, 72076 Tübingen, Germany

**Abstract.** The present paper is the second of a series of publications that aim at investigating relevant directions to turn the nuclear energy density functional (EDF) method as an effective field theory (EFT). The EDF approach has known numerous successes in nuclear theory over the past decades [1] and is currently the only microscopic technique that can be applied to all atomic nuclei. However, the phenomenological character of the EDF method also comes with important limitations, such as the lack of an explicit connection with quantum chromodynamics (QCD). As was argued in the first paper of this series [2], reformulating the EDF framework as an EFT would enable us to overcome these limitations. In particular, path-integral (PI) techniques are suited to achieve such a purpose as they allow us to design numerous non-perturbative approximations and can take Lagrangians possibly derived from EFTs of QCD as inputs. In our previous paper [2], we have illustrated such technical features for diagrammatic PI techniques in a study of the (0+0)-D $O(N)$-symmetric $\varphi^4$-theory. In the present work, we consider another class of PI techniques, i.e. functional renormalization group (FRG) approaches, that we apply to the same toy model. Despite our explicit interest for the nuclear many-body problem, the presented study is also directed towards FRG practitioners from various fields: technical details are provided for FRG techniques based on 1-particle-irreducible (1PI), 2-particle-irreducible (2PI) and 2-particle-point-irreducible (2PPI) effective actions (EAs), coined respectively as 1PI-, 2PI- and 2PPI-FRGs, and the treatment of the $O(N)$ symmetry is also addressed thoroughly. Connections between these various FRG methods are identified as well.

## 1 Introduction

### 1.1 Context: Energy density functional method and path-integral formalism

The energy density functional (EDF) method is arguably one of the most exploited microscopic approach in theoretical nuclear physics. This has actually been the case for the past several decades. As a matter of fact, nuclear physicists have estimated ground state (gs) observables and excited state properties throughout the whole nuclear chart thanks to EDF calculations [1, 3]. In the first paper of this series, coined as paper I, we have outlined the main features of the EDF method: i) The EDF approach relies on functionals of the nucleonic density that depend on free parameters fitted on nuclear data and the analytical forms of these functionals are postulated consistently with the symmetries of the two-body nucleon interaction; ii) The recipe underlying EDF calculations is essentially split into two steps: the single-reference implementation used as a first step, which is basically a mean-field approach obtained through a variational treatment of the aforementioned functional, and the multi-reference scheme, where all symmetries spontaneously broken at the single-reference level are restored, owing to the fact that mesoscopic systems such as nuclei can not exhibit spontaneous symmetry breakings (SSBs). Despite the presence of density functionals in the EDF formalism according to point i), the EDF approach should not be confused with density functional theory (DFT), as we have argued in more detail in paper I. Actually, the difference between EDF and DFT is notably due to point ii) and the presence of SSB in the EDF framework since DFT, and more precisely the Hohenberg-Kohn theorems [4] and the Kohn-Sham scheme [5, 6], are only formulated in a symmetry-conserving framework [7].

The phenomenological nature of the functionals (point i)) and the spontaneous breakdown and subsequent restoration of symmetries (point ii)) both allow for describing collective properties across the whole nuclear chart at low cost within the EDF framework. Naturally, this also comes with a price as this phenomenological character induces problematic limitations of the EDF approach as well, such as the lack of possibility to systematically improve the obtained results or the absence of explicit connection with quantum chromodynamics (QCD). In paper I, we explained technically that all of these limitations could be overcome by turning the EDF approach into an effective field theory (EFT) [8–20]. Although we pointed out other parallel directions aiming at designing more reliable approaches also based on density functionals (like the density-matrix expansion [21, 22]), we argued in addition that a framework of choice to achieve the reformulation of the EDF method as an EFT is the path-integral (PI) formulation of quantum field theory. This is due to the fact that PI techniques take Lagrangians (or classical actions), potentially derived from EFTs of QCD, as inputs. Most PI techniques are also themselves systematically improvable, as was illustrated in paper I. For the nuclear models we aim at constructing, this implies that both the nuclear interaction (i.e. the Lagrangian resulting from an EFT) and the many-body treatment (i.e. the PI technique) rely on an expansion and can therefore be the subject of systematically improvable approximations. Furthermore, there are several ways to introduce collective degrees of freedom (dofs) within the PI framework whereas such dofs also play a pivotal role for the EDF method, either at the level of the functional through densities or via order parameters introduced at the multi-reference level. On the side of the PI formalism, collective dofs can be introduced by means of a Hubbard-Stratonovich transformation (HST) and/or using higher-order effective actions (EAs), i.e. $n$-particle-(point-)irreducible ($n$P(P)I) EAs with $n > 1$. More precisely, 2PI EAs are functionals of propagators whereas their local versions, i.e. 2PPI EAs, are by definition functionals of the diagonal parts of these propagators, which can be identified as densities.

## 1.2 Recall from paper I

Let us then recall what was done in the first part of our study within paper I. In the latter, we have investigated several PI techniques: the loop expansion (LE) which notably implements perturbation theory in the PI formalism, optimized perturbation theory (OPT) and self-consistent perturbation theory (SCPT). Some of these methods were also combined with various resummation techniques when relevant. The playground chosen for these applications was the (0+0)-D $O(N)$-symmetric $\varphi^4$-theory, also called (0+0)-D $O(N)$ model, which is specified by the classical action:

$$S\left(\vec{\widetilde{\varphi}}\right) = \frac{m^2}{2}\vec{\widetilde{\varphi}}^2 + \frac{\lambda}{4!}\left(\vec{\widetilde{\varphi}}^2\right)^2 , \tag{1}$$

and relies e.g. on the generating functional[1]:

$$Z\left(\vec{J},\boldsymbol{K}\right) = e^{W\left(\vec{J},\boldsymbol{K}\right)} = \int_{\mathbb{R}^N} d^N\vec{\widetilde{\varphi}}\; e^{-S_{JK}\left(\vec{\widetilde{\varphi}}\right)} , \tag{2}$$

with[2]

$$S_{JK}\left(\vec{\widetilde{\varphi}}\right) \equiv S\left(\vec{\widetilde{\varphi}}\right) - \sum_{a=1}^{N} J_a\widetilde{\varphi}_a - \frac{1}{2}\sum_{a_1,a_2=1}^{N} \widetilde{\varphi}_{a_1}\boldsymbol{K}_{a_1a_2}\widetilde{\varphi}_{a_2} , \tag{3}$$

$\vec{J}$ and $\boldsymbol{K}$ being the sources and the integration in Eq. (2) runs over the $N$ components of the field $\vec{\widetilde{\varphi}}$, i.e.:

$$\vec{\widetilde{\varphi}} \equiv \begin{pmatrix} \widetilde{\varphi}_1 \\ \vdots \\ \widetilde{\varphi}_N \end{pmatrix} . \tag{4}$$

In paper I, we have investigated the effects of a HST on the performances of the aforementioned PI techniques by considering three mathematically equivalent representations of the (0+0)-D $O(N)$ model. Firstly, there is the *original representation* based on classical action (1), involving only the original dofs (4). Then, the *mixed representation* can be obtained by applying a HST to $Z\big(\vec{J}=\vec{0},\boldsymbol{K}=\boldsymbol{0}\big)$, thus yielding:

$$Z_{\text{mix}}(\mathcal{J},\mathcal{K}) = e^{W_{\text{mix}}(\mathcal{J},\mathcal{K})} = \int_{\mathbb{R}^{N+1}} d^N\vec{\widetilde{\varphi}}d\widetilde{\sigma}\; e^{-S_{\text{mix},\mathcal{J}\mathcal{K}}\left(\vec{\widetilde{\varphi}},\widetilde{\sigma}\right)} , \tag{5}$$

with

$$\begin{split} S_{\text{mix},\mathcal{J}\mathcal{K}}\left(\widetilde{\Psi}\right) \equiv {} & S_{\text{mix}}\left(\widetilde{\Psi}\right) - \sum_{b=1}^{N+1}\mathcal{J}_b\widetilde{\Psi}_b \\ & - \frac{1}{2}\sum_{b_1,b_2=1}^{N+1}\widetilde{\Psi}_{b_1}\mathcal{K}_{b_1b_2}\widetilde{\Psi}_{b_2} , \end{split} \tag{6}$$

$$S_{\text{mix}}\left(\widetilde{\Psi}\right) = \frac{m^2}{2}\vec{\widetilde{\varphi}}^2 + \frac{1}{2}\widetilde{\sigma}^2 + i\sqrt{\frac{\lambda}{12}}\widetilde{\sigma}\vec{\widetilde{\varphi}}^2 , \tag{7}$$

and

$$\widetilde{\Psi} \equiv \begin{pmatrix} \vec{\widetilde{\varphi}} \\ \widetilde{\sigma} \end{pmatrix} , \tag{8}$$

$$\mathcal{J} \equiv \begin{pmatrix} \vec{J} \\ j \end{pmatrix} , \tag{9}$$

[1] We have set $\hbar = 1$ in Eq. (2) and follow this convention throughout this entire paper.

[2] We slightly change our notations as compared to paper I, labeling color (supercolor) indices by $a_1$, $a_2$, $a_3$, ... ($b_1$, $b_2$, $b_3$, ...) instead of $a$, $b$, $c$, ... ($\alpha$, $\beta$, $\gamma$, ...) for convenience. Recall that color (supercolor) indices run by definition from 1 to $N$ ($N+1$) and label e.g. the components of $\vec{\widetilde{\varphi}}$ ($\widetilde{\Psi}$).

$$\mathcal{K} \equiv \begin{pmatrix} \boldsymbol{K} & \vec{0} \\ \vec{0}^{\mathrm{T}} & k \end{pmatrix} . \tag{10}$$

The theory now involves also a collective dof under the form of the fluctuating auxiliary field $\widetilde{\sigma}$. The original dofs can also be integrated out within the mixed representation by means of Gaussian integration, which gives us the *collective representation* of the system, involving the collective dof $\widetilde{\sigma}$ only. A generating functional of this collective representation is:

$$Z_{\mathrm{col}}(\mathcal{J}) = e^{W_{\mathrm{col}}(\mathcal{J})} = \int_{\mathbb{R}} d\widetilde{\sigma} \; e^{-S_{\mathrm{col},\mathcal{J}}(\widetilde{\sigma})} , \tag{11}$$

with

$$S_{\mathrm{col},\mathcal{J}}(\widetilde{\sigma}) = S_{\mathrm{col}}(\widetilde{\sigma}) - j\widetilde{\sigma} - \frac{1}{2} \sum_{a_1,a_2=1}^{N} J_{a_1} \boldsymbol{G}_{\widetilde{\sigma},a_1 a_2} J_{a_2} , \tag{12}$$

$$S_{\mathrm{col}}(\widetilde{\sigma}) = \frac{1}{2}\widetilde{\sigma}^2 - \frac{1}{2}\mathrm{Tr}_a \left[\ln(2\pi \boldsymbol{G}_{\widetilde{\sigma}})\right] , \tag{13}$$

and

$$\boldsymbol{G}^{-1}_{\widetilde{\sigma},a_1 a_2} = \left( m^2 + i\sqrt{\frac{\lambda}{3}}\widetilde{\sigma} \right) \delta_{a_1 a_2} , \tag{14}$$

and $\mathrm{Tr}_a$ denotes the trace over color indices.[3] The introduction of a collective dof via HST turned out to be particularly fruitful in our study of paper I. In particular, the LE in the collective representation and SCPT based on a 2PI EA in the mixed representation both yielded much better performances than their counterparts implemented within the original representation.

Among the different families of approaches tested in paper I, only SCPT is implemented in the EA framework. We have thus also investigated the effect of introducing collective dofs via 2PI EAs within SCPT. More precisely, SCPT has the peculiarity to rely on optimized expansions of EAs, where the optimization is carried out by extremizing the EA under consideration. The variational equations thus obtained are actually self-consistent in finite dimensions which, combined with the fact that some EAs are functionals of densities, enable us to say that SCPT is the PI technique closest in spirit to the EDF method (although it should be stressed that SCPT overcomes the aforementioned limitations of the EDF approach, enabling for instance to systematically improve Hartree-Fock(-Bogoliubov) results, as was illustrated in paper I).

## 1.3 Focus on paper II

In the present paper, i.e. paper II, we investigate other means to treat EAs by exploiting FRG approaches, which rely instead on (integro-)differential equations. The (0+0)-D $O(N)$-symmetric $\varphi^4$-theory will remain our playground and our study will therefore benefit once again from various features of this model. Firstly, its simplicity (and more specifically its zero-dimensional nature) allows for pushing our approaches to quite high truncation orders, to identify more readily connections between the different tested techniques and provides us with exact (analytical) solutions for benchmarking. Secondly, despite its simplicity, the classical action of this toy model possesses a symmetry, i.e. the $O(N)$ symmetry which is continuous for $N > 1$: this can also be seen as an advantage considering the pivotal role of symmetries within the EDF method. We stress that the exact solution of this zero-dimensional toy model always conserves the $O(N)$ symmetry[4] (explicitly and spontaneously), which constitutes an interesting link (illustrated in section 2 of paper I) with mesoscopic systems such as nuclei that can not spontaneously break symmetries either. However, we point out that the analysis of the possible occurrence of spontaneous breakdown of the $O(N)$ symmetry will be much more concise here as compared to that of paper I, notably because most of the tested FRG implementations (all those based on 2PI or 2PPI EAs) can not break the $O(N)$ symmetry by construction.

Regarding the tested FRG methods actually, we will also follow the steps of paper I by exploiting the aforementioned techniques to introduce collective (bosonic) dofs in the theory and examining their impacts on the performances of those methods. When possible, we will notably discuss the effect of a HST by considering the mixed and/or collective representation(s) of the studied model. Higher-order EAs will be exploited as well since we will implement FRG techniques formulated from 1PI, 2PI and 2PPI EAs, respectively referred to as 1PI-, 2PI- and 2PPI-FRGs. Although higher-order EAs occupy a rather small place in the FRG community as most FRG applications are based on 1PI EAs, this is certainly not the case in nuclear physics where a particular interest has been manifested for 2PPI EAs. Although PI techniques remain clearly marginal in the landscape of nuclear theory, several works exploring the 2PPI EA framework are to be mentioned like the contributions of Furnstahl and coworkers (already discussed in paper I) who pointed out the possibility and advantages to turn the EDF framework into an EFT in several of their papers [27–32]. A particular reason to consider this direction is the connection between DFT and 2PPI EAs, as it has been shown that the latter are density functionals in the sense of DFT [33], even in the sense of Kohn-Sham DFT [34]. The approach exploited in this way by Furnstahl and collaborators is thus coined as *ab initio* DFT for nuclei [30] but it should be stressed that it treats 2PPI EAs with SCPT, which has also been discussed within our (0+0)-D $O(N)$ model in paper I.[5]

[3] We refer to paper I for the expressions of the generating functionals (2), (5) and (11) in arbitrary dimensions.

[4] This feature of the exact solution of the (0+0)-D $O(N)$ model can be traced back to Mermin-Wagner theorem [23–25], as was shown in Ref. [26].

[5] In paper I, we have more precisely implemented SCPT based on 2PI EAs for the (0+0)-D $O(N)$ model and found that the optimal solutions of the corresponding gap equations

We will investigate in the present paper the possibility to design such an *ab initio* DFT using 2PPI-FRG instead of SCPT, thus following the direction set notably by Kemler and Braun [35–37] as well as Yokota and collaborators [38–43] in the framework of different models. However, the present study differentiates itself from the latter works by comparing 2PPI-FRG techniques with other FRG approaches, for various truncation orders. This will notably enable us to identify connections between 2PPI-FRG implementations and the 2PI-FRG formalism.

In what follows, sections 2, 3 and 4 contain our analysis based on 1PI-, 2PI- and 2PPI-FRG approaches, respectively. As in paper I, the results obtained for the gs energy and density of the (0+0)-D $O(N)$ model from the tested PI techniques will be benchmarked against the corresponding exact solutions, $E_{\text{gs}}^{\text{exact}}$ and $\rho_{\text{gs}}^{\text{exact}}$, whose analytical representations are already given in paper I. Furthermore, once again following the steps of paper I, those numerical applications will be performed at $N = 1$ or 2 (mostly at $N = 2$ for the sake of conciseness) with $\mathcal{Re}(\lambda) \geq 0$ and $\mathcal{Im}(\lambda) = 0$, investigating the two following phases: the phase or regime with $m^2 > 0$ coined as unbroken-symmetry phase and that with $m^2 < 0$ referred to as broken-symmetry phase. However, one must keep in mind that the exact solution of the studied (0+0)-D model actually does not exhibit any symmetry breaking, as mentioned previously, and, in our so-called broken-symmetry phase, the $O(N)$ symmetry is broken only *spontaneously* and only at the *classical* level. Finally, section 5 contains our conclusion, which is twofold. We first conclude on the FRG study of the present paper, comparing the most performing tested 1PI-, 2PI- and 2PPI-FRG approaches. Then, we conclude on our whole comparative study made of papers I and II, comparing this time the most performing FRG method of paper II with the best LE, OPT and SCPT implementations identified in paper I. In particular, since 2PI- and 2PPI-FRG approaches are significantly less present in the literature than their 1PI counterparts, we will keep the discussion general throughout most of sections 3 and 4, presenting notably the corresponding formalisms in arbitrary dimensions. In any case, technical details will be provided for all tested FRG techniques, in the main part of the text as well as in appendices:

- Appendix A: an introduction to the bosonic index formalism used in the 2PI EA framework.
- Appendix B: a discussion on two specific implementations of the 2PI-FRG, coined as C-flow [44] and CU-flow [45], illustrating notably, with numerical results for the (0+0)-D $O(N)$-symmetric $\varphi^4$-theory, that the C-flow is not suited to construct reliable approximation schemes.
- Appendix C: an explanation on how to simplify the flow equations within the 1PI- and 2PI-FRGs using the conservation of the $O(N)$ symmetry.
- Appendix D: a presentation of additional flow equations (with complementary derivations when considered relevant) that were used to obtain the numerical results presented in this paper.

This study has been carried out as part of a PhD project and we refer to the corresponding PhD thesis [46] for readers interested in further technical details.

## 2 1PI functional renormalization group

### 2.1 Generalities

As already mentioned, the most widespread FRG approach is the 1PI-FRG, that was proposed by Wetterich in Refs. [47–50] alongside with others [51–56]. It remains an active area in numerous fields of physics, as e.g. in QCD [57–69], quantum gravity [70–79], condensed matter physics [80–90] and even in out-of-equilibrium physics [91–101]. A few applications in the context of nuclear physics have been performed as well [102, 103]. Various toy model studies based on the 1PI-FRG are to be noted too, like those of the quantum anharmonic oscillator, i.e. the (0+1)-D $\varphi^4$-theory [104, 105]. In particular, the 1PI-FRG has already been used to treat the (0+0)-D $O(N)$ model [26, 106–108], notably in the framework of a recent project designing novel implementations of the 1PI-FRG in connection with fluid dynamics [26, 107–109]. We do not aim at presenting here all FRG implementations based on 1PI EAs but rather refer to reviews [110–112] or textbooks [113, 114] for complete overviews on this topic. We discuss instead the main ingredients of the 1PI-FRG approach that we will use throughout this entire study, which is still presently the most widespread FRG implementation for fermionic systems.

To define those main ingredients, we consider a local source, that we denote as $J_\alpha$, and a Schwinger functional $W[J]$, that we express as:

$$Z[J] = e^{W[J]} = \int \mathcal{D}\widetilde{\psi}\ e^{-S\left[\widetilde{\psi}\right]+\int_\alpha J_\alpha \widetilde{\psi}_\alpha} , \qquad (15)$$

where $\widetilde{\psi}$ is either a bosonic or a Grassmann field. The index $\alpha$ encompasses all types of internal indices $i$ as well as coordinates in Euclidean $D$-dimensional spacetime, such that the corresponding integration reads:

$$\int_\alpha \equiv \sum_i \int_0^{1/T} d\tau \int d^{D-1}r , \qquad (16)$$

with $T$ the temperature. The 1PI EA $\Gamma^{(\text{1PI})}[\psi]$ is related to $W[J]$ via the Legendre transform:

$$\Gamma^{(\text{1PI})}[\psi] = -W[J] + \int_\alpha J_\alpha \psi_\alpha , \qquad (17)$$

with

$$\psi_\alpha = \frac{\delta W[J]}{\delta J_\alpha} . \qquad (18)$$

always fully conserve the $O(N)$ symmetry. Furthermore, in the absence of SSB, the 2PI and 2PPI EAs of our zero-dimensional toy model coincide, which is why we have also tested SCPT based on 2PPI EAs in this comparative study.

Within the FRG framework, one introduces a scale-dependence into the generating functionals of interest via a cutoff function. For the Schwinger functional and the 1PI EA introduced in Eqs. (15) and (17), this can be done as follows:

$$Z_k[J] = e^{W_k[J]} = \int \mathcal{D}\widetilde{\psi}\ e^{-S\left[\widetilde{\psi}\right]-\Delta S_k\left[\widetilde{\psi}\right]+\int_\alpha J_\alpha \widetilde{\psi}_\alpha} \ , \quad (19)$$

$$\Gamma^{(1\mathrm{PI})}_k[\psi] = -W_k[J] + \int_\alpha J_\alpha \psi_\alpha - \Delta S_k[\psi] \ , \quad (20)$$

where

$$\psi_\alpha \equiv \psi_{k,\alpha}[J] = \frac{\delta W_k[J]}{\delta J_\alpha} \ , \quad (21)$$

and $\Delta S_k$ is defined in terms of the cutoff function $R_k$ as:

$$\Delta S_k\left[\widetilde{\psi}\right] = \frac{1}{2}\int_{\alpha_1,\alpha_2} \widetilde{\psi}_{\alpha_1} R_{k,\alpha_1\alpha_2} \widetilde{\psi}_{\alpha_2} \ . \quad (22)$$

By then differentiating Eq. (20) with respect to the scale $k$, one can derive the following *exact* differential equation for the effective average action or flowing EA $\Gamma^{(1\mathrm{PI})}_k[\psi]$:

$$\begin{aligned}\dot{\Gamma}^{(1\mathrm{PI})}_k[\psi] &\equiv \partial_k \Gamma^{(1\mathrm{PI})}_k[\psi] \\ &= \frac{1}{2}\mathrm{STr}\left[\dot{R}_k\left(\Gamma^{(1\mathrm{PI})(2)}_k[\psi] + R_k\right)^{-1}\right] \ ,\end{aligned} \quad (23)$$

which is the celebrated Wetterich equation [48]. The functional matrix $\Gamma^{(1\mathrm{PI})(2)}_k[\psi] \equiv \frac{\delta^2 \Gamma^{(1\mathrm{PI})}_k[\psi]}{\delta\psi\delta\psi}$ is the Hessian of $\Gamma^{(1\mathrm{PI})}_k$ and STr denotes the supertrace with respect to $\alpha$-indices. In practice, the Wetterich equation can not be solved directly (except for some toy models, such as the $O(N)$ model considered in this study [106]) but must rather be turned into a *truncated* set of coupled differential equations using a given expansion scheme. In the present study, the Wetterich equation is always treated within one expansion scheme, namely the vertex expansion [55, 115]. The latter consists in expanding all sides of the Wetterich equation (23) in powers of the field $\psi$. In particular, such an expansion for the left-hand side of Eq. (23) directly follows from the following Taylor expansion of $\Gamma^{(1\mathrm{PI})}_k$:

$$\begin{aligned}\Gamma^{(1\mathrm{PI})}_k[\psi] = \overline{\Gamma}^{(1\mathrm{PI})}_k & \\ + \sum^{\infty}_{n=2} \frac{1}{n!} & \int_{\alpha_1,\cdots,\alpha_n} \overline{\Gamma}^{(1\mathrm{PI})(n)}_{k,\alpha_1\cdots\alpha_n} \\ \times & \left(\psi-\overline{\psi}_k\right)_{\alpha_1}\cdots\left(\psi-\overline{\psi}_k\right)_{\alpha_n} \ ,\end{aligned} \quad (24)$$

where $\overline{\psi}_{k,\alpha} = \left.\frac{\delta W_k[J]}{\delta J_\alpha}\right|_{J=0}$, $\overline{\Gamma}^{(1\mathrm{PI})}_k \equiv \Gamma^{(1\mathrm{PI})}_k[\psi=\overline{\psi}_k]$, $\overline{\Gamma}^{(1\mathrm{PI})(n)}_{k,\alpha_1\cdots\alpha_n} \equiv \left.\frac{\delta^n \Gamma^{(1\mathrm{PI})}_k[\psi]}{\delta\psi_{\alpha_1}\cdots\delta\psi_{\alpha_n}}\right|_{\psi=\overline{\psi}_k}$ and $\overline{\psi}_k$ must extremize the flowing EA, i.e.:

$$\left.\frac{\delta\Gamma^{(1\mathrm{PI})}_k[\psi]}{\delta\psi_\alpha}\right|_{\psi=\overline{\psi}_k} = 0 \quad \forall\alpha, k \ . \quad (25)$$

The final step of the vertex expansion amounts to identifying the terms of identical powers in the field in the expanded Wetterich equation to infer an infinite tower or hierarchy of coupled integro-differential equations for the 1PI vertices $\overline{\Gamma}^{(1\mathrm{PI})(n)}_k$. This tower is of course always truncated in practice: in current treatments of fermionic systems for instance [116–121], one typically uses the so-called level-2 truncation in which ($\overline{\Gamma}^{(1\mathrm{PI})}_k$,) $\overline{\Gamma}^{(1\mathrm{PI})(2)}_k$ and $\overline{\Gamma}^{(1\mathrm{PI})(4)}_k$ are the only flowing vertices.

Although this approach based on the vertex expansion can be seen as the current canonical 1PI-FRG implementation to describe fermions, it is not suited to tackle strongly-coupled purely fermionic systems since the expansion (24) can only be performed around the trivial configuration $\overline{\psi}_{k,\alpha} = 0\ \forall\alpha$ if $\psi$ is a Grassmann field. This implies that, in this situation, the FRG reduces basically to a LE and is therefore perturbative in essence. One can however directly remedy this by e.g. bosonizing the system with a HST and then perform the vertex expansion around a non-trivial configuration for the Hubbard-Stratonovich field, the resulting approach being henceforth non-perturbative. This highlights the importance of our 1PI-FRG investigations within the mixed and collective representations of the (0+0)-D $O(N)$ model: indeed, even if the latter model is originally purely bosonic, we will often discuss the relevance of the tested methods for (realistic) fermionic problems and analyze the possibility to extend the conclusions of our comparative study towards such problems. Note also that there are other very performing expansion schemes of the Wetterich equation besides the vertex expansion, especially to treat $O(N)$ models. In that respect, one can quote in particular the derivative expansion (DE) [122–127] and the Blaizot-Méndez-Galain-Wschebor (BMW) approximation[6] [129–137]. However, it should be noted that, in (0+0)-D, the DE amounts to rewriting exactly the Wetterich equation at lowest order, i.e. the DE is no longer an approximate method in (0+0)-D, as discussed in Ref. [106]. Therefore, we will not consider the DE in the present toy model study. The BMW approximation will not be investigated in our analysis either and we will content ourselves with the vertex expansion.

Finally, we want to address a key aspect of FRG approaches, i.e. the *starting point of the flow*. These approaches indeed all rely on coupled differential equations that require initial conditions to be solved numerically. Before discussing the matter of initial conditions further, we would like to stress that the original formulation of the 1PI-FRG based on the Wetterich equation implements the *Wilsonian momentum-shell integration*, namely a step-by-step integration with respect to the momentum scale, which allows for an efficient description of critical phenomena [49, 112, 138–144]. Despite this, one should keep in mind that there are still fundamental differences be-tween the 1PI-FRG and Wilson's renormalization group (RG): pedagogical explanations on this point can be found e.g. in Ref. [128]. The flow parameter $k$ introduced via $R_k$ in Eq. (19) thus corresponds to a momentum scale in the 1PI-FRG considered here. More precisely, the differential equations underlying this FRG approach are solved by evolving the flow parameter $k$ from a chosen ultraviolet cutoff $k_\mathrm{i} = \Lambda$ down to $k_\mathrm{f} = 0$. Although we will not discuss in detail the different possible relevant choices for the analytical expressions of $R_k$ in finite dimensions,[7] we will just point out that this cutoff function must satisfy in the present case:

$$\begin{cases} R_{k=k_\mathrm{i},\alpha_1\alpha_2} = \infty \quad \forall \alpha_1,\alpha_2 \; . & \text{(26a)} \\ R_{k=k_\mathrm{f},\alpha_1\alpha_2} = 0 \quad \forall \alpha_1,\alpha_2 \; . & \text{(26b)} \end{cases}$$

[6] See Refs. [112,128] for pedagogical presentations of the DE and the BMW approximation, and Ref. [112] for e.g. a comparison of the performances of those various expansion schemes in the determination of critical exponents.

Condition (26b) directly follows from the fact that we must recover the theory under consideration, specified by Eq. (15), at the end of the flow. Furthermore, it can also be shown that the divergence of the cutoff function imposed by Eq. (26a) at $k = k_\mathrm{i}$ imposes that the starting point of the flow coincides with the classical theory, as a consequence of the modification of the Legendre transform in Eq. (20). This translates into:

$$\begin{cases} \Gamma^{(\mathrm{1PI})}_{k=k_\mathrm{i}}[\psi] = S\left[\widetilde{\psi} = \psi\right] \; . & \text{(27a)} \\ \Gamma^{(\mathrm{1PI})}_{k=k_\mathrm{f}}[\psi] = \Gamma^{(\mathrm{1PI})}[\psi] \; . & \text{(27b)} \end{cases}$$

Therefore, the 1PI-FRG procedure originally formulated by Wetterich starts from the classical theory, incorporates progressively quantum correlations on top of it throughout the flow by solving integro-differential equations, so as to reach the corresponding quantum theory. It should be noted however that several drawbacks of this 1PI-FRG formulation, based on both the Wetterich equation and on the Wilsonian momentum-shell integration, have been identified along the years [148], such as violations of Ward identities at finite values of $R_k$ [149, 150]. For this reason notably, other flow schemes, still treating the Wetterich equation but using different initial conditions and possibly other flow parameters (such as the temperature), became popular notably in the condensed matter theory community: we can mention for example the temperature flow [151], the interaction flow [152] and the $\Omega$-flow [153,154] in that respect. Once again, the framework chosen for the present study is certainly not relevant to compare all these flow schemes,[8] owing to the vanishing of momentum and frequency dependences in (0+0)-D in particular. Moreover, the aforementioned drawbacks do not apply to the (0+0)-D model studied here either and we will therefore only consider the FRG approach described so far, which uses the classical theory as starting point, in our 1PI-FRG analysis.

Regardless of the chosen flow scheme, one only reaches an approximated version of the quantum theory specified by $\Gamma^{(\mathrm{1PI})}[\psi]$ at the end of the flow in practice, i.e. equality (27b) is only approximatively fulfilled in actual calculations owing to the truncation of the infinite hierarchy of equations resulting from the treatment of the Wetterich equation. One might actually say that this is where the power of this approach lies since such a truncation implies that approximations within the FRG scheme do not rely on the smallness of coupling constants (although one should keep in mind that this does not hold for the 1PI-FRG based on a vertex expansion of the Wetterich equation for purely fermionic systems, as explained earlier) and can therefore provide us with the possibility to design various non-perturbative approaches. This will be thoroughly illustrated in the following sections. We can however already appreciate such a potential for our aim of reformulating the EDF method, considering the strongly-coupled nature of the nuclear many-body problem.

## 2.2 Original 1PI functional renormalization group

We now turn to our 1PI-FRG applications. As opposed to the other sections dedicated to the 2PI- and 2PPI-FRGs, we mainly consider the formalism in (0+0)-D for our 1PI-FRG analysis, notably since the corresponding developments at finite dimensions are already exhaustively discussed in the available literature [110–114]. We thus consider the (0+0)-D limit from now on until the end of section 2. In this framework, the Wetterich equation for a 1PI EA $\Gamma^{(\mathrm{1PI})}_k\left(\vec{\phi}\right)$ reads:

$$\begin{aligned} \dot{\Gamma}^{(\mathrm{1PI})}_k\left(\vec{\phi}\right) &= \frac{1}{2}\mathrm{STr}\left[\dot{\boldsymbol{R}}_k\left(\Gamma^{(\mathrm{1PI})(2)}_k\left(\vec{\phi}\right)+\boldsymbol{R}_k\right)^{-1}\right] \\ &= \frac{1}{2}\sum^{N}_{a_1,a_2=1}\dot{\boldsymbol{R}}_{k,a_1a_2}\left(\Gamma^{(\mathrm{1PI})(2)}_k\left(\vec{\phi}\right)+\boldsymbol{R}_k\right)^{-1}_{a_2a_1} \; , \end{aligned} \tag{28}$$

with

$$\Gamma^{(\mathrm{1PI})(2)}_{k,a_1a_2}\left(\vec{\phi}\right) \equiv \frac{\partial^2\Gamma^{(\mathrm{1PI})}_k\left(\vec{\phi}\right)}{\partial\phi_{a_1}\partial\phi_{a_2}} \; . \tag{29}$$

As can be deduced from the general formalism presented in section 2.1, the vertex expansion procedure applied to Eq. (28) starts from the Taylor expansion of $\Gamma^{(\mathrm{1PI})}_k\left(\vec{\phi}\right)$ around its flowing extremum at $\vec{\phi} = \vec{\overline{\phi}}_k$, which reduces in (0+0)-D to:

$$\begin{aligned} \Gamma^{(\mathrm{1PI})}_k\left(\vec{\phi}\right) &= \overline{\Gamma}^{(\mathrm{1PI})}_k \\ &+ \sum^{\infty}_{n=2}\frac{1}{n!}\sum^{N}_{a_1,\cdots,a_n=1}\overline{\Gamma}^{(\mathrm{1PI})(n)}_{k,a_1\cdots a_n} \\ &\times\left(\vec{\phi}-\vec{\overline{\phi}}_k\right)_{a_1}\cdots\left(\vec{\phi}-\vec{\overline{\phi}}_k\right)_{a_n} \; . \end{aligned} \tag{30}$$

---

[7] Widespread choices for $R_k$ are the exponential regulator or the Litim regulator [145,146]. See Refs. [110,112,147] for more exhaustive discussions on this topic.

[8] The $\Omega$- and temperature flows have already been compared in the framework of a (2+1)-D Hubbard model [155].

We have used in Eq. (30) the definitions:

$$\overline{\Gamma}^{(1\mathrm{PI})}_{k} \equiv \Gamma^{(1\mathrm{PI})}_{k}\left(\vec{\phi}=\vec{\overline{\phi}}_{k}\right) \quad \forall k \;, \tag{31}$$

$$\overline{\Gamma}^{(1\mathrm{PI})(n)}_{k,a_1\cdots a_n} \equiv \left.\frac{\partial^n \Gamma^{(1\mathrm{PI})}_{k}(\vec{\phi})}{\partial\phi_{a_1}\cdots\partial\phi_{a_n}}\right|_{\vec{\phi}=\vec{\overline{\phi}}_{k}} \quad \forall a_1,\cdots,a_n,k \;, \tag{32}$$

and

$$\overline{\Gamma}^{(1\mathrm{PI})(1)}_{k,a} = 0 \quad \forall a,k \;, \tag{33}$$

by construction. We will then distinguish two situations to pursue the vertex expansion further. On the one hand, in the unbroken-symmetry regime (i.e. in the phase with $m^2>0$) where $\vec{\overline{\phi}}_k = \vec{0}\ \forall k$, the infinite set of differential equations resulting from the vertex expansion includes[9]:

$$\dot{\overline{\Gamma}}^{(1\mathrm{PI})}_{k} = \frac{N}{2}\dot{R}_k\left(\overline{G}_k - \overline{G}^{(0)}_k\right) \;, \tag{34}$$

$$\dot{\overline{\Gamma}}^{(1\mathrm{PI})(2)}_{k} = -\frac{N+2}{6}\dot{R}_k\overline{G}^2_k\overline{\Gamma}^{(1\mathrm{PI})(4)}_{k} \;, \tag{35}$$

$$\begin{aligned}\dot{\overline{\Gamma}}^{(1\mathrm{PI})(4)}_{k} =& \frac{N+8}{3}\dot{R}_k\overline{G}^3_k\left(\overline{\Gamma}^{(1\mathrm{PI})(4)}_{k}\right)^2 \\ &-\frac{N+4}{10}\dot{R}_k\overline{G}^2_k\overline{\Gamma}^{(1\mathrm{PI})(6)}_{k} \;,\end{aligned} \tag{36}$$

$$\begin{aligned}\dot{\overline{\Gamma}}^{(1\mathrm{PI})(6)}_{k} =& -\frac{5N+130}{3}\dot{R}_k\overline{G}^4_k\left(\overline{\Gamma}^{(1\mathrm{PI})(4)}_{k}\right)^3 \\ &+(N+14)\,\dot{R}_k\overline{G}^3_k\overline{\Gamma}^{(1\mathrm{PI})(4)}_{k}\overline{\Gamma}^{(1\mathrm{PI})(6)}_{k} \\ &-\frac{N+6}{14}\dot{R}_k\overline{G}^2_k\overline{\Gamma}^{(1\mathrm{PI})(8)}_{k} \;,\end{aligned} \tag{37}$$

where, as shown later by Eqs. (45) and (46), $\overline{G}_k$ and $\overline{G}^{(0)}_k$ are respectively the diagonal parts of the propagators $\boldsymbol{\overline{G}}_k$ and $\boldsymbol{\overline{G}}^{(0)}_k$ defined by:

$$\boldsymbol{\overline{G}}^{-1}_{k,a_1a_2} \equiv \overline{\Gamma}^{(1\mathrm{PI})(2)}_{k,a_1a_2} + \boldsymbol{R}_{k,a_1a_2} \;, \tag{38}$$

$$\left(\boldsymbol{\overline{G}}^{(0)}_k\right)^{-1}_{a_1a_2} \equiv \overline{\Gamma}^{(1\mathrm{PI})(2)}_{k=k_\mathrm{i},a_1a_2} + \boldsymbol{R}_{k,a_1a_2} \;, \tag{39}$$

with

$$\boldsymbol{R}_{k,a_1a_2} = R_k\ \delta_{a_1a_2} \;. \tag{40}$$

In Eqs. (34) to (37), we have also used the following relations resulting from the $O(N)$ symmetry:

$$\overline{\Gamma}^{(1\mathrm{PI})(2)}_{k,a_1a_2} = \overline{\Gamma}^{(1\mathrm{PI})(2)}_{k}\ \delta_{a_1a_2} \quad \forall a_1,a_2 \;, \tag{41}$$

$$\begin{aligned}\overline{\Gamma}^{(1\mathrm{PI})(4)}_{k,a_1a_2a_3a_4} =& \frac{\overline{\Gamma}^{(1\mathrm{PI})(4)}_{k}}{3}\left(\delta_{a_1a_2}\delta_{a_3a_4}+\delta_{a_1a_3}\delta_{a_2a_4}\right. \\ &\left.+\delta_{a_1a_4}\delta_{a_2a_3}\right) \quad \forall a_1,a_2,a_3,a_4 \;,\end{aligned} \tag{42}$$

$$\begin{aligned}\overline{\Gamma}^{(1\mathrm{PI})(6)}_{k,a_1a_2a_3a_4a_5a_6} =& \frac{\overline{\Gamma}^{(1\mathrm{PI})(6)}_{k}}{45}\left(\delta_{a_1a_2}\delta_{a_3a_4}\delta_{a_5a_6}\right. \\ &+\delta_{a_1a_2}\delta_{a_3a_5}\delta_{a_4a_6}+\delta_{a_1a_2}\delta_{a_3a_6}\delta_{a_4a_5} \\ &+\delta_{a_1a_3}\delta_{a_2a_4}\delta_{a_5a_6}+\delta_{a_1a_3}\delta_{a_2a_5}\delta_{a_4a_6} \\ &+\delta_{a_1a_3}\delta_{a_2a_6}\delta_{a_4a_5}+\delta_{a_1a_4}\delta_{a_2a_3}\delta_{a_5a_6} \\ &+\delta_{a_1a_4}\delta_{a_2a_5}\delta_{a_3a_6}+\delta_{a_1a_4}\delta_{a_2a_6}\delta_{a_3a_5} \\ &+\delta_{a_1a_5}\delta_{a_2a_3}\delta_{a_4a_6}+\delta_{a_1a_5}\delta_{a_2a_4}\delta_{a_3a_6} \\ &+\delta_{a_1a_5}\delta_{a_2a_6}\delta_{a_3a_4}+\delta_{a_1a_6}\delta_{a_2a_3}\delta_{a_4a_5} \\ &\left.+\delta_{a_1a_6}\delta_{a_2a_4}\delta_{a_3a_5}+\delta_{a_1a_6}\delta_{a_2a_5}\delta_{a_3a_4}\right) \\ &\forall a_1,\cdots,a_6 \;,\end{aligned} \tag{43}$$

$$\overline{\Gamma}^{(1\mathrm{PI})(n)}_{k,a_1\cdots a_n} = 0 \quad \forall a_1,\cdots,a_n,\ \forall n \text{ odd} \;. \tag{44}$$

As a consequence of Eq. (41), the propagators $\boldsymbol{\overline{G}}_k$ and $\boldsymbol{\overline{G}}^{(0)}_k$ satisfy:

$$\boldsymbol{\overline{G}}_{k,a_1a_2} = \overline{G}_k\ \delta_{a_1a_2} = \left(\overline{\Gamma}^{(1\mathrm{PI})(2)}_{k} + R_k\right)^{-1}\delta_{a_1a_2} \;, \tag{45}$$

$$\boldsymbol{\overline{G}}^{(0)}_{k,a_1a_2} = \overline{G}^{(0)}_k\ \delta_{a_1a_2} = \left(\overline{\Gamma}^{(1\mathrm{PI})(2)}_{k=k_\mathrm{i}} + R_k\right)^{-1}\delta_{a_1a_2} \;. \tag{46}$$

Note that the differential equations (34) to (37) are already given in Ref. [106] which indeed already treats the unbroken-symmetry phase of the (0+0)-D $O(N)$ model in its original representation with the 1PI-FRG considered here, i.e. that based on the vertex expansion of the Wetterich equation. This 1PI-FRG implementation is however not exploited to tackle the broken-symmetry phase of our toy model in the study of Ref. [106].[10] In this phase (i.e.

[9] We have performed the substitution $\overline{G}_k \rightarrow \overline{G}_k - \overline{G}^{(0)}_k$ in the output of the vertex expansion procedure so as to obtain Eq. (34) (the same remark applies to Eq. (47) given afterwards), which is also the procedure followed in Refs. [105,106]. Physical observables are not affected by this shift and the necessity of it can be seen by the fact that no quantum corrections must be added to $\overline{\Gamma}^{(1\mathrm{PI})}_{k}$ throughout the flow if $\lambda=0$, i.e. the relation $\overline{\Gamma}^{(1\mathrm{PI})}_{k=k_\mathrm{f}} = \overline{\Gamma}^{(1\mathrm{PI})}_{k=k_\mathrm{i}}$ must hold in the free case, which is indeed satisfied thanks to this operation. We note also that this induces a redefinition of the 1PI EA, which we nonetheless still denote as $\Gamma^{(1\mathrm{PI})}$ in what follows.

[10] Although Ref. [106] discusses the performances of several *approximated* PI techniques (perturbation theory, $1/N$-expansion, 1PI-FRG based on the vertex expansion of the Wetterich equation) only in the framework of the unbroken-symmetry phase of the (0+0)-D $O(N)$ model, it also reports a calculation of the effective potential by an *exact* resolution of the Wetterich equation in the broken-symmetry phase. It shows in this way that the 1PI-FRG flow restores the $O(N)$ symmetry broken by its classical starting point, which is consistent with Fig. 1 of paper I together with the Mermin-Wagner theorem.

in the phase with $m^2 < 0$), the infinite tower resulting from the vertex expansion contains the following differential equations at $N = 1$:

$$\dot{\overline{\Gamma}}_k^{(\mathrm{1PI})} = \frac{1}{2}\dot{R}_k\left(\overline{G}_k - \overline{G}_k^{(0)}\right) , \tag{47}$$

$$\dot{\overline{\phi}}_k = \frac{1}{2\overline{\Gamma}_k^{(\mathrm{1PI})(2)}}\dot{R}_k\overline{G}_k^2\overline{\Gamma}_k^{(\mathrm{1PI})(3)} , \tag{48}$$

$$\begin{aligned}\dot{\overline{\Gamma}}_k^{(\mathrm{1PI})(2)} &= \dot{\overline{\phi}}_k\overline{\Gamma}_k^{(\mathrm{1PI})(3)} + \dot{R}_k\overline{G}_k^3\left(\overline{\Gamma}_k^{(\mathrm{1PI})(3)}\right)^2 \\ &\quad - \frac{1}{2}\dot{R}_k\overline{G}_k^2\overline{\Gamma}_k^{(\mathrm{1PI})(4)} ,\end{aligned} \tag{49}$$

$$\begin{aligned}\dot{\overline{\Gamma}}_k^{(\mathrm{1PI})(3)} &= \dot{\overline{\phi}}_k\overline{\Gamma}_k^{(\mathrm{1PI})(4)} - 3\dot{R}_k\overline{G}_k^4\left(\overline{\Gamma}_k^{(\mathrm{1PI})(3)}\right)^3 \\ &\quad + 3\dot{R}_k\overline{G}_k^3\overline{\Gamma}_k^{(\mathrm{1PI})(3)}\overline{\Gamma}_k^{(\mathrm{1PI})(4)} \\ &\quad - \frac{1}{2}\dot{R}_k\overline{G}_k^2\overline{\Gamma}_k^{(\mathrm{1PI})(5)} ,\end{aligned} \tag{50}$$

$$\begin{aligned}\dot{\overline{\Gamma}}_k^{(\mathrm{1PI})(4)} &= \dot{\overline{\phi}}_k\overline{\Gamma}_k^{(\mathrm{1PI})(5)} + 12\dot{R}_k\overline{G}_k^5\left(\overline{\Gamma}_k^{(\mathrm{1PI})(3)}\right)^4 \\ &\quad - 18\dot{R}_k\overline{G}_k^4\left(\overline{\Gamma}_k^{(\mathrm{1PI})(3)}\right)^2\overline{\Gamma}_k^{(\mathrm{1PI})(4)} \\ &\quad + 4\dot{R}_k\overline{G}_k^3\overline{\Gamma}_k^{(\mathrm{1PI})(3)}\overline{\Gamma}_k^{(\mathrm{1PI})(5)} \\ &\quad + 3\dot{R}_k\overline{G}_k^3\left(\overline{\Gamma}_k^{(\mathrm{1PI})(4)}\right)^2 - \frac{1}{2}\dot{R}_k\overline{G}_k^2\overline{\Gamma}_k^{(\mathrm{1PI})(6)} ,\end{aligned} \tag{51}$$

with $\overline{\Gamma}_{k,1\cdots 1}^{(\mathrm{1PI})(n)} \equiv \overline{\Gamma}_k^{(\mathrm{1PI})(n)}$. Moreover, the propagators $\overline{G}_k \equiv \overline{\boldsymbol{G}}_{k,11}$ and $\overline{G}_k^{(0)} \equiv \overline{\boldsymbol{G}}_{k,11}^{(0)}$ are still given by Eqs. (38) and (39), respectively.

At the present stage, we still have not specified the model under consideration: all we know is that we are dealing with a 1PI EA depending on a single field $\vec{\phi}$ which is a vector in color space and lives in a (0+0)-D spacetime. In the framework of FRG approaches, the model (i.e. the classical action under consideration) is often only specified via the initial conditions used to solve the underlying set of differential equations. The initial conditions used to solve the above two sets of differential equations for our (0+0)-D $O(N)$ model can be obtained by assuming that the $O(N)$ symmetry can only be spontaneously broken in the direction set by $a = N$ in color space (i.e. by assuming that $\vec{\phi}^2 = \phi_N^2$), thus following the convention already used in paper I. In this case, they are given by[11]:

$$\overline{\phi}_{k=k_{\mathrm{i}},a} = \overline{\varphi}_{\mathrm{cl},a} = \begin{cases} 0 \quad \forall a,\ \forall m^2 > 0 , \\ \pm\sqrt{-\dfrac{6m^2}{\lambda}}\ \delta_{aN} \quad \forall a,\ \forall m^2 < 0 \text{ and } \lambda \neq 0 , \end{cases} \tag{52}$$

[11] The logarithm term in Eq. (53) was added to shift the calculated gs energy $E_{\mathrm{gs}}$ so that the latter coincides with the corresponding exact solution given in paper I for $\lambda = 0$ and $m^2 > 0$.

$$\overline{\Gamma}_{k=k_{\mathrm{i}}}^{(\mathrm{1PI})} = \begin{cases} -\dfrac{N}{2}\ln\left(\dfrac{2\pi}{m^2}\right) + S\left(\vec{\widetilde{\varphi}} = \vec{\overline{\phi}}_{k=k_{\mathrm{i}}}\right) = -\dfrac{N}{2}\ln\left(\dfrac{2\pi}{m^2}\right) \quad \forall m^2 > 0 , \\ S\left(\vec{\widetilde{\varphi}} = \vec{\overline{\phi}}_{k=k_{\mathrm{i}}}\right) = \dfrac{m^2}{2}\vec{\overline{\phi}}_{k=k_{\mathrm{i}}}^2 + \dfrac{\lambda}{4!}\left(\vec{\overline{\phi}}_{k=k_{\mathrm{i}}}^2\right)^2 \quad \forall m^2 < 0 , \end{cases} \tag{53}$$

$$\overline{\Gamma}_{k=k_{\mathrm{i}},a_1\cdots a_n}^{(\mathrm{1PI})(n)} = \left.\frac{\partial^n S\left(\vec{\widetilde{\varphi}}\right)}{\partial\widetilde{\varphi}_{a_1}\cdots\partial\widetilde{\varphi}_{a_n}}\right|_{\vec{\widetilde{\varphi}}=\vec{\overline{\phi}}_{k=k_{\mathrm{i}}}} \quad \forall a_1,\cdots,a_n . \tag{54}$$

From Eq. (54), we then directly deduce the initial conditions for the symmetric part $\overline{\Gamma}_k^{(\mathrm{1PI})(n)}$ of the 1PI vertices of even order $n$ introduced via Eqs. (41) to (43) for $m^2 > 0$:

$$\overline{\Gamma}_{k=k_{\mathrm{i}}}^{(\mathrm{1PI})(2)} = m^2 , \tag{55}$$

$$\overline{\Gamma}_{k=k_{\mathrm{i}}}^{(\mathrm{1PI})(4)} = \lambda , \tag{56}$$

$$\overline{\Gamma}_{k=k_{\mathrm{i}}}^{(\mathrm{1PI})(n)} = 0 \quad \forall n \geq 6 . \tag{57}$$

The truncation of the infinite tower of differential equations containing either Eqs. (34) to (37) (for all $N$ and $m^2 > 0$) or Eqs. (47) to (51) (for $N = 1$ and $m^2 < 0$) is implemented by the condition:

$$\overline{\Gamma}_k^{(\mathrm{1PI})(n)} = \overline{\Gamma}_{k=k_{\mathrm{i}}}^{(\mathrm{1PI})(n)} \quad \forall k,\ \forall n > N_{\mathrm{max}} , \tag{58}$$

where $\overline{\Gamma}_k^{(\mathrm{1PI})(n)}$ corresponds to: i) the symmetric part of the 1PI vertices of (even) order $n$ (as defined via Eqs. (41) to (43) up to $n = 6$) for all $N$ and $m^2 > 0$; ii) the 1PI vertices themselves for $N = 1$ and $m^2 < 0$ according to the definition $\overline{\Gamma}_k^{(\mathrm{1PI})(n)} \equiv \overline{\Gamma}_{k,1\cdots 1}^{(\mathrm{1PI})(n)}$. We finally infer the gs energy from $\overline{\Gamma}_k^{(\mathrm{1PI})}$ at the end of the flow using the equality:

$$E_{\mathrm{gs}}^{\mathrm{1PI\text{-}FRG;orig}} = \overline{\Gamma}_{k=k_{\mathrm{f}}}^{(\mathrm{1PI})} . \tag{59}$$

Furthermore, the gs density $\rho_{\mathrm{gs}}$ is determined at $m^2 > 0$ from the relation:

$$\begin{aligned}\rho_{\mathrm{gs}}^{\mathrm{1PI\text{-}FRG;orig}} &= \frac{1}{N}\sum_{a=1}^{N}\left.\frac{\partial^2 W_{k=k_{\mathrm{f}}}\left(\vec{J}\right)}{\partial J_a^2}\right|_{\vec{J}=\vec{0}} \\ &= \frac{1}{N}\sum_{a=1}^{N}\left(\overline{\Gamma}_{k=k_{\mathrm{f}}}^{(\mathrm{1PI})(2)}\right)_{aa}^{-1} \\ &= \left(\overline{\Gamma}_{k=k_{\mathrm{f}}}^{(\mathrm{1PI})(2)}\right)^{-1} ,\end{aligned} \tag{60}$$

which results from Eq. (41). We will actually neither calculate $E_{\mathrm{gs}}$ nor $\rho_{\mathrm{gs}}$ in the regime with $m^2 < 0$, as explained below. Finally, the chosen cutoff function for both $m^2 < 0$ and $m^2 > 0$ is:

$$\boldsymbol{R}_{k,a_1a_2} = R_k\ \delta_{a_1a_2} = \left(k^{-1} - 1\right)\delta_{a_1a_2} \quad \forall a_1, a_2 , \tag{61}$$

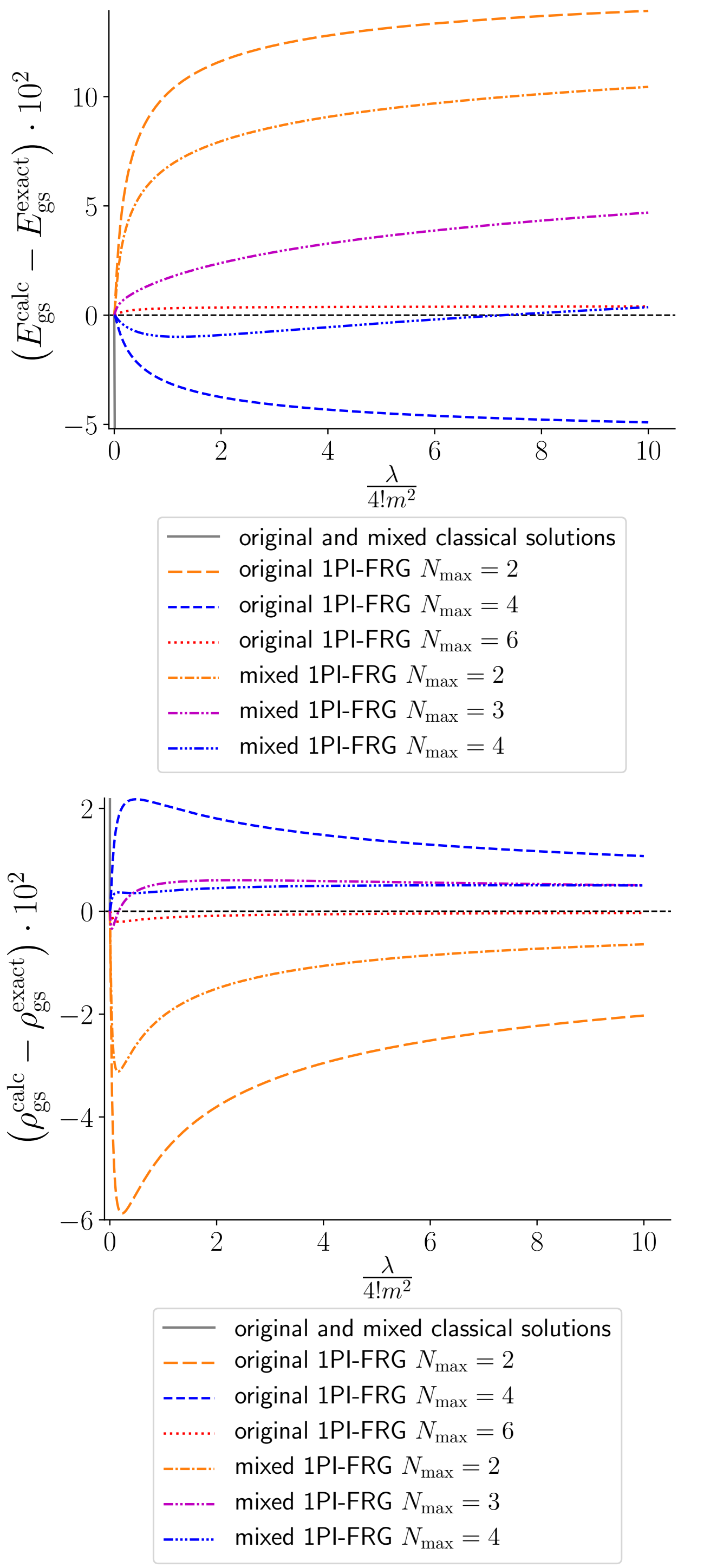

**Fig. 1.** Difference between the calculated gs energy $E_{\mathrm{gs}}^{\mathrm{calc}}$ (upper panel) or density $\rho_{\mathrm{gs}}^{\mathrm{calc}}$ (lower panel) and the corresponding exact solution $E_{\mathrm{gs}}^{\mathrm{exact}}$ or $\rho_{\mathrm{gs}}^{\mathrm{exact}}$ at $m^2 = +1$ and $N = 2$.

which indeed satisfies the boundary conditions (26) (i.e. we have indeed $R_{k=k_{\mathrm{i}}} = \infty$ and $R_{k=k_{\mathrm{f}}} = 0$) as the flow parameter $k$ runs from $k_{\mathrm{i}} = 0$ to $k_{\mathrm{f}} = 1$ in the present case.

Regarding the regime with $m^2 > 0$, we solve the differential equations (34) to (37) up to $N_{\mathrm{max}} = 6$, with the initial conditions (55) to (57) and the cutoff function (61). Due to the symmetry constraint (44), this enables us to determine the first three non-trivial orders of the 1PI-FRG, set by the truncations $N_{\mathrm{max}} = 2$, 4 and 6. The results thus obtained are displayed in Fig. 1 for both $E_{\mathrm{gs}}$ and $\rho_{\mathrm{gs}}$ with $N = 2$. While the first two non-trivial order results all lie within a few percents away from the corresponding exact solution over the whole range of tested values for the coupling constant (i.e. $\lambda/4! \in [0, 10]$ as for most figures of paper I), those obtained at $N_{\mathrm{max}} = 6$ are hardly distinguishable from their exact solutions. The performances of this FRG approach are also barely affected as the coupling constant $\lambda/4!$ evolves, hence the non-perturbative character of this approach.

However, the resolution of the system given by Eqs. (47) to (51) with $m^2 < 0$, using the initial conditions (52) to (54) and the cutoff function (61), is prevented, the system being too stiff, at least for the numerical tools used in this study.[12] Note that we have checked that this problem still occurs at $N = 2$ and for $N_{\mathrm{max}}$ up to 6 [46]. Hence, we only display original 1PI-FRG results for the unbroken-symmetry phase.

In conclusion, we can directly see from Fig. 1 that, especially at second and third non-trivial orders (i.e. at $N_{\mathrm{max}} = 4$ and 6), the 1PI-FRG based on a vertex expansion of the Wetterich equation does manage to describe reliably the strongly-coupled regime of our *bosonic* toy model in its original representation. We have however discussed earlier that purely *fermionic* systems, and therefore nuclear systems in principle,[13] can not be treated at strong couplings via this 1PI-FRG implementation. We have also pointed out that the use of HSTs within the fermionic 1PI-FRG framework can drastically change the situation and give us access to non-perturbative physics, as a result of vertex expansions around non-trivial minima for Hubbard-Stratonovich fields. With this in mind, we thus address in what follows the mixed and collective representations of the (0+0)-D $O(N)$ model. We will notably check if the stiffness of the equation systems to solve within the 1PI-FRG is sufficiently reduced in these situations to tackle the regime with $m^2 < 0$.

### 2.3 Mixed 1PI functional renormalization group

We have stressed before that the 1PI-FRG implementation based on a vertex expansion of the Wetterich equation has already been applied to the (0+0)-D $O(N)$ model in its unbroken-symmetry phase and in its original representation. This is to our knowledge not the case for the mixed

[12] In the present study, all sets of differential equations underpinning FRG approaches are solved numerically using the `NDSolve` function of `Mathematica 12.1`.

[13] One should stress however that nuclear Lagrangians are not necessarily based on fermionic dofs: one can for instance simply exploit $\alpha$-particles as dofs for the description of $^{16}O$. The choice of relevant dofs in nuclear EFTs is still a very active debate [156].

or the collective representation, regardless of the values of $m^2$ and $\lambda$. Let us thus now consider the 1PI-FRG in the mixed representation, i.e. the mixed 1PI-FRG. The Wetterich equation for our zero-dimensional $O(N)$ model in this representation reads:

$$\dot{\Gamma}^{(\mathrm{1PI})}_{\mathrm{mix},k}\left(\vec{\phi},\eta\right) = \frac{1}{2}\mathcal{STr}\left[\dot{\mathcal{R}}_k\left(\Gamma^{(\mathrm{1PI})(2)}_{\mathrm{mix},k}\left(\vec{\phi},\eta\right)+\mathcal{R}_k\right)^{-1}\right] = \frac{1}{2}\sum_{b_1,b_2=1}^{N+1}\dot{\mathcal{R}}_{k,b_1 b_2}\left(\Gamma^{(\mathrm{1PI})(2)}_{\mathrm{mix},k}\left(\vec{\phi},\eta\right)+\mathcal{R}_k\right)^{-1}_{b_2 b_1} , \quad (62)$$

where the Hessian of the flowing EA is now given by:

$$\Gamma^{(\mathrm{1PI})(2)}_{\mathrm{mix},k}\left(\vec{\phi},\eta\right) = \begin{pmatrix} \frac{\partial^2\Gamma^{(\mathrm{1PI})}_{\mathrm{mix},k}}{\partial\vec{\phi}\partial\vec{\phi}} & \frac{\partial^2\Gamma^{(\mathrm{1PI})}_{\mathrm{mix},k}}{\partial\vec{\phi}\partial\eta} \\ \frac{\partial^2\Gamma^{(\mathrm{1PI})}_{\mathrm{mix},k}}{\partial\eta\partial\vec{\phi}} & \frac{\partial^2\Gamma^{(\mathrm{1PI})}_{\mathrm{mix},k}}{\partial\eta\partial\eta} \end{pmatrix} \equiv \begin{pmatrix} \Gamma^{(\mathrm{1PI})(2\phi)}_{\mathrm{mix},k} & \Gamma^{(\mathrm{1PI})(1\phi,1\eta)}_{\mathrm{mix},k} \\ \Gamma^{(\mathrm{1PI})(1\phi,1\eta)}_{\mathrm{mix},k} & \Gamma^{(\mathrm{1PI})(2\eta)}_{\mathrm{mix},k} \end{pmatrix} , \quad (63)$$

and the cutoff function $\mathcal{R}_k$ exhibits the following matrix structure in extended color space:

$$\mathcal{R}_k = \begin{pmatrix} \boldsymbol{R}^{(\phi)}_k & \vec{0} \\ \vec{0}^{\mathrm{T}} & R^{(\eta)}_k \end{pmatrix} = \begin{pmatrix} R_k\mathbb{I}_N & \vec{0} \\ \vec{0}^{\mathrm{T}} & R_k \end{pmatrix} = R_k\mathbb{I}_{N+1} , \quad (64)$$

with $\mathbb{I}_D$ the $D$-dimensional identity matrix. We then apply the vertex expansion procedure to the Wetterich equation in the form (62), starting from the Taylor series:

$$\Gamma^{(\mathrm{1PI})}_{\mathrm{mix},k}\left(\vec{\phi},\eta\right) = \overline{\Gamma}^{(\mathrm{1PI})}_{\mathrm{mix},k} + \sum_{n=2}^{\infty}\frac{1}{n!}\sum_{m=0}^{n}\begin{pmatrix} n \\ m \end{pmatrix} \times \sum_{a_1,\cdots,a_m=1}^{N}\overline{\Gamma}^{(\mathrm{1PI})(m\phi,(n-m)\eta)}_{\mathrm{mix},k,a_1\cdots a_m} \times \left(\vec{\phi}-\vec{\overline{\phi}}_k\right)_{a_1}\cdots\left(\vec{\phi}-\vec{\overline{\phi}}_k\right)_{a_m} \times \left(\eta-\overline{\eta}_k\right)^{n-m} , \quad (65)$$

with

$$\overline{\Gamma}^{(\mathrm{1PI})}_{\mathrm{mix},k} \equiv \Gamma^{(\mathrm{1PI})}_{\mathrm{mix},k}\left(\vec{\phi}=\vec{\overline{\phi}}_k,\eta=\overline{\eta}_k\right) \quad \forall k , \quad (66)$$

$$\overline{\Gamma}^{(\mathrm{1PI})(n\phi,m\eta)}_{\mathrm{mix},k,a_1\cdots a_n} \equiv \left.\frac{\partial^{n+m}\Gamma^{(\mathrm{1PI})}_{\mathrm{mix},k}(\vec{\phi},\eta)}{\partial\phi_{a_1}\cdots\partial\phi_{a_n}\partial\eta^m}\right|_{\substack{\vec{\phi}=\vec{\overline{\phi}}_k \\ \eta=\overline{\eta}_k}} \quad \forall a_1,\cdots,a_n,k , \quad (67)$$

and[14]

$$\overline{\Gamma}^{(\mathrm{1PI})(1\phi)}_{\mathrm{mix},k,a} = \overline{\Gamma}^{(\mathrm{1PI})(1\eta)}_{\mathrm{mix},k} = 0 \quad \forall a,k , \quad (68)$$

since the flowing EA is now extremal at $\left(\vec{\phi}\;\eta\right) = \left(\vec{\overline{\phi}}_k\;\overline{\eta}_k\right)$. Furthermore, as this EA now depends on several fields, it will be convenient to use the following definition:

$$\Gamma^{(\mathrm{1PI})(2)}_{\mathrm{mix},k} + \mathcal{R}_k = \mathcal{P}_k + \mathcal{F}_k , \quad (69)$$

where the fluctuation matrix $\mathcal{F}_k$ contains all the field dependence, i.e.:

$$\mathcal{F}_k = \begin{pmatrix} \Delta\Gamma^{(\mathrm{1PI})(2\phi)}_{\mathrm{mix},k} & \Delta\Gamma^{(\mathrm{1PI})(1\phi,1\eta)}_{\mathrm{mix},k} \\ \Delta\Gamma^{(\mathrm{1PI})(1\phi,1\eta)}_{\mathrm{mix},k} & \Delta\Gamma^{(\mathrm{1PI})(2\eta)}_{\mathrm{mix},k} \end{pmatrix} , \quad (70)$$

with

$$\Delta\Gamma^{(\mathrm{1PI})\mathrm{X}}_{\mathrm{mix},k} = \Gamma^{(\mathrm{1PI})\mathrm{X}}_{\mathrm{mix},k} - \overline{\Gamma}^{(\mathrm{1PI})\mathrm{X}}_{\mathrm{mix},k} , \quad (71)$$

which imposes that $\mathcal{P}_k$ satisfies:

$$\mathcal{P}_k = \begin{pmatrix} \boldsymbol{R}^{(\phi)}_k + \overline{\Gamma}^{(\mathrm{1PI})(2\phi)}_{\mathrm{mix},k} & \overline{\Gamma}^{(\mathrm{1PI})(1\phi,1\eta)}_{\mathrm{mix},k} \\ \overline{\Gamma}^{(\mathrm{1PI})(1\phi,1\eta)}_{\mathrm{mix},k} & R^{(\eta)}_k + \overline{\Gamma}^{(\mathrm{1PI})(2\eta)}_{\mathrm{mix},k} \end{pmatrix} . \quad (72)$$

For $m^2 > 0$, we will use the counterparts of Eqs. (41) to (44) introducing the symmetric parts of 1PI vertices:

$$\overline{\Gamma}^{(\mathrm{1PI})(2\phi,n\eta)}_{\mathrm{mix},k,a_1 a_2} = \overline{\Gamma}^{(\mathrm{1PI})(2\phi,n\eta)}_{\mathrm{mix},k}\;\delta_{a_1 a_2} \quad \forall a_1,a_2,n , \quad (73)$$

$$\overline{\Gamma}^{(\mathrm{1PI})(4\phi,n\eta)}_{\mathrm{mix},k,a_1 a_2 a_3 a_4} = \overline{\Gamma}^{(\mathrm{1PI})(4\phi,n\eta)}_{\mathrm{mix},k}\left(\delta_{a_1 a_2}\delta_{a_3 a_4}+\delta_{a_1 a_3}\delta_{a_2 a_4}+\delta_{a_1 a_4}\delta_{a_2 a_3}\right) \quad \forall a_1,a_2,a_3,a_4,n , \quad (74)$$

$$\overline{\Gamma}^{(\mathrm{1PI})(n\phi,m\eta)}_{\mathrm{mix},k,a_1\cdots a_n} = 0 \quad \forall a_1,\cdots,a_n,m,\;\forall n \text{ odd} . \quad (75)$$

The next step of the vertex expansion consists in carrying out matrix products between $\mathcal{P}^{-1}_k$ and $\mathcal{F}_k$ and the rest of the recipe towards the set of coupled differential equations to solve is detailed in appendix D.1. Although we only give our derived expressions for these flow equations in the latter appendix, we would like to point out here, for the needs of our discussion, that the underlying derivations require to define the following (inverse) propagators:

$$\left(\boldsymbol{G}^{(\phi)}_k\right)^{-1}_{a_1 a_2} = \overline{\Gamma}^{(2\phi)}_{\mathrm{mix},k,a_1 a_2} + \boldsymbol{R}^{(\phi)}_{k,a_1 a_2} = \left(\overline{G}^{(\phi)}_k\right)^{-1}\delta_{a_1 a_2} = \left(\overline{\Gamma}^{(2\phi)}_{\mathrm{mix},k}+R_k\right)\delta_{a_1 a_2} \quad \forall a_1,a_2 , \quad (76)$$

$$\left(\overline{G}^{(\eta)}_k\right)^{-1} = \overline{\Gamma}^{(2\eta)}_{\mathrm{mix},k} + R_k , \quad (77)$$

which are the mixed counterparts of Eq. (38) or (45) and the last line of Eq. (76) is only valid in the unbroken-symmetry phase. Hence, as a first level of approximation which is basically mean-field theory (MFT), we can set all bosonic entries of $\mathcal{P}^{-1}_k$ equal to zero [157]. For the unbroken-symmetry regime of the toy model under consideration, this amounts to neglecting the bottom-right component of $\mathcal{P}^{-1}_k$, i.e. this amounts to setting $\mathcal{P}^{-1}_{k,N+1\,N+1} =$

[14] The relations $\overline{\Gamma}^{(\mathrm{1PI})(n\phi)}_{\mathrm{mix},k} \equiv \overline{\Gamma}^{(\mathrm{1PI})(n\phi,0\eta)}_{\mathrm{mix},k}$ and $\overline{\Gamma}^{(\mathrm{1PI})(n\eta)}_{\mathrm{mix},k} \equiv \overline{\Gamma}^{(\mathrm{1PI})(0\phi,n\eta)}_{\mathrm{mix},k}$ are assumed for all $n$ in Eq. (68) as well as in subsequent equations.

$\overline{G}_{k}^{(\eta)} = 0$. The sets of differential equations to solve in the framework of MFT can therefore be directly inferred from those resulting from the standard vertex expansion procedure and given in appendix D.1.

Whether we restrict ourselves to MFT or not, the initial conditions used to solve the differential equations within the mixed 1PI-FRG are directly inferred for all $N$ from the classical action $S_{\mathrm{mix}}$ expressed in Eq. (7), which gives us[15]:

$$\begin{pmatrix}\vec{\overline{\phi}}_{k=k_{\mathrm{i}}}\\ \overline{\eta}_{k=k_{\mathrm{i}}}\end{pmatrix} = \begin{pmatrix}\vec{\widetilde{\varphi}}_{\mathrm{cl}}\\ \widetilde{\sigma}_{\mathrm{cl}}\end{pmatrix} = \begin{cases}\begin{pmatrix}\vec{0}\\ 0\end{pmatrix} & \forall m^{2}>0\;,\\[2ex] \begin{pmatrix}0\\ \vdots\\ 0\\ \pm\sqrt{\frac{-6m^{2}}{\lambda}}\\ im^{2}\sqrt{\frac{3}{\lambda}}\end{pmatrix} & \forall m^{2}<0 \text{ and } \lambda\neq 0\;,\end{cases} \tag{78}$$

$$\overline{\Gamma}_{\mathrm{mix},k=k_{\mathrm{i}}}^{(\mathrm{1PI})} = \begin{cases} -\frac{N}{2}\ln\left(\frac{2\pi}{m^{2}}\right) + S_{\mathrm{mix}}\left(\vec{\widetilde{\varphi}}=\vec{\overline{\phi}}_{k=k_{\mathrm{i}}},\widetilde{\sigma}=\overline{\eta}_{k=k_{\mathrm{i}}}\right) = -\frac{N}{2}\ln\left(\frac{2\pi}{m^{2}}\right) & \forall m^{2}>0\;,\\[2ex] S_{\mathrm{mix}}\left(\vec{\widetilde{\varphi}}=\vec{\overline{\phi}}_{k=k_{\mathrm{i}}},\widetilde{\sigma}=\overline{\eta}_{k=k_{\mathrm{i}}}\right) = \frac{m^{2}}{2}\vec{\overline{\phi}}_{k=k_{\mathrm{i}}}^{2} + \frac{1}{2}\overline{\eta}_{k=k_{\mathrm{i}}}^{2} + i\sqrt{\frac{\lambda}{12}}\,\overline{\eta}_{k=k_{\mathrm{i}}}\vec{\overline{\phi}}_{k=k_{\mathrm{i}}}^{2} & \forall m^{2}<0\;,\end{cases} \tag{79}$$

$$\overline{\Gamma}_{\mathrm{mix},k=k_{\mathrm{i}},a_{1}\cdots a_{n}}^{(\mathrm{1PI})(n\phi,m\eta)} = \left.\frac{\partial^{n+m}S_{\mathrm{mix}}\left(\vec{\widetilde{\varphi}},\widetilde{\sigma}\right)}{\partial\widetilde{\varphi}_{a_{1}}\cdots\partial\widetilde{\varphi}_{a_{n}}\partial\widetilde{\sigma}^{m}}\right|_{\substack{\vec{\widetilde{\varphi}}=\vec{\overline{\phi}}_{k=k_{\mathrm{i}}}\\ \widetilde{\sigma}=\overline{\eta}_{k=k_{\mathrm{i}}}}} \quad \forall a_{1},\cdots,a_{n}\;, \tag{80}$$

where the latter relation translates in the unbroken-symmetry regime to:

$$\overline{\Gamma}_{\mathrm{mix},k=k_{\mathrm{i}}}^{(\mathrm{1PI})(2\phi)} = m^{2}\;, \tag{81}$$

$$\overline{\Gamma}_{\mathrm{mix},k=k_{\mathrm{i}}}^{(\mathrm{1PI})(2\eta)} = 1\;, \tag{82}$$

$$\overline{\Gamma}_{\mathrm{mix},k=k_{\mathrm{i}}}^{(\mathrm{1PI})(2\phi,1\eta)} = i\sqrt{\frac{\lambda}{3}}\;, \tag{83}$$

$$\overline{\Gamma}_{\mathrm{mix},k=k_{\mathrm{i}}}^{(\mathrm{1PI})(3\eta)} = 0\;, \tag{84}$$

$$\overline{\Gamma}_{\mathrm{mix},k=k_{\mathrm{i}}}^{(\mathrm{1PI})(m\phi,n\eta)} = 0 \quad \forall\; m+n\geq 4\;. \tag{85}$$

The truncation of the infinite set of differential equations resulting from the vertex expansion is now imposed by:

$$\overline{\Gamma}_{\mathrm{mix},k}^{(\mathrm{1PI})(n\phi,m\eta)} = \overline{\Gamma}_{\mathrm{mix},k=k_{\mathrm{i}}}^{(\mathrm{1PI})(n\phi,m\eta)} \qquad \forall k,\;\forall\; n+m>N_{\mathrm{max}}\;. \tag{86}$$

Finally, the gs energy is deduced in the present case from:

$$E_{\mathrm{gs}}^{\text{1PI-FRG;mix}} = \overline{\Gamma}_{\mathrm{mix},k=k_{\mathrm{f}}}^{(\mathrm{1PI})}\;, \tag{87}$$

whereas the gs density is determined from:

$$\rho_{\mathrm{gs}}^{\text{1PI-FRG;mix}} = \frac{1}{N}\sum_{a=1}^{N}\left(\overline{\Gamma}_{\mathrm{mix},k=k_{\mathrm{f}}}^{(\mathrm{1PI})(2\phi)}\right)_{aa}^{-1} = \left(\overline{\Gamma}_{\mathrm{mix},k=k_{\mathrm{f}}}^{(\mathrm{1PI})(2\phi)}\right)^{-1}\;, \tag{88}$$

where $\overline{\Gamma}_{\mathrm{mix},k}^{(\mathrm{1PI})(2\phi)}$ is introduced in the right-hand side using Eq. (73) (at $n=0$) and assuming that $m^{2}>0$. For all our mixed 1PI-FRG calculations, we have also exploited the cutoff function $R_{k}$ of Eq. (61) for both the original and auxiliary field sectors.

Let us first concentrate our discussion on the regime with $m^{2}>0$. Without the MFT approximation, our mixed 1PI-FRG results exhibit a distinct convergence towards the exact solution, as can be seen in Fig. 1 for both $E_{\mathrm{gs}}$ and $\rho_{\mathrm{gs}}$ with $N=2$. More specifically, this figure shows that the mixed 1PI-FRG outperforms the original 1PI-FRG at $N_{\mathrm{max}}=2$ and 4.[16] The superiority of the mixed 1PI-FRG as compared to the original one for a given $N_{\mathrm{max}}$ can notably be attributed to the 1-point correlation function of the auxiliary field taking non-trivial values, as illustrated by Fig. 2. This echoes very clearly our comparison between SCPT based on the original and mixed 2PI EAs in paper I where the 1-point correlation function of the auxiliary field was also put forward to explain the difference between the bare vertex approximation and the original Hartree-Fock result.

We have also implemented the MFT by setting the propagator $\overline{G}_{k}^{(\eta)}$ equal to zero for all $k$. This amounts to setting the mass of the bosonic field $\widetilde{\sigma}$ (or equivalently the associated cutoff function $R_{k}^{(\eta)}$ introduced in Eq. (64)) to infinity. This completely freezes the fluctuations of this field. In other words, the MFT can not capture radiative corrections associated with the auxiliary field. This notably excludes all contributions beyond the leading order of the collective LE. Therefore, the MFT can not, by construction, outperform the leading order of the collective LE, which coincides with the auxiliary field LE (LOAF) approximation [158–167] for the studied model. This means that the MFT should tend to the LOAF approximation as the truncation order $N_{\rm max}$ increases, which is illustrated by Fig. 3. Finally, although Fig. 4 shows that the approximation underlying MFT induces a significant loss in the accuracy of mixed 1PI-FRG results, they also illustrate that its efficiency is not affected in the strongly-coupled regime: the MFT can therefore be considered as a first level of non-perturbative approximations.

[15] The remark of footnote 11 also applies to the logarithm term in Eq. (79).

[16] We note however the specific case of the determination of $E_{\mathrm{gs}}$ at $N=1$ and $N_{\mathrm{max}}=2$ reported in Ref. [46], where the original 1PI-FRG outperforms its mixed counterpart.

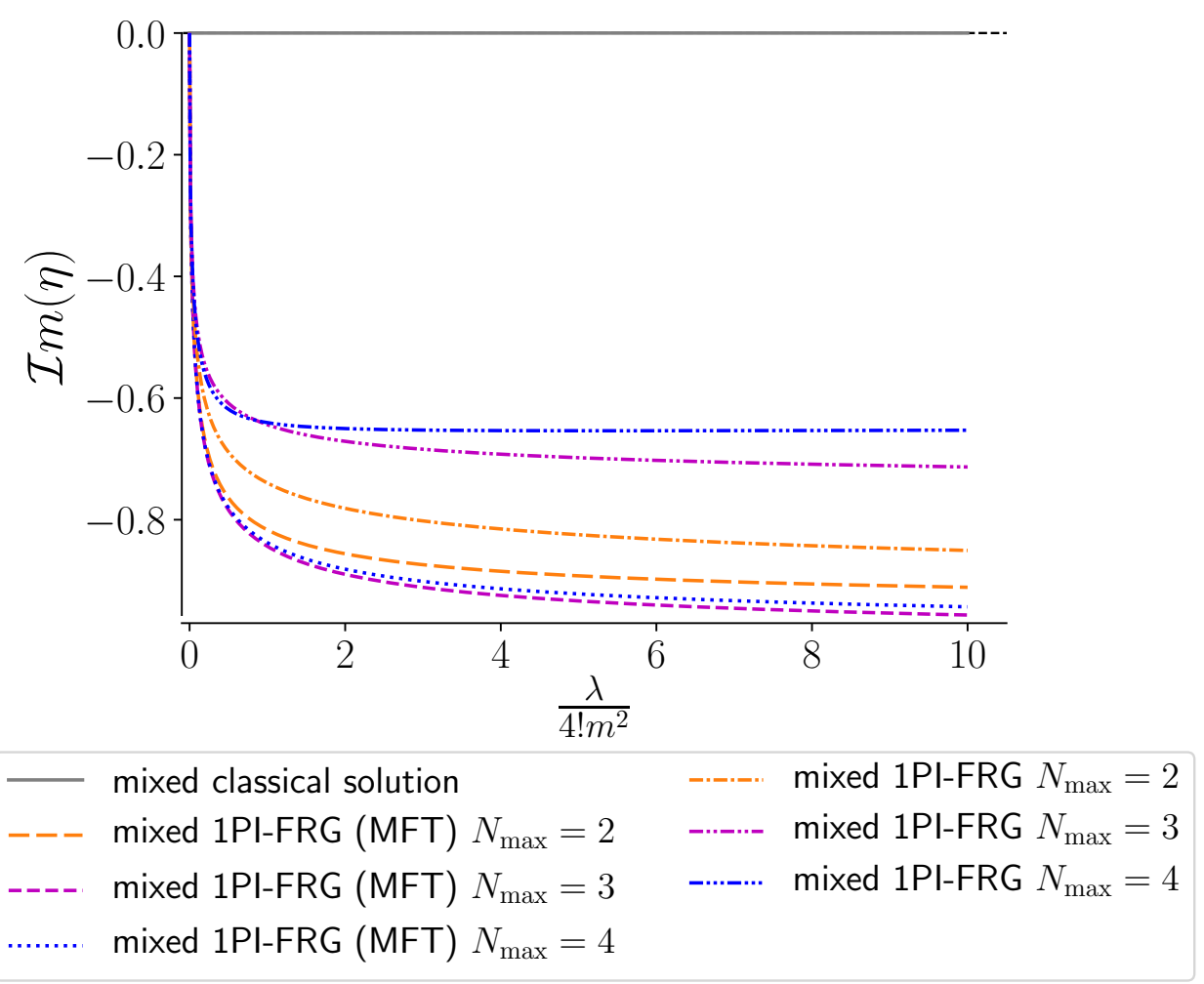

**Fig. 2.** Imaginary part of the 1-point correlation function of the auxiliary field in the framework of the mixed representation at $N=2$ and $m^2=+1$.

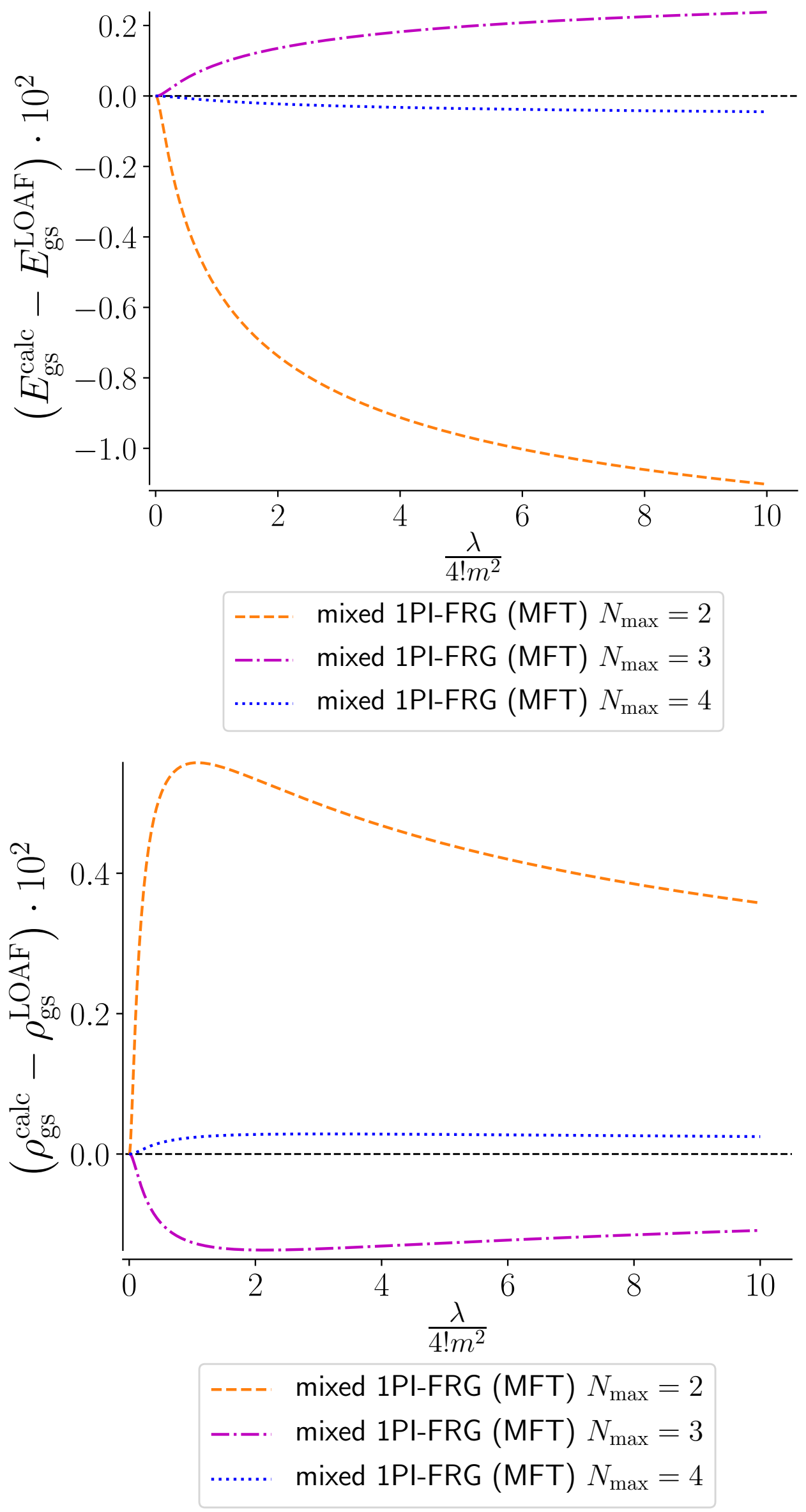

**Fig. 3.** Difference between the calculated gs energy $E_{\rm gs}^{\rm calc}$ (upper panel) or density $\rho_{\rm gs}^{\rm calc}$ (lower panel) and the corresponding LOAF approximation results $E_{\rm gs}^{\rm LOAF}$ or $\rho_{\rm gs}^{\rm LOAF}$ at $m^2=+1$ and $N=2$.

Regarding the phase with $m^2<0$, we encounter the same limitation as in the original representation: the set of differential equations resulting from the vertex expansion procedure applied to Eq. (62) is too stiff to be solved from $k_{\rm i}=0$ to $k_{\rm f}=1$ with the chosen numerical tools[12]. We will therefore turn to the collective representation as our last attempt to describe the broken-symmetry phase with the 1PI-FRG.

### 2.4 Collective 1PI functional renormalization group

For the collective representation of the (0+0)-D $O(N)$ model, the Wetterich equation reduces to:

$$\dot{\Gamma}^{(\rm 1PI)}_{{\rm col},k}\left(\vec{\phi}\right)=\frac{1}{2}\dot{R}_k\left(\Gamma^{(\rm 1PI)(2)}_{{\rm col},k}(\eta)+R_k\right)^{-1}, \tag{89}$$

with

$$\Gamma^{(\rm 1PI)(2)}_{{\rm col},k}(\eta)\equiv\frac{\partial^2\Gamma^{(\rm 1PI)}_{{\rm col},k}(\eta)}{\partial\eta^2}. \tag{90}$$

Hence, we are basically in the same situation as when treating the broken-symmetry phase of the original representation with $N=1$. The output of the vertex expansion procedure applied to Eq. (89) can therefore be directly deduced from the set of differential equations containing notably Eqs. (47) to (51), which yields:

$$\dot{\overline{\Gamma}}^{(\rm 1PI)}_{{\rm col},k}=\frac{1}{2}\dot{R}_k\left(\overline{D}_k-\overline{D}_k^{(0)}\right), \tag{91}$$

$$\dot{\overline{\eta}}_k=\frac{1}{2\overline{\Gamma}^{(\rm 1PI)(2)}_{{\rm col},k}}\dot{R}_k\overline{D}_k^2\overline{\Gamma}^{(\rm 1PI)(3)}_{{\rm col},k}, \tag{92}$$

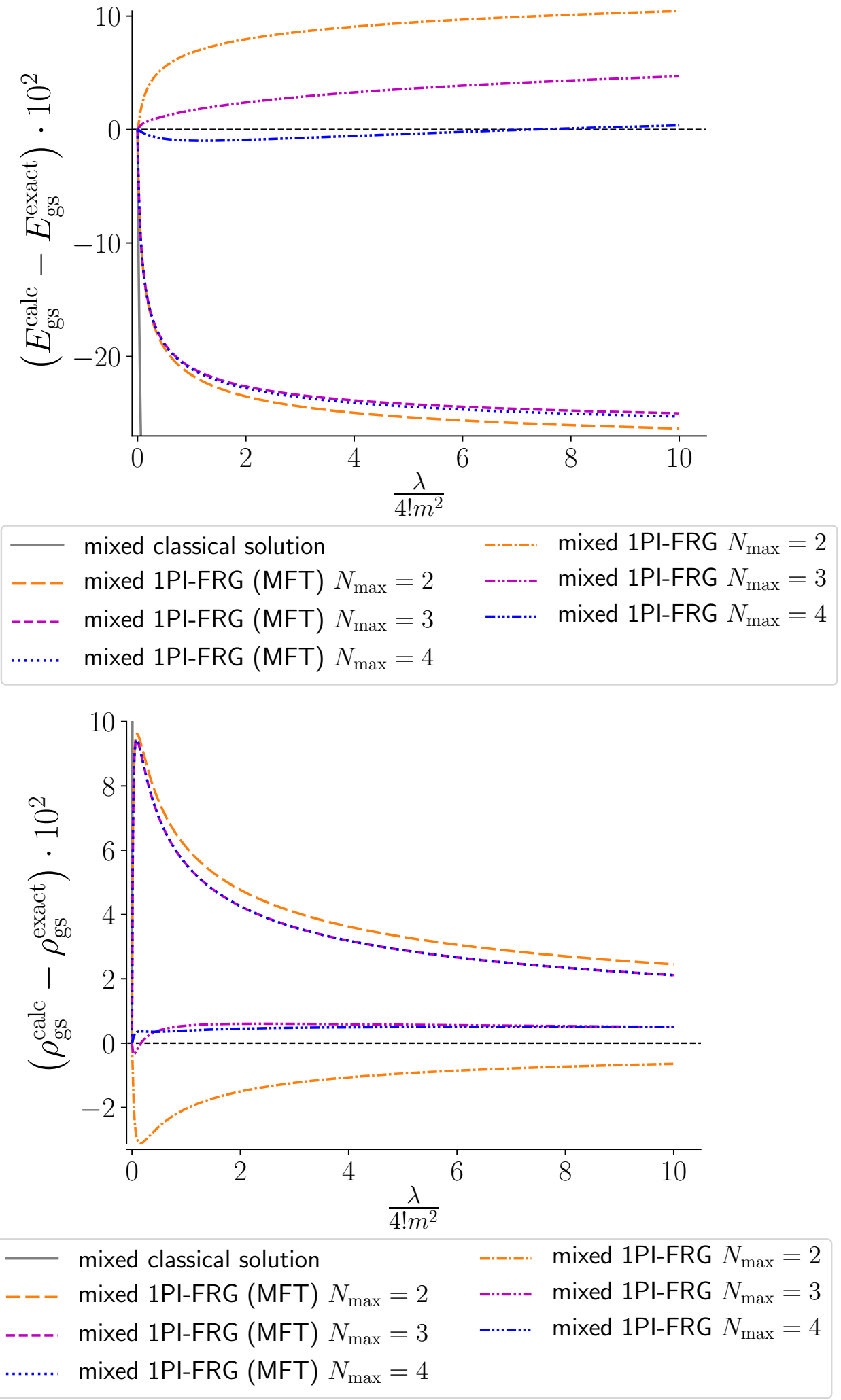

**Fig. 4.** Difference between the calculated gs energy $E_{\rm gs}^{\rm calc}$ (upper panel) or density $\rho_{\rm gs}^{\rm calc}$ (lower panel) and the corresponding exact solution $E_{\rm gs}^{\rm exact}$ or $\rho_{\rm gs}^{\rm exact}$ at $m^2 = +1$ and $N = 2$.

$$\begin{split}\dot{\overline{\Gamma}}^{(\rm 1PI)(2)}_{{\rm col},k} = {} & \dot{\overline{\eta}}_k \overline{\Gamma}^{(\rm 1PI)(3)}_{{\rm col},k} + \dot{R}_k \overline{D}_k^3 \left(\overline{\Gamma}^{(\rm 1PI)(3)}_{{\rm col},k}\right)^2 \\ & - \frac{1}{2}\dot{R}_k \overline{D}_k^2 \overline{\Gamma}^{(\rm 1PI)(4)}_{{\rm col},k} \,,\end{split} \quad (93)$$

$$\begin{split}\dot{\overline{\Gamma}}^{(\rm 1PI)(3)}_{{\rm col},k} = {} & \dot{\overline{\eta}}_k \overline{\Gamma}^{(\rm 1PI)(4)}_{{\rm col},k} - 3\dot{R}_k \overline{D}_k^4 \left(\overline{\Gamma}^{(\rm 1PI)(3)}_{{\rm col},k}\right)^3 \\ & + 3\dot{R}_k \overline{D}_k^3 \overline{\Gamma}^{(\rm 1PI)(3)}_{{\rm col},k} \overline{\Gamma}^{(\rm 1PI)(4)}_{{\rm col},k} \\ & - \frac{1}{2}\dot{R}_k \overline{D}_k^2 \overline{\Gamma}^{(\rm 1PI)(5)}_{{\rm col},k} \,,\end{split} \quad (94)$$

$$\begin{split}\dot{\overline{\Gamma}}^{(\rm 1PI)(4)}_{{\rm col},k} = {} & \dot{\overline{\eta}}_k \overline{\Gamma}^{(\rm 1PI)(5)}_{{\rm col},k} + 12\dot{R}_k \overline{D}_k^5 \left(\overline{\Gamma}^{(\rm 1PI)(3)}_{{\rm col},k}\right)^4 \\ & - 18\dot{R}_k \overline{D}_k^4 \left(\overline{\Gamma}^{(\rm 1PI)(3)}_{{\rm col},k}\right)^2 \overline{\Gamma}^{(\rm 1PI)(4)}_{{\rm col},k} \\ & + 4\dot{R}_k \overline{D}_k^3 \overline{\Gamma}^{(\rm 1PI)(3)}_{{\rm col},k} \overline{\Gamma}^{(\rm 1PI)(5)}_{{\rm col},k} \\ & + 3\dot{R}_k \overline{D}_k^3 \left(\overline{\Gamma}^{(\rm 1PI)(4)}_{{\rm col},k}\right)^2 - \frac{1}{2}\dot{R}_k \overline{D}_k^2 \overline{\Gamma}^{(\rm 1PI)(6)}_{{\rm col},k} \,,\end{split} \quad (95)$$

with

$$\overline{\Gamma}^{(\rm 1PI)}_{{\rm col},k} \equiv \Gamma^{(\rm 1PI)}_{{\rm col},k}(\eta = \overline{\eta}_k) \quad \forall k \,, \quad (96)$$

$$\overline{\Gamma}^{(\rm 1PI)(n)}_{{\rm col},k} \equiv \left.\frac{\partial^n \Gamma^{(\rm 1PI)}_{{\rm col},k}(\eta)}{\partial \eta^n}\right|_{\eta = \overline{\eta}_k} \quad \forall k \,, \quad (97)$$

and

$$\overline{D}_k^{-1} \equiv \overline{\Gamma}^{(\rm 1PI)(2)}_{{\rm col},k} + R_k \,, \quad (98)$$

$$\left(\overline{D}_k^{(0)}\right)^{-1} \equiv \overline{\Gamma}^{(\rm 1PI)(2)}_{{\rm col},k=k_{\rm i}} + R_k \,. \quad (99)$$

The corresponding initial conditions are:

$$\overline{\eta}_{k=k_{\rm i}} = \overline{\sigma}_{\rm cl} = i\left(\frac{\sqrt{3m^2 - \sqrt{3m^4 + 2N\lambda}}}{2\sqrt{\lambda}}\right) \,, \quad (100)$$

$$\begin{split}\overline{\Gamma}^{(\rm 1PI)}_{{\rm col},k=k_{\rm i}} & = S_{\rm col}\big(\widetilde{\sigma} = \overline{\eta}_{k=k_{\rm i}}\big) \\ & = \frac{1}{2}\left(\overline{\eta}_{k=k_{\rm i}}\right)^2 - \frac{N}{2}\ln\left(\frac{2\pi}{m^2 + i\sqrt{\frac{\lambda}{3}}\overline{\eta}_{k=k_{\rm i}}}\right) \,,\end{split} \quad (101)$$

$$\begin{split}\overline{\Gamma}^{(\rm 1PI)(n)}_{{\rm col},k=k_{\rm i}} & = \left.\frac{\partial^n S_{\rm col}(\widetilde{\sigma})}{\partial \widetilde{\sigma}^n}\right|_{\widetilde{\sigma} = \overline{\eta}_{k=k_{\rm i}}} \\ & = \delta_{n2} + (-1)^{n+1}\frac{N}{2}(n-1)!\left(\frac{i\sqrt{\frac{\lambda}{3}}}{m^2 + i\sqrt{\frac{\lambda}{3}}\overline{\eta}_{k=k_{\rm i}}}\right)^n \\ & \quad \forall n \geq 2 \,.\end{split} \quad (102)$$

In addition, the infinite tower of differential equations including Eqs. (91) to (95) is truncated by imposing:

$$\overline{\Gamma}^{(\rm 1PI)(n)}_{{\rm col},k} = \overline{\Gamma}^{(\rm 1PI)(n)}_{{\rm col},k=k_{\rm i}} \quad \forall k,\ \forall n > N_{\rm max} \,. \quad (103)$$

Furthermore, the gs energy is deduced from:

$$E_{\rm gs}^{\rm 1PI\text{-}FRG;col} = \overline{\Gamma}^{(\rm 1PI)}_{{\rm col},k=k_{\rm f}} \,, \quad (104)$$

whereas the gs density is estimated with the relation:

$$\rho_{\rm gs}^{\rm 1PI\text{-}FRG;col} = \frac{i}{N}\sqrt{\frac{12}{\lambda}}\,\overline{\eta}_{k=k_{\rm f}} \,. \quad (105)$$

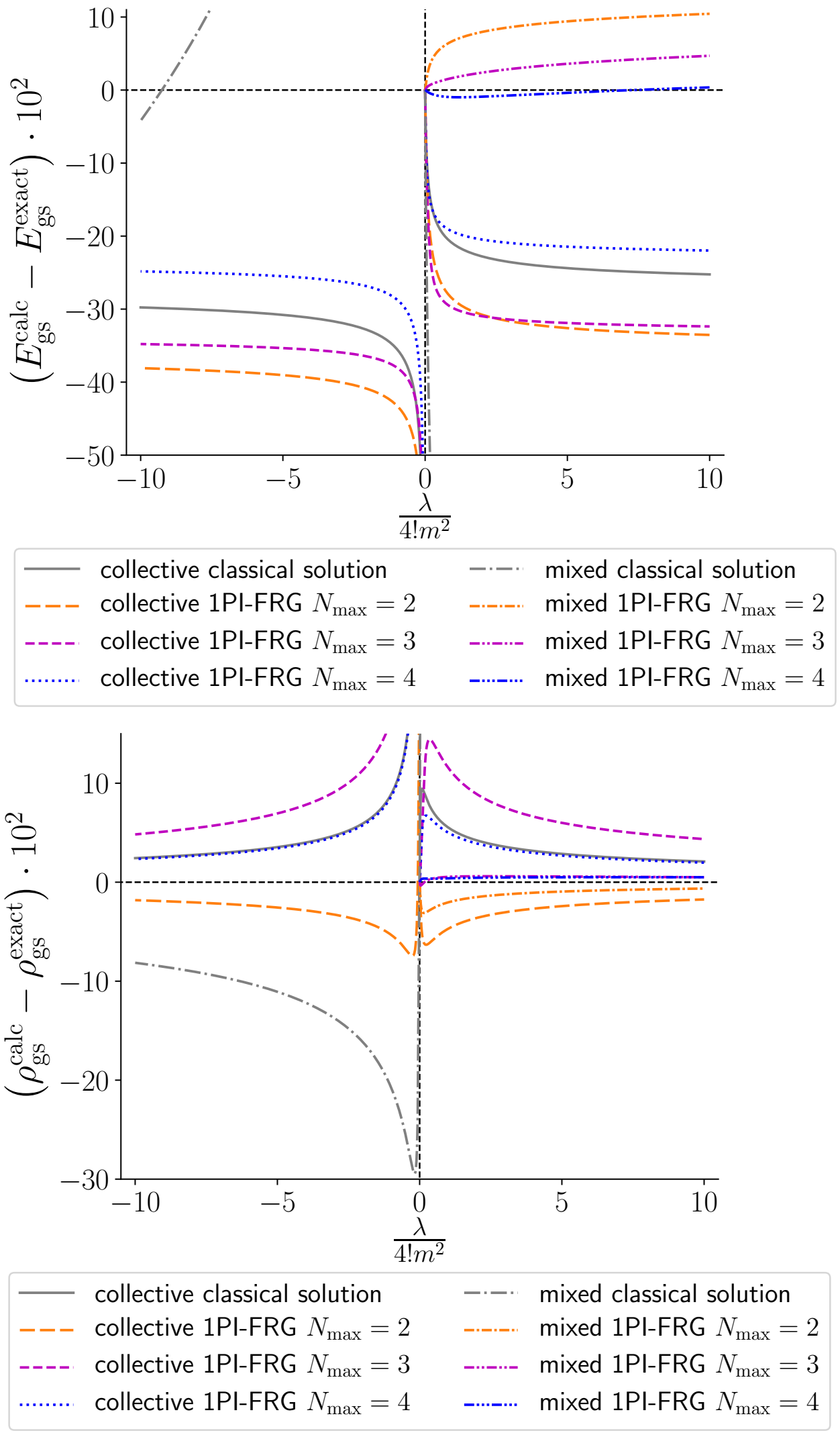

**Fig. 5.** Difference between the calculated gs energy $E_{\mathrm{gs}}^{\mathrm{calc}}$ (upper panel) or density $\rho_{\mathrm{gs}}^{\mathrm{calc}}$ (lower panel) and the corresponding exact solution $E_{\mathrm{gs}}^{\mathrm{exact}}$ or $\rho_{\mathrm{gs}}^{\mathrm{exact}}$ at $m^2 = \pm 1$ and $N = 2$.

Finally, we still use the cutoff function $R_k$ given by Eq. (61). Note that all analytical results given since Eq. (89) are valid for both the unbroken- and broken-symmetry regimes. This follows from the fact that the $O(N)$ symmetry does not constrain the auxiliary field as it does for the original field via Eqs. (41) to (44). Hence, as opposed to the original situation, there is no additional difficulty in treating the phase with $m^2 < 0$ instead of $m^2 > 0$ in the framework of the collective representation.

From this implementation of the collective 1PI-FRG, we are able to calculate $E_{\mathrm{gs}}$ and $\rho_{\mathrm{gs}}$ for all signs of $m^2$, which yields notably our first 1PI-FRG results for $m^2 < 0$. These results, shown in Fig. 5 at $N = 2$, are however disappointing in the sense that, for $m^2 > 0$, they are all outperformed by the original and mixed 1PI-FRG approaches at a given $N_{\mathrm{max}}$. Actually, for both $m^2 < 0$ and $m^2 > 0$, Fig. 5 shows for $E_{\mathrm{gs}}$ at $N = 2$ that the collective 1PI-FRG must be pushed beyond $N_{\mathrm{max}} = 4$ to yield an accuracy below 10%, whereas this is already achieved by the mixed 1PI-FRG at $N_{\mathrm{max}} = 2$ for $m^2 > 0$.

The connection between the collective 1PI-FRG and MFT is also clear. The starting point of the collective 1PI-FRG procedure coincides with the collective classical action, i.e. with the LOAF approximation towards which the MFT tends. Hence, the collective 1PI-FRG incorporates quantum corrections (which correspond to the bosonic fluctuations neglected by the MFT) on top of the LOAF approximation throughout the flow: it is therefore by construction more performing than the MFT version of the mixed 1PI-FRG.

## 3 2PI functional renormalization group

### 3.1 Generalities

We have argued in the introduction (especially in section 1.3) that higher-order EAs such as 2PI and 2PPI EAs play an important role in our aim of reformulating the nuclear EDF approach with PI techniques. We thus continue our FRG study by considering the 2PI EA framework here before treating the 2PPI EA one in the next section. Formulations of FRG approaches for 2PI EAs have started since the 2000s [44, 147, 168, 169]. For example, the detailed discussion of Ref. [147] outlines the recipe to construct flow equations for any $n$PI EA (i.e. for $n$PI EAs with $n \geq 1$) by interpreting cutoff functions as shifts for the sources. Moreover, in Ref. [168], the ideas of the work of Alexandre, Polonyi and Sailer [170,171] deriving a generalization of the Callan-Symanzik equation [172–174] via the 1PI EA are exploited to determine a flow equation for the 2PI EA. The resulting approach, called internal space RG, has been compared with other RG methods, including the standard Callan-Symanzik RG, the Wegner-Houghton RG [175], the local potential approximation [176, 177] of the DE within the 1PI-FRG, in the framework of a comparative study of a (0+1)-D $O(N)$-symmetric $\varphi^4$-theory [178].

We will rather focus in this study on the more recent 2PI-FRG formalism put forward by Dupuis in Refs. [44, 45], and more specifically on its different versions called C-flow [44], U-flow [45] and CU-flow [45] that we will define further below. The U-flow and CU-flow can be both formulated via a modification of the Legendre transform defining the 2PI EA, in the same way as for the 1PI-FRG with the extra term $\varDelta S_k[\phi]$ in Eq. (20). In any case, the aim remains to obtain a starting point as convenient as possible for the flow: the presence of $\varDelta S_k[\phi]$ in Eq. (20) enables us to start the 1PI-FRG procedure at the classical theory whereas the 2PI-FRG flow can begin at the result of SCPT in this way. In fact, Wetterich also developed a 2PI-FRG approach based on a modified Legendre transform as well [169]. However, as opposed to this work, Dupuis' 2PI-FRG is based on flow equations for the Luttinger-Ward (LW) functional and not for the 2PI

EA itself, which significantly improves its convergence.[17] Moreover, both Wetterich and Dupuis ignore the field dependence of the 2PI EA in their 2PI-FRG formulations, i.e. they consider $\Gamma^{(2\mathrm{PI})}[\phi=0,G]$, which is also referred to as bosonic EA [169]. Some appealing features of Dupuis' 2PI-FRG[18] are listed below, stressing some advantages of this 2PI-FRG as compared to its 1PI counterpart:

- Contrary to a 1PI EA depending only on Grassmann fields, expansions of $\Gamma^{(2\mathrm{PI})}[\phi=0,G]$ around *non-trivial minima* are possible since the propagator $G$ is a bosonic variable. As a result, the 2PI-FRG is well suited to grasp non-perturbative physics in fermionic systems, even without HSTs.
- When the 2PI-FRG flow is designed to take results of SCPT as inputs, the 2PI-FRG offers the possibility to *start the flow in a broken-symmetry phase*, which enables us to avoid phase transitions (and the associated problematic divergences encountered in the 1PI-FRG) during the flow.
- Besides its convenient starting points, the 2PI-FRG is designed itself to avoid the undesirable divergences from which the 1PI-FRG suffers. This is simply because the quantities calculated during the flow are different: one calculates the 1PI vertices (i.e. derivatives of the 1PI EA) during the 1PI-FRG flow and the *2PI vertices* (i.e. derivatives of the LW functional) during the 2PI-FRG flow. A concrete example is given in Ref. [44] for an application of the C-flow implementation of the 2PI-FRG to the BCS theory.

However, the most performing implementations of the 2PI-FRG, i.e. the U-flow and the CU-flow, both require to solve the Bethe-Salpeter equation at each step of the flow, as will be explained technically later in this section. This increases significantly the numerical cost of these FRG approaches, although we will argue throughout this study that such a cost is worth paying in various situations (when treating strongly-coupled fermions notably) and some ideas can actually be put forward to lower this cost (see section 5.1). Nonetheless, this is probably due to this important numerical weight that very few applications of the 2PI-FRG have been carried out so far. Among these, we can mention the work of Rentrop, Jakobs and Meden on the (0+1)-D $\varphi^4$-theory [179] and on the (0+1)-D Anderson impurity model[19] [180]. Note in addition the study of Ref. [181] that discusses connections between the 2PI-FRG C-flow and the 1PI-FRG in the context of the Fermi-edge singularity, which is a model used to describe optical excitations in electronic systems. Notably, the 2PI-FRG has also been designed by Katanin so as to take the 2PI vertices calculated from another non-perturbative method, i.e. dynamical mean-field theory (DMFT) [182, 183], as inputs [184], thus developing a 2PI counterpart for the so-called DMF$^2$RG already developed in the 1PI framework [120, 185–187]. Most importantly, the tower of differential equations resulting from this approach is tractable enough to perform a quantitative description of a (2+1)-D extended[20] Hubbard model [184]. Besides the 2PI-FRG study of the (3+1)-D $\varphi^4$-theory reported in Ref. [188], this is to our knowledge the only 2PI-FRG application to a model with finite space dimensions, which illustrates that this promising 2PI approach has barely been beyond the stage of toy model applications.

Before presenting the 2PI-FRG formalism, we first define some notations and review the symmetry properties of the main objects involved in the 2PI EA framework. All 2PI-FRG implementations investigated in the present study are based on the generating functional:

$$Z[K]=e^{W[K]}=\int\mathcal{D}\widetilde{\psi}\;e^{-S\left[\widetilde{\psi}\right]+\frac{1}{2}\int_{\alpha,\alpha'}\widetilde{\psi}_\alpha K_{\alpha\alpha'}\widetilde{\psi}_{\alpha'}}\;, \tag{106}$$

where $\widetilde{\psi}$ is either a real bosonic field ($\zeta=+1$) or a real Grassmann field ($\zeta=-1$). The index $\alpha\equiv(i,x)$ used in Eq. (106) combines this time all types of internal indices $i$ with $x\equiv(r,\tau,c)$ including the space coordinate $r$, the imaginary time $\tau$ and a charge index $c$ if necessary. The latter index is defined as follows[21]:

$$\widetilde{\psi}_\alpha=\widetilde{\psi}_{i,x}=\begin{cases}\widetilde{\psi}_i(r,\tau) & \text{for } c=-\;.\\ \widetilde{\psi}^\dagger_i(r,\tau) & \text{for } c=+\;.\end{cases} \tag{107}$$

The shorthand notation for integration exploited in Eq. (106) is set by:

$$\int_\alpha\equiv\sum_{i,c}\int_0^{1/T}d\tau\int d^{D-1}r\;, \tag{108}$$

assuming that the studied system lives in a $D$-dimensional spacetime. It will be also most convenient to group $\alpha$-indices by pairs via a bosonic index:

$$\gamma\equiv(\alpha,\alpha')\;, \tag{109}$$

and the main features of the 2PI EA formalism based on such indices are presented in appendix A. For instance, the connected correlation functions can be expressed in terms of such indices:

$$W^{(n)}_{\gamma_1\cdots\gamma_n}[K]\equiv\frac{\delta^n W[K]}{\delta K_{\gamma_1}\cdots\delta K_{\gamma_n}}=\frac{\delta^n W[K]}{\delta K_{\alpha_1\alpha_1'}\cdots\delta K_{\alpha_n\alpha_n'}}\;, \tag{110}$$

[17] This improvement will be thoroughly illustrated via the numerical applications presented in section 4.3 on the 2PPI-FRG.

[18] In what follows, 2PI-FRG refers implicitly to the 2PI-FRG à la Dupuis developed in Refs. [44, 45].

[19] The work of Refs. [179,180] echoes that of Ref. [104] which treats the same models with the 1PI-FRG.

[20] The adjective "extended" for the Hubbard model indicates that the latter involves a non-local Coulomb interaction, i.e. a Coulomb interaction between fermions on different lattice sites.

[21] Therefore, with this notation based on a charge index, $\widetilde{\psi}_\alpha$ is always mathematically treated as a real field with an extra index, instead of a complex one.

which defines the connected propagator:

$$G_{\gamma} = W^{(1)}_{\gamma}[K] \;, \tag{111}$$

for $n = 1$. As the components of the source $K$ satisfy $K_{\alpha\alpha'} = \zeta K_{\alpha'\alpha}$, the correlation functions of Eq. (110) possess the symmetry properties:

$$\begin{cases} W^{(n)}_{\gamma_1\cdots(\alpha_i,\alpha_i')\cdots\gamma_n}[K] = \zeta W^{(n)}_{\gamma_1\cdots(\alpha_i',\alpha_i)\cdots\gamma_n}[K] \;, & \text{(112a)} \\ W^{(n)}_{\gamma_1\cdots\gamma_n}[K] = W^{(n)}_{\gamma_{P(1)}\cdots\gamma_{P(n)}}[K] \;, & \text{(112b)} \end{cases}$$

with $P$ denoting an arbitrary element of the permutation group of order $n$, and especially:

$$G_{\alpha\alpha'} = \zeta G_{\alpha'\alpha} \;, \tag{113}$$

at $n = 1$. With this bosonic index notation, the Legendre transform defining the 2PI EA under consideration reads:

$$\Gamma^{(\mathrm{2PI})}[G] = -W[K] + \frac{1}{2}\int_{\gamma} K_{\gamma} G_{\gamma} \;, \tag{114}$$

where an integration over bosonic indices amounts to integrating (or just summing) over all its constituent indices, i.e.:

$$\int_{\gamma} \equiv \int_{\alpha,\alpha'} \;. \tag{115}$$

In what follows, we will use a DeWitt-like notation for the integration over bosonic indices. For $n$ arbitrary bosonic matrices $M_m$ (with $m = 1,\ldots,n$), it takes the following form:

$$M_{1,\gamma_1\hat{\gamma}_1}\cdots M_{n,\hat{\gamma}_{n-1}\gamma_2} = \frac{1}{2^{n-1}}\int_{\gamma_3,\cdots,\gamma_{n+1}} M_{1,\gamma_1\gamma_3}\cdots M_{n,\gamma_{n+1}\gamma_2} \;, \tag{116}$$

where the hatted indices are all dummy and the non-hatted ones are all free by convention. In the 2PI EA framework, it is also natural to consider the LW functional $\Phi[G]$, which corresponds to the interaction part of $\Gamma^{(\mathrm{2PI})}[G]$, i.e.:

$$\Phi[G] \equiv \Gamma^{(\mathrm{2PI})}[G] - \Gamma^{(\mathrm{2PI})}_{0}[G] \;, \tag{117}$$

where the non-interacting part of the 2PI EA can be expressed from Gaussian integration, thus yielding[22]:

$$\Gamma^{(\mathrm{2PI})}_{0}[G] = -\frac{\zeta}{2}\mathrm{STr}\left[\ln(G)\right] + \frac{\zeta}{2}\mathrm{STr}\left[GC^{-1} - \mathbb{I}\right] \;, \tag{118}$$

with $\mathbb{I}$ denoting the identity with respect to $\alpha$-indices (i.e. $\mathbb{I}_{\alpha_1\alpha_2} = \delta_{\alpha_1\alpha_2} = \delta_{i_1 i_2}\delta_{c_1 c_2}\delta^{(D-1)}(r_1 - r_2)\delta^{(1)}(\tau_1 - \tau_2)$) and $C$ being the free propagator, i.e.:

$$C^{-1}_{\alpha\alpha'} = \left.\frac{\delta^2 S\big[\widetilde{\psi}\big]}{\delta\widetilde{\psi}_{\alpha}\delta\widetilde{\psi}_{\alpha'}}\right|_{\widetilde{\psi}=0} \;. \tag{119}$$

[22] Since we have now two types of indices ($\alpha$- and bosonic indices), we stress that STr still denotes the trace with respect to $\alpha$-indices in the whole section 3, as in the rest of this paper.

Recall that the LW functional is the sum of 2PI diagrams, with propagator lines corresponding to the full propagator $G$. The so-called 2PI vertices correspond to its derivatives, i.e. $\Phi^{(n)}_{\gamma_1\cdots\gamma_n}[G] \equiv \frac{\delta^n \Phi[G]}{\delta G_{\gamma_1}\cdots\delta G_{\gamma_n}}$. Another important relation of the 2PI formalism is the Dyson equation that follows from Eq. (117), i.e.:

$$\frac{\delta\Gamma^{(\mathrm{2PI})}[G]}{\delta G_{\gamma}} = K_{\gamma} = -G^{-1}_{\gamma} + C^{-1}_{\gamma} - \Sigma_{\gamma}[G] \;, \tag{120}$$

where the self-energy $\Sigma$ is directly connected to the LW functional:

$$\Sigma_{\gamma}[G] \equiv -\frac{\delta\Phi[G]}{\delta G_{\gamma}} \;. \tag{121}$$

All equations introduced since the beginning of section 3.1 underlie the 2PI EA framework, but are not inherent to the 2PI-FRG formalism. All 2PI-FRG approaches treated in this study can be applied to any system whose classical action can be put in the form:

$$S\Big[\widetilde{\psi}\Big] = \frac{1}{2}\int_{\alpha_1,\alpha_2} \widetilde{\psi}_{\alpha_1} C^{-1}_{\alpha_1\alpha_2}\widetilde{\psi}_{\alpha_2} + \frac{1}{4!}\int_{\alpha_1,\alpha_2,\alpha_3,\alpha_4} U_{\alpha_1\alpha_2\alpha_3\alpha_4}\widetilde{\psi}_{\alpha_1}\widetilde{\psi}_{\alpha_2}\widetilde{\psi}_{\alpha_3}\widetilde{\psi}_{\alpha_4} \;, \tag{122}$$

with the two-body interaction $U$ satisfying the symmetry property:

$$U_{\alpha_1\alpha_2\alpha_3\alpha_4} = \zeta^{N(P)} U_{\alpha_{P(1)}\alpha_{P(2)}\alpha_{P(3)}\alpha_{P(4)}} \;, \tag{123}$$

$N(P)$ being the number of inversions in the permutation $P$. Hence, in the present 2PI-FRG study, we are only treating systems with two-body interactions at most. However, it is straightforward to generalize the formalism presented in the whole section 3 to interactions which are three-body or more after including in $S\big[\widetilde{\psi}\big]$ terms being sextic in $\widetilde{\psi}_{\alpha}$ or more. This is a particularly important remark to prepare the ground for applications to nuclear systems as it is acknowledged that a treatment of three-body interactions is necessary for many-body techniques grounded on EFTs of QCD and aiming at a quantitative description of atomic nuclei. Regarding the 2PI-FRG considered here, three main implementations can be defined from action (122) depending on the way cutoff functions $R_{\mathfrak{s}}$ are introduced: either to dress the free propagator via e.g. $C^{-1} \rightarrow C^{-1}_{\mathfrak{s}} = C^{-1} + R_{\mathfrak{s}}$, or to dress the two-body interaction via e.g. $U \rightarrow U_{\mathfrak{s}} = U + R_{\mathfrak{s}}$, or both. These three options define respectively the *C-flow*, the *U-flow* and the *CU-flow* of the 2PI-FRG. We stress that, in principle, all these three versions can implement the Wilsonian momentum-shell integration, but this is not always the case depending on the choice of cutoff function(s). The flow parameter can thus be dimensionless and we will therefore denote it by $\mathfrak{s}$ instead of $k$ in what follows.

After performing the substitution(s) $C^{-1} \rightarrow C^{-1}_{\mathfrak{s}}$ and/or $U \rightarrow U_{\mathfrak{s}}$, all relevant functionals become dependent on $\mathfrak{s}$, including the 2PI EA defined by Eq. (114) that

becomes in this way:

$$\Gamma_{\mathfrak{s}}^{(2\mathrm{PI})}[G] = -W_{\mathfrak{s}}[K] + \frac{1}{2}\int_{\gamma} K_{\gamma} G_{\gamma} \;, \tag{124}$$

with

$$G_{\gamma} \equiv G_{\mathfrak{s},\gamma}[K] = W_{\mathfrak{s},\gamma}^{(1)}[K] \;. \tag{125}$$

In particular, for any approach based on the EA (124) including the entire 2PI-FRG framework, the 1-point correlation function $\phi = \langle\widetilde{\psi}\rangle$ is imposed to vanish and spontaneous breakdowns of the $O(N)$ symmetry can therefore not be produced. An exact flow equation can be derived for $\Gamma_{\mathfrak{s}}^{(2\mathrm{PI})}[G]$, i.e.:

$$\begin{aligned}\dot{\Gamma}_{\mathfrak{s}}^{(2\mathrm{PI})}[G] &\equiv \partial_{\mathfrak{s}}\Gamma_{\mathfrak{s}}^{(2\mathrm{PI})}[G] \\ &= \underbrace{\dot{C}_{\mathfrak{s},\hat{\gamma}}^{-1} G_{\hat{\gamma}}}_{C^{-1}\rightarrow C_{\mathfrak{s}}^{-1}} + \underbrace{\frac{1}{6}\dot{U}_{\mathfrak{s},\hat{\gamma}_1\hat{\gamma}_2}\left(W_{\mathfrak{s},\hat{\gamma}_2\hat{\gamma}_1}^{(2)}[K] + \frac{1}{2}\Pi_{\hat{\gamma}_2\hat{\gamma}_1}[G]\right)}_{U\rightarrow U_{\mathfrak{s}}} \;,\end{aligned} \tag{126}$$

which is the counterpart of the Wetterich equation for the 2PI-FRG. The underbraces indicate which terms are generated by the flow dependence of either $C_{\mathfrak{s}}$ or $U_{\mathfrak{s}}$, which means that the term induced by $U \rightarrow U_{\mathfrak{s}}$ ($C^{-1} \rightarrow C_{\mathfrak{s}}^{-1}$) is to be discarded for the $C$-flow ($U$-flow). Eq. (126) also involves the pair propagator $\Pi[G]$ that can be expressed as:

$$\Pi_{\gamma_1\gamma_2}[G] \equiv \frac{\delta^2 W_0[K(G)]}{\delta K_{\gamma_1}\delta K_{\gamma_2}} = G_{\alpha_1\alpha_2'}G_{\alpha_1'\alpha_2} + \zeta G_{\alpha_1\alpha_2}G_{\alpha_1'\alpha_2'} \;, \tag{127}$$

where $W_0[K]$ is by definition $W[K]$ evaluated at vanishing $U$, i.e. $W_0[K]$ is the non-interacting version of the Schwinger functional defined by Eq. (106). As in the fermionic 1PI-FRG, the conventional treatment of the exact flow equation (126) is the vertex expansion, which now consists in Taylor expanding both sides of Eq. (126) with respect to $G$. In particular, the corresponding expansion for $\dot{\Gamma}_{\mathfrak{s}}^{(2\mathrm{PI})}[G]$ is directly deduced from:

$$\begin{aligned}\Gamma_{\mathfrak{s}}^{(2\mathrm{PI})}[G] = {} & \overline{\Gamma}_{\mathfrak{s}}^{(2\mathrm{PI})} \\ & + \sum_{n=2}^{\infty}\frac{1}{n!}\int_{\gamma_1,\cdots,\gamma_n}\overline{\Gamma}_{\mathfrak{s},\gamma_1\cdots\gamma_n}^{(2\mathrm{PI})(n)} \\ & \times \left(G-\overline{G}_{\mathfrak{s}}\right)_{\gamma_1}\cdots\left(G-\overline{G}_{\mathfrak{s}}\right)_{\gamma_n} \;,\end{aligned} \tag{128}$$

with $\overline{\Gamma}_{\mathfrak{s}}^{(2\mathrm{PI})} \equiv \Gamma_{\mathfrak{s}}^{(2\mathrm{PI})}[G = \overline{G}_{\mathfrak{s}}]$, $\overline{\Gamma}_{\mathfrak{s},\gamma_1\cdots\gamma_n}^{(2\mathrm{PI})(n)} \equiv \left.\frac{\delta^n\Gamma_{\mathfrak{s}}^{(2\mathrm{PI})}[G]}{\delta G_{\gamma_1}\cdots\delta G_{\gamma_n}}\right|_{G=\overline{G}_{\mathfrak{s}}}$ and $\overline{G}_{\mathfrak{s}}$ must satisfy:

$$\left.\frac{\delta\Gamma_{\mathfrak{s}}^{(2\mathrm{PI})}[G]}{\delta G_{\gamma}}\right|_{G=\overline{G}_{\mathfrak{s}}} = 0 \quad \forall\gamma,\mathfrak{s} \;. \tag{129}$$

An infinite tower of coupled differential equations can then be derived for the 2PI vertices $\overline{\Phi}_{\mathfrak{s}}^{(n)} \equiv \Phi_{\mathfrak{s}}^{(n)}[G = \overline{G}_{\mathfrak{s}}]$ by comparing the two expanded sides of Eq. (126) and equalizing the terms with identical powers of $G - \overline{G}_{\mathfrak{s}}$. The 2PI vertices are introduced in those equations after replacing the derivatives $\overline{\Gamma}_{\mathfrak{s}}^{(2\mathrm{PI})(n)}$ with $\overline{\Phi}_{\mathfrak{s}}^{(n)}$ using splitting (117). In what follows, we will deduce these coupled flow equations for 2PI vertices by directly using the functional chain rule, which is equivalent to exploiting the vertex expansion outlined here. This equivalence is actually illustrated in section 4 by comparing numerical results obtained from 2PI- and 2PPI-FRGs.[23]

All versions of the 2PI-FRG can be applied to the (0+0)-D $O(N)$ model in its original representation. Indeed, after setting:

$$\widetilde{\psi}_{\alpha} = \widetilde{\varphi}_{a} \;, \tag{130}$$

$$\int_{\gamma} = \sum_{a,a'=1}^{N} \;, \tag{131}$$

$$\gamma \equiv (\alpha,\alpha') = (a,a') \;, \tag{132}$$

one can see that the generic classical action (122) reduces to $S(\vec{\widetilde{\varphi}})$ expressed by Eq. (1), i.e. we obtain in this way:

$$\begin{aligned}S\left[\widetilde{\psi}\right] &= S\left(\vec{\widetilde{\varphi}}\right) \\ &= \frac{1}{2}\sum_{a_1,a_2=1}^{N}\widetilde{\varphi}_{a_1}C_{a_1a_2}^{-1}\widetilde{\varphi}_{a_2} \\ &+ \frac{1}{4!}\sum_{a_1,a_2,a_3,a_4=1}^{N}U_{a_1a_2a_3a_4}\widetilde{\varphi}_{a_1}\widetilde{\varphi}_{a_2}\widetilde{\varphi}_{a_3}\widetilde{\varphi}_{a_4} \;,\end{aligned} \tag{133}$$

with the free inverse propagator:

$$C_{a_1a_2}^{-1} = m^2\delta_{a_1a_2} \;, \tag{134}$$

and the two-body interaction:

$$U_{a_1a_2a_3a_4} = \frac{\lambda}{3}\left(\delta_{a_1a_2}\delta_{a_3a_4} + \delta_{a_1a_3}\delta_{a_2a_4} + \delta_{a_1a_4}\delta_{a_2a_3}\right) \;, \tag{135}$$

which is consistent with Eq. (123) at $\zeta = +1$. However, some implementations of the 2PI-FRG (the U-flow and the CU-flow) are not suited to treat the mixed and collective representations of our toy model, because the interaction part of the classical action can no longer be put in the form $\int_{\alpha_1,\alpha_2,\alpha_3,\alpha_4} U_{\alpha_1\alpha_2\alpha_3\alpha_4}\widetilde{\psi}_{\alpha_1}\widetilde{\psi}_{\alpha_2}\widetilde{\psi}_{\alpha_3}\widetilde{\psi}_{\alpha_4}$ in those situations. We will discuss the mixed case in more detail in section 3.3. In the rest of section 3, we will further develop the 2PI-FRG formalism in a generic $D$-dimensional framework instead of specifying to the (0+0)-D $O(N)$ model, as opposed to the 1PI-FRG section. The equations for the studied toy model can be directly inferred from the generic equations thus derived by imposing equalities (130), (131), (132), (134) and (135) with $\zeta = +1$. Before moving to section 4 on the 2PPI-FRG, we will actually only treat the U-flow implementation of the 2PI-FRG, in its standard version, i.e. the plain U-flow (pU-flow), and its improved version, i.e. the modified U-flow (mU-flow). The underpinning reason is that we want to illustrate that the 2PI-FRG provides us with excellent correlated starting points given by SCPT based on 2PI EAs, which is particularly appealing in our aim to treat strongly correlated systems. We will discuss the price to pay (in terms of numerical cost) for such powerful starting points and all of this can be addressed in the U-flow (and more specifically in the mU-flow) framework. In addition, there already exists studies of 2PI-FRG methods comparing numerical results of C-flow, U-flow and CU-flow approaches [179,180], with Ref. [180] notably discussing how to construct $\Phi$-derivable approximations from some of these 2PI-FRG implementations. From this, we can expect the mU-flow to be the most performing version of the 2PI-FRG (in general and for the studied $O(N)$ model in particular), alongside with the CU-flow which yields similar performances. Moreover, the C-flow is not suited to design reliable approximation schemes. The last remarks on the CU-flow and the C-flow are thoroughly illustrated in appendix B. Finally, we stress that the present work is to our knowledge the first to apply the 2PI-FRG developed in Refs. [44,45] to an $O(N)$ model with $N > 1$ (see appendix C.2 in particular) and we will also exploit the simplicity of our zero-dimensional toy model to reach higher truncation orders than those investigated in previous studies [179,180], allowing notably the 2PI three-particle vertex $\overline{\Phi}^{(3)}_{\mathfrak{s}}$ to flow.

[23] See Fig. 11 more specifically.

### 3.2 Plain U-flow

The U-flow scheme, developed in Refs. [45,179], is based on the substitution $U \rightarrow U_{\mathfrak{s}} = R_{\mathfrak{s}} U$ or equivalently $U \rightarrow U_{\mathfrak{s}} = U + R_{\mathfrak{s}}$. The U-flow implements in principle the Wilsonian momentum-shell integration, with $\mathfrak{s}$ being connected to the momentum scale. This is notably the case with the cutoff function $R_{\mathfrak{s}}$ put forward in section IV.C. of Ref. [45] which plays the role of an infrared regulator for (low-energy) collective fluctuations, thus preventing problematic divergences during the flow. However, a perfectly valid choice for $R_{\mathfrak{s}}$, used for all our 2PI-FRG applications in (0+0)-D with the U-flow, could be set by $U_{\mathfrak{s}} = R_{\mathfrak{s}} U = \mathfrak{s} U$, with $\mathfrak{s}$ a dimensionless parameter running from $\mathfrak{s}_{\mathrm{i}} = 0$ to $\mathfrak{s}_{\mathrm{f}} = 1$. This follows the philosophy of the 2PPI-FRG discussed in section 4 and the resulting 2PI-FRG implementation does not carry out the momentum-shell integration à la Wilson. However, we will explain later that such a choice for $R_{\mathfrak{s}}$ can enable us to start the flow in broken-symmetry phases, which could also be advantageous in the description of critical physics. The boundary conditions for $U_{\mathfrak{s}}$ are:

$$\begin{cases} U_{\mathfrak{s}=\mathfrak{s}_{\mathrm{i}},\gamma_1\gamma_2} = 0 \quad \forall \gamma_1,\gamma_2 \; . & (136a) \\ U_{\mathfrak{s}=\mathfrak{s}_{\mathrm{f}}} = U \; . & (136b) \end{cases}$$

Hence, the starting point of the flow corresponds to the free theory, at least for the standard implementation of the U-flow called the pU-flow. We will see in section 3.3 that, after a suitable transformation of the LW functional, condition (136a) implies that the 2PI-FRG flow starts from SCPT results, which defines the mU-flow.

Regarding the infinite hierarchy of differential equations to solve within the pU-flow, one can start by differentiating the Dyson equation (120) with respect to $\mathfrak{s}$ at vanishing source $K$:

$$\dot{\overline{G}}_{\mathfrak{s},\alpha_1\alpha_1'} = \int_{\alpha_2,\alpha_2'} \overline{G}_{\mathfrak{s},\alpha_1\alpha_2} \dot{\overline{\Sigma}}_{\mathfrak{s},\alpha_2\alpha_2'} \overline{G}_{\mathfrak{s},\alpha_2'\alpha_1'} \; , \tag{137}$$

where we have used $\dot{C}^{-1}_{\gamma} = \dot{C}^{-1}_{\mathfrak{s},\gamma} = 0 \; \forall\gamma$, which holds by definition of the U-flow. Moreover, a differential equation expressing $\dot{\overline{\Gamma}}^{(\mathrm{2PI})}_{\mathfrak{s}}$ is directly found from the master equation (126):

$$\dot{\overline{\Gamma}}^{(\mathrm{2PI})}_{\mathfrak{s}} = \frac{1}{6} \dot{U}_{\mathfrak{s},\hat{\gamma}_1\hat{\gamma}_2} \left( \overline{W}^{(2)}_{\mathfrak{s}} + \frac{1}{2} \overline{\Pi}_{\mathfrak{s}} \right)_{\hat{\gamma}_2\hat{\gamma}_1} \; . \tag{138}$$

Following the derivations of Refs. [45, 179], the corresponding flow equations for the 2PI vertices can be directly inferred from the chain rule based on bosonic indices (see appendix A):

$$\begin{cases} \dot{\overline{\Sigma}}_{\mathfrak{s},\gamma} = -\dot{\overline{\Phi}}^{(1)}_{\mathfrak{s},\gamma} - \dot{\overline{G}}_{\mathfrak{s},\hat{\gamma}} \overline{\Phi}^{(2)}_{\mathfrak{s},\hat{\gamma}\gamma} = \dot{\overline{\Sigma}}_{\mathfrak{s},\gamma} - \dot{\overline{G}}_{\mathfrak{s},\hat{\gamma}} \overline{\Phi}^{(2)}_{\mathfrak{s},\hat{\gamma}\gamma} \; , & (139a) \\ \dot{\overline{\Phi}}^{(n)}_{\mathfrak{s},\gamma_1\cdots\gamma_n} = \dot{\overline{\Phi}}^{(n)}_{\mathfrak{s},\gamma_1\cdots\gamma_n} + \dot{\overline{G}}_{\mathfrak{s},\hat{\gamma}} \overline{\Phi}^{(n+1)}_{\mathfrak{s},\hat{\gamma}\gamma_1\cdots\gamma_n} \quad \forall n \geq 2 \; , & (139b) \end{cases}$$

with $\dot{\overline{\Phi}}^{(n)}_{\mathfrak{s}} \equiv \dot{\Phi}^{(n)}_{\mathfrak{s}}[G = \overline{G}_{\mathfrak{s}}]$. The derivatives $\dot{\overline{\Phi}}^{(n)}_{\mathfrak{s}}$ are then expressed by differentiating the flow equation for the 2PI EA (126), combined with the definition of the LW functional given by Eqs. (117) and (118). Up to $n = 3$, this leads to:

$$\begin{aligned} \dot{\overline{\Sigma}}_{\mathfrak{s},\gamma} = & - \frac{1}{3} \left( \mathcal{I} + \overline{\Pi}_{\mathfrak{s}} \overline{\Phi}^{(2)}_{\mathfrak{s}} \right)^{\mathrm{inv}}_{\gamma\hat{\gamma}_1} \left[ 2 \left( \mathcal{I} + \overline{\Pi}_{\mathfrak{s}} \overline{\Phi}^{(2)}_{\mathfrak{s}} \right)^{\mathrm{inv}} \dot{U}_{\mathfrak{s}} \left( \mathcal{I} + \overline{\Pi}_{\mathfrak{s}} \overline{\Phi}^{(2)}_{\mathfrak{s}} \right)^{\mathrm{inv}} + \dot{U}_{\mathfrak{s}} \right]_{\hat{\alpha}_1\hat{\alpha}_2\hat{\alpha}_2'\hat{\alpha}_1'} \overline{G}_{\mathfrak{s},\hat{\gamma}_2} \\ & + \frac{1}{6} \left( \mathcal{I} + \overline{\Pi}_{\mathfrak{s}} \overline{\Phi}^{(2)}_{\mathfrak{s}} \right)^{\mathrm{inv}}_{\gamma\hat{\gamma}_1} \dot{U}_{\mathfrak{s},\hat{\gamma}_2\hat{\gamma}_3} \overline{W}^{(2)}_{\mathfrak{s},\hat{\gamma}_3\hat{\gamma}_4} \overline{\Phi}^{(3)}_{\mathfrak{s},\hat{\gamma}_1\hat{\gamma}_4\hat{\gamma}_5} \overline{W}^{(2)}_{\mathfrak{s},\hat{\gamma}_5\hat{\gamma}_2} \; , \end{aligned} \tag{140}$$

$$\begin{aligned}
\dot{\overline{\Phi}}^{(2)}_{\mathfrak{s},\gamma_1\gamma_2} = \frac{1}{3}\dot{U}_{\mathfrak{s},\hat{\gamma}_1\hat{\gamma}_2}\Bigg[ & \overline{W}^{(2)}_{\mathfrak{s},\hat{\gamma}_2\hat{\gamma}_3}\left(\overline{\Pi}^{\mathrm{inv}}_{\mathfrak{s},\hat{\gamma}_3\hat{\gamma}_4}\frac{\delta\overline{\Pi}_{\mathfrak{s},\hat{\gamma}_4\hat{\gamma}_5}}{\delta\overline{G}_{\mathfrak{s},\gamma_1}}\overline{\Pi}^{\mathrm{inv}}_{\mathfrak{s},\hat{\gamma}_5\hat{\gamma}_6} - \overline{\Phi}^{(3)}_{\mathfrak{s},\gamma_1\hat{\gamma}_3\hat{\gamma}_6}\right)\overline{W}^{(2)}_{\mathfrak{s},\hat{\gamma}_6\hat{\gamma}_7} \\
& \times\left(\overline{\Pi}^{\mathrm{inv}}_{\mathfrak{s},\hat{\gamma}_7\hat{\gamma}_8}\frac{\delta\overline{\Pi}_{\mathfrak{s},\hat{\gamma}_8\hat{\gamma}_9}}{\delta\overline{G}_{\mathfrak{s},\gamma_2}}\overline{\Pi}^{\mathrm{inv}}_{\mathfrak{s},\hat{\gamma}_9\hat{\gamma}_{10}} - \overline{\Phi}^{(3)}_{\mathfrak{s},\gamma_2\hat{\gamma}_7\hat{\gamma}_{10}}\right)\overline{W}^{(2)}_{\mathfrak{s},\hat{\gamma}_{10}\hat{\gamma}_1} \\
& - \overline{W}^{(2)}_{\mathfrak{s},\hat{\gamma}_2\hat{\gamma}_3}\overline{\Pi}^{\mathrm{inv}}_{\mathfrak{s},\hat{\gamma}_3\hat{\gamma}_4}\frac{\delta\overline{\Pi}_{\mathfrak{s},\hat{\gamma}_4\hat{\gamma}_5}}{\delta\overline{G}_{\mathfrak{s},\gamma_1}}\overline{\Pi}^{\mathrm{inv}}_{\mathfrak{s},\hat{\gamma}_5\hat{\gamma}_6}\frac{\delta\overline{\Pi}_{\mathfrak{s},\hat{\gamma}_6\hat{\gamma}_7}}{\delta\overline{G}_{\mathfrak{s},\gamma_2}}\overline{\Pi}^{\mathrm{inv}}_{\mathfrak{s},\hat{\gamma}_7\hat{\gamma}_8}\overline{W}^{(2)}_{\mathfrak{s},\hat{\gamma}_8\hat{\gamma}_1} \\
& + \frac{1}{2}\overline{W}^{(2)}_{\mathfrak{s},\hat{\gamma}_2\hat{\gamma}_3}\left(\overline{\Pi}^{\mathrm{inv}}_{\mathfrak{s},\hat{\gamma}_3\hat{\gamma}_4}\frac{\delta^2\overline{\Pi}_{\mathfrak{s},\hat{\gamma}_4\hat{\gamma}_5}}{\delta\overline{G}_{\mathfrak{s},\gamma_1}\delta\overline{G}_{\mathfrak{s},\gamma_2}}\overline{\Pi}^{\mathrm{inv}}_{\mathfrak{s},\hat{\gamma}_5\hat{\gamma}_6} - \overline{\Phi}^{(4)}_{\mathfrak{s},\gamma_1\gamma_2\hat{\gamma}_3\hat{\gamma}_6}\right)\overline{W}^{(2)}_{\mathfrak{s},\hat{\gamma}_6\hat{\gamma}_1} + \frac{1}{4}\frac{\delta^2\overline{\Pi}_{\mathfrak{s},\hat{\gamma}_2\hat{\gamma}_1}}{\delta\overline{G}_{\mathfrak{s},\gamma_1}\delta\overline{G}_{\mathfrak{s},\gamma_2}}\Bigg] + \dot{\overline{G}}_{\mathfrak{s},\hat{\gamma}}\overline{\Phi}^{(3)}_{\mathfrak{s},\hat{\gamma}\gamma_1\gamma_2} \, ,
\end{aligned} \tag{141}$$

$$\begin{aligned}
\dot{\overline{\Phi}}^{(3)}_{\mathfrak{s},\gamma_1\gamma_2\gamma_3} = \frac{1}{3}\dot{U}_{\mathfrak{s},\hat{\gamma}_1\hat{\gamma}_2}\Bigg[ & 3\overline{W}^{(2)}_{\mathfrak{s},\hat{\gamma}_2\hat{\gamma}_3}\left(\overline{\Pi}^{\mathrm{inv}}_{\mathfrak{s},\hat{\gamma}_3\hat{\gamma}_4}\frac{\delta\overline{\Pi}_{\mathfrak{s},\hat{\gamma}_4\hat{\gamma}_5}}{\delta\overline{G}_{\mathfrak{s},\gamma_1}}\overline{\Pi}^{\mathrm{inv}}_{\mathfrak{s},\hat{\gamma}_5\hat{\gamma}_6} - \overline{\Phi}^{(3)}_{\mathfrak{s},\gamma_1\hat{\gamma}_3\hat{\gamma}_6}\right)\overline{W}^{(2)}_{\mathfrak{s},\hat{\gamma}_6\hat{\gamma}_7}\Bigg(\overline{\Pi}^{\mathrm{inv}}_{\mathfrak{s},\hat{\gamma}_7\hat{\gamma}_8}\frac{\delta\overline{\Pi}_{\mathfrak{s},\hat{\gamma}_8\hat{\gamma}_9}}{\delta\overline{G}_{\mathfrak{s},\gamma_2}}\overline{\Pi}^{\mathrm{inv}}_{\mathfrak{s},\hat{\gamma}_9\hat{\gamma}_{10}} \\
& - \overline{\Phi}^{(3)}_{\mathfrak{s},\gamma_2\hat{\gamma}_7\hat{\gamma}_{10}}\Bigg)\overline{W}^{(2)}_{\mathfrak{s},\hat{\gamma}_{10}\hat{\gamma}_{11}}\left(\overline{\Pi}^{\mathrm{inv}}_{\mathfrak{s},\hat{\gamma}_{11}\hat{\gamma}_{12}}\frac{\delta\overline{\Pi}_{\mathfrak{s},\hat{\gamma}_{12}\hat{\gamma}_{13}}}{\delta\overline{G}_{\mathfrak{s},\gamma_3}}\overline{\Pi}^{\mathrm{inv}}_{\mathfrak{s},\hat{\gamma}_{13}\hat{\gamma}_{14}} - \overline{\Phi}^{(3)}_{\mathfrak{s},\gamma_3\hat{\gamma}_{11}\hat{\gamma}_{14}}\right)\overline{W}^{(2)}_{\mathfrak{s},\hat{\gamma}_{14}\hat{\gamma}_1} \\
& + \Bigg(\overline{W}^{(2)}_{\mathfrak{s},\hat{\gamma}_2\hat{\gamma}_3}\left(\overline{\Pi}^{\mathrm{inv}}_{\mathfrak{s},\hat{\gamma}_3\hat{\gamma}_4}\frac{\delta\overline{\Pi}_{\mathfrak{s},\hat{\gamma}_4\hat{\gamma}_5}}{\delta\overline{G}_{\mathfrak{s},\gamma_1}}\overline{\Pi}^{\mathrm{inv}}_{\mathfrak{s},\hat{\gamma}_5\hat{\gamma}_6} - \overline{\Phi}^{(3)}_{\mathfrak{s},\gamma_1\hat{\gamma}_3\hat{\gamma}_6}\right)\overline{W}^{(2)}_{\mathfrak{s},\hat{\gamma}_6\hat{\gamma}_7} \\
& \times\left(\overline{\Pi}^{\mathrm{inv}}_{\mathfrak{s},\hat{\gamma}_7\hat{\gamma}_8}\frac{\delta^2\overline{\Pi}_{\mathfrak{s},\hat{\gamma}_8\hat{\gamma}_9}}{\delta\overline{G}_{\mathfrak{s},\gamma_2}\delta\overline{G}_{\mathfrak{s},\gamma_3}}\overline{\Pi}^{\mathrm{inv}}_{\mathfrak{s},\hat{\gamma}_9\hat{\gamma}_{10}} - \overline{\Phi}^{(4)}_{\mathfrak{s},\gamma_2\gamma_3\hat{\gamma}_7\hat{\gamma}_{10}}\right)\overline{W}^{(2)}_{\mathfrak{s},\hat{\gamma}_{10}\hat{\gamma}_1} + (\gamma_2,\gamma_1,\gamma_3) + (\gamma_3,\gamma_1,\gamma_2)\Bigg) \\
& - 2\Bigg(\overline{W}^{(2)}_{\mathfrak{s},\hat{\gamma}_2\hat{\gamma}_3}\left(\overline{\Pi}^{\mathrm{inv}}_{\mathfrak{s},\hat{\gamma}_3\hat{\gamma}_4}\frac{\delta\overline{\Pi}_{\mathfrak{s},\hat{\gamma}_4\hat{\gamma}_5}}{\delta\overline{G}_{\mathfrak{s},\gamma_1}}\overline{\Pi}^{\mathrm{inv}}_{\mathfrak{s},\hat{\gamma}_5\hat{\gamma}_6} - \overline{\Phi}^{(3)}_{\mathfrak{s},\gamma_1\hat{\gamma}_3\hat{\gamma}_6}\right)\overline{W}^{(2)}_{\mathfrak{s},\hat{\gamma}_6\hat{\gamma}_7} \\
& \times\overline{\Pi}^{\mathrm{inv}}_{\mathfrak{s},\hat{\gamma}_7\hat{\gamma}_8}\frac{\delta\overline{\Pi}_{\mathfrak{s},\hat{\gamma}_8\hat{\gamma}_9}}{\delta\overline{G}_{\mathfrak{s},\gamma_2}}\overline{\Pi}^{\mathrm{inv}}_{\mathfrak{s},\hat{\gamma}_9\hat{\gamma}_{10}}\frac{\delta\overline{\Pi}_{\mathfrak{s},\hat{\gamma}_{10}\hat{\gamma}_{11}}}{\delta\overline{G}_{\mathfrak{s},\gamma_3}}\overline{\Pi}^{\mathrm{inv}}_{\mathfrak{s},\hat{\gamma}_{11}\hat{\gamma}_{12}}\overline{W}^{(2)}_{\mathfrak{s},\hat{\gamma}_{12}\hat{\gamma}_1} + (\gamma_2,\gamma_1,\gamma_3) + (\gamma_3,\gamma_1,\gamma_2)\Bigg) \\
& + 3\overline{W}^{(2)}_{\mathfrak{s},\hat{\gamma}_2\hat{\gamma}_3}\overline{\Pi}^{\mathrm{inv}}_{\mathfrak{s},\hat{\gamma}_3\hat{\gamma}_4}\frac{\delta\overline{\Pi}_{\mathfrak{s},\hat{\gamma}_4\hat{\gamma}_5}}{\delta\overline{G}_{\mathfrak{s},\gamma_1}}\overline{\Pi}^{\mathrm{inv}}_{\mathfrak{s},\hat{\gamma}_5\hat{\gamma}_6}\frac{\delta\overline{\Pi}_{\mathfrak{s},\hat{\gamma}_6\hat{\gamma}_7}}{\delta\overline{G}_{\mathfrak{s},\gamma_2}}\overline{\Pi}^{\mathrm{inv}}_{\mathfrak{s},\hat{\gamma}_7\hat{\gamma}_8}\frac{\delta\overline{\Pi}_{\mathfrak{s},\hat{\gamma}_8\hat{\gamma}_9}}{\delta\overline{G}_{\mathfrak{s},\gamma_3}}\overline{\Pi}^{\mathrm{inv}}_{\mathfrak{s},\hat{\gamma}_9\hat{\gamma}_{10}}\overline{W}^{(2)}_{\mathfrak{s},\hat{\gamma}_{10}\hat{\gamma}_1} \\
& - \Bigg(\overline{W}^{(2)}_{\mathfrak{s},\hat{\gamma}_2\hat{\gamma}_3}\overline{\Pi}^{\mathrm{inv}}_{\mathfrak{s},\hat{\gamma}_3\hat{\gamma}_4}\frac{\delta\overline{\Pi}_{\mathfrak{s},\hat{\gamma}_4\hat{\gamma}_5}}{\delta\overline{G}_{\mathfrak{s},\gamma_1}}\overline{\Pi}^{\mathrm{inv}}_{\mathfrak{s},\hat{\gamma}_5\hat{\gamma}_6}\frac{\delta^2\overline{\Pi}_{\mathfrak{s},\hat{\gamma}_6\hat{\gamma}_7}}{\delta\overline{G}_{\mathfrak{s},\gamma_2}\delta\overline{G}_{\mathfrak{s},\gamma_3}}\overline{\Pi}^{\mathrm{inv}}_{\mathfrak{s},\hat{\gamma}_7\hat{\gamma}_8}\overline{W}^{(2)}_{\mathfrak{s},\hat{\gamma}_8\hat{\gamma}_1} + (\gamma_2,\gamma_1,\gamma_3) + (\gamma_3,\gamma_1,\gamma_2)\Bigg) \\
& - \frac{1}{2}\overline{W}^{(2)}_{\mathfrak{s},\hat{\gamma}_2\hat{\gamma}_3}\overline{\Phi}^{(5)}_{\mathfrak{s},\gamma_1\gamma_2\gamma_3\hat{\gamma}_3\hat{\gamma}_4}\overline{W}^{(2)}_{\mathfrak{s},\hat{\gamma}_4\hat{\gamma}_1}\Bigg] + \dot{\overline{G}}_{\mathfrak{s},\hat{\gamma}}\overline{\Phi}^{(4)}_{\mathfrak{s},\hat{\gamma}\gamma_1\gamma_2\gamma_3} \, ,
\end{aligned} \tag{142}$$

where the following shorthand notations were used:

$$\frac{\delta^n\overline{\Pi}_{\mathfrak{s}}}{\delta\overline{G}_{\mathfrak{s},\gamma_1}\cdots\delta\overline{G}_{\mathfrak{s},\gamma_n}} \equiv \left.\frac{\delta^n\Pi[G]}{\delta G_{\gamma_1}\cdots\delta G_{\gamma_n}}\right|_{G=\overline{G}_{\mathfrak{s}}} \, , \tag{143}$$

and

$$\mathcal{F}_{\gamma_1\gamma_2\gamma_3} + (\gamma_2,\gamma_1,\gamma_3) + (\gamma_3,\gamma_1,\gamma_2) \equiv \mathcal{F}_{\gamma_1\gamma_2\gamma_3} + \mathcal{F}_{\gamma_2\gamma_1\gamma_3} + \mathcal{F}_{\gamma_3\gamma_1\gamma_2} \, , \tag{144}$$

valid for any functional $\mathcal{F}$. Eqs. (140) to (142) also involve the inverse pair propagator $\Pi^{\mathrm{inv}}$ (and more specifically $\overline{\Pi}^{\mathrm{inv}}_{\mathfrak{s}} \equiv \Pi^{\mathrm{inv}}[G=\overline{G}_{\mathfrak{s}}]$), which satisfies:

$$\Pi^{\mathrm{inv}}_{\gamma_1\gamma_2}[G] = G^{-1}_{\alpha_1\alpha'_2}G^{-1}_{\alpha'_1\alpha_2} + \zeta G^{-1}_{\alpha_1\alpha_2}G^{-1}_{\alpha'_1\alpha'_2} \, , \tag{145}$$

as "inv" labels inverses with respect to bosonic indices[24] by definition, i.e.:

$$\mathcal{I}_{\gamma_1\gamma_2} = M_{\gamma_1\hat{\gamma}}M^{\mathrm{inv}}_{\hat{\gamma}\gamma_2} \, , \tag{146}$$

with $M$ an arbitrary bosonic matrix and $\mathcal{I}$ the bosonic identity matrix defined via:

$$\mathcal{I}_{\gamma_1\gamma_2} \equiv \frac{\delta G_{\gamma_1}}{\delta G_{\gamma_2}} = \delta_{\alpha_1\alpha_2}\delta_{\alpha'_1\alpha'_2} + \zeta\delta_{\alpha_1\alpha'_2}\delta_{\alpha'_1\alpha_2} \, . \tag{147}$$

Hence, Eqs. (137) and (138) together with the hierarchy made of Eqs. (140) to (142) constitute the set of differential equations to solve within the pU-flow.[25] To solve these equations, the derivative $\overline{W}^{(2)}_{\mathfrak{s}}$ must be determined from the flowing 2PI vertex $\overline{\Phi}^{(2)}_{\mathfrak{s}}$ at each value taken by $\mathfrak{s}$ throughout the flow. This can be achieved by determining $\overline{W}^{(2)}_{\mathfrak{s}}$ self-consistently from the relation:

$$\overline{W}^{(2)}_{\mathfrak{s},\gamma_1\gamma_2} = \overline{\Pi}_{\mathfrak{s},\gamma_1\gamma_2} - \overline{\Pi}_{\mathfrak{s},\gamma_1\hat{\gamma}_1}\overline{\Phi}^{(2)}_{\mathfrak{s},\hat{\gamma}_1\hat{\gamma}_2}\overline{W}^{(2)}_{\mathfrak{s},\hat{\gamma}_2\gamma_2} \; , \tag{148}$$

or by inverting a bosonic matrix according to:

$$\overline{W}^{(2)}_{\mathfrak{s}} = \left(\overline{\Gamma}^{(\text{2PI})(2)}_{\mathfrak{s}}\right)^{\text{inv}} = \left(\overline{\Pi}^{\text{inv}}_{\mathfrak{s}} + \overline{\Phi}^{(2)}_{\mathfrak{s}}\right)^{\text{inv}} \; . \tag{149}$$

[24] Inverses with respect to $\alpha$-indices are on the other hand always indicated by a "$-1$" as exponent in our notations.

[25] In the framework of the (0+0)-D $O(N)$ model, these equations can be further simplified using the $O(N)$ symmetry, which is discussed in detail in appendix D.2.1.

Eqs. (148) and (149), which can be derived from each other, are two equivalent versions of the *Bethe-Salpeter equation*. This is why the latter equation must be solved at each step of the flow in the framework of the U-flow implementation of the 2PI-FRG. It remains in general a challenging task for finite-dimensional models but some recent progress in the treatment of the Bethe-Salpeter equation should also be pointed out [189,190], especially since they have led to the successful 2PI-FRG application to the (2+1)-D Hubbard model mentioned earlier [184]. One can of course apply drastic approximations to circumvent the need for repeating the resolution of this equation at each step of the flow but this leads in principle to significant losses in the quality of the obtained results. One can mention for example the truncated U-flow (tU-flow) which is a version of the pU-flow approximating Eq. (149) as:

$$\overline{W}^{(2)}_{\mathfrak{s}} = \overline{\Pi}_{\mathfrak{s}} + \mathcal{O}\left(\overline{\Phi}^{(2)}_{\mathfrak{s}}\right) \simeq \overline{\Pi}_{\mathfrak{s}} \; . \tag{150}$$

We will not investigate the tU-flow further in this study since it has already been illustrated that it is a perturbative approach in essence [179] and we are mainly interested in methods suited for quantitative descriptions of strongly-coupled systems. Before going any further, we also point out that the resolution of the Bethe-Salpeter equation in the form of Eq. (149) and in the framework of the (0+0)-D $O(N)$ model is discussed in detail in appendix D.2.

At the starting point of the pU-flow, the flowing objects are simply set equal to their non-interacting versions. In particular, the LW functional and all its derivatives vanish in the framework of the free theory by construction. This yields:

$$\overline{G}_{\mathfrak{s}=\mathfrak{s}_{\text{i}}} = C \; , \tag{151}$$

$$\overline{\Gamma}^{(\text{2PI})}_{\mathfrak{s}=\mathfrak{s}_{\text{i}}} = \Gamma^{(\text{2PI})}_{0}\left[G = \overline{G}_{\mathfrak{s}=\mathfrak{s}_{\text{i}}}\right] = -\frac{\zeta}{2}\text{STr}\left[\ln(C)\right] \; , \tag{152}$$

$$\overline{\Sigma}_{\mathfrak{s}=\mathfrak{s}_{\text{i}},\gamma} = 0 \quad \forall\gamma \; , \tag{153}$$

$$\overline{\Phi}^{(n)}_{\mathfrak{s}=\mathfrak{s}_{\text{i}},\gamma_1,\cdots,\gamma_n} = 0 \quad \forall\gamma_1,\cdots,\gamma_n,\ \forall n \geq 2 \; . \tag{154}$$

Moreover, the truncation of the infinite hierarchy underpinning the pU-flow is set by choosing an integer $N_{\text{max}}$ such that:

$$\overline{\Phi}^{(n)}_{\mathfrak{s}} = \overline{\Phi}^{(n)}_{\mathfrak{s}=\mathfrak{s}_{\text{i}}} \quad \forall\mathfrak{s},\ \forall n > N_{\text{max}} \; . \tag{155}$$

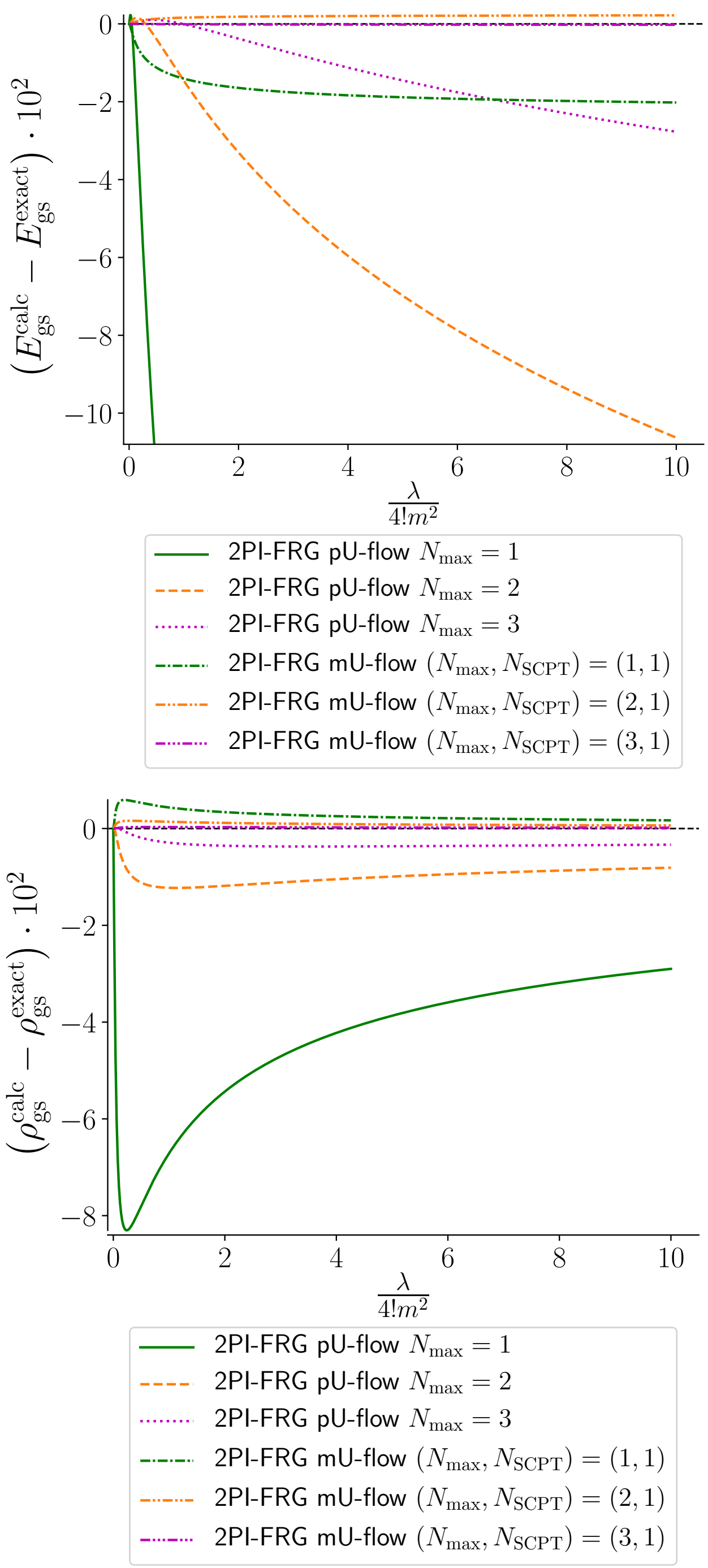

**Fig. 6.** Difference between the calculated gs energy $E^{\text{calc}}_{\text{gs}}$ or density $\rho^{\text{calc}}_{\text{gs}}$ and the corresponding exact solution $E^{\text{exact}}_{\text{gs}}$ or $\rho^{\text{exact}}_{\text{gs}}$ at $m^2 = +1$ and $N = 2$.

Finally, for our numerical applications to the (0+0)-D $O(N)$ model, the gs energy and density are directly deduced from $\overline{\Gamma}^{(\text{2PI})}_{\mathfrak{s}}$ and $\overline{G}_{\mathfrak{s}}$ at the end of the flow, i.e.:

$$E^{\text{2PI-FRG;pU-flow}}_{\text{gs}} = \overline{\Gamma}^{(\text{2PI})}_{\mathfrak{s}=\mathfrak{s}_{\text{f}}} \; , \tag{156}$$

and

$$\rho_{\mathrm{gs}}^{\text{2PI-FRG;pU-flow}} = \frac{1}{N}\sum_{a=1}^{N}\overline{G}_{\mathfrak{s}=\mathfrak{s}_{\mathrm{f}},aa} \; . \tag{157}$$

Our pU-flow results for the (0+0)-D $O(N)$ model at $N = 2$ are reported in Fig. 6. Only the unbroken-symmetry regime is shown in this figure since the starting point of the pU-flow, namely the free theory, is ill-defined in the phase with $m^2 < 0$ as the generating functional (2) diverges at $\lambda = 0$, $m^2 < 0$ and at vanishing sources $\vec{J}$ and $\boldsymbol{K}$. Fig. 6 clearly illustrates the convergence of the pU-flow results towards the exact solutions for both $E_{\mathrm{gs}}$ and $\rho_{\mathrm{gs}}$ as the truncation order $N_{\mathrm{max}}$ increases from $N_{\mathrm{max}} = 1$ to $N_{\mathrm{max}} = 3$, thus reaching an accuracy of about 2% or less for $E_{\mathrm{gs}}$ and $\rho_{\mathrm{gs}}$ at $N_{\mathrm{max}} = 3$ and $\lambda/4! = 10$. However, the truncation order $N_{\mathrm{max}} = 3$ requires to calculate the 2PI three-particle vertex $\overline{\Phi}_{\mathfrak{s}}^{(3)}$ throughout the flow, which is very involved for realistic theories. We will thus examine in what follows how we can reach such an accuracy at lowest truncation orders by using correlated starting points for the 2PI-FRG flow.

### 3.3 Modified U-flow

As was already mentioned, the mU-flow is an implementation of the U-flow version of the 2PI-FRG based on correlated starting points in the form of SCPT results. Its formulation relies on a diagrammatic expression of the LW functional, which reads for the theory based on the classical action (122):

$$\Phi[G] = \Phi_{\mathrm{SCPT}}[U,G]$$

$$\equiv \frac{1}{8}\; \text{(diagram)} \; - \frac{1}{48}\; \text{(diagram)} \; + \mathcal{O}\big(U^3\big)$$

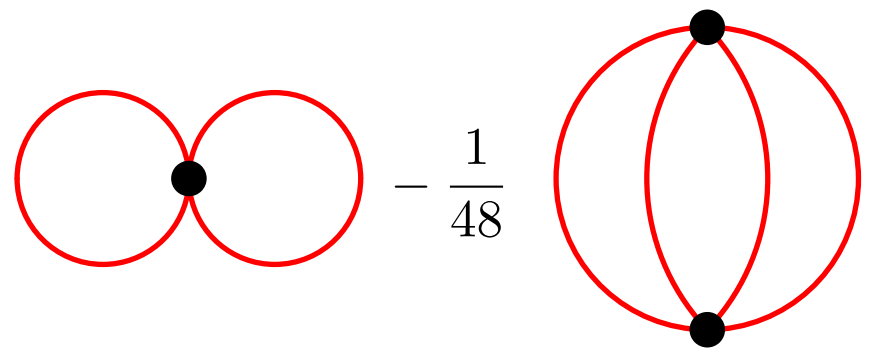

$$= \frac{1}{8}\int_{\gamma_1,\gamma_2} U_{\gamma_1\gamma_2} G_{\gamma_1} G_{\gamma_2} - \frac{1}{48}\int_{\gamma_1,\gamma_2,\gamma_3,\gamma_4} U_{\alpha_1\alpha_2\alpha_3\alpha_4} U_{\alpha_1'\alpha_2'\alpha_3'\alpha_4'} G_{\gamma_1} G_{\gamma_2} G_{\gamma_3} G_{\gamma_4} + \mathcal{O}\big(U^3\big) \; . \tag{158}$$

We have introduced in this way the functional $\Phi_{\mathrm{SCPT}}[U,G]$, which is identical to the LW functional, although we stress its dependence with respect to the interaction $U$ (usually left implicit) and the subscript "SCPT" indicates that we consider its expression (158) usually taken as input for the variational procedure underlying SCPT. From this functional, one can define the transformation of the LW functional on which the mU-flow is based, i.e.:

$$\boldsymbol{\Phi}_{\mathfrak{s}}[G] \equiv \Phi_{\mathfrak{s}}[G] + \Phi_{\mathrm{SCPT},N_{\mathrm{SCPT}}}[U,G] - \Phi_{\mathrm{SCPT},N_{\mathrm{SCPT}}}[U_{\mathfrak{s}},G] \; , \tag{159}$$

where $\boldsymbol{\Phi}_{\mathfrak{s}}[G]$ will be referred to as the modified LW functional and the functional $\Phi_{\mathrm{SCPT},N_{\mathrm{SCPT}}}[U,G]$ only contains the terms of order $\mathcal{O}\big(U^{N_{\mathrm{SCPT}}}\big)$ or less by definition. According to definitions (117) and (159), we can also define a modified 2PI EA $\Gamma_{\mathfrak{s}}^{(\mathrm{2PI})}[G]$ as:

$$\begin{split}\boldsymbol{\Gamma}_{\mathfrak{s}}^{(\mathrm{2PI})}[G] \equiv \Gamma_{\mathfrak{s}}^{(\mathrm{2PI})}[G] + \Phi_{\mathrm{SCPT},N_{\mathrm{SCPT}}}[U,G] \\ - \Phi_{\mathrm{SCPT},N_{\mathrm{SCPT}}}[U_{\mathfrak{s}},G] \; .\end{split} \tag{160}$$

In the same way, the modified 2PI vertices $\boldsymbol{\Phi}_{\mathfrak{s},\gamma_1\cdots\gamma_n}^{(n)}[G] \equiv \frac{\delta^n \boldsymbol{\Phi}_{\mathfrak{s}}[G]}{\delta G_{\gamma_1}\cdots\delta G_{\gamma_n}}$ are related to the derivatives $\Phi_{\mathrm{SCPT},N_{\mathrm{SCPT}},\gamma_1\cdots\gamma_n}^{(n)}[U,G] \equiv \frac{\delta^n \Phi_{\mathrm{SCPT},N_{\mathrm{SCPT}}}[U,G]}{\delta G_{\gamma_1}\cdots\delta G_{\gamma_n}}$ via:

$$\begin{split}\boldsymbol{\Phi}_{\mathfrak{s}}^{(n)}[G] \equiv \Phi_{\mathfrak{s}}^{(n)}[G] + \Phi_{\mathrm{SCPT},N_{\mathrm{SCPT}}}^{(n)}[U,G] \\ - \Phi_{\mathrm{SCPT},N_{\mathrm{SCPT}}}^{(n)}[U_{\mathfrak{s}},G] \quad \forall n \in \mathbb{N}^* \; ,\end{split} \tag{161}$$

which, at $n = 1$, gives us the self-energy:

$$\boldsymbol{\Sigma}_{\mathfrak{s},\gamma}[G] \equiv -\frac{\delta \boldsymbol{\Phi}_{\mathfrak{s}}[G]}{\delta G_\gamma} \; . \tag{162}$$

Furthermore, we introduce the configuration $\overline{\boldsymbol{G}}_{\mathfrak{s}}$ of the propagator $G$ which extremizes the bold 2PI EA of Eq. (160) according to:

$$\left.\frac{\delta \boldsymbol{\Gamma}_{\mathfrak{s}}^{(\mathrm{2PI})}[G]}{\delta G_\gamma}\right|_{G=\overline{\boldsymbol{G}}_{\mathfrak{s}}} = 0 \quad \forall \gamma, \mathfrak{s} \; . \tag{163}$$

Similarly to Eq. (120) at vanishing source $K$ and at $G = \overline{G}_{\mathfrak{s}}$, definition (163) can also be rewritten in the form of a Dyson equation, i.e.:

$$\overline{\boldsymbol{G}}_{\mathfrak{s}}^{-1} = C^{-1} - \boldsymbol{\Sigma}_{\mathfrak{s}}\big[G = \overline{\boldsymbol{G}}_{\mathfrak{s}}\big] \; , \tag{164}$$

as follows from Eq. (162). Hence, we have just assigned bold counterparts to all flowing objects of the 2PI-FRG framework.

Let us then consider the boundary conditions (136) for $U_{\mathfrak{s}}$ which still hold within the mU-flow. From Eq. (136a), one can directly deduce that $\Phi_{\mathfrak{s}}[G]$ and all its derivatives $\Phi_{\mathfrak{s}}^{(n)}[G]$ vanish at the starting point of the flow. According to definitions (159) and (160), it also implies that, at $\mathfrak{s} = \mathfrak{s}_{\mathrm{i}}$, the modified LW functional and its derivatives coincide with their perturbative expression inferred from Eq. (158), i.e.:

$$\boldsymbol{\Phi}_{\mathfrak{s}=\mathfrak{s}_{\mathrm{i}}}[G] = \Phi_{\mathrm{SCPT},N_{\mathrm{SCPT}}}[U,G] \; , \tag{165}$$

$$\boldsymbol{\Phi}_{\mathfrak{s}=\mathfrak{s}_{\mathrm{i}}}^{(n)}[G] = \Phi_{\mathrm{SCPT},N_{\mathrm{SCPT}}}^{(n)}[U,G] \quad \forall n \in \mathbb{N}^* \; , \tag{166}$$

whereas, according to Eq. (136b), they reduce to their original counterparts at $\mathfrak{s} = \mathfrak{s}_{\mathrm{f}}$, namely:

$$\boldsymbol{\Phi}_{\mathfrak{s}=\mathfrak{s}_{\mathrm{f}}}[G] = \Phi_{\mathfrak{s}=\mathfrak{s}_{\mathrm{f}}}[G] \; , \tag{167}$$

$$\boldsymbol{\Phi}_{\mathfrak{s}=\mathfrak{s}_{\mathrm{f}}}^{(n)}[G] = \Phi_{\mathfrak{s}=\mathfrak{s}_{\mathrm{f}}}^{(n)}[G] \quad \forall n \in \mathbb{N}^* \; . \tag{168}$$

Still regarding the initial conditions of the mU-flow, inserting Eq. (166) at $n=1$ into the Dyson equation (164) at $\mathfrak{s}=\mathfrak{s}_{\mathrm{i}}$ yields the relation:

$$\overline{\boldsymbol{G}}^{-1}_{\mathfrak{s}=\mathfrak{s}_{\mathrm{i}}} = C^{-1} + \Phi^{(1)}_{\mathrm{SCPT},N_{\mathrm{SCPT}}}\left[U, G=\overline{\boldsymbol{G}}_{\mathfrak{s}=\mathfrak{s}_{\mathrm{i}}}\right] , \quad (169)$$

which is nothing other than the gap equation underlying SCPT based on $\Phi_{\mathrm{SCPT},N_{\mathrm{SCPT}}}[U,G]$, with $\overline{G}_{\mathrm{SCPT},N_{\mathrm{SCPT}}}$ being a solution for the propagator $G$. The initial conditions of the mU-flow are thus fully determined from SCPT solutions. For a given value of $N_{\mathrm{SCPT}}$, this translates into:

$$\overline{\boldsymbol{G}}_{\mathfrak{s}=\mathfrak{s}_{\mathrm{i}}} = \overline{G}_{\mathrm{SCPT},N_{\mathrm{SCPT}}} , \quad (170)$$

$$\begin{aligned}\overline{\boldsymbol{\Gamma}}^{(\mathrm{2PI})}_{\mathfrak{s}=\mathfrak{s}_{\mathrm{i}}} = {} & \Gamma^{(\mathrm{2PI})}_{0}\left[G=\overline{G}_{\mathrm{SCPT},N_{\mathrm{SCPT}}}\right] \\ & + \Phi_{\mathrm{SCPT},N_{\mathrm{SCPT}}}\left[U, G=\overline{G}_{\mathrm{SCPT},N_{\mathrm{SCPT}}}\right] ,\end{aligned} \quad (171)$$

$$\overline{\boldsymbol{\Sigma}}_{\mathfrak{s}=\mathfrak{s}_{\mathrm{i}}} = -\Phi^{(1)}_{\mathrm{SCPT},N_{\mathrm{SCPT}}}\left[U, G=\overline{G}_{\mathrm{SCPT},N_{\mathrm{SCPT}}}\right] , \quad (172)$$

$$\overline{\boldsymbol{\Phi}}^{(n)}_{\mathfrak{s}=\mathfrak{s}_{\mathrm{i}}} = \Phi^{(n)}_{\mathrm{SCPT},N_{\mathrm{SCPT}}}\left[U, G=\overline{G}_{\mathrm{SCPT},N_{\mathrm{SCPT}}}\right] \quad \forall n \geq 2 , \quad (173)$$

and the truncation of the infinite tower of flow equations is set in the same spirit as in the pU-flow by:

$$\overline{\boldsymbol{\Phi}}^{(n)}_{\mathfrak{s}} = \overline{\boldsymbol{\Phi}}^{(n)}_{\mathfrak{s}=\mathfrak{s}_{\mathrm{i}}} \quad \forall \mathfrak{s},\ \forall n > N_{\mathrm{max}} . \quad (174)$$

In summary, the quantities $\overline{\boldsymbol{G}}_{\mathfrak{s}}$, $\overline{\boldsymbol{\Gamma}}^{(\mathrm{2PI})}_{\mathfrak{s}} \equiv \boldsymbol{\Gamma}^{(\mathrm{2PI})}_{\mathfrak{s}}\left[G=\overline{\boldsymbol{G}}_{\mathfrak{s}}\right]$, $\overline{\boldsymbol{\Sigma}}_{\mathfrak{s}} \equiv \boldsymbol{\Sigma}_{\mathfrak{s}}\left[G=\overline{\boldsymbol{G}}_{\mathfrak{s}}\right]$ and $\overline{\boldsymbol{\Phi}}^{(n)}_{\mathfrak{s}} \equiv \boldsymbol{\Phi}^{(n)}_{\mathfrak{s}}\left[G=\overline{\boldsymbol{G}}_{\mathfrak{s}}\right]$ (with $n\geq 2$) are the flowing objects in the framework of the mU-flow, in contrast with $\overline{G}_{\mathfrak{s}}$, $\Gamma^{(\mathrm{2PI})}_{\mathfrak{s}}\left[G=\overline{G}_{\mathfrak{s}}\right]$, $\Sigma_{\mathfrak{s}}\left[G=\overline{G}_{\mathfrak{s}}\right]$ and $\Phi^{(n)}_{\mathfrak{s}}\left[G=\overline{G}_{\mathfrak{s}}\right]$ (with $n\geq 2$) for the pU-flow. To clarify, in this section 3.3 and as opposed to the notations used in section 3.2, the upper bars always label functionals evaluated at $G=\overline{\boldsymbol{G}}_{\mathfrak{s}}$ (with $\overline{\boldsymbol{G}}_{\mathfrak{s}}$ defined by Eq. (163)) instead of $G=\overline{G}_{\mathfrak{s}}$ (with $\overline{G}_{\mathfrak{s}}$ defined by Eq. (129)). Furthermore, according to our previous explanation based on Eqs. (167) and (168), physical quantities are recovered at the end of the flow in the present case in the same manner as in the pU-flow. For the gs energy and density of the (0+0)-D $O(N)$ model, this gives us the following relations from their pU-flow counterparts (i.e. Eqs. (156) and (157)):

$$E^{\text{2PI-FRG;mU-flow}}_{\mathrm{gs}} = \overline{\boldsymbol{\Gamma}}^{(\mathrm{2PI})}_{\mathfrak{s}=\mathfrak{s}_{\mathrm{f}}} , \quad (175)$$

and

$$\rho^{\text{2PI-FRG;mU-flow}}_{\mathrm{gs}} = \frac{1}{N}\sum_{a=1}^{N} \overline{\boldsymbol{G}}_{\mathfrak{s}=\mathfrak{s}_{\mathrm{f}},aa} . \quad (176)$$

Before going further with discussing our numerical results obtained from the mU-flow, we address three additional points related to the starting points of this approach:

– Starting the flow at the *Hartree-Fock* result:
According to our notations, the Hartree-Fock result for SCPT based on the 2PI EA is obtained at $N_{\mathrm{SCPT}}=1$, in which case the LW functional is approximated as follows:

$$\Phi_{\mathrm{SCPT},N_{\mathrm{SCPT}}=1}[U,G] = \frac{1}{2}U_{\hat{\gamma}_1\hat{\gamma}_2}G_{\hat{\gamma}_1}G_{\hat{\gamma}_2} , \quad (177)$$

consistently with Eq. (158). Plugging expression (177) into definition (160) of the modified 2PI EA leads to:

$$\boldsymbol{\Gamma}^{(\mathrm{2PI})}_{\mathfrak{s}}[G] = \Gamma^{(\mathrm{2PI})}_{\mathfrak{s}}[G] + \frac{1}{2}\left(U-U_{\mathfrak{s}}\right)_{\hat{\gamma}_1\hat{\gamma}_2} G_{\hat{\gamma}_1}G_{\hat{\gamma}_2} . \quad (178)$$

Choosing an additive cutoff function $U_{\mathfrak{s}} = U + R_{\mathfrak{s}}$ for the sake of our comparison, the rightmost term of Eq. (178) becomes:

$$\frac{1}{2}\left(U-U_{\mathfrak{s}}\right)_{\hat{\gamma}_1\hat{\gamma}_2} G_{\hat{\gamma}_1}G_{\hat{\gamma}_2} = -\frac{1}{2}G_{\hat{\gamma}_1}R_{\mathfrak{s},\hat{\gamma}_1\hat{\gamma}_2}G_{\hat{\gamma}_2} . \quad (179)$$

We can thus see that the rightmost term of Eq. (178) has a very similar expression to that of $-\Delta S_k[\widetilde{\psi}]$ (given by Eq. (22)) that modifies the Legendre transform defining the 1PI EA (in Eq. (20)) within the FRG based on the Wetterich equation. Hence, both in the 1PI-FRG based on the Wetterich equation and in the mU-flow version of the 2PI-FRG, the Legendre transform defining the underlying flowing EA is modified to obtain a convenient starting point for the flow: the classical theory in the former case and the Hartree-Fock theory (for $N_{\mathrm{SCPT}}=1$) or higher orders of SCPT (for $N_{\mathrm{SCPT}}>1$) in the latter one. Before moving to the next point, we also refer to appendix D.2.2 for the expressions of flow equations underpinning the mU-flow at $N_{\mathrm{SCPT}}=1$ at arbitrary dimensions as well as for the specific case of the (0+0)-D $O(N)$ model.

– Starting the flow in *broken-symmetry* phases:
Let us consider the situation where the choice for the flow-dependent interaction $U_{\mathfrak{s}}$ is such that the Wilsonian momentum-shell integration is not implemented through the U-flow. At first sight, this could be taken as a severe limitation of this approach for the purpose of describing critical phenomena. However, this issue is remarkably circumvented in the framework of the mU-flow which can be designed to start in a broken-symmetry phase if necessary (or more generally in the phase that we aim to describe) so as to avoid undesirable phase transitions during the flow, depending on the solution $\overline{G}_{\mathrm{SCPT},N_{\mathrm{SCPT}}}$ chosen for the initial conditions. This remark only applies if the different solutions of the Dyson equation (169) give us access to each phase that we seek to describe, which works in principle for the $U(1)$ symmetry notably, and therefore for the description of superfluid systems, as the relevant order parameters can be identified as components of the propagator $G$ in such a situation.[26]

[26] However, the mU-flow does not provide us with a similar freedom to tackle an $O(N)$ symmetry as neither $\Gamma^{(\mathrm{2PI})}_{\mathfrak{s}}[G]$ nor $\boldsymbol{\Gamma}^{(\mathrm{2PI})}_{\mathfrak{s}}[G]$ is capable of spontaneously breaking such a symmetry, as was already explained right below Eq. (125).

– Starting the flow with SCPT based on *mixed* 2PI EAs: In paper I, SCPT based on a mixed 2PI EA clearly stood out among all tested PI techniques. We have illustrated in particular, still in paper I, that SCPT based on a mixed 2PI EA manages to grasp further correlations, especially in the strongly-coupled regime of our toy model, as compared to its counterpart formulated without HST (i.e. in the original representation). This enhancement of performance can be attributed to the role of the 1-point correlation function of the Hubbard-Stratonovich field efficiently used as variational parameter in this implementation of SCPT. For that reason, it is clearly appealing to design a mU-flow exploiting SCPT based on mixed 2PI EAs as starting point for the flow. However, for the $O(N)$ model treated in this study and for models based on a two-body interaction in general, the mixed 2PI EA will be defined from a classical action with a Yukawa interaction, which causes problems in the derivations of the (m)U-flow equations. Such problems actually already arise in the derivation of the master equation for the mixed 2PI EA and can be simply illustrated for the studied toy model based on the mixed generating functional (5). In the derivation of the master equation for the original 2PI EA $\Gamma_{\mathfrak{s}}^{(2\mathrm{PI})}$ for the U-flow, one needs to introduce the bosonic matrix $W_{\mathfrak{s}}^{(2)}$ that must be connected to the 2PI vertices to solve the resulting coupled flow equations. As explained earlier, this connection is achieved by solving the Bethe-Salpeter equation which amounts to inverting a bosonic matrix according to definition (146). For our (0+0)-D $O(N)$ model, the latter translates into:

$$\mathcal{I}_{(a_1,a_1')(a_2,a_2')} = \frac{1}{2} \sum_{a_3,a_3'=1}^{N} M_{(a_1,a_1')(a_3,a_3')} M^{\mathrm{inv}}_{(a_3,a_3')(a_2,a_2')} \; . \tag{180}$$

By following the same procedure in the mixed representation of this model, we would have to invert instead matrices with components of the form $\mathcal{M}_{b_1(b_2,b_2')}$ (labeled by a supercolor index $b$ and a bosonic index $(b,b')$), as a result of the Yukawa interaction $\mathcal{Y}_{b_1(b_2,b_2')}$ introduced at the expense of the two-body one $U_{(a_1,a_1')(a_2,a_2')} \equiv U_{a_1 a_1' a_2 a_2'}$. Unfortunately, it is not straightforward to find inverses for such matrices. For instance, the relations:

$$\mathcal{I}_{(b_1,b_1')(b_2,b_2')} = \sum_{b_3=1}^{N+1} \mathcal{M}_{(b_1,b_1')b_3} \mathcal{M}^{\mathrm{inv}}_{b_3(b_2,b_2')} \; , \tag{181}$$

and

$$\delta_{b_1 b_2} = \frac{1}{2} \sum_{b_3,b_3'=1}^{N+1} \mathcal{M}_{b_1(b_3,b_3')} \mathcal{M}^{\mathrm{inv}}_{(b_3,b_3')b_2} \; , \tag{182}$$

do not provide the right number of conditions to fix the components of $\mathcal{M}^{\mathrm{inv}}$ in an unambiguous manner. Hence, extending the mU-flow implementation of the 2PI-FRG in the framework of the mixed representation of $O(N)$ models (in (0+0)-D as well as in finite dimensions), and more generally to theories based on Yukawa interactions, still requires a consequent work on the side of the formalism, that we postpone to subsequent projects.

With Figs. 6, 7 and 8, we then discuss our mU-flow results exploiting SCPT based on the original 2PI EA as starting point. We can already appreciate from Fig. 6 the gain in accuracy achieved by the mU-flow with its simplest correlated starting point (i.e. the Hartree-Fock result) as compared to the pU-flow starting from the free theory. For instance, the mU-flow results for $E_{\mathrm{gs}}$ in Fig. 6 are already less than 2% away from the exact solution at $N_{\mathrm{max}} = 1$ over the whole tested range for the coupling constant (i.e. $\lambda/4! \in [0,10]$) whereas such a performance is barely reached for $\lambda/4! > 1$ (i.e. beyond the weakly-coupled regime) at $N_{\mathrm{max}} < 3$ for the pU-flow. Actually, Fig. 6 shows that, over most of the range defined by $\lambda/4! \in [0,10]$ and at $m^2 = 1$, the pU-flow at $N_{\mathrm{max}} = 2$ (where the two-particle vertex $\Phi_{\mathfrak{s}}^{(2)}[G = \overline{G}_{\mathfrak{s}}]$ is flowing) is outperformed by the mU-flow at $N_{\mathrm{max}} = 1$ (where the two-particle vertex $\overline{\boldsymbol{\Phi}}_{\mathfrak{s}}^{(2)}$ is *not* flowing). Put differently, we can achieve, even in strongly-coupled regimes, fairly satisfactory quantitative performances (accuracy below 2% for $E_{\mathrm{gs}}$ and below 1% for $\rho_{\mathrm{gs}}$ in Fig. 6) with the mU-flow at $N_{\mathrm{max}} = 1$, i.e. by discarding the flow of $\overline{\boldsymbol{\Phi}}_{\mathfrak{s}}^{(2)}$. These statements have very profound consequences as the 2PI two-particle vertex $\Phi_{\mathfrak{s}}^{(2)}[G = \overline{G}_{\mathfrak{s}}]$ (or its modified version $\overline{\boldsymbol{\Phi}}_{\mathfrak{s}}^{(2)}$ equivalently) is already significantly demanding to calculate for realistic theories, just like the 1PI two-particle vertex $\overline{\Gamma}_k^{(1\mathrm{PI})(4)} \equiv \Gamma_k^{(1\mathrm{PI})(4)}[\psi = \overline{\psi}_k]$ in the 1PI-FRG framework. We have thus illustrated in this way the power and usefulness of FRG implementations based on correlated starting points.

We stress nonetheless that the above remarks based on the results of Fig. 6 concern the mU-flow at $N_{\mathrm{SCPT}} = 1$. It is then quite natural to test the mU-flow for more involved starting points, i.e. for $N_{\mathrm{SCPT}} > 1$, which has never been done so far to our knowledge. We exploit the simplicity of our toy model at $N = 1$ to achieve this (see appendix D.2.2 for the expressions of the underlying flow equations), which will be sufficient to make our point. Hence, Fig. 7 shows mU-flow results up to $N_{\mathrm{max}} = 3$ with $N_{\mathrm{SCPT}} = 1, 2$ and 3 for both $E_{\mathrm{gs}}$ and $\rho_{\mathrm{gs}}$ in the unbroken- and broken-symmetry regimes of our toy model. We point out first of all the appearance of the same stiffness issues as before with the used numerical tools[12] for the mU-flow at $N_{\mathrm{SCPT}} = 2$ and 3 with $N_{\mathrm{max}} = 2$, hence explaining the absence of the corresponding curves in Fig. 7. This does not prevent us from noticing in this figure that, besides a few exceptions, the mU-flow at $N_{\mathrm{SCPT}} = 1$ is more performing than at $N_{\mathrm{SCPT}} = 2$ or 3 for a given truncation order $N_{\mathrm{max}}$, for $E_{\mathrm{gs}}$ as well as for $\rho_{\mathrm{gs}}$. Although the starting point contains more and more information about the system to describe as $N_{\mathrm{SCPT}}$ increases, we recall that it is the bare SCPT results (i.e. without resummation) that the mU-flow procedure takes as inputs. Such results take the form of diverging asymptotic series for the original 2PI EA $\Gamma^{(2\mathrm{PI})}(G)$ (as was shown in paper I notably) and the corresponding estimates for $E_{\mathrm{gs}}$ and $\rho_{\mathrm{gs}}$ all worsen as the truncation order $N_{\mathrm{SCPT}}$ increases, except for particularly small values of the coupling constant $\lambda$. It was indeed only after applying a resummation procedure that we managed to turn SCPT into a systematically improvable technique in paper I. In the light of the latter comments, we can conclude that, unlike resummation procedures, the mU-flow approach is not suited to extract the full information from the asymptotic series representing the 2PI EA (and the corresponding 2PI vertices) taken as input(s). The mU-flow is thus most efficient at $N_{\mathrm{SCPT}} = 1$, which is why we have also tested this approach in the unbroken-and broken-symmetry regimes for $N=2$, as reported by Fig. 8. Our qualitative conclusions on the latter figure are particularly positive and thus echo those inferred from the mU-flow results of Fig. 6: for $E_{\text{gs}}$ and $\rho_{\text{gs}}$ and for both signs of $m^2$, we see that the mU-flow procedure at $N_{\text{max}}=1$ clearly improves the Hartree-Fock curve representing its starting point, and this mU-flow result is itself improved by increasing the truncation order $N_{\text{max}}$ until the curve corresponding to $N_{\text{max}}=3$ becomes barely distinguishable from the exact solution. To conclude on section 3, the mU-flow with $N_{\text{SCPT}}=1$ clearly stands out from our 2PI-FRG study.

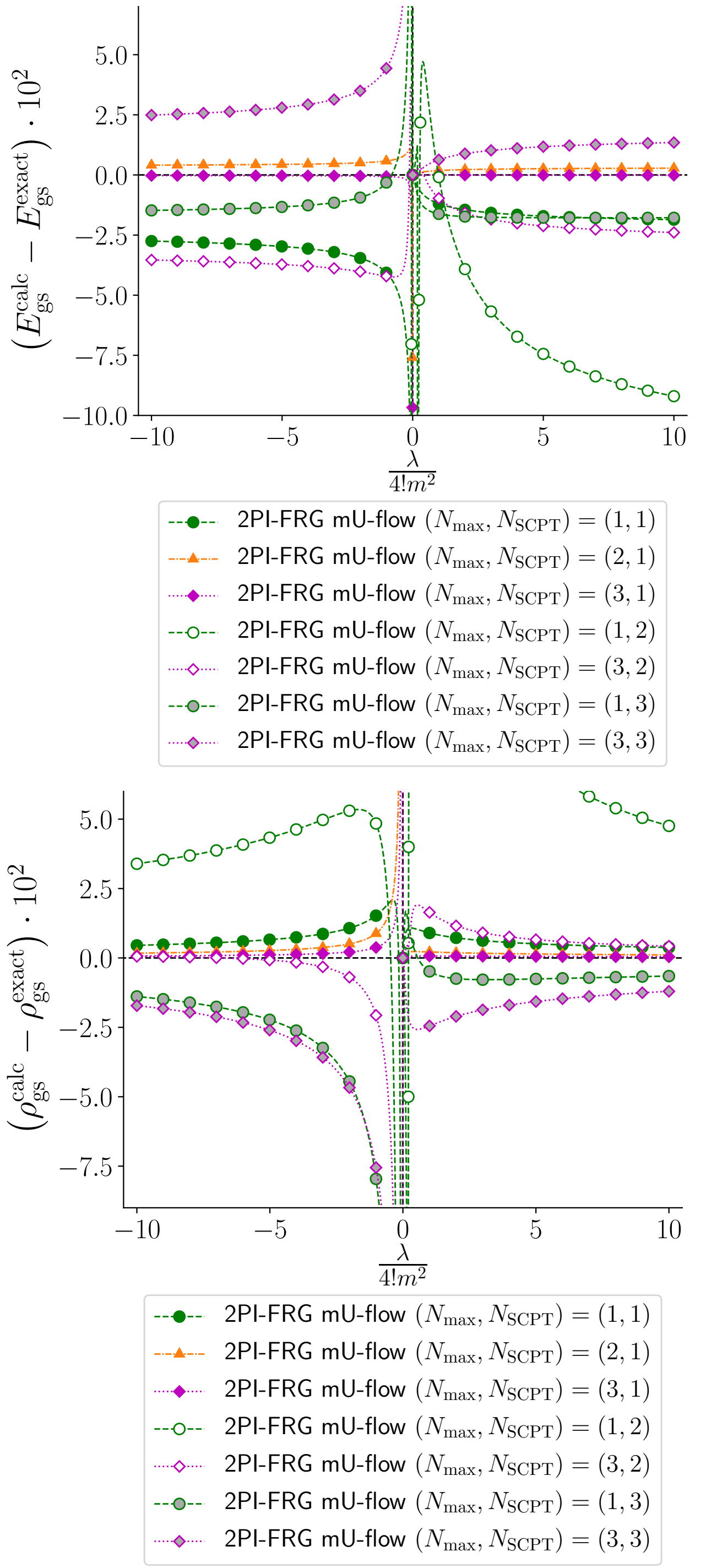

**Fig. 7.** Difference between the calculated gs energy $E_{\mathrm{gs}}^{\mathrm{calc}}$ or density $\rho_{\mathrm{gs}}^{\mathrm{calc}}$ and the corresponding exact solution $E_{\mathrm{gs}}^{\mathrm{exact}}$ or $\rho_{\mathrm{gs}}^{\mathrm{exact}}$ at $m^2 = \pm 1$ and $N = 1$.

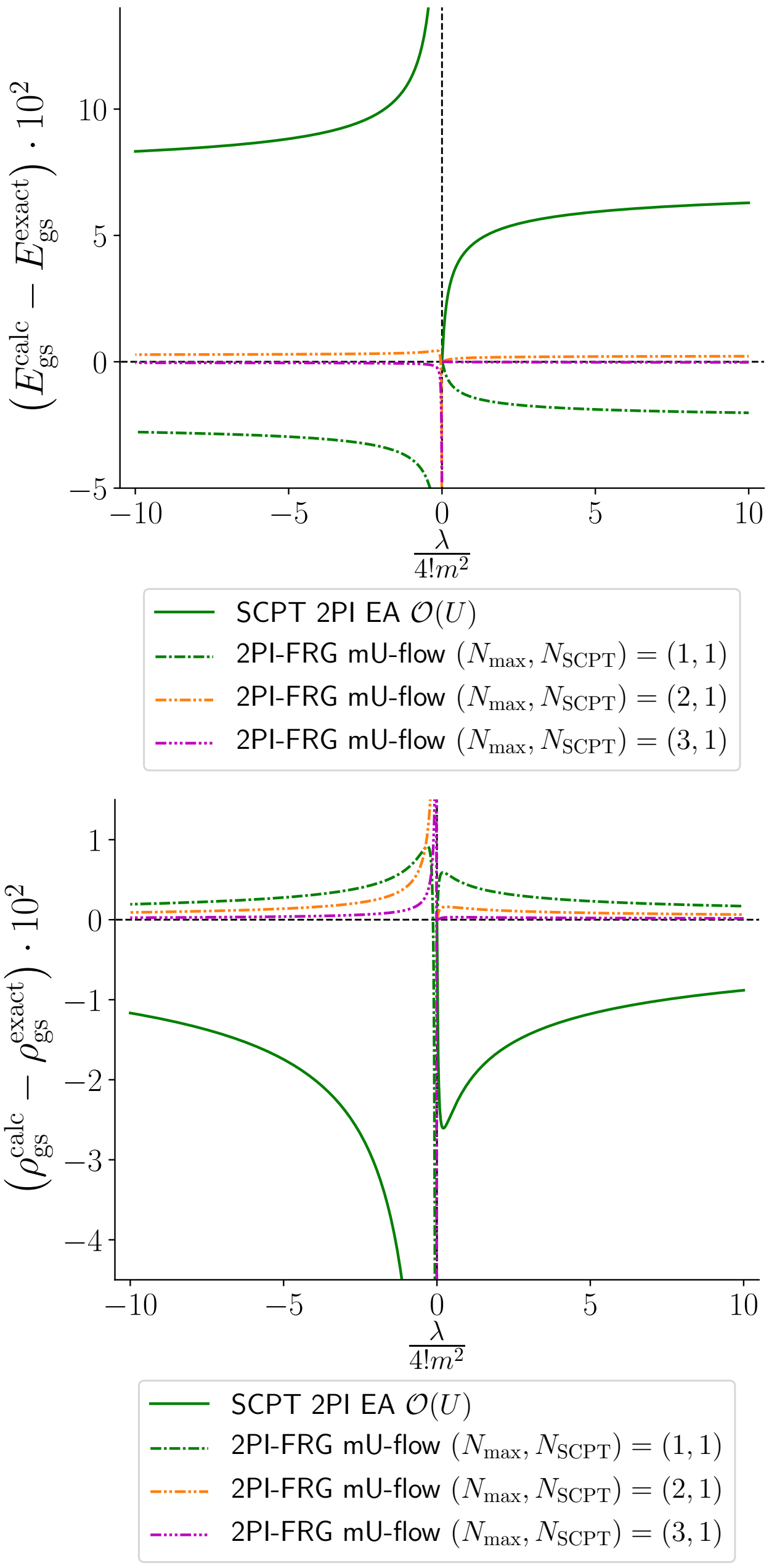

**Fig. 8.** Difference between the calculated gs energy $E_{\mathrm{gs}}^{\mathrm{calc}}$ or density $\rho_{\mathrm{gs}}^{\mathrm{calc}}$ and the corresponding exact solution $E_{\mathrm{gs}}^{\mathrm{exact}}$ or $\rho_{\mathrm{gs}}^{\mathrm{exact}}$ at $m^2 = \pm 1$ and $N = 2$. The SCPT result labeled "SCPT 2PI EA $\mathcal{O}(U)$" is identical to that corresponding to the legend "SCPT original 2PI EA $\mathcal{O}(\hbar^2)$" in the figures of paper I.

We have thus illustrated that FRG approaches designed by dressing interaction parts of classical actions with cutoff functions can exploit very efficiently correlated starting points. However, we have also shown why such an introduction of cutoff functions leads to complications of the flow equations, translating into the appearance of the Bethe-Salpeter equation in the U-flow formalism, but this has to be contrasted with the possibility to achieve satisfactory performances by ignoring the flow of complicated objects such as 2PI two-particle vertices.

# 4 2PPI functional renormalization group

## 4.1 Generalities

We now consider FRG methods based on density functionals in the form of 2PPI EAs. The explicit presence of the density as dof brings us one step closer to the nuclear EDF formalism, although we recall that we still rely on completely different types of equations (differential equations for FRG and self-consistent equations for EDF). Since 2PPI EAs can be viewed as simplified versions of 2PI EAs, the 2PPI-FRG approaches designed so far rely on flow equations that are less demanding to treat numerically, as compared to those presented in the latter section for the 2PI-FRG. The question is then whether one can implement a 2PPI-FRG approach as performing as the mU-flow version of the 2PI-FRG, but at a lower numerical cost. This will be addressed in this section 4 where we will once again present FRG formalisms in a general setting before focusing on the (0+0)-D $O(N)$ model for our numerical applications. Before doing so, we make a state of play of the 2PPI-FRG approaches and corresponding applications achieved hitherto. The 2PPI-FRG was first developed in the early 2000s by Polonyi and Sailer [191] and discussed in the context of quantum electrodynamics. Links between this approach and DFT were also emphasized in Ref. [191], as well as in Ref. [192] which puts forward the 2PPI-FRG as a means for calculating properties of nuclear systems in a systematic manner. The first numerical application of this method came out almost a decade later with the work of Kemler and Braun [35], who took as theoretical laboratories the (0+0)-D $\varphi^4$-theory in its unbroken-symmetry regime (i.e. the studied toy model with $N=1$ and $m^2>0$) and the (0+1)-D $\varphi^4$-theory, still in the phase without SSB. Corrections of the application to the latter toy model were pointed out subsequently by Rentrop and collaborators in Ref. [179]. A few years later, an extension of the 2PPI-FRG formalism, coined as Kohn-Sham FRG (KS-FRG) due to its connection with the Kohn-Sham scheme of DFT, was developed by Liang, Niu and Hatsuda [193]. To our knowledge, the KS-FRG has only been applied to the toy model considered in the present study with $N=1$ and in its unbroken-symmetry regime [193].

The 2PPI-FRG practitioners have also managed to treat a (1+1)-D model [36–40], called the Alexandrou-Negele nuclei [194]. Such a model reproduces some basic properties of the nuclear force (short-range repulsive and long-range attractive). As any other toy model, the Alexandrou-Negele nuclei have been used to benchmark different theoretical approaches (see Ref. [195] for the similarity renormalization group) but they have also been exploited to describe real physical systems such as ultracold fermionic atoms interacting via a dipolar interaction [196]. For that reason in addition to the technical difficulties related to the inclusion of a space dimension, the work of Kemler and Braun presented in Ref. [36] can be considered as a pioneering work for the 2PPI-FRG community. More specifically, it presents results obtained for the gs energies (in comparison with Monte Carlo results [194]), intrinsic densities and density correlation functions. Other subsequent applications were carried out for an infinite number of particles by Yokota and collaborators in order to study spinless nuclear matter with this model: this led to the determination of the nuclear saturation curve and other gs properties on the one hand [38] and to the calculation of spectral functions for the study of excited states on the other hand [39]. Note also that other 2PPI-FRG results for a (1+1)-D fermionic system are reported in Ref. [197]. Then, applications to higher-dimensional systems were performed recently, still by Yokota and collaborators, on (2+1)-D [40, 198] and (3+1)-D [42] homogeneous electron gases, thus achieving the first two-dimensional and three-dimensional applications of the 2PPI-FRG. The work of Ref. [43], which designs a 2PPI-FRG approach to describe classical liquids, can be mentioned as well. Finally, the 2PPI-FRG formalism has also been generalized to treat superfluid systems [41], thus marking a significant step towards the description of systems with competing instabilities. The resulting approach can actually be considered as a DFT for systems with pairing correlations and echoes the work of Furnstahl and collaborators with SCPT based on 2PPI EAs [29].

As a next step, we outline the basic ingredients of the 2PPI EA formalism. The generating functional underlying this FRG approach is given by[27]:

$$Z[K]=e^{W[K]}=\int\mathcal{D}\widetilde{\psi}^\dagger\mathcal{D}\widetilde{\psi}\;e^{-S\left[\widetilde{\psi}^\dagger,\widetilde{\psi}\right]+\int_\alpha K_\alpha\widetilde{\psi}^\dagger_\alpha\widetilde{\psi}_\alpha}\;, \tag{183}$$

[27] Although the notations for the generating functionals $Z[K]$ and $W[K]$ are identical to those of Eq. (106) related to the 2PI-FRG, we stress that $Z[K]$ and $W[K]$ are always defined by Eq. (183) in all our discussions related to the 2PPI-FRG.

where we consider here a complex field $\widetilde{\psi}$ which is either bosonic or fermionic. The different configurations of $\widetilde{\psi}$ are now specified by an $\alpha$-index which is essentially the same as that used in section 3 in our presentation of the 2PI-FRG, with the exception of the charge index $c$. Namely, we now have $\alpha \equiv (i,x)$ with $x \equiv (r,\tau)$ and the following shorthand notation holds throughout the entire section 4:

$$\int_\alpha \equiv \sum_i \int_0^{1/T} d\tau \int d^{D-1}r \;. \tag{184}$$

The connected correlation functions deduced from Eq. (183) are:

$$W^{(n)}_{\alpha_1\cdots\alpha_n}[K] \equiv \frac{\delta^n W[K]}{\delta K_{\alpha_1}\cdots\delta K_{\alpha_n}} \;, \tag{185}$$

which yields the density:

$$\rho_\alpha = W^{(1)}_\alpha[K] \;, \tag{186}$$

at $n=1$. As opposed to the 2PI-FRG, the source $K$ is now local and does not exhibit antisymmetric properties, even if $\widetilde{\psi}$ is a Grassmann field. This also translates into simpler symmetry properties for the underlying correlation functions, with e.g. for those of Eq. (185):

$$W^{(n)}_{\alpha_1\cdots\alpha_n}[K] = W^{(n)}_{\alpha_{P(1)}\cdots\alpha_{P(n)}}[K] \;, \tag{187}$$

to be compared with Eqs. (112) for the 2PI-FRG formalism. Finally, the following 2PPI EA is defined from the Schwinger functional $W[K]$ introduced in Eq. (183):

$$\Gamma^{(\mathrm{2PPI})}[\rho] = -W[K] + \int_\alpha K_\alpha \rho_\alpha \;. \tag{188}$$

All implementations of the 2PPI-FRG treated in this paper are applicable to any system whose classical action can be written as:

$$\begin{split} S\left[\widetilde{\psi}^\dagger,\widetilde{\psi}\right] = & \int_\alpha \widetilde{\psi}^\dagger_\alpha \left(\hat{\mathcal{O}}_{\mathrm{kin},\alpha} + V_\alpha - \mu\right)\widetilde{\psi}_\alpha \\ & + \frac{1}{2}\int_{\alpha_1,\alpha_2} \widetilde{\psi}^\dagger_{\alpha_1}\widetilde{\psi}^\dagger_{\alpha_2} U_{\alpha_1\alpha_2}\widetilde{\psi}_{\alpha_2}\widetilde{\psi}_{\alpha_1} \;, \end{split} \tag{189}$$

where

$$U_{\alpha_1\alpha_2} = U_{\alpha_2\alpha_1} \;, \tag{190}$$

$\hat{\mathcal{O}}_{\mathrm{kin}}$ contains the kinetic operator[28] and we place ourselves in the grand canonical ensemble here by using a chemical potential $\mu$ to monitor the particle number.[29] In the framework of the 2PPI-FRG, the flow parameter $\mathfrak{s}$ is always dimensionless and the resulting approach thus does not implement the Wilsonian momentum-shell integration.[30] It is also typically the one-body potential $V$ and the two-body interaction $U$ which are rendered flow-dependent by introducing cutoff functions through the substitutions $U \rightarrow U_\mathfrak{s}$ and $V \rightarrow V_\mathfrak{s}$ applied to Eq. (189). After doing so, definition (188) becomes:

$$\Gamma^{(\mathrm{2PPI})}_\mathfrak{s}[\rho] = -W_\mathfrak{s}[K] + \int_\alpha K_\alpha \rho_\alpha \;, \tag{191}$$

where

$$\rho_\alpha \equiv \rho_{\mathfrak{s},\alpha}[K] = W^{(1)}_{\mathfrak{s},\alpha}[K] \;. \tag{192}$$

Moreover, although the classical action (189) only contains a two-body interaction, the 2PPI-FRG approaches discussed in this study can be straightforwardly extended to treat three-body and even higher-body interactions by including higher powers of $\widetilde{\psi}^\dagger\widetilde{\psi}$ into Eq. (189), similarly to Eq. (122) for the 2PI-FRG.

We can also draw a parallel with our discussions of section 3 on the 2PI-FRG by noticing that setting:

$$\widetilde{\psi}_\alpha = \widetilde{\psi}^\dagger_\alpha = \widetilde{\varphi}_a \;, \tag{193}$$

$$\int_\alpha = \sum_{a=1}^N \;, \tag{194}$$

and imposing all kinetic terms to vanish enable us to get back the classical action $S(\vec{\widetilde{\varphi}})$ of the (0+0)-D $O(N)$ model as follows:

$$\begin{split} S\left[\widetilde{\psi}^\dagger,\widetilde{\psi}\right] = & \, S\left(\vec{\widetilde{\varphi}}\right) \\ = & \sum_{a=1}^N \widetilde{\varphi}_a V_a \widetilde{\varphi}_a + \frac{1}{2}\sum_{a_1,a_2=1}^N \widetilde{\varphi}_{a_1}\widetilde{\varphi}_{a_2} U_{a_1 a_2}\widetilde{\varphi}_{a_2}\widetilde{\varphi}_{a_1} \;, \end{split} \tag{195}$$

with the one-body potential:

$$V_a = \frac{m^2}{2} \quad \forall a \;, \tag{196}$$

and the two-body interaction:

$$U_{a_1a_2} = \frac{\lambda}{12} \quad \forall a_1,a_2 \;. \tag{197}$$

Therefore, as for the 2PI-FRG, the 2PPI-FRG formalism can be directly applied to the studied toy model in its original representation, as opposed to its mixed and collective counterparts. The 2PPI-FRG flow equations to treat for the (0+0)-D $O(N)$ model can thus be straightforwardly inferred from the generic ones given in this section 4 after exploiting Eqs. (194), (196) and (197). We finally stress that, to our knowledge, the present study is the first presenting an application of the 2PPI-FRG to an $O(N)$ model with $N>1$.

[28] For instance, $\hat{\mathcal{O}}_{\mathrm{kin},\alpha} = \partial_\tau - \frac{\boldsymbol{\nabla}^2}{2m}$ for a non-relativistic system of mass $m$.

[29] Alternatively, one can also simply impose a given particle number at the initial conditions of the 2PPI-FRG procedure since the particle number is conserved during the flow, as shown in Ref. [36].

[30] To be more specific, none of the 2PPI-FRG implementations exploited so far, i.e. the approach of Polonyi and Sailer [191] and the KS-FRG of Liang and collaborators [193], implements the Wilsonian momentum-shell integration, although this is in principle possible via suitable choices of cutoff functions, as for the 2PI-FRG.

### 4.2 Standard 2PPI functional renormalization group

We start by discussing the standard version of the 2PPI-FRG, as proposed by Polonyi and Sailer [191] and then exploited e.g. by Kemler and Braun [35, 36]. After performing the substitutions $V \rightarrow V_{\mathfrak{s}}$ and $U \rightarrow U_{\mathfrak{s}}$ into the classical action (189), we can show that the corresponding flow-dependent 2PPI EA $\Gamma^{(\mathrm{2PPI})}_{\mathfrak{s}}[\rho]$, defined by Eq. (191), satisfies the master equation:

$$\dot{\Gamma}^{(\mathrm{2PPI})}_{\mathfrak{s}}[\rho] = \int_{\alpha} \dot{V}_{\mathfrak{s},\alpha}\rho_{\alpha} + \frac{1}{2}\mathrm{STr}\left[\dot{U}_{\mathfrak{s}}\left(\Gamma^{(\mathrm{2PPI})(2)}_{\mathfrak{s}}[\rho]\right)^{-1}\right] + \frac{1}{2}\int_{\alpha_1,\alpha_2} \dot{U}_{\mathfrak{s},\alpha_1\alpha_2}\rho_{\alpha_1}\rho_{\alpha_2} \;, \tag{198}$$

where the 2PPI vertices satisfy:

$$\Gamma^{(\mathrm{2PPI})(n)}_{\mathfrak{s},\alpha_1\cdots\alpha_n}[\rho] \equiv \frac{\delta^n \Gamma^{(\mathrm{2PPI})}_{\mathfrak{s}}[\rho]}{\delta\rho_{\alpha_1}\cdots\delta\rho_{\alpha_n}} \;. \tag{199}$$

Result (198) is thus the counterpart of the Wetterich equation for the standard 2PPI-FRG. The flow equation (198) is then treated via a vertex expansion procedure based on the following expression of $\Gamma^{(\mathrm{2PPI})}_{\mathfrak{s}}[\rho]$:

$$\Gamma^{(\mathrm{2PPI})}_{\mathfrak{s}}[\rho] = \overline{\Gamma}^{(\mathrm{2PPI})}_{\mathfrak{s}} + \sum^{\infty}_{n=2}\frac{1}{n!}\int_{\alpha_1,\cdots,\alpha_n} \overline{\Gamma}^{(\mathrm{2PPI})(n)}_{\mathfrak{s},\alpha_1\cdots\alpha_n} \times \left(\rho-\overline{\rho}_{\mathfrak{s}}\right)_{\alpha_1}\cdots\left(\rho-\overline{\rho}_{\mathfrak{s}}\right)_{\alpha_n} \;, \tag{200}$$

where $\overline{\Gamma}^{(\mathrm{2PPI})}_{\mathfrak{s}} \equiv \Gamma^{(\mathrm{2PPI})}_{\mathfrak{s}}[\rho=\overline{\rho}_{\mathfrak{s}}]$, $\overline{\Gamma}^{(\mathrm{2PPI})(n)}_{\mathfrak{s},\alpha_1\cdots\alpha_n} \equiv \left.\frac{\delta^n \Gamma^{(\mathrm{2PPI})}_{\mathfrak{s}}[\rho]}{\delta\rho_{\alpha_1}\cdots\delta\rho_{\alpha_n}}\right|_{\rho=\overline{\rho}_{\mathfrak{s}}}$ and the flowing density $\overline{\rho}_{\mathfrak{s}}$ extremizes the flowing 2PPI EA:

$$\left.\frac{\delta\Gamma^{(\mathrm{2PPI})}_{\mathfrak{s}}[\rho]}{\delta\rho_{\alpha}}\right|_{\rho=\overline{\rho}_{\mathfrak{s}}} = 0 \quad \forall\alpha,\mathfrak{s} \;, \tag{201}$$

similarly to Eq. (25) for the 1PI-FRG. Expanding both sides of Eq. (198) around the configuration $\rho=\overline{\rho}_{\mathfrak{s}}$ and then identifying the terms involving identical powers of $\rho-\overline{\rho}_{\mathfrak{s}}$ gives us the infinite set of differential equations underlying the standard 2PPI-FRG. The first equations of this hierarchy are[31]:

$$\dot{\overline{\Gamma}}^{(\mathrm{2PPI})}_{\mathfrak{s}} = \int_{\alpha}\dot{V}_{\mathfrak{s},\alpha}\overline{\rho}_{\mathfrak{s},\alpha} + \frac{1}{2}\int_{\alpha_1,\alpha_2}\dot{U}_{\mathfrak{s},\alpha_1\alpha_2}\left(\overline{\boldsymbol{G}}_{\mathfrak{s},\alpha_2\alpha_1}+\overline{\rho}_{\mathfrak{s},\alpha_2}\overline{\rho}_{\mathfrak{s},\alpha_1}\right) \;, \tag{202}$$

$$\dot{\overline{\rho}}_{\mathfrak{s},\alpha_1} = \int_{\alpha_2}\overline{\boldsymbol{G}}_{\mathfrak{s},\alpha_1\alpha_2}\left(-\dot{V}_{\mathfrak{s},\alpha_2}+\frac{1}{2}\int_{\alpha_3,\cdots,\alpha_6}\dot{U}_{\mathfrak{s},\alpha_3\alpha_4}\overline{\boldsymbol{G}}_{\mathfrak{s},\alpha_3\alpha_5}\overline{\Gamma}^{(\mathrm{2PPI})(3)}_{\mathfrak{s},\alpha_2\alpha_5\alpha_6}\overline{\boldsymbol{G}}_{\mathfrak{s},\alpha_6\alpha_4} - \int_{\alpha_3}\dot{U}_{\mathfrak{s},\alpha_2\alpha_3}\overline{\rho}_{\mathfrak{s},\alpha_3}\right) \;, \tag{203}$$

$$\begin{split}\dot{\overline{\Gamma}}^{(\mathrm{2PPI})(2)}_{\mathfrak{s},\alpha_1\alpha_2} = {} & \int_{\alpha_3}\dot{\overline{\rho}}_{\mathfrak{s},\alpha_3}\overline{\Gamma}^{(\mathrm{2PPI})(3)}_{\mathfrak{s},\alpha_3\alpha_1\alpha_2} + \dot{U}_{\mathfrak{s},\alpha_1\alpha_2} + \int_{\alpha_3,\cdots,\alpha_8}\dot{U}_{\mathfrak{s},\alpha_3\alpha_4}\overline{\boldsymbol{G}}_{\mathfrak{s},\alpha_3\alpha_5}\overline{\Gamma}^{(\mathrm{2PPI})(3)}_{\mathfrak{s},\alpha_1\alpha_5\alpha_6}\overline{\boldsymbol{G}}_{\mathfrak{s},\alpha_6\alpha_7}\overline{\Gamma}^{(\mathrm{2PPI})(3)}_{\mathfrak{s},\alpha_2\alpha_7\alpha_8}\overline{\boldsymbol{G}}_{\mathfrak{s},\alpha_8\alpha_4} \\ & - \frac{1}{2}\int_{\alpha_3,\cdots,\alpha_6}\dot{U}_{\mathfrak{s},\alpha_3\alpha_4}\overline{\boldsymbol{G}}_{\mathfrak{s},\alpha_3\alpha_5}\overline{\Gamma}^{(\mathrm{2PPI})(4)}_{\mathfrak{s},\alpha_1\alpha_2\alpha_5\alpha_6}\overline{\boldsymbol{G}}_{\mathfrak{s},\alpha_6\alpha_4} \;,\end{split} \tag{204}$$

$$\begin{split}\dot{\overline{\Gamma}}^{(\mathrm{2PPI})(3)}_{\mathfrak{s},\alpha_1\alpha_2\alpha_3} = {} & \int_{\alpha_4}\dot{\overline{\rho}}_{\mathfrak{s},\alpha_4}\overline{\Gamma}^{(\mathrm{2PPI})(4)}_{\mathfrak{s},\alpha_4\alpha_1\alpha_2\alpha_3} \\ & - \left(\int_{\alpha_4,\cdots,\alpha_{11}}\dot{U}_{\mathfrak{s},\alpha_4\alpha_5}\overline{\boldsymbol{G}}_{\mathfrak{s},\alpha_4\alpha_6}\overline{\Gamma}^{(\mathrm{2PPI})(3)}_{\mathfrak{s},\alpha_1\alpha_6\alpha_7}\overline{\boldsymbol{G}}_{\mathfrak{s},\alpha_7\alpha_8}\overline{\Gamma}^{(\mathrm{2PPI})(3)}_{\mathfrak{s},\alpha_2\alpha_8\alpha_9}\overline{\boldsymbol{G}}_{\mathfrak{s},\alpha_9\alpha_{10}}\overline{\Gamma}^{(\mathrm{2PPI})(3)}_{\mathfrak{s},\alpha_3\alpha_{10}\alpha_{11}}\overline{\boldsymbol{G}}_{\mathfrak{s},\alpha_{11}\alpha_5}\right. \\ & \left.+(\alpha_2,\alpha_1,\alpha_3)+(\alpha_1,\alpha_3,\alpha_2)\right) + \left(\int_{\alpha_4,\cdots,\alpha_9}\dot{U}_{\mathfrak{s},\alpha_4\alpha_5}\overline{\boldsymbol{G}}_{\mathfrak{s},\alpha_4\alpha_6}\overline{\Gamma}^{(\mathrm{2PPI})(4)}_{\mathfrak{s},\alpha_1\alpha_2\alpha_6\alpha_7}\overline{\boldsymbol{G}}_{\mathfrak{s},\alpha_7\alpha_8}\overline{\Gamma}^{(\mathrm{2PPI})(3)}_{\mathfrak{s},\alpha_3\alpha_8\alpha_9}\overline{\boldsymbol{G}}_{\mathfrak{s},\alpha_9\alpha_5}\right. \\ & \left.+(\alpha_1,\alpha_3,\alpha_2)+(\alpha_2,\alpha_3,\alpha_1)\right) - \frac{1}{2}\int_{\alpha_4,\cdots,\alpha_7}\dot{U}_{\mathfrak{s},\alpha_4\alpha_5}\overline{\boldsymbol{G}}_{\mathfrak{s},\alpha_4\alpha_6}\overline{\Gamma}^{(\mathrm{2PPI})(5)}_{\mathfrak{s},\alpha_1\alpha_2\alpha_3\alpha_6\alpha_7}\overline{\boldsymbol{G}}_{\mathfrak{s},\alpha_7\alpha_5} \;,\end{split} \tag{205}$$

where we have used the definition:

$$\overline{\boldsymbol{G}}^{-1}_{\mathfrak{s}} \equiv \overline{\Gamma}^{(\mathrm{2PPI})(2)}_{\mathfrak{s}} \;, \tag{206}$$

together with the notation set by Eq. (144) for $\alpha$-indices. Regarding then the choice of cutoff functions, $U_{\mathfrak{s}}$ must satisfy the same boundary conditions as those encountered

[31] In principle, the chemical potential $\mu$ also depends on the flow parameter $\mathfrak{s}$ in the present situation, which translates into an extra flow equation in the infinite hierarchies underpinning 2PPI-FRG approaches. However, we discard this flow equation here as it is not relevant for our (0+0)-D applications.

in the U-flow implementation of the 2PI-FRG, i.e.:

$$\begin{cases} U_{\mathfrak{s}=\mathfrak{s}_{\mathrm{i}},\alpha_1\alpha_2} = 0 \quad \forall\alpha_1,\alpha_2 \;. & \text{(207a)} \\ U_{\mathfrak{s}=\mathfrak{s}_{\mathrm{f}}} = U \;. & \text{(207b)} \end{cases}$$

For the analytical form of the one-body potential $V_{\mathfrak{s}}$, most aforementioned 2PPI-FRG studies simply use $V_{\mathfrak{s}} = V$ $\forall\mathfrak{s}$,[32] for which the starting point of the flow coincides with the free theory[33] (according to Eq. (207a)), as for the pU-flow version of the 2PI-FRG. We will therefore also follow this path to obtain all numerical results presented in this section 4, using also the same cutoff function for $U_{\mathfrak{s}}$ as for our 2PI-FRG study of section 3 (i.e. $U_{\mathfrak{s}} = \mathfrak{s}U$ with $\mathfrak{s}$ still running from $\mathfrak{s}_{\mathrm{i}} = 0$ to $\mathfrak{s}_{\mathrm{f}} = 1$).

However, the initial conditions for the 2PPI-FRG are not as easily determined from the free theory as it was for the pU-flow implementation of the 2PI-FRG. The underlying reason is that the Legendre transforms defining 2PPI EAs can not be done explicitly,[34] which implies that we can not express $\Gamma^{(\mathrm{2PPI})}[\rho]$, or its non-interacting version $\Gamma_0^{(\mathrm{2PPI})}[\rho]$, explicitly in terms of $\rho$ at finite dimensions. As a result, one must express the 2PPI vertices:

$$\overline{\Gamma}^{(\mathrm{2PPI})(n)}_{\mathfrak{s}=\mathfrak{s}_{\mathrm{i}},\alpha_1\cdots\alpha_n} = \left.\frac{\delta^n \Gamma_0^{(\mathrm{2PPI})}[\rho]}{\delta\rho_{\alpha_1}\cdots\delta\rho_{\alpha_n}}\right|_{\rho=\overline{\rho}_{\mathfrak{s}=\mathfrak{s}_{\mathrm{i}}}} \;, \tag{208}$$

in terms of derivatives of the non-interacting version $W_0[K]$ of the Schwinger functional expressed by Eq. (183) (with Eq. (189) as classical action), i.e.:

$$W_0[K] = \zeta \mathrm{STr}[\ln(G_K)] \;, \tag{209}$$

where

$$G^{-1}_{K,\alpha_1\alpha_2} = \left(\hat{O}_{\mathrm{kin},\alpha_1} + V_{\alpha_1} - \mu - K_{\alpha_1}\right)\delta_{\alpha_1\alpha_2} \;, \tag{210}$$

and $\zeta = \pm 1$ to distinguish the situations where $\widetilde{\psi}$ is a bosonic or a Grassmann field, as in section 3. The connection between the 2PPI vertices $\overline{\Gamma}^{(\mathrm{2PPI})(n)}_{\mathfrak{s}=\mathfrak{s}_{\mathrm{i}}}$ and the derivatives of $W_0[K]$ can be achieved from the chain rule:

$$\frac{\delta}{\delta\rho_{\alpha_1}} = \int_{\alpha_2} \frac{\delta K_{\alpha_2}}{\delta\rho_{\alpha_1}} \frac{\delta}{\delta K_{\alpha_2}} = \int_{\alpha_2} \left(W^{(2)}_{\mathfrak{s}}[K]\right)^{-1}_{\alpha_1\alpha_2} \frac{\delta}{\delta K_{\alpha_2}} \;, \tag{211}$$

which follows from Eq. (192), and we have used $W^{(n)}_{\mathfrak{s},\alpha_1\cdots\alpha_n}[K] \equiv \frac{\delta^n W_{\mathfrak{s}}[K]}{\delta K_{\alpha_1}\cdots\delta K_{\alpha_n}}$. From this, we can indeed infer the relations:

$$\overline{\Gamma}^{(\mathrm{2PPI})(2)}_{\mathfrak{s}=\mathfrak{s}_{\mathrm{i}},\alpha_1\alpha_2} = \left(W^{(2)}_0[K=0]\right)^{-1}_{\alpha_1\alpha_2} \;, \tag{212}$$

$$\begin{split} \overline{\Gamma}^{(\mathrm{2PPI})(n)}_{\mathfrak{s}=\mathfrak{s}_{\mathrm{i}},\alpha_1\cdots\alpha_n} = & \int_{\alpha_{2n-2}} \left(W^{(2)}_0[K]\right)^{-1}_{\alpha_n\alpha_{2n-2}} \frac{\delta}{\delta K_{\alpha_{2n-2}}} \cdots \\ & \times \int_{\alpha_{n+1}} \left(W^{(2)}_0[K]\right)^{-1}_{\alpha_3\alpha_{n+1}} \\ & \times \frac{\delta}{\delta K_{\alpha_{n+1}}} \left.\left(W^{(2)}_0[K]\right)^{-1}_{\alpha_1\alpha_2}\right|_{K=0} \quad \forall n \geq 3 \;, \end{split} \tag{213}$$

whereas the initial condition for the flowing density is simply given by:

$$\overline{\rho}_{\mathfrak{s}=\mathfrak{s}_{\mathrm{i}},\alpha} = W^{(1)}_{0,\alpha}[K=0] \;. \tag{214}$$

Some 2PPI-FRG studies like that of Ref. [36] exploit the master equation expressing $\dot{W}_{\mathfrak{s}}[K]$ (instead of Eq. (198) expressing $\dot{\Gamma}^{(\mathrm{2PPI})}_{\mathfrak{s}}[\rho]$) and treat both sides of this equation with an expansion around $K_\alpha = 0$ $\forall\alpha$ (instead of the vertex expansion around $\rho = \overline{\rho}_{\mathfrak{s}}$). In this way, such a lengthy determination of initial conditions can be avoided since the flowing objects are $W^{(n)}_{\mathfrak{s}}[K=0]$ and the 2PPI vertices can be recovered at the end of the flow if necessary through relations such as Eqs. (212) and (213). Fortunately, in (0+0)-D, the 2PPI EA $\Gamma^{(\mathrm{2PPI})}(\rho)$ (defined by Eq. (188)) coincides with the 2PI one $\Gamma^{(\mathrm{2PI})}(G)$ (defined by Eq. (114)) for which we have an explicit expression in terms of $G$ in the non-interacting case, as given by Eq. (118) in arbitrary dimensions. This enables us to determine the initial conditions for the 2PPI vertices of the (0+0)-D $O(N)$ model as follows:

$$\overline{\Gamma}^{(\mathrm{2PPI})(n)}_{\mathfrak{s}=\mathfrak{s}_{\mathrm{i}},a_1\cdots a_n} = -\frac{1}{2}\left.\frac{\partial^n}{\partial\rho_{a_1}\cdots\partial\rho_{a_n}} \sum_{a_{n+1}=1}^{N} \ln(\rho_{a_{n+1}})\right|_{\rho=\overline{\rho}_{\mathfrak{s}=\mathfrak{s}_{\mathrm{i}}}} \quad \forall n \geq 2 \;, \tag{215}$$

and the corresponding flowing density satisfies:

$$\overline{\rho}_{\mathfrak{s}=\mathfrak{s}_{\mathrm{i}},a} = \frac{1}{m^2} \quad \forall a \;. \tag{216}$$

Then, to truncate the infinite hierarchy of equations underpinning the standard 2PPI-FRG, one can design the three following conditions:

[32] However, even when one imposes that $V_{\mathfrak{s}} = V$ $\forall\mathfrak{s}$, the evolution of the chemical potential throughout the flow (already pointed out in footnote 31) can also be considered as a flow-dependent shift to the one-body potential $V$, and therefore as a renormalization of $V$.

[33] Ref. [192] also puts forward the alternative $V_{\mathfrak{s}} = (1-\mathfrak{s})V_{\mathrm{KS}}$ $\forall\mathfrak{s}$ (with $\mathfrak{s}_{\mathrm{i}} = 0$ and $\mathfrak{s}_{\mathrm{f}} = 1$), for which the starting point of the flow corresponds to the Kohn-Sham system specified by the Kohn-Sham potential $V_{\mathrm{KS}}$. That said, it should also be pointed out that, even in the situation where $V_{\mathfrak{s}} = V$ $\forall\mathfrak{s}$, it is possible to incorporate information on $V_{\mathrm{KS}}$ already at the starting point of the 2PPI-FRG flow by adjusting the chemical potential so that it coincides with that of the Kohn-Sham system at $\mathfrak{s} = \mathfrak{s}_{\mathrm{i}}$, as is done e.g. in Ref. [198]. These 2PPI-FRG approaches with correlated starting points are however not discussed further in the present study.

[34] See Refs. [46,199] for technical explanations on this point.

– For the standard U-flow (sU-flow):

$$\overline{\Gamma}^{(\mathrm{2PPI})(n)}_{\mathfrak{s}} = 0 \quad \forall \mathfrak{s},\ \forall n > N_{\mathrm{max}}\ . \tag{217}$$

– For the plain U-flow (pU-flow):

$$\overline{\Gamma}^{(\mathrm{2PPI})(n)}_{\mathfrak{s}} = \overline{\Gamma}^{(\mathrm{2PPI})(n)}_{\mathfrak{s}=\mathfrak{s}_{\mathrm{i}}} \quad \forall \mathfrak{s},\ \forall n > N_{\mathrm{max}}\ . \tag{218}$$

– For the improved U-flow (iU-flow):

$$\overline{\Gamma}^{(\mathrm{2PPI})(n)}_{\mathfrak{s}} = \overline{\Gamma}^{(\mathrm{2PPI})(n)}_{\mathfrak{s}=\mathfrak{s}_{\mathrm{i}}}\Big|_{\overline{\rho}_{\mathfrak{s}=\mathfrak{s}_{\mathrm{i}}}\rightarrow\overline{\rho}_{\mathfrak{s}}} \quad \forall \mathfrak{s},\ \forall n > N_{\mathrm{max}}\ . \tag{219}$$

Although the standard 2PPI-FRG has already been applied to the (0+0)-D model considered here (only in its unbroken-symmetry regime with $N=1$) [35], it should be stressed at the present stage that only the simplest truncation scheme, i.e. the sU-flow, was exploited in this case. The sU-flow truncation is indeed clearly the most drastic whereas the iU-flow one is *a priori* the most refined. The condition underpinning the sU-flow enables us to content ourselves with the determination of the initial conditions $\overline{\Gamma}^{(\mathrm{2PPI})(n)}_{\mathfrak{s}=\mathfrak{s}_{\mathrm{i}}}$ up to $n=N_{\mathrm{max}}$. However, the pU-flow requires us to pursue this procedure up to $n = N_{\mathrm{max}}+2$ since the differential equations (resulting from the vertex expansion) expressing $\dot{\overline{\Gamma}}^{(\mathrm{2PPI})(n)}_{\mathfrak{s}}$ (with $2 \leq n \leq N_{\mathrm{max}}$) depend on other 2PPI vertices $\overline{\Gamma}^{(\mathrm{2PPI})(m)}_{\mathfrak{s}}$ of order up to $m=n+2$. Finally, the iU-flow was first introduced in Ref. [35] under the name "RG improvement" or "RGi".[35] Regarding then the quantities of interest for the present study, i.e. the gs energy and density of the (0+0)-D $O(N)$ model, they are obtained at the end of the standard 2PPI-FRG flow from the relations:

$$E^{\mathrm{s2PPI\text{-}FRG}}_{\mathrm{gs}} = \overline{\Gamma}^{(\mathrm{2PPI})}_{\mathfrak{s}=\mathfrak{s}_{\mathrm{f}}}\ , \tag{220}$$

$$\rho^{\mathrm{s2PPI\text{-}FRG}}_{\mathrm{gs}} = \frac{1}{N}\sum_{a=1}^{N}\overline{\rho}_{\mathfrak{s}=\mathfrak{s}_{\mathrm{f}},a}\ . \tag{221}$$

Our numerical results for the standard 2PPI-FRG up to $N_{\mathrm{max}}=3$ and at $N=2$ are reported in Fig. 9. The sU-flow can not be implemented at $N_{\mathrm{max}}=1$ since condition (217) implies that $\boldsymbol{G}^{-1}_{\mathfrak{s},a_1 a_2} = \overline{\Gamma}^{(\mathrm{2PPI})(2)}_{\mathfrak{s},a_1 a_2} = 0\ \forall a_1, a_2, \mathfrak{s}$ at this truncation order, which renders the associated flow equations (202) and (203) ill-defined. Moreover, in accordance with a previous remark on the pU-flow implementation of the 2PI-FRG, Fig. 9 only displays results in the unbroken-symmetry regime as the free theory, used as starting point for the flow, is ill-defined at $m^2 < 0$.

[35] The substitution $\overline{\rho}_{\mathfrak{s}=\mathfrak{s}_{\mathrm{i}}} \rightarrow \overline{\rho}_{\mathfrak{s}}$ can not be directly carried out in the framework of finite-dimensional theories for which it is not possible to express $\overline{\Gamma}^{(\mathrm{2PPI})(n)}_{\mathfrak{s}=\mathfrak{s}_{\mathrm{i}}}$ explicitly in terms of $\overline{\rho}_{\mathfrak{s}=\mathfrak{s}_{\mathrm{i}}}$ (despite this, Ref. [35] exploited an alternative definition of the iU-flow allowing for an application to a (0+1)-D $\varphi^4$-theory). This is why the iU-flow has not been fully worked out for finite-dimensional theories up to now.

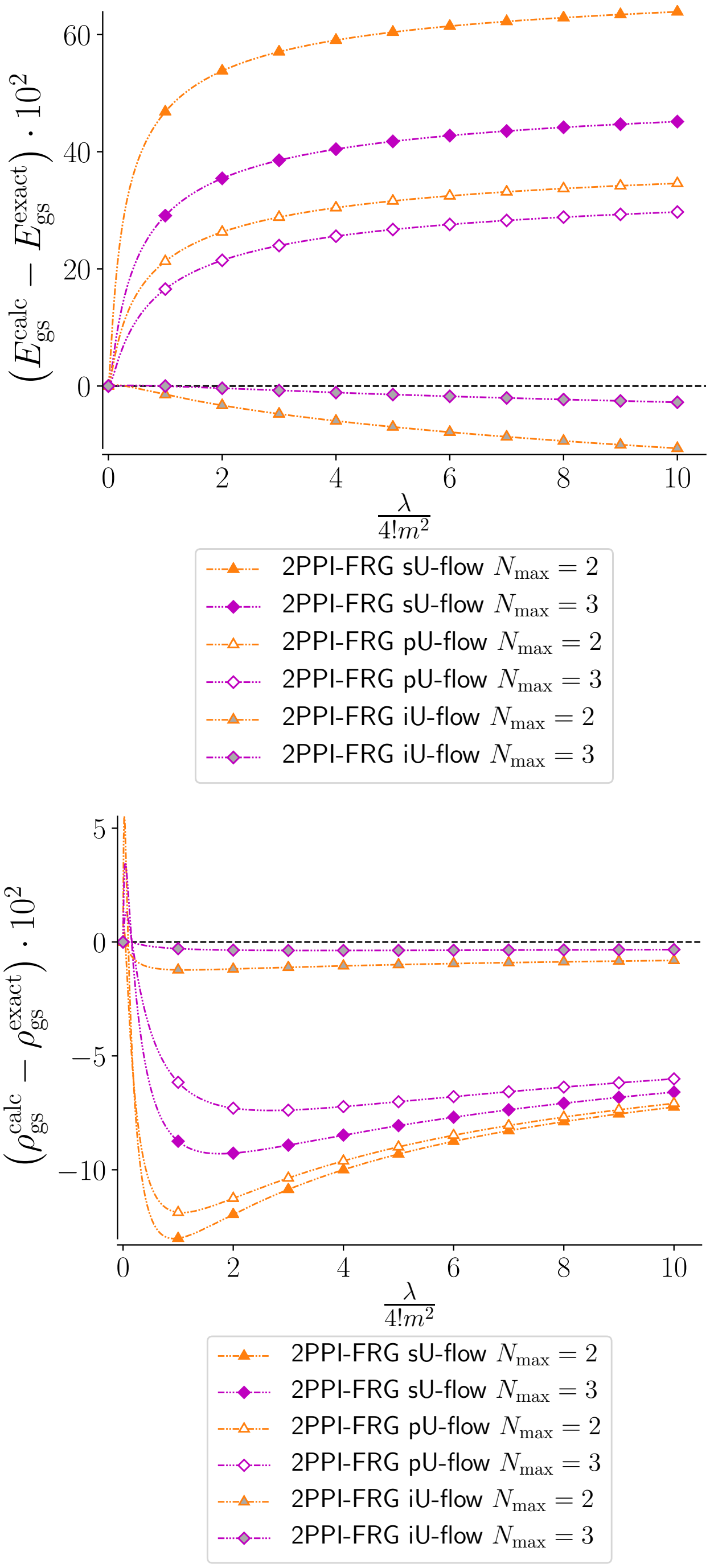

**Fig. 9.** Difference between the calculated gs energy $E^{\mathrm{calc}}_{\mathrm{gs}}$ or density $\rho^{\mathrm{calc}}_{\mathrm{gs}}$ and the corresponding exact solution $E^{\mathrm{exact}}_{\mathrm{gs}}$ or $\rho^{\mathrm{exact}}_{\mathrm{gs}}$ at $m^2=+1$ and $N=2$.

From this figure and for both $E_{\mathrm{gs}}$ and $\rho_{\mathrm{gs}}$, it can clearly be seen that, for a given truncation order $N_{\mathrm{max}}$, the sU-flow is outperformed by the pU-flow which is itself clearly less performing than the iU-flow, as expected. As a next step, we will examine what are the connections between these various versions of the standard 2PPI-FRG and the KS-FRG.

## 4.3 Kohn-Sham functional renormalization group

The KS-FRG was put forward as a "novel optimization theory of FRG with faster convergence" [193], as compared to other 2PPI-FRG approaches (especially the sU-flow and pU-flow of the standard 2PPI-FRG). It is based on the splitting of the total 2PPI EA $\Gamma^{(\mathrm{2PPI})}_{\mathfrak{s}}[\rho] \equiv \Gamma^{(\mathrm{2PPI})}[\rho; V_{\mathfrak{s}}, U_{\mathfrak{s}}]$ (introduced in Eq. (191)) into the mean-field part $\Gamma_{\mathrm{KS},\mathfrak{s}}[\rho] \equiv \Gamma^{(\mathrm{2PPI})}[\rho; V_{\mathfrak{s}} = V_{\mathrm{KS},\mathfrak{s}}, U_{\mathfrak{s}} = 0]$ (specified by a flow-dependent Kohn-Sham potential $V_{\mathrm{KS},\mathfrak{s}}$) and the correlation part $\gamma_{\mathfrak{s}}[\rho]$, i.e.:

$$\Gamma^{(\mathrm{2PPI})}_{\mathfrak{s}}[\rho] = \Gamma_{\mathrm{KS},\mathfrak{s}}[\rho] + \gamma_{\mathfrak{s}}[\rho] \; . \tag{222}$$

Plugging this decomposition into the exact flow equation for $\Gamma^{(\mathrm{2PPI})}_{\mathfrak{s}}[\rho]$, expressed by Eq. (198), gives us:

$$\begin{split}\dot{\gamma}_{\mathfrak{s}}[\rho] = & \int_{\alpha_1} \rho_{\alpha_1} \left( \dot{V}_{\mathfrak{s},\alpha_1} + \int_{\alpha_2} \overline{\Gamma}^{(2)}_{\mathrm{KS},\mathfrak{s},\alpha_1\alpha_2} \dot{\overline{\rho}}_{\mathfrak{s},\alpha_2} \right) \\ & + \frac{1}{2} \mathrm{STr}\left[ \dot{U}_{\mathfrak{s}} \left( \Gamma^{(2)}_{\mathrm{KS},\mathfrak{s}}[\rho] + \gamma^{(2)}_{\mathfrak{s}}[\rho] \right)^{-1} \right] \\ & + \frac{1}{2} \int_{\alpha_1,\alpha_2} \dot{U}_{\mathfrak{s},\alpha_1\alpha_2} \rho_{\alpha_1} \rho_{\alpha_2} \; ,\end{split} \tag{223}$$

after using the following chain rule:

$$\begin{split}\dot{\Gamma}_{\mathrm{KS},\mathfrak{s}}[\rho] = & \int_{\alpha_1,\alpha_2} \frac{\delta \Gamma_{\mathrm{KS},\mathfrak{s}}[\rho]}{\delta V_{\mathrm{KS},\mathfrak{s},\alpha_1}} \frac{\delta V_{\mathrm{KS},\mathfrak{s},\alpha_1}}{\delta \overline{\rho}_{\mathfrak{s},\alpha_2}} \dot{\overline{\rho}}_{\mathfrak{s},\alpha_2} \\ = & - \int_{\alpha_1,\alpha_2} \rho_{\alpha_1} \overline{\Gamma}^{(2)}_{\mathrm{KS},\mathfrak{s},\alpha_1\alpha_2} \dot{\overline{\rho}}_{\mathfrak{s},\alpha_2} \; ,\end{split} \tag{224}$$

which assumes that $\frac{\delta^n \overline{\Gamma}_{\mathrm{KS},\mathfrak{s}}}{\delta \overline{\rho}_{\mathfrak{s},\alpha_1} \cdots \delta \overline{\rho}_{\mathfrak{s},\alpha_n}} = \left. \frac{\delta^n \Gamma_{\mathrm{KS},\mathfrak{s}}[\rho]}{\delta \rho_{\alpha_1} \cdots \delta \rho_{\alpha_n}} \right|_{\rho=\overline{\rho}_{\mathfrak{s}}} \equiv \overline{\Gamma}^{(n)}_{\mathrm{KS},\mathfrak{s},\alpha_1\cdots\alpha_n}$ $\forall \alpha_1, \cdots, \alpha_n, \mathfrak{s}$. Eq. (223) is the master equation of the KS-FRG. It is also turned into an infinite hierarchy of coupled differential equations through a vertex expansion procedure, which relies in this situation on the expansion of the correlation part $\gamma_{\mathfrak{s}}[\rho]$:

$$\begin{split}\gamma_{\mathfrak{s}}[\rho] = & \, \overline{\gamma}_{\mathfrak{s}} \\ & + \sum^{\infty}_{n=2} \int_{\alpha_1,\cdots,\alpha_n} \overline{\gamma}^{(n)}_{\mathfrak{s},\alpha_1\cdots\alpha_n} \\ & \times \left(\rho - \overline{\rho}_{\mathfrak{s}}\right)_{\alpha_1} \cdots \left(\rho - \overline{\rho}_{\mathfrak{s}}\right)_{\alpha_n} \; ,\end{split} \tag{225}$$

with $\overline{\gamma}_{\mathfrak{s}} \equiv \gamma_{\mathfrak{s}}[\rho = \overline{\rho}_{\mathfrak{s}}]$, $\overline{\gamma}^{(n)}_{\mathfrak{s},\alpha_1\cdots\alpha_n} \equiv \left. \frac{\delta^n \gamma_{\mathfrak{s}}[\rho]}{\delta \rho_{\alpha_1} \cdots \delta \rho_{\alpha_n}} \right|_{\rho=\overline{\rho}_{\mathfrak{s}}}$ and $\overline{\rho}_{\mathfrak{s}}$ now verifies the equality:

$$\left. \frac{\delta \gamma_{\mathfrak{s}}[\rho]}{\delta \rho_{\alpha}} \right|_{\rho=\overline{\rho}_{\mathfrak{s}}} = 0 \quad \forall \alpha, \mathfrak{s} \; . \tag{226}$$

The derivatives $\overline{\Gamma}^{(n)}_{\mathrm{KS},\mathfrak{s}}$ involved in the coupled differential equations, obtained from this vertex expansion and given in appendix D.3, are specified from the flow-dependent Kohn-Sham potential $V_{\mathrm{KS},\mathfrak{s}}$ which is itself determined by solving, at each step of the flow, the Kohn-Sham equation [5,6] in the form[36]:

$$\left. \frac{\delta \Gamma_{\mathrm{KS},\mathfrak{s}}[\rho]}{\delta \rho_{\alpha}} \right|_{\rho=\overline{\rho}_{\mathfrak{s}}} = 0 \quad \forall \alpha, \mathfrak{s} \; . \tag{227}$$

Therefore, the enhanced convergence of the KS-FRG, as compared to some standard 2PPI-FRG approaches, comes at the price of solving the Kohn-Sham equation at each step of the flow. Although this extra cost is negligible for the studied (0+0)-D toy model (see appendix D.3), one should stress that it might substantially burden the numerical resolution of the KS-FRG flow equations for more realistic models.

The KS-FRG is designed to use the same starting points as the standard 2PPI-FRG. The free theory[33] will thus still be exploited to initialize the flow for our KS-FRG applications. In any case, the initial conditions for the flowing objects $\overline{\gamma}_{\mathfrak{s}}$ and $\overline{\gamma}^{(n)}_{\mathfrak{s}}$ are trivial and read:

$$\overline{\gamma}_{\mathfrak{s}=\mathfrak{s}_{\mathrm{i}}} = 0 \; , \tag{228}$$

$$\overline{\gamma}^{(n)}_{\mathfrak{s}=\mathfrak{s}_{\mathrm{i}},\alpha_1,\cdots,\alpha_n} = 0 \quad \forall \alpha_1, \cdots, \alpha_n, \; \forall n \geq 2 \; , \tag{229}$$

whereas the truncation of the infinite tower of differential equations generated by the vertex expansion of Eq. (223) is simply implemented by:

$$\overline{\gamma}^{(n)}_{\mathfrak{s}} = \overline{\gamma}^{(n)}_{\mathfrak{s}=\mathfrak{s}_{\mathrm{i}}} \quad \forall \mathfrak{s}, \; \forall n > N_{\mathrm{max}} \; , \tag{230}$$

for a given truncation order $N_{\mathrm{max}}$. The gs energy and density of the (0+0)-D $O(N)$ model are estimated from the flowing objects of the KS-FRG via the relations (see appendix D.3 for the determination of $\Gamma_{\mathrm{KS},\mathfrak{s}}(\rho)$):

$$\begin{split}E^{\mathrm{KS\text{-}FRG}}_{\mathrm{gs}} = & \, \overline{\Gamma}^{(\mathrm{2PPI})}_{\mathfrak{s}=\mathfrak{s}_{\mathrm{f}}} = \overline{\Gamma}_{\mathrm{KS},\mathfrak{s}=\mathfrak{s}_{\mathrm{f}}} + \overline{\gamma}_{\mathfrak{s}=\mathfrak{s}_{\mathrm{f}}} \\ = & - \frac{1}{2} \sum^{N}_{a=1} \ln\left(2\pi \overline{\rho}_{\mathfrak{s}=\mathfrak{s}_{\mathrm{f}},a}\right) + \overline{\gamma}_{\mathfrak{s}=\mathfrak{s}_{\mathrm{f}}} \; ,\end{split} \tag{231}$$

$$\rho^{\mathrm{KS\text{-}FRG}}_{\mathrm{gs}} = \frac{1}{N} \sum^{N}_{a=1} \overline{\rho}_{\mathfrak{s}=\mathfrak{s}_{\mathrm{f}},a} \; . \tag{232}$$

Finally, we analyze numerical results for the KS-FRG at $N = 2$ with Figs. 10 and 11. The KS-FRG was introduced to improve the convergence of the standard 2PPI-FRG in its most basic implementations, i.e. with the sU-flow or pU-flow as truncation schemes. This improvement is illustrated by Fig. 10, which clearly shows that, at the same truncation order $N_{\mathrm{max}}$, the KS-FRG outperforms the pU-flow version of the standard 2PPI-FRG (which itself outperforms its sU-flow counterpart according to our previous discussion on Fig. 9) over the whole range of tested values for the coupling constant (i.e. for $\lambda/4! \in [0,10]$) for both $E_{\mathrm{gs}}$ and $\rho_{\mathrm{gs}}$ up to $N_{\mathrm{max}} = 3$.

[36] From Eqs. (226) and (227) together with splitting (222), one can see that the flowing density $\overline{\rho}_{\mathfrak{s}}$ of the KS-FRG still fulfills condition (201) exploited within the standard 2PPI-FRG.

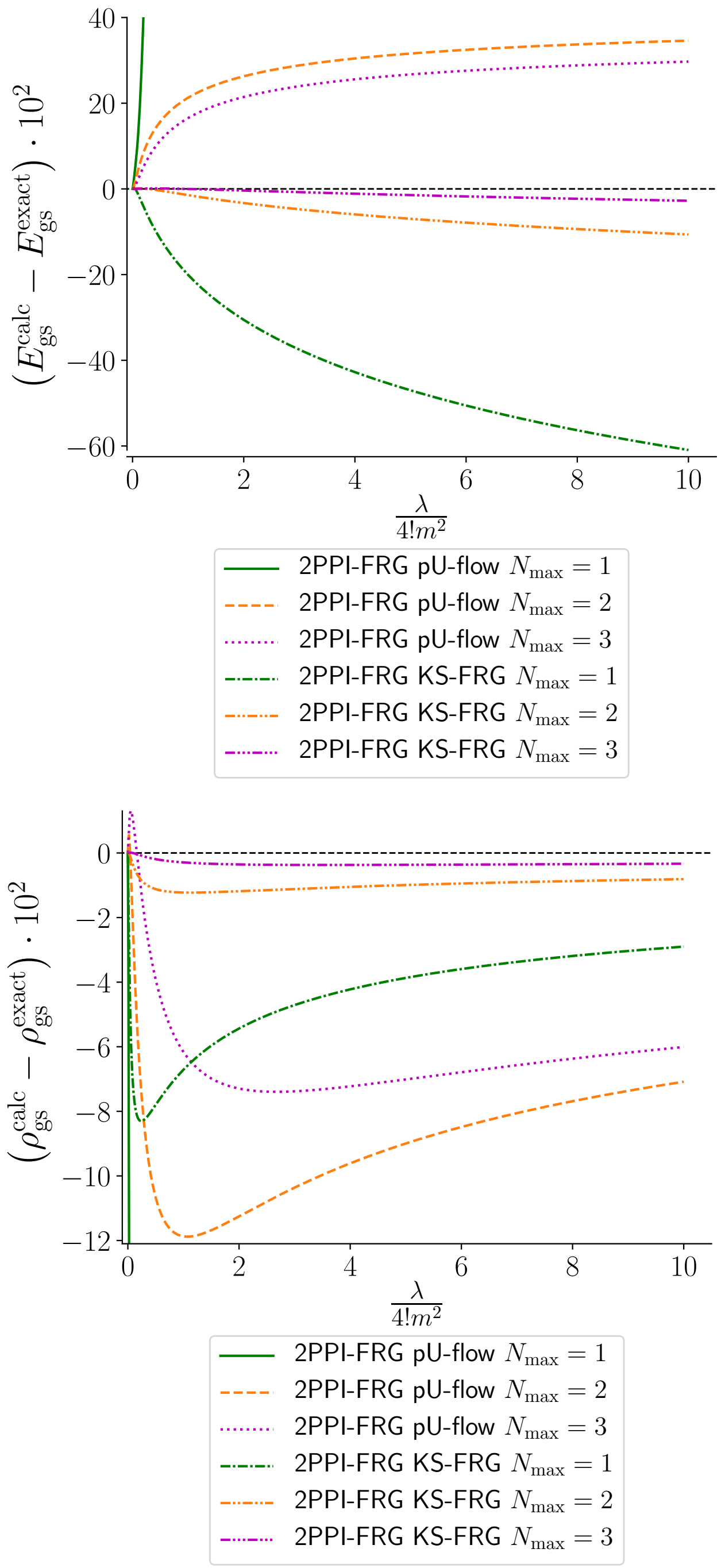

**Fig. 10.** Difference between the calculated gs energy $E_{\mathrm{gs}}^{\mathrm{calc}}$ or density $\rho_{\mathrm{gs}}^{\mathrm{calc}}$ and the corresponding exact solution $E_{\mathrm{gs}}^{\mathrm{exact}}$ or $\rho_{\mathrm{gs}}^{\mathrm{exact}}$ at $m^2 = +1$ and $N = 2$.

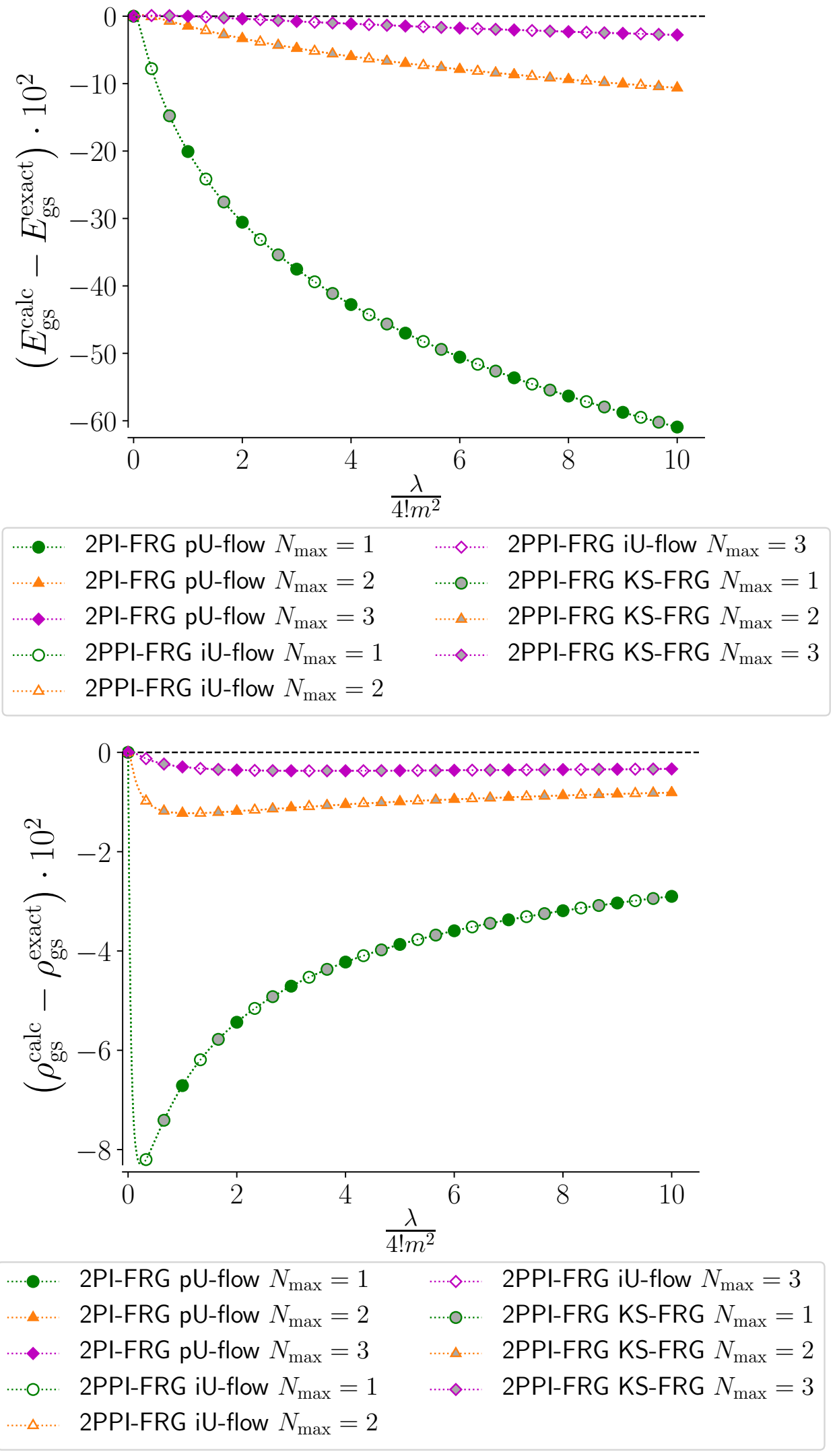

**Fig. 11.** Difference between the calculated gs energy $E_{\mathrm{gs}}^{\mathrm{calc}}$ or density $\rho_{\mathrm{gs}}^{\mathrm{calc}}$ and the corresponding exact solution $E_{\mathrm{gs}}^{\mathrm{exact}}$ or $\rho_{\mathrm{gs}}^{\mathrm{exact}}$ at $m^2 = +1$ and $N = 2$.

We also point out interesting connections between different FRG approaches tested so far: Fig. 11 illustrates that the pU-flow version of the 2PI-FRG, the iU-flow version of the 2PPI-FRG and the KS-FRG lead to identical results for the studied toy model. The equivalence between the pU-flow of the 2PI-FRG on the one hand and the latter two 2PPI-FRG approaches on the other hand would no longer be valid at finite dimensions since the 2PI EA $\Gamma^{(\mathrm{2PI})}(G)$ and the 2PPI EA $\Gamma^{(\mathrm{2PPI})}(\rho)$ only coincide in (0+0)-D. However, the connection between the iU-flow and the KS-FRG remains unaffected as dimension increases since both of these approaches are based on the 2PPI EA $\Gamma^{(\mathrm{2PPI})}(\rho)$. The latter remark enables us to characterize the KS-FRG as a more easily implementable version of the iU-flow, notably since it is in general not straightforward (if not impossible) to generalize the substitution $\overline{\rho}_{\mathfrak{s}=\mathfrak{s}_{\mathrm{i}}} \rightarrow \overline{\rho}_{\mathfrak{s}}$ defining the iU-flow in Eq. (219) to finite-dimensional theories[35]. Furthermore, the KS-FRG completely bypasses the cumbersome determination of the initial conditions $\overline{\Gamma}^{(\mathrm{2PPI})(n)}_{\mathfrak{s}=\mathfrak{s}_{\mathrm{i}}}$ that must be performed up to $n = N_{\mathrm{max}} + 2$ within the iU-flow but this should be put in contrast with the evaluation of the derivatives $\overline{\Gamma}^{(n)}_{\mathrm{KS},\mathfrak{s}}$

through the KS-FRG flow, requiring notably the resolution of the Kohn-Sham equation at each step of the flow.

To conclude this section on the 2PPI-FRG, we get back to the equivalence between the most performing 2PPI-FRG approaches tested in this study and the pU-flow version of the 2PI-FRG illustrated with Fig. 11. Although this only holds in (0+0)-D, it certainly also highlights the relevance of a mU-flow implementation of the 2PPI-FRG even for finite-dimensional theories, considering the excellent performances of the 2PI-FRG mU-flow reported in section 3. Such an implementation of the mU-flow in the 2PPI EA framework would indeed be significantly less demanding to implement *numerically* (notably since it would not rely on the Bethe-Salpeter equation, as opposed to its 2PI counterpart). However, on the side of the *formalism*, the initial conditions can not be as readily determined as in the 2PI case, still because the 2PPI EA $\Gamma^{(\mathrm{2PPI})}_{\mathfrak{s}}[\rho]$ (and the 2PPI vertices) can not be explicitly expressed in terms of $\rho$ in general. Although such a feature might be handled with the inversion method (developed by Fukuda and collaborators [33,200]) to successfully construct a mU-flow for finite-dimensional systems within the 2PPI EA framework, it clearly complexifies the underlying formalism as compared to the 2PI case. This illustrates that the 2PPI-FRG framework provides us with *numerically* less demanding alternatives to 2PI-FRG approaches, but the underlying *formalism* is also less flexible, notably to exploit correlated starting points.

# 5 Conclusions

## 5.1 Conclusion on the functional renormalization group study

In this comparative study, we have investigated three FRG schemes, the 1PI-FRG, the 2PI-FRG and the 2PPI-FRG, formulated respectively from 1PI, 2PI and 2PPI EAs. The results of sections 2, 3 and 4 were presented mostly for a single value of $N$ (i.e. at $N = 2$) notably since, as opposed to the original LE with the divergences produced by the Goldstone modes at $N \geq 2$ (see paper I), the FRG formalisms considered here do not discriminate specific values of $N$. One can actually check in Ref. [46] that the performances of the tested FRG methods are hardly modified at $N = 1$ instead of 2.

As the KS-FRG stands out as the most efficient 2PPI-FRG approach tested in the present study, we compare the three first non-trivial orders of this method in Fig. 12, still at $N = 2$, with those of the most performing 1PI-FRG and 2PI-FRG techniques tested in previous sections, i.e. the mixed 1PI-FRG and the mU-flow version of the 2PI-FRG at $N_{\mathrm{SCPT}} = 1$. This comparison is done for both the unbroken- and broken-symmetry regimes of the (0+0)-D $O(N)$ model, except for the mixed 1PI-FRG which is limited to the phase with $m^2 > 0$ owing to the stiffness issues reported in section 2. Since the pU-flow version of the 2PI-FRG is equivalent to the KS-FRG for the toy model under consideration, Fig. 12 basically compares the mU-flow and pU-flow implementations of the 2PI-FRG. As was pointed out in section 3, the better performances of the mU-flow can be attributed to the quality of its starting point: the Hartree-Fock theory for the mU-flow and the free theory (at squared mass $|m^2|$) for the pU-flow or the KS-FRG. Furthermore, one can see in Fig. 12 (and for the regime with $m^2 > 0$) that the mU-flow of the 2PI-FRG clearly outperforms the mixed 1PI-FRG whereas the KS-FRG yields a similar or a slightly better accuracy than the latter. In addition to the absence of mixed 1PI-FRG results in the broken-symmetry phase, it can be said that our best 1PI-FRG results are rather disappointing as compared to the 2PI-FRG and 2PPI-FRG ones. However, it can not be concluded from this study that the 1PI-FRG is less efficient or performing than the 2PI- and 2PPI-FRGs on a general setting, especially considering that the full-fledged FRG machinery can not be completely exploited in the chosen (0+0)-D framework. We will actually come back to this last point later in this concluding section. That said, we do expect the excellent performances of the mU-flow version of the 2PI-FRG to hold in the framework of more realistic (fermionic) systems, notably because, like many other FRG implementations, this approach is designed to tackle both particle-hole and particle-particle channels on an equal footing.

Besides the numerical results, some striking features of the formalisms underlying the tested FRG methods should also be highlighted. All FRG implementations exploited in this study, whether they are formulated in the 1PI, the 2PI or the 2PPI framework, are based on a vertex expansion of the exact flow equation for the corresponding EA (although this may not appear clearly in the 2PI-FRG formalism). For purely fermionic systems and in the 1PI framework, we are somehow limited to the vertex expansion since functionals of Grassmann variables can only be exploited via their Taylor expansions, unless of course one studies an auxiliary bosonic system by making use of e.g. a HST. This highlights an advantage of the 2PI- and 2PPI-FRGs over their 1PI counterpart: the 2PI and 2PPI EAs are functionals of propagators or densities, which are not Grassmann variables even for fermionic systems. This implies that, even for purely fermionic systems (and without HST), one can directly exploit vertex expansions around non-trivial minima and thus describe non-perturbative phenomena. However, to exploit 2PI EAs with FRG in a reliable and performing manner, the limitations of the C-flow scheme within the 2P(P)I-FRG (see appendix B.1) forces us to introduce a cutoff function in the interaction part of the classical action taken as input. This results in consequent complications of the underlying flow equations, notably with the resolution of the Bethe-Salpeter equation required at each step of the flow within the U-flow and CU-flow schemes of the 2PI-FRG. In that respect, the 1PI-FRG based on the Wetterich equation, which does not suffer from such complications, is certainly appealing, with or without HST. There are however some developments, currently available in the literature, that could help reducing significantly the complexity of U-flow and CU-flow equations, as will be discussed further below.

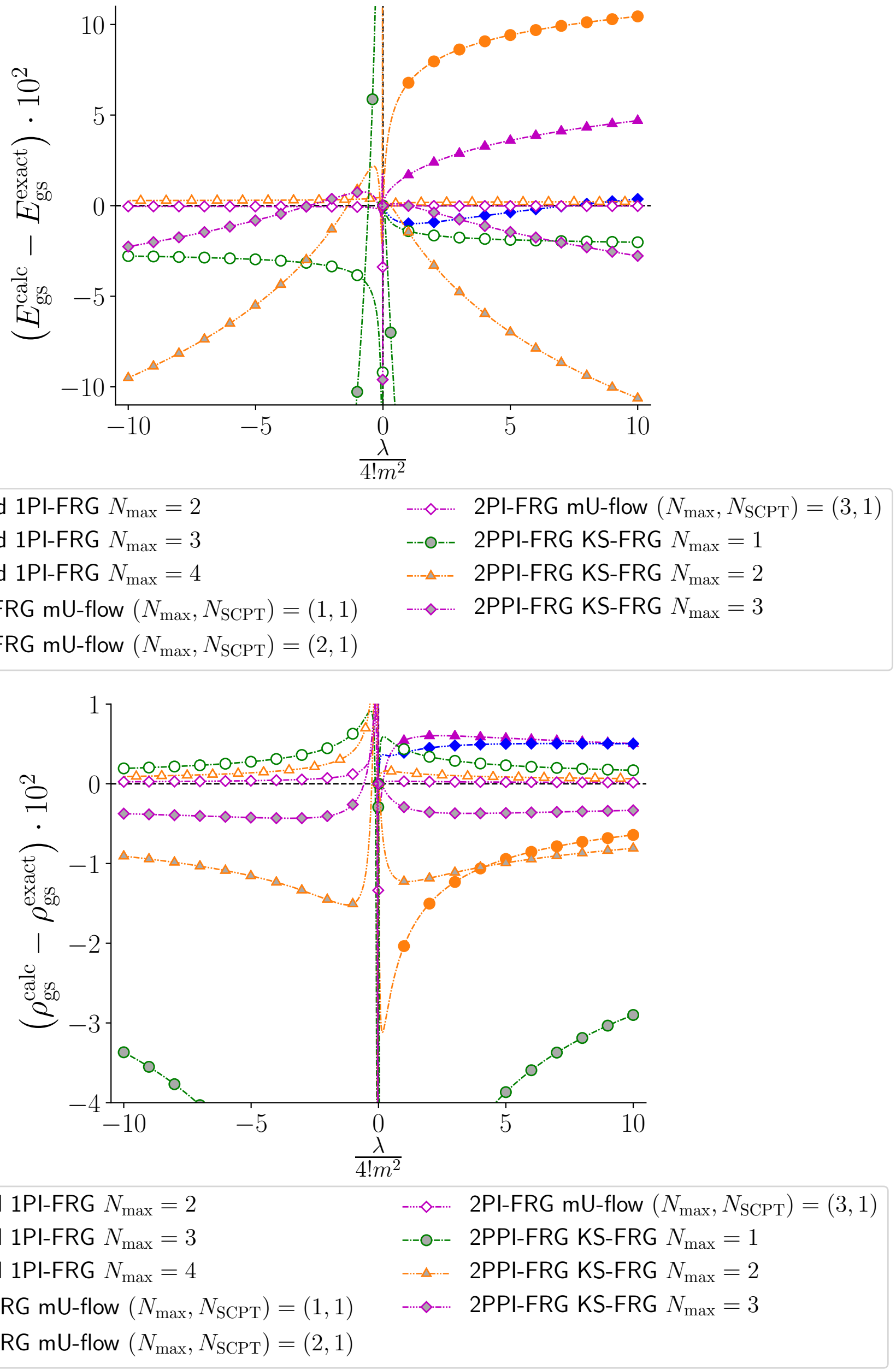

**Fig. 12.** Difference between the calculated gs energy $E_{\mathrm{gs}}^{\mathrm{calc}}$ or density $\rho_{\mathrm{gs}}^{\mathrm{calc}}$ and the corresponding exact solution $E_{\mathrm{gs}}^{\mathrm{exact}}$ or $\rho_{\mathrm{gs}}^{\mathrm{exact}}$ at $m^2 = \pm 1$ and $N = 2$. For the sake of comparison with the 2PI-FRG mU-flow results, the KS-FRG was extended to the broken-symmetry phase by exploiting the flow-dependent one-body potential $V_{\mathfrak{s}} = (2\mathfrak{s} - 1)V = (2\mathfrak{s} - 1)\frac{m^2}{2}$ $\forall \mathfrak{s}$ for $m^2 < 0$, alongside with the usual two-body interaction $U_{\mathfrak{s}} = \mathfrak{s}U$ $\forall \mathfrak{s}$. In this way, the starting point of the KS-FRG corresponds to the free theory with $\left|m^2\right|$ as squared mass.

Another remark should be made concerning a drawback that one might assign to the 2PI and 2PPI frameworks. One should have in mind that the 1PI vertices correspond to the physical interactions in the studied quantum system and are therefore, in many contexts, the objects of interest that we want to calculate in the end (see Refs. [44, 45] for the relations which allow for the determination of 1PI vertices from the 2PI ones). In condensed matter physics for instance, physical susceptibilities are usually extracted from the two-particle vertex $\overline{\Gamma}^{(1\mathrm{PI})(4)}$ at the end of the flow (see Refs. [117, 201–203] for examples in the context of the 1PI-FRG). However, in nuclear physics and more specifically in nuclear structure, one rather focuses e.g. on gs energies or densities, or maybe on the propagators of the theory and these objects are all directly accessible in the 2PI scheme (i.e. their determination does not necessitate an intermediate calculation of 1PI vertices), as we have illustrated throughout the present study.

Hence, we have just concluded on the results presented in sections 2 to 4 as well as on the formalisms of the FRG techniques that led to these results. Before going further with a general conclusion on papers I and II, we believe that it is also important to conclude on what could not be discussed in this paper. In particular, the 1PI-FRG is very widespread and has been subject to numerous theoretical developments, even very recently. Although we have exploited its conventional implementation for fermionic systems (based on a vertex expansion of the Wetterich equation), there are several state-of-the-art extensions or refinements of the 1PI-FRG that were not discussed in this study, either because their implementation in the framework of our zero-dimensional toy model was not relevant[37] or simply for the sake of conciseness. In particular, the following features of the 1PI-FRG can all be relevant in the treatment of (finite-dimensional) fermionic systems:

- Flowing bosonization:
  A very fruitful FRG implementation is the so-called scale-dependent or flowing bosonization that was developed by combining the 1PI-FRG approach with HSTs in the following manners: either by exploiting a HST that is scale-dependent itself [204, 205], or rather by simply letting the expectation value of the Hubbard-Stratonovich field depend on the momentum scale[38] [206–208]. For example, phase transitions are typically signaled by diverging vertices, which renders the flow equations untractable and simply forces the FRG solver to stop. However, flowing bosonization could enable us to distort the theory at each step of the flow in such a way that these divergences do not occur, thus allowing us for continuing the flow into a broken-symmetry phase that might have been inaccessible otherwise. Flowing bosonization can thus be extremely relevant to tackle critical physics. Other relevant applications of this approach could be mentioned, like the treatment of the Fierz ambiguity which is known to spoil FRG approximations combined with partial bosonization for fermionic systems [157].
- DMF$^2$RG:
  We have already mentioned several times that the 1PI-FRG based on a vertex expansion of the Wetterich equation is not suited to describe strongly-coupled fermions and that such a limitation can be overcome with a HST. For fermionic models that can be reliably treated with DMFT, another option to tackle strongly-coupled regimes is provided by a recent extension of the 1PI-FRG, developed notably for the Hubbard model and coined as DMF$^2$RG [120, 185–187]. In the latter approach, the flow exploits a correlated (non-perturbative) starting point in the form of DMFT results and the FRG procedure then grasps (perturbative) non-local correlations on top of the local ones captured by DMFT.
- Truncated unity FRG (TUFRG):
  The TUFRG was introduced to capture more efficiently the momentum dependences of the 1PI vertices in the context of the fermionic 1PI-FRG [116, 153]. The underlying idea is to introduce truncated completeness relations, involving sums over form factors, in the FRG equations, hence the name "truncated unity". In this way, some momentum dependences of the vertices can be represented by a set of coordinates associated to a few form factors instead of relying on the discretization of the full momentum space (or the Brillouin zone). Moreover, another version of the TUFRG, called TU$^2$FRG [119], was put forward recently to unify the original TU formulation in both real and momentum spaces and exhibits a much better numerical scaling for a certain class of models breaking translational symmetry.
- Multiloop FRG:
  For our original 1PI-FRG applications in the unbroken-symmetry phase, we have identified the first three non-trivial orders as those associated to the truncation orders $N_{\mathrm{max}} = 2$, $N_{\mathrm{max}} = 4$ and $N_{\mathrm{max}} = 6$, respectively. The truncation at $N_{\mathrm{max}} = 4$, known as the level-2 truncation, already relies on the calculation of the two-particle vertex $\overline{\Gamma}_k^{(1\mathrm{PI})(4)}$ throughout the flow, which is already quite involved, whereas the next truncation order (i.e. that at $N_{\mathrm{max}} = 6$) includes as flowing object the three-particle vertex $\overline{\Gamma}_k^{(1\mathrm{PI})(6)}$, whose direct calculation is currently out of reach for most finite-dimensional models. Before discussing very briefly some state-of-the-art treatments of $\overline{\Gamma}^{(1\mathrm{PI})(4)}$ within the FRG framework, we would like to point out another recently developed expansion scheme for the FRG in the 1PI framework, coined as multiloop FRG [201–203, 209–211], which consists in taking into account effects of the three-particle vertex $\overline{\Gamma}^{(1\mathrm{PI})(6)}$ in a systematic manner on top of the level-2 truncation of the vertex expansion.
- Single-boson exchange (SBE) decomposition:
  A conventional treatment of the two-particle vertex $\overline{\Gamma}^{(1\mathrm{PI})(4)}$ is based on the parquet approximation combined with a decomposition of its 2-particle-reducible (2PR) part in terms of high-frequency asymptotics [212]. There is also the SBE decomposition of $\overline{\Gamma}^{(1\mathrm{PI})(4)}$ [213], which rather suggests to retain the $U$-reducible (i.e. 1-vertex-reducible) diagrams contributing to $\overline{\Gamma}^{(1\mathrm{PI})(4)}$ instead of the 2PR ones. The flow equations to solve within the SBE scheme at the level-2 truncation can actually also be derived from a multi-channel HST [120]. In the 1PI-FRG frame-work, the implementation of the SBE decomposition has recently shown some significant computational advantages as compared to its conventional counterpart based on high-frequency asymptotics [121].

[37] We have notably mentioned alternatives to the vertex expansion, such as the DE which is an exact method for the (0+0)-D model under consideration.

[38] This formulation based on scale-dependent expectation values has been generalized [147] such that only the fluctuating composite operators introduced via HSTs (and not their expectation values) are scale-dependent (see Ref. [67] for a concrete example of application in the context of QCD). A further generalization aiming at treating explicit symmetry breaking is developed in Ref. [69].

For instance, regarding the most recent developments of these various techniques, a merging of the SBE formalism with the DMF$^2$RG has already been achieved successfully in the framework of the (2+1)-D Hubbard model [120] and the multiloop FRG formalism based on the SBE decomposition has already been published [214].

Some of the above features of the 1PI-FRG could be relevant to improve the 2PI- and 2PPI-FRG approaches as well. As mentioned in section 3.1, the DMF$^2$RG has also been exploited within the 2PI-FRG framework [184]. An advantage of this DMF$^2$RG formulation over its counterpart based on the 1PI framework and the Wetterich equation [120,185–187] is the possibility to properly treat models with non-local two-body interactions. Furthermore, the idea of TUFRG could also be very relevant in the 2PI framework, as was already pointed out by Katanin [215]. Notably, this would drastically simplify the resolution of the Bethe-Salpeter equation within the U-flow and CU-flow versions of the 2PI-FRG, and therefore render the 2PI-FRG scheme more amenable to realistic physical systems.

## 5.2 Conclusion on the whole comparative study of path-integral techniques

To conclude on the study of papers I and II, we finally consider Figs. 13 and 14, where respectively the first and third non-trivial orders of the most performing tested PI techniques (combined with resummation theory when relevant) are compared, namely the LE in the collective representation (i.e. the collective LE), OPT based on the principle of minimal sensitivity (PMS), SCPT for the mixed 2PI EA and the mU-flow version of the 2PI-FRG with Hartree-Fock starting point (i.e. with $N_{\text{SCPT}} = 1$). Before comparing their numerical results, we point out that, to our knowledge, the study of paper I is the first to push the collective LE and SCPT based on the mixed 2PI EA up to their third non-trivial orders (with or without combination with resummation theory) whereas paper II discusses first applications of 2PI-FRG and 2PPI-FRG methods to an $O(N)$ model with $N > 1$, pushing notably the 2PI-FRG mU-flow up to truncation orders not investigated so far.

The estimates of both $E_{\text{gs}}$ and $\rho_{\text{gs}}$ determined from those four PI techniques depicted in Figs. 13 and 14 are all fairly close, in both the unbroken- and broken-symmetry regimes of the (0+0)-D $O(N)$ model. That said, one should stress that all results presented in those figures are obtained at $N = 2$ and, as opposed to the other three approaches, the performances of the collective LE are expected to vary substancially as $N$ varies, and more specifically to improve as $N$ increases[39] owing to its connection with the $1/N$-expansion identified in paper I. One might also notice from the results of Figs. 13 and 14 that OPT based on the PMS is slightly less performing than the collective LE, SCPT for the mixed 2PI EA and the mU-flow version of the 2PI-FRG. However, this should be put in contrast with a certain ease of implementation underlying the OPT approach. Indeed, its variational parameter (the auxiliary classical field) is in principle local at finite dimensions, which contrasts with the bilocal objects (the propagators) involved in the 2PI EA formalism. We have nonetheless pointed out in section 5.1 some relevant direc-tions to reduce the cost of the numerical implementation of FRG approaches for finite-dimensional models. The same can be done for other PI techniques: see e.g. recent applications of the truncated unity approximation [116, 153] to the resolution of parquet equations in Refs. [216, 217].

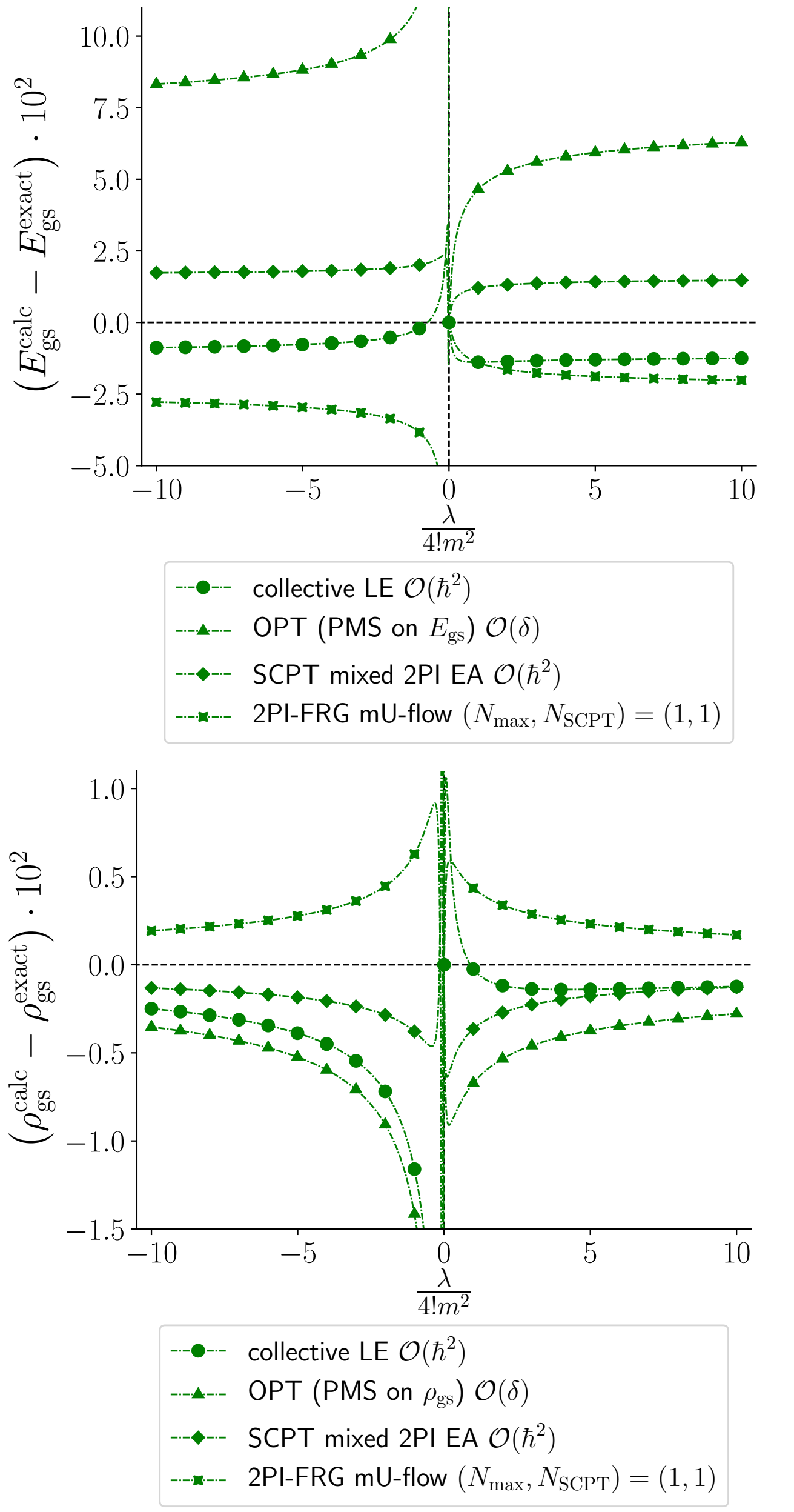

**Fig. 13.** Difference between the calculated gs energy $E_{\text{gs}}^{\text{calc}}$ or density $\rho_{\text{gs}}^{\text{calc}}$ and the corresponding exact solution $E_{\text{gs}}^{\text{exact}}$ or $\rho_{\text{gs}}^{\text{exact}}$ at $m^2 = \pm 1$ and $N = 2$.

[39] This can already be observed in paper I by comparing results at $N = 1$ and 2.

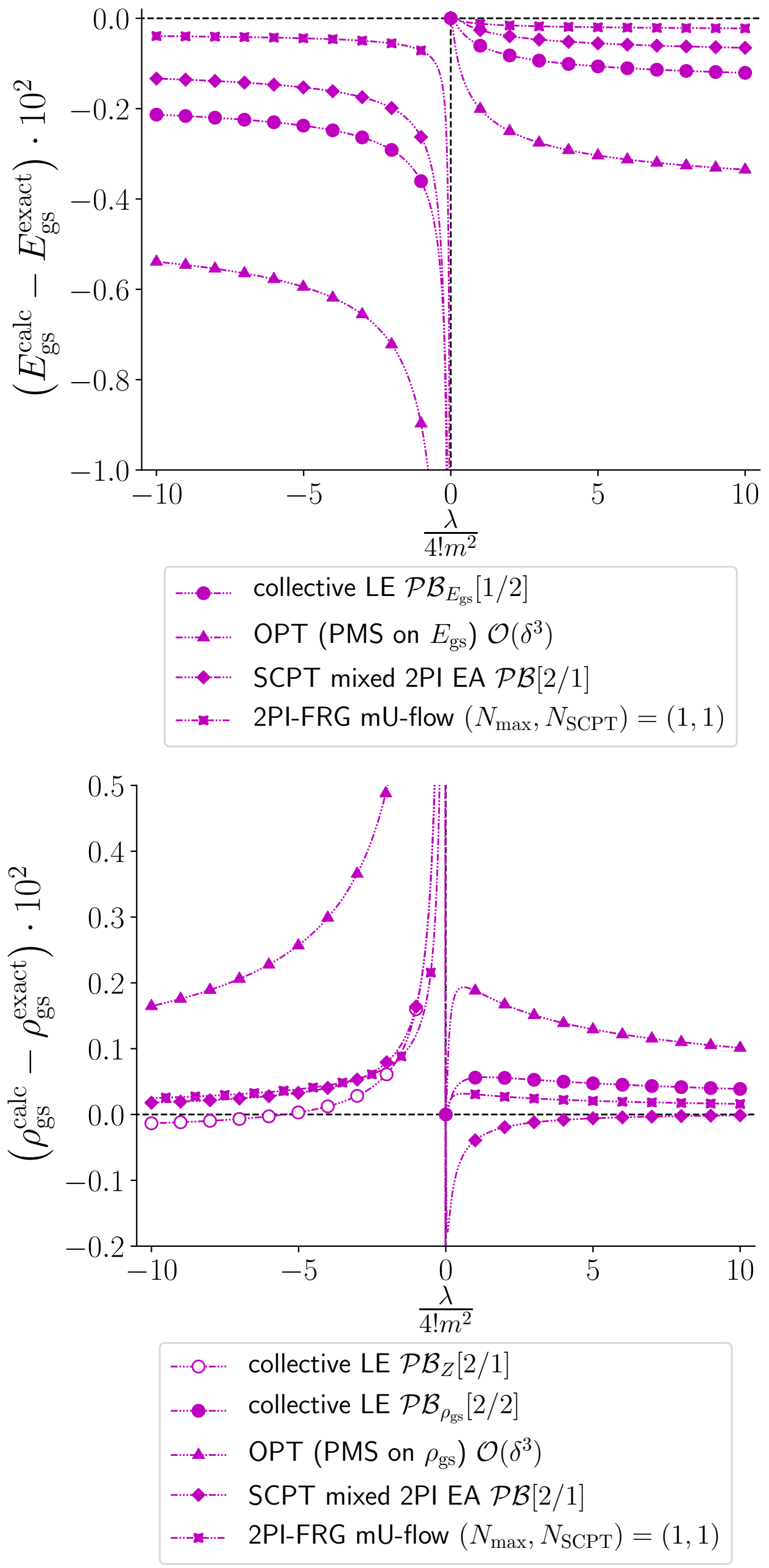

**Fig. 14.** Difference between the calculated gs energy $E_{\text{gs}}^{\text{calc}}$ or density $\rho_{\text{gs}}^{\text{calc}}$ and the corresponding exact solution $E_{\text{gs}}^{\text{exact}}$ or $\rho_{\text{gs}}^{\text{exact}}$ at $m^2 = \pm 1$ and $N = 2$. The legend $\mathcal{PB}[U/V]$ or $\mathcal{PB}_X[U/V]$ indicates that the Padé-Borel resummation was exploited with a $[U/V]$ Padé approximant, whereas $X$ in "$\mathcal{PB}_X[U/V]$" specifies that the Padé approximant is calculated from the series representing $X$ (see paper I for more details on the implementation of Padé-Borel resummation in the framework of the LE or SCPT).

On the side of the formalism, all four performing PI methods discussed here were implemented by introducing various bosonic dofs in the problem: i) the Hubbard-Stratonovich field for the collective LE and SCPT based on the mixed 2PI EA; ii) the classical auxiliary field for OPT; iii) the propagator(s) for the 2PI-FRG and SCPT based on the mixed 2PI EA. This illustrates the richness of the PI framework as well as its relevance to reformulate the nuclear EDF framework, whose efficiency also relies on bosonic dofs in the form of the nucleonic density. We also recall that PI techniques are actually suited to turn the EDF method into a more reliable approach owing to their ability of treating classical actions (derived e.g. from an EFT of QCD) as inputs, thus potentially providing an explicit connection with the dynamics of quarks and gluons which can not be accessed via current EDF calculations. Furthermore, we have also highlighted several links between PI techniques and DFT, which is also of special interest for those aiming at improving the EDF framework (or at finding more reliable alternatives) [27–32, 36, 37, 192]. In particular, all PI approaches relying on 2PPI EAs, such as the 2PPI-FRG treated in section 4, can be viewed as an implementation of DFT [33]. One can also mention the self-consistent condition of OPT, which was tested in paper I and is in spirit quite close to the Kohn-Sham scheme. The above remarks, alongside with the whole studies of papers I and II, emphasize the power and richness of the PI formalism and its ability to describe strongly-coupled quantum systems (whether nuclear or not) with its various methods, sometimes clearly connected to celebrated approaches of quantum many-body theory. Further connections with the nuclear EDF method might still be identified: for example, one could investigate if the mU-flow implementation of the 2PI-FRG has the power to describe collective phenomena by restoring the symmetries broken down by the SCPT results used as its starting point, which is precisely what is achieved within the multi-reference scheme of the EDF framework.

## Acknowledgments

The authors thank Thomas Duguet, Vittorio Somà, Nicolas Dupuis and Fabian Rennecke for stimulating discussions. K. F. also acknowledges fruitful exchanges with Takeru Yokota as well as with colleagues at the Institut für Theoretische Physik in Tübingen (Aiman Al-Eryani, Sarah Heinzelmann and Sabine Andergassen) and particularly thanks Sabine Andergassen for her patience while writing the manuscript.

### Data Availability

The datasets generated during and/or analyzed during the current study are available from the corresponding author on reasonable request.

# A Introduction to the bosonic index formalism

The formalism related to bosonic indices involves additional subtleties as compared to that based solely on the fermionic ones, denoted by $\alpha$ in our notations. This stems from the symmetry properties (expressed by Eqs. (112) and (113)) of objects like $G_{\alpha\alpha'}$ or $K_{\alpha\alpha'}$. In particular, this bosonic index formalism relies on an identity matrix $\mathcal{I}$ constructed so as to exhibit such symmetries, i.e.:

$$\begin{aligned}\mathcal{I}_{\gamma_1\gamma_2} = \mathcal{I}_{\gamma_2\gamma_1} &\equiv \frac{\delta G_{\gamma_1}}{\delta G_{\gamma_2}} = \frac{\delta K_{\gamma_1}}{\delta K_{\gamma_2}} \\ &= \delta_{\alpha_1\alpha_2}\delta_{\alpha'_1\alpha'_2} + \zeta\delta_{\alpha_1\alpha'_2}\delta_{\alpha'_1\alpha_2} \ .\end{aligned} \tag{233}$$

An expression for the functional derivative $\frac{\delta W[K]}{\delta K_\gamma}$ in terms of the propagator $G_\gamma$ directly follows from definition (233) and from the generating functional expressed by Eq. (106):

$$\begin{aligned}\frac{\delta W[K]}{\delta K_{\alpha_1\alpha'_1}} &= \frac{1}{Z[K]}\int \mathcal{D}\widetilde{\psi}\left[\frac{1}{2}\int_{\alpha_2,\alpha'_2}\widetilde{\psi}_{\alpha_2}\frac{\delta K_{\alpha_2\alpha'_2}}{\delta K_{\alpha_1\alpha'_1}}\widetilde{\psi}_{\alpha'_2}\right] \\ &\quad\times e^{-S\left[\widetilde{\psi}\right]+\frac{1}{2}\int_{\alpha_2,\alpha'_2}\widetilde{\psi}_{\alpha_2}K_{\alpha_2\alpha'_2}\widetilde{\psi}_{\alpha'_2}} \\ &= \frac{1}{Z[K]}\int \mathcal{D}\widetilde{\psi}\left[\frac{1}{2}\int_{\alpha_2,\alpha'_2}\widetilde{\psi}_{\alpha_2}\big(\delta_{\alpha_2\alpha_1}\delta_{\alpha'_2\alpha'_1}\right. \\ &\quad\left.+\zeta\delta_{\alpha_2\alpha'_1}\delta_{\alpha'_2\alpha_1}\big)\widetilde{\psi}_{\alpha'_2}\right] \\ &\quad\times e^{-S\left[\widetilde{\psi}\right]+\frac{1}{2}\int_{\alpha_2,\alpha'_2}\widetilde{\psi}_{\alpha_2}K_{\alpha_2\alpha'_2}\widetilde{\psi}_{\alpha'_2}} \\ &= \frac{1}{Z[K]}\int \mathcal{D}\widetilde{\psi}\ \frac{1}{2}\Big(\widetilde{\psi}_{\alpha_1}\widetilde{\psi}_{\alpha'_1}+\zeta\underbrace{\widetilde{\psi}_{\alpha'_1}\widetilde{\psi}_{\alpha_1}}_{\zeta\widetilde{\psi}_{\alpha_1}\widetilde{\psi}_{\alpha'_1}}\Big) \\ &\quad\times e^{-S\left[\widetilde{\psi}\right]+\frac{1}{2}\int_{\alpha_2,\alpha'_2}\widetilde{\psi}_{\alpha_2}K_{\alpha_2\alpha'_2}\widetilde{\psi}_{\alpha'_2}} \\ &= \frac{1}{Z[K]}\int \mathcal{D}\widetilde{\psi}\ \widetilde{\psi}_{\alpha_1}\widetilde{\psi}_{\alpha'_1} \\ &\quad\times e^{-S\left[\widetilde{\psi}\right]+\frac{1}{2}\int_{\alpha_2,\alpha'_2}\widetilde{\psi}_{\alpha_2}K_{\alpha_2\alpha'_2}\widetilde{\psi}_{\alpha'_2}} \\ &= G_{\alpha_1\alpha'_1} \ ,\end{aligned} \tag{234}$$

or, in terms of bosonic indices,

$$\frac{\delta W[K]}{\delta K_\gamma} = G_\gamma \ . \tag{235}$$

Another important point is the matrix multiplication with respect to bosonic indices. For two bosonic matrices $M$ and $N$, this gives us:

$$\begin{aligned}(MN)_{\gamma_1\gamma_2} &= \frac{1}{2}\int_{\gamma_3} M_{\gamma_1\gamma_3}N_{\gamma_3\gamma_2} \\ &= \frac{1}{2}\int_{\alpha_3,\alpha'_3} M_{\gamma_1(\alpha_3,\alpha'_3)}N_{(\alpha_3,\alpha'_3)\gamma_2} \\ &= M_{\gamma_1\hat{\gamma}}N_{\hat{\gamma}\gamma_2} \ ,\end{aligned} \tag{236}$$

which is generalized to $n$ bosonic matrices by Eq. (116). Hence, it involves an extra $1/2$ factor as compared to the standard matrix multiplication with respect to $\alpha$-indices, which is convenient considering the symmetry properties discussed right above Eq. (233). For instance, the bosonic identity matrix involving two terms in its definition (233) so as to satisfy those symmetries, such a $1/2$ factor is usually canceled out as follows:

$$\begin{aligned}\frac{\delta}{\delta G_{\gamma_1}}G_{\hat{\gamma}}M_{\hat{\gamma}\gamma_2} &= \frac{1}{2}\int_{\gamma_3}\underbrace{\frac{\delta G_{\gamma_3}}{\delta G_{\gamma_1}}}_{\mathcal{I}_{\gamma_1\gamma_3}}M_{\gamma_3\gamma_2} \\ &= \frac{1}{2}\int_{\gamma_3}\left(\delta_{\alpha_1\alpha_3}\delta_{\alpha'_1\alpha'_3}+\zeta\delta_{\alpha_1\alpha'_3}\delta_{\alpha'_1\alpha_3}\right)M_{\gamma_3\gamma_2} \\ &= \frac{1}{2}\Big(M_{\gamma_1\gamma_2}+\zeta\underbrace{M_{(\alpha'_1,\alpha_1)\gamma_2}}_{\zeta M_{(\alpha_1,\alpha'_1)\gamma_2}}\Big) \\ &= M_{\gamma_1\gamma_2} \ ,\end{aligned} \tag{237}$$

with $M$ being an arbitrary bosonic matrix (independent of $M$) possessing also the symmetry properties of Eq. (112). Note in addition that specific realizations of Eq. (236) are the chain rule:

$$\frac{\delta W[K]}{\delta G_{\gamma_1}} = \frac{1}{2}\int_{\gamma_2}\frac{\delta K_{\gamma_2}}{\delta G_{\gamma_1}}\frac{\delta W[K]}{\delta K_{\gamma_2}} \ , \tag{238}$$

and the definition of an inverse $M^{\text{inv}}$ of a bosonic matrix $M$:

$$\mathcal{I}_{\gamma_1\gamma_2} = \left(MM^{\text{inv}}\right)_{\gamma_1\gamma_2} = \frac{1}{2}\int_{\gamma_3}M_{\gamma_1\gamma_3}M^{\text{inv}}_{\gamma_3\gamma_2} \ , \tag{239}$$

which is already given by Eq. (146). We can also evaluate the derivative $\frac{\delta\Gamma^{(\text{2PI})}[G]}{\delta G_\gamma}$ with the help of the chain rule expressed by Eq. (238). This is achieved by differentiating both sides of Eq. (114) with respect to $G$, thus yielding:

$$\begin{aligned}\frac{\delta\Gamma^{(\text{2PI})}[G]}{\delta G_{\alpha_1\alpha'_1}} &= -\frac{\delta W[K]}{\delta G_{\alpha_1\alpha'_1}}+\frac{1}{2}\int_{\gamma_2}\frac{\delta K_{\gamma_2}}{\delta G_{\alpha_1\alpha'_1}}G_{\gamma_2} \\ &\quad+\frac{1}{2}\int_{\alpha_2,\alpha'_2}K_{\alpha_2\alpha'_2}\frac{\delta G_{\alpha_2\alpha'_2}}{\delta G_{\alpha_1\alpha'_1}} \\ &= -\frac{1}{2}\int_{\gamma_2}\frac{\delta K_{\gamma_2}}{\delta G_{\alpha_1\alpha'_1}}\underbrace{\frac{\delta W[K]}{\delta K_{\gamma_2}}}_{G_{\gamma_2}}+\frac{1}{2}\int_{\gamma_2}\frac{\delta K_{\gamma_2}}{\delta G_{\alpha_1\alpha'_1}}G_{\gamma_2} \\ &\quad+\frac{1}{2}\Big(K_{\alpha_1\alpha'_1}+\zeta\underbrace{K_{\alpha'_1\alpha_1}}_{\zeta K_{\alpha_1\alpha'_1}}\Big) \\ &= K_{\alpha_1\alpha'_1} \ ,\end{aligned} \tag{240}$$

and, in terms of bosonic indices,

$$\frac{\delta\Gamma^{(\text{2PI})}[G]}{\delta G_\gamma} = K_\gamma \ , \tag{241}$$

which is consistent with the leftmost equality in result (120).

# B Additional numerical results for the 2PI functional renormalization group

## B.1 C-flow

The underlying idea of the C-flow version of the 2PI-FRG [44] remains to implement the momentum-shell integration à la Wilson. By definition, the C-flow consists in considering a flow-dependent free propagator $C_{\mathfrak{s}}$. This amounts to inserting a cutoff function $R_{\mathfrak{s}}$ in the classical action (122) via the substitution $C^{-1} \rightarrow C_{\mathfrak{s}}^{-1} = R_{\mathfrak{s}} C^{-1}$ or equivalently $C^{-1} \rightarrow C_{\mathfrak{s}}^{-1} = C^{-1} + R_{\mathfrak{s}}$, which is exactly what is done for the 1PI-FRG based on the Wetterich equation by introducing the term $\varDelta S_k$ in Eq. (19). The C-flow is therefore close in spirit to Wetterich's approach. Owing to such a connection, we can deduce the required values for $C_{\mathfrak{s}=\mathfrak{s}_{\mathrm{i}}}$ and $C_{\mathfrak{s}=\mathfrak{s}_{\mathrm{f}}}$ from the boundary conditions for $R_k$ set by Eqs. (26):

$$\begin{cases} C_{\mathfrak{s}=\mathfrak{s}_{\mathrm{i}},\gamma} = 0 \quad \forall \gamma \ , & (242a) \\ C_{\mathfrak{s}=\mathfrak{s}_{\mathrm{f}}} = C \ , & (242b) \end{cases}$$

with $\mathfrak{s}_{\mathrm{i}}$ and $\mathfrak{s}_{\mathrm{f}}$ still denoting the values of the flow parameter at the beginning and at the end of the flow, respectively. Moreover, the LW functional $\varPhi[G]$, defined as the interacting part of the 2PI EA $\varGamma^{(2\mathrm{PI})}[G]$ through Eq. (117), does not depend on the free propagator $C$, and therefore not on $\mathfrak{s}$ for the C-flow. Consequently, it is an invariant of the flow in this case, which translates into:

$$\dot{\varPhi}_{\mathfrak{s}}[G] = 0 \quad \forall \mathfrak{s} \ . \tag{243}$$

In particular, Eq. (243) implies that all components of $\dot{\overline{\varSigma}}_{\mathfrak{s}}$ and $\dot{\overline{\varPhi}}{}_{\mathfrak{s}}^{(n)}$ (with $n \geq 2$) vanish, which enables us to simplify the chain rules (139) for the C-flow:

$$\begin{cases} \dot{\overline{\varSigma}}_{\mathfrak{s},\gamma} = -\dot{\overline{G}}_{\mathfrak{s},\hat{\gamma}} \overline{\varPhi}{}^{(2)}_{\mathfrak{s},\hat{\gamma}\gamma} \ . & (244a) \\ \dot{\overline{\varPhi}}{}^{(n)}_{\mathfrak{s},\gamma_1\cdots\gamma_n} = \dot{\overline{G}}_{\mathfrak{s},\hat{\gamma}} \overline{\varPhi}{}^{(n+1)}_{\mathfrak{s},\hat{\gamma}\gamma_1\cdots\gamma_n} \quad \forall n \geq 2 \ . & (244b) \end{cases}$$

The flow dependence of $\overline{G}_{\mathfrak{s}}$, involved in both Eqs. (244a) to (244b), follows from that of $C_{\mathfrak{s}}$ according to:

$$\dot{\overline{G}}_{\mathfrak{s},\alpha_1\alpha_1'} = -\int_{\alpha_2,\alpha_2'} \overline{G}_{\mathfrak{s},\alpha_1\alpha_2} \left( \dot{C}_{\mathfrak{s}}^{-1} - \dot{\overline{\varSigma}}_{\mathfrak{s}} \right)_{\alpha_2\alpha_2'} \overline{G}_{\mathfrak{s},\alpha_2'\alpha_1'} \ . \tag{245}$$

Result (245) can be derived by differentiating the Dyson equation (120) with respect to $\mathfrak{s}$ at vanishing source $K$, similarly to Eq. (137) for the U-flow. A flow equation for $\overline{\varGamma}{}^{(2\mathrm{PI})}_{\mathfrak{s}}$ can also be directly inferred from the master equation (126) expressing $\dot{\varGamma}{}^{(2\mathrm{PI})}_{\mathfrak{s}}[G]$, which gives us:

$$\dot{\overline{\varGamma}}{}^{(2\mathrm{PI})}_{\mathfrak{s}} = \dot{C}^{-1}_{\mathfrak{s},\hat{\gamma}} \overline{G}_{\mathfrak{s},\hat{\gamma}} \ , \tag{246}$$

for the C-flow. However, Eq. (246) is actually not tractable. To see this, one can consider the Dyson equation for $\overline{G}_{\mathfrak{s}}$, i.e.:

$$\overline{G}_{\mathfrak{s}} = \left( C_{\mathfrak{s}}^{-1} - \overline{\varSigma}_{\mathfrak{s}} \right)^{-1} \ , \tag{247}$$

to notice that $\overline{G}_{\mathfrak{s}}$ must satisfy $\overline{G}_{\mathfrak{s}=\mathfrak{s}_{\mathrm{i}},\gamma} = 0$ $\forall \gamma$, as a result of condition (242a). According to Eq. (118), this induces a divergence of $\frac{\zeta}{2}\mathrm{STr}\left[\ln(\overline{G}_{\mathfrak{s}=\mathfrak{s}_{\mathrm{i}}})\right]$ in $\varGamma^{(2\mathrm{PI})}_{0,\mathfrak{s}=\mathfrak{s}_{\mathrm{i}}}$, and in $\overline{\varGamma}{}^{(2\mathrm{PI})}_{\mathfrak{s}=\mathfrak{s}_{\mathrm{i}}}$ as a result. Therefore, we will calculate during the flow the following quantity:

$$\begin{aligned} \varDelta\overline{\varGamma}{}^{(2\mathrm{PI})}_{\mathfrak{s}} &\equiv \overline{\varGamma}{}^{(2\mathrm{PI})}_{\mathfrak{s}} - \varGamma^{(2\mathrm{PI})}_{0,\mathfrak{s}}[G = C_{\mathfrak{s}}] \\ &= \overline{\varGamma}{}^{(2\mathrm{PI})}_{\mathfrak{s}} + \frac{\zeta}{2}\mathrm{STr}\left[\ln(C_{\mathfrak{s}})\right] \ , \end{aligned} \tag{248}$$

instead of $\overline{\varGamma}{}^{(2\mathrm{PI})}_{\mathfrak{s}}$. The extra term $\varGamma^{(2\mathrm{PI})}_{0,\mathfrak{s}}[G = C_{\mathfrak{s}}]$ eliminates the aforementioned divergence problem, thanks to condition (242a). In conclusion, the tower of differential equations underlying the C-flow version of the 2PI-FRG is thus given by:

$$\dot{\overline{G}}_{\mathfrak{s},\alpha_1\alpha_1'} = -\int_{\alpha_2,\alpha_2'} \overline{G}_{\mathfrak{s},\alpha_1\alpha_2} \left( \dot{C}_{\mathfrak{s}}^{-1} - \dot{\overline{\varSigma}}_{\mathfrak{s}} \right)_{\alpha_2\alpha_2'} \overline{G}_{\mathfrak{s},\alpha_2'\alpha_1'} \ , \tag{249}$$

$$\varDelta\dot{\overline{\varGamma}}{}^{(2\mathrm{PI})}_{\mathfrak{s}} = \dot{C}^{-1}_{\mathfrak{s},\hat{\gamma}} \left( \overline{G}_{\mathfrak{s}} - C_{\mathfrak{s}} \right)_{\hat{\gamma}} \ , \tag{250}$$

$$\dot{\overline{\varSigma}}_{\mathfrak{s},\gamma} = -\dot{\overline{G}}_{\mathfrak{s},\hat{\gamma}} \overline{\varPhi}{}^{(2)}_{\mathfrak{s},\hat{\gamma}\gamma} \ , \tag{251}$$

$$\dot{\overline{\varPhi}}{}^{(n)}_{\mathfrak{s},\gamma_1\cdots\gamma_n} = \dot{\overline{G}}_{\mathfrak{s},\hat{\gamma}} \overline{\varPhi}{}^{(n+1)}_{\mathfrak{s},\hat{\gamma}\gamma_1\cdots\gamma_n} \quad \forall n \geq 2 \ . \tag{252}$$

The quantity $\overline{\varGamma}{}^{(2\mathrm{PI})}_{\mathfrak{s}=\mathfrak{s}_{\mathrm{f}}}$ is readily obtained at the end of the flow of $\varDelta\overline{\varGamma}{}^{(2\mathrm{PI})}_{\mathfrak{s}}$ through definition (248) in the form $\overline{\varGamma}{}^{(2\mathrm{PI})}_{\mathfrak{s}=\mathfrak{s}_{\mathrm{f}}} = \varDelta\overline{\varGamma}{}^{(2\mathrm{PI})}_{\mathfrak{s}=\mathfrak{s}_{\mathrm{f}}} - \frac{\zeta}{2}\mathrm{STr}\left[\ln(C)\right]$.

We then discuss the initial conditions for the C-flow. It was explained above that the following relation follows from both Dyson equation (in the form of Eq. (247)) and the initial condition for $C_{\mathfrak{s}}$ (given by (242a)):

$$\overline{G}_{\mathfrak{s}=\mathfrak{s}_{\mathrm{i}},\gamma} = 0 \quad \forall \gamma \ . \tag{253}$$

After combining Eq. (253) with the diagrammatic expression of $\varPhi[G]$ given by result (158), one can find the initial conditions for all 2PI vertices in the framework of the C-flow according to:

$$\overline{\varPhi}{}^{(n)}_{\mathfrak{s}=\mathfrak{s}_{\mathrm{i}},\gamma_1\cdots\gamma_n} = \left.\frac{\delta^n \varPhi_{\mathrm{SCPT}}[U,G]}{\delta G_{\gamma_1}\cdots\delta G_{\gamma_n}}\right|_{G=\overline{G}_{\mathfrak{s}=\mathfrak{s}_{\mathrm{i}}}} \ , \tag{254}$$

which translates for specific values of $n$ into:

$$\overline{\varSigma}_{\mathfrak{s}=\mathfrak{s}_{\mathrm{i}},\gamma} = 0 \quad \forall \gamma \ , \tag{255}$$

$$\overline{\varPhi}{}^{(2)}_{\mathfrak{s}=\mathfrak{s}_{\mathrm{i}},\gamma_1\gamma_2} = U_{\gamma_1\gamma_2} \ , \tag{256}$$

$$\overline{\Phi}^{(4)}_{\mathfrak{s}=\mathfrak{s}_{\mathrm{i}},\gamma_1\gamma_2\gamma_3\gamma_4} = -\frac{1}{2}\Big[\Big(\Big\{\Big[U_{\alpha_1\alpha_2\alpha_3\alpha_4}U_{\alpha'_1\alpha'_2\alpha'_3\alpha'_4} + \zeta\left(\alpha_1 \leftrightarrow \alpha'_1\right)\Big] + \zeta\left(\alpha_2 \leftrightarrow \alpha'_2\right)\Big\} + \zeta\left(\alpha_3 \leftrightarrow \alpha'_3\right)\Big) + \zeta\left(\alpha_4 \leftrightarrow \alpha'_4\right)\Big] , \tag{257}$$

$$\overline{\Phi}^{(n)}_{\mathfrak{s}=\mathfrak{s}_{\mathrm{i}},\gamma_1\cdots\gamma_n} = 0 \quad \forall \gamma_1,\cdots,\gamma_n,\ \forall n \text{ odd} . \tag{258}$$

Even though Eqs. (255) to (258) are derived from a truncated result in practice, these equations are exact at $\mathfrak{s}=\mathfrak{s}_{\mathrm{i}}$ because of condition (253). It remains to determine the initial condition for $\Delta\overline{\Gamma}^{(\mathrm{2PI})}_{\mathfrak{s}}$. Combining definition (248) with Eqs. (117) and (118) at $\mathfrak{s}=\mathfrak{s}_{\mathrm{i}}$ gives us:

$$\begin{aligned}\Delta\overline{\Gamma}^{(\mathrm{2PI})}_{\mathfrak{s}=\mathfrak{s}_{\mathrm{i}}} &= \overline{\Gamma}^{(\mathrm{2PI})}_{\mathfrak{s}=\mathfrak{s}_{\mathrm{i}}} - \Gamma^{(\mathrm{2PI})}_{0,\mathfrak{s}=\mathfrak{s}_{\mathrm{i}}}\left[G = C_{\mathfrak{s}=\mathfrak{s}_{\mathrm{i}}}\right] \\ &= -\frac{\zeta}{2}\mathrm{STr}\left[\ln\!\left(\overline{G}_{\mathfrak{s}=\mathfrak{s}_{\mathrm{i}}}\right)\right] + \frac{\zeta}{2}\mathrm{STr}\left[\overline{G}_{\mathfrak{s}=\mathfrak{s}_{\mathrm{i}}}C^{-1}_{\mathfrak{s}=\mathfrak{s}_{\mathrm{i}}} - \mathbb{I}\right] \\ &\quad + \overline{\Phi}_{\mathfrak{s}=\mathfrak{s}_{\mathrm{i}}} + \frac{\zeta}{2}\mathrm{STr}\left[\ln(C_{\mathfrak{s}=\mathfrak{s}_{\mathrm{i}}})\right] ,\end{aligned} \tag{259}$$

which, according to the latter results, simply reduces to:

$$\Delta\overline{\Gamma}^{(\mathrm{2PI})}_{\mathfrak{s}=\mathfrak{s}_{\mathrm{i}}} = 0 . \tag{260}$$

As a next step, we address the truncation of the infinite hierarchy of flow equations for the 2PI vertices within the C-flow. Two truncation schemes have been designed for this version of the 2PI-FRG [44]. We discuss both of them in what follows:

- Truncated C-flow (tC-flow):
  The tC-flow is a specific implementation of the C-flow in which the infinite tower of differential equations given by Eqs. (249) to (252) is rendered finite according to the condition:

$$\overline{\Phi}^{(n)}_{\mathfrak{s}} = \overline{\Phi}^{(n)}_{\mathfrak{s}=\mathfrak{s}_{\mathrm{i}}} \quad \forall \mathfrak{s},\ \forall n > N_{\mathrm{max}} . \tag{261}$$

  In particular, it is shown in Ref. [179] that the tC-flow scheme with truncation order $N_{\mathrm{max}} = 2N_{\mathrm{SCPT}} - 1$ or $2N_{\mathrm{SCPT}}$ (with $N_{\mathrm{SCPT}} \in \mathbb{N}^*$) is equivalent to SCPT carried out up to order $\mathcal{O}\big(U^{N_{\mathrm{SCPT}}}\big)$, e.g. the tC-flow with $N_{\mathrm{max}} = 1$ or 2 reproduces Hartree-Fock results.
- Modified C-flow (mC-flow):
  The mC-flow is an alternative to the tC-flow and relies on the following condition:

$$\overline{\Phi}^{(n)}_{\mathfrak{s}} = \overline{\Phi}^{(n)}_{\mathrm{SCPT},N_{\mathrm{SCPT}},\mathfrak{s}}\Big|_{U\rightarrow\overline{\Phi}^{(2)}_{\mathrm{sym},\mathfrak{s}}} \quad \forall \mathfrak{s},\ \forall n > N_{\mathrm{max}} , \tag{262}$$

  with

$$\overline{\Phi}^{(2)}_{\mathrm{sym},\mathfrak{s},\alpha_1\alpha_2\alpha_3\alpha_4} = \frac{1}{3}\Big(\overline{\Phi}^{(2)}_{\mathfrak{s},\alpha_1\alpha_2\alpha_3\alpha_4} + \overline{\Phi}^{(2)}_{\mathfrak{s},\alpha_2\alpha_3\alpha_1\alpha_4} + \overline{\Phi}^{(2)}_{\mathfrak{s},\alpha_3\alpha_1\alpha_2\alpha_4}\Big) , \tag{263}$$

  instead of Eq. (261). The truncation condition (262) is an ansatz based on the perturbative expression of the LW functional $\Phi_{\mathrm{SCPT}}[U,G]$ given by Eq. (158), and involves its truncated version $\Phi_{\mathrm{SCPT},N_{\mathrm{SCPT}}}[U,G]$ already defined between Eqs. (159) and (160). From this functional, we introduce the following quantities:

$$\overline{\Phi}_{\mathrm{SCPT},N_{\mathrm{SCPT}},\mathfrak{s}} \equiv \Phi_{\mathrm{SCPT},N_{\mathrm{SCPT}}}\left[U, G = \overline{G}_{\mathfrak{s}}\right] , \tag{264}$$

  and

$$\overline{\Phi}^{(n)}_{\mathrm{SCPT},N_{\mathrm{SCPT}},\mathfrak{s},\gamma_1\cdots\gamma_n} \equiv \left.\frac{\delta^n\Phi_{\mathrm{SCPT},N_{\mathrm{SCPT}}}[U,G]}{\delta G_{\gamma_1}\cdots\delta G_{\gamma_n}}\right|_{G=\overline{G}_{\mathfrak{s}}} . \tag{265}$$

  Hence, we have for example at $N_{\mathrm{SCPT}} = 1$:

$$\begin{aligned}&\overline{\Phi}_{\mathrm{SCPT},N_{\mathrm{SCPT}}=1,\mathfrak{s}}\Big|_{U\rightarrow\overline{\Phi}^{(2)}_{\mathrm{sym},\mathfrak{s}}} \\ &\equiv \frac{1}{8}\int_{\gamma_1,\gamma_2}\overline{\Phi}^{(2)}_{\mathrm{sym},\mathfrak{s},\gamma_1\gamma_2}\overline{G}_{\mathfrak{s},\gamma_1}\overline{G}_{\mathfrak{s},\gamma_2} .\end{aligned} \tag{266}$$

  Furthermore, the motivation for replacing $U$ by $\overline{\Phi}^{(2)}_{\mathfrak{s}}$ is set by the initial condition for $\overline{\Phi}^{(2)}_{\mathfrak{s}}$, i.e. Eq. (256). Generalizing the latter relation to all $\mathfrak{s}$ indeed suggests to substitute $U$ by $\overline{\Phi}^{(2)}_{\mathfrak{s}}$ in Eq. (262) but the problem is that $U$ and $\overline{\Phi}^{(2)}_{\mathfrak{s}}$ have different symmetry properties. Indeed, condition (123), i.e.:

$$U_{\alpha_1\alpha_2\alpha_3\alpha_4} = \zeta^{N(P)}U_{\alpha_{P(1)}\alpha_{P(2)}\alpha_{P(3)}\alpha_{P(4)}} , \tag{267}$$

  imposes an invariance of $U$ (up to a sign) under $4! = 24$ permutations, which is not equivalent to the conditions set by Eqs. (112) for $\overline{\Phi}^{(2)}_{\mathfrak{s}}$ involving also:

$$\overline{\Phi}^{(2)}_{\mathfrak{s},(\alpha_1,\alpha'_1)(\alpha_2,\alpha'_2)} = \zeta\overline{\Phi}^{(2)}_{\mathfrak{s},(\alpha'_1,\alpha_1)(\alpha_2,\alpha'_2)} , \tag{268}$$

  but not

$$\overline{\Phi}^{(2)}_{\mathfrak{s},(\alpha_1,\alpha'_1)(\alpha_2,\alpha'_2)} = \zeta\overline{\Phi}^{(2)}_{\mathfrak{s},(\alpha_1,\alpha_2)(\alpha'_1,\alpha'_2)} , \tag{269}$$

  for instance. This brings us to the relevance of $\overline{\Phi}^{(2)}_{\mathrm{sym},\mathfrak{s}}$ which is constructed so as to possess the same symmetry properties as $U$ [179]:

$$\overline{\Phi}^{(2)}_{\mathrm{sym},\mathfrak{s},\alpha_1\alpha_2\alpha_3\alpha_4} = \zeta^{N(P)}\overline{\Phi}^{(2)}_{\mathrm{sym},\mathfrak{s},\alpha_{P(1)}\alpha_{P(2)}\alpha_{P(3)}\alpha_{P(4)}} , \tag{270}$$

  hence the substitution $U \rightarrow \overline{\Phi}^{(2)}_{\mathrm{sym},\mathfrak{s}}$ in Eq. (262). Finally, we note that, according to expression (158) of the LW functional, a choice $N_{\mathrm{SCPT}} \leq N_{\mathrm{max}}/2$ induces that $\overline{\Phi}^{(N_{\mathrm{max}}+1)}_{\mathrm{SCPT},N_{\mathrm{SCPT}},\mathfrak{s}}[U,G]$ vanishes for all $\mathfrak{s}$, which implies that the mC-flow reduces to the tC-flow in this case.

Let us then focus on our numerical applications of the C-flow to the (0+0)-D $O(N)$ model. We will more particularly focus on the situation where $N = 1$, which will enable us to investigate more readily high truncation orders (up to $N_{\mathrm{max}} = 10$ to be specific), thus illustrating clearly why the C-flow implementation of the 2PI-FRG is not suited to construct reliable approximation schemes. For the (0+0)-D $O(N)$ model at $N=1$, the flow equations (249) to (252) reduce to:

$$\dot{\overline{G}}_{\mathfrak{s}} = -\overline{G}_{\mathfrak{s}}^{2}\left(\dot{C}_{\mathfrak{s}}^{-1} - \dot{\overline{\Sigma}}_{\mathfrak{s}}\right) , \tag{271}$$

$$\Delta\dot{\Gamma}_{\mathfrak{s}}^{(\mathrm{2PI})} = \frac{1}{2}\dot{C}_{\mathfrak{s}}^{-1}\left(\overline{G}_{\mathfrak{s}} - C_{\mathfrak{s}}\right) , \tag{272}$$

$$\dot{\overline{\Sigma}}_{\mathfrak{s}} = -\frac{1}{2}\dot{\overline{G}}_{\mathfrak{s}}\overline{\Phi}_{\mathfrak{s}}^{(2)} , \tag{273}$$

$$\dot{\overline{\Phi}}_{\mathfrak{s}}^{(n)} = \frac{1}{2}\dot{\overline{G}}_{\mathfrak{s}}\overline{\Phi}_{\mathfrak{s}}^{(n+1)} \quad \forall n \geq 2 , \tag{274}$$

where we have introduced the shorthand notations $\overline{G}_{\mathfrak{s}} \equiv \overline{G}_{\mathfrak{s},11}$, $\overline{\Sigma}_{\mathfrak{s}} \equiv \overline{\Sigma}_{\mathfrak{s},11}$ and $\overline{\Phi}_{\mathfrak{s}}^{(n)} \equiv \overline{\Phi}_{\mathfrak{s},(1,1)\cdots(1,1)}^{(n)}$ $\forall n$. To implement the tC-flow up to $N_{\mathrm{max}}=10$, we will need to consider the perturbative expression of the LW functional up to order 5 in $\lambda$, which reads according to Ref. [218]:

$$\begin{aligned}\Phi_{\mathrm{SCPT}}(U,G) = {}& \frac{1}{8}\lambda G^{2} - \frac{1}{48}\lambda^{2}G^{4} + \frac{1}{48}\lambda^{3}G^{6} - \frac{5}{128}\lambda^{4}G^{8} \\ & + \frac{101}{960}\lambda^{5}G^{10} + \mathcal{O}\left(\lambda^{6}\right) ,\end{aligned} \tag{275}$$

still at $N=1$. From this, we infer the following initial conditions for the 2PI vertices:

$$\overline{\Phi}_{\mathfrak{s}=\mathfrak{s}_{\mathrm{i}}}^{(2)} = \lambda , \tag{276}$$

$$\overline{\Phi}_{\mathfrak{s}=\mathfrak{s}_{\mathrm{i}}}^{(4)} = -8\lambda^{2} , \tag{277}$$

$$\overline{\Phi}_{\mathfrak{s}=\mathfrak{s}_{\mathrm{i}}}^{(6)} = 960\lambda^{3} , \tag{278}$$

$$\overline{\Phi}_{\mathfrak{s}=\mathfrak{s}_{\mathrm{i}}}^{(8)} = -403200\lambda^{4} , \tag{279}$$

$$\overline{\Phi}_{\mathfrak{s}=\mathfrak{s}_{\mathrm{i}}}^{(10)} = 390942720\lambda^{5} , \tag{280}$$

where we have taken into account that the identity matrix of the bosonic index formalism (given by Eq. (147)) reduces to 2 (and not 1) in the (0+0)-D limit at $N=1$, i.e. Eq. (147) becomes in the (0+0)-D limit:

$$\mathcal{I}_{(a_{1},a_{1}')(a_{2},a_{2}')} \equiv \frac{\partial G_{a_{1}a_{1}'}}{\partial G_{a_{2}a_{2}'}} = \delta_{a_{1}a_{2}}\delta_{a_{1}'a_{2}'} + \delta_{a_{1}a_{2}'}\delta_{a_{1}'a_{2}} , \tag{281}$$

which yields at $N=1$:

$$\mathcal{I} \equiv \frac{\partial G}{\partial G} = 2 , \tag{282}$$

as opposed to standard derivation rules.[40] The truncation conditions underpinning the mC-flow can also be derived from expression (275) alongside with Eq. (282) as derivation rule. We have notably investigated the following combinations for $(N_{\mathrm{max}},N_{\mathrm{SCPT}})$[41,42]:

- At $N_{\mathrm{max}}=2$:
  - At $N_{\mathrm{SCPT}}=2$:

$$\begin{aligned}\overline{\Phi}_{\mathfrak{s}}^{(3)} = {}& \left.\overline{\Phi}_{\mathrm{SCPT},N_{\mathrm{SCPT}}=2,\mathfrak{s}}^{(3)}\right|_{\lambda\rightarrow\overline{\Phi}_{\mathfrak{s}}^{(2)}} \\ = {}& -4\left(\overline{\Phi}_{\mathfrak{s}}^{(2)}\right)^{2}\overline{G}_{\mathfrak{s}} .\end{aligned} \tag{283}$$

  - At $N_{\mathrm{SCPT}}=3$:

$$\begin{aligned}\overline{\Phi}_{\mathfrak{s}}^{(3)} = {}& \left.\overline{\Phi}_{\mathrm{SCPT},N_{\mathrm{SCPT}}=3,\mathfrak{s}}^{(3)}\right|_{\lambda\rightarrow\overline{\Phi}_{\mathfrak{s}}^{(2)}} \\ = {}& -4\left(\overline{\Phi}_{\mathfrak{s}}^{(2)}\right)^{2}\overline{G}_{\mathfrak{s}} + 20\left(\overline{\Phi}_{\mathfrak{s}}^{(2)}\right)^{3}\overline{G}_{\mathfrak{s}}^{3} .\end{aligned} \tag{284}$$

- At $N_{\mathrm{max}}=3$:
  - At $N_{\mathrm{SCPT}}=2$:

$$\begin{aligned}\overline{\Phi}_{\mathfrak{s}}^{(4)} = {}& \left.\overline{\Phi}_{\mathrm{SCPT},N_{\mathrm{SCPT}}=2,\mathfrak{s}}^{(4)}\right|_{\lambda\rightarrow\overline{\Phi}_{\mathfrak{s}}^{(2)}} \\ = {}& -8\left(\overline{\Phi}_{\mathfrak{s}}^{(2)}\right)^{2} .\end{aligned} \tag{285}$$

  - At $N_{\mathrm{SCPT}}=3$:

$$\begin{aligned}\overline{\Phi}_{\mathfrak{s}}^{(4)} = {}& \left.\overline{\Phi}_{\mathrm{SCPT},N_{\mathrm{SCPT}}=3,\mathfrak{s}}^{(4)}\right|_{\lambda\rightarrow\overline{\Phi}_{\mathfrak{s}}^{(2)}} \\ = {}& -8\left(\overline{\Phi}_{\mathfrak{s}}^{(2)}\right)^{2} + 120\left(\overline{\Phi}_{\mathfrak{s}}^{(2)}\right)^{3}\overline{G}_{\mathfrak{s}}^{2} .\end{aligned} \tag{286}$$

- At $N_{\mathrm{max}}=4$:
  - At $N_{\mathrm{SCPT}}=3$:

$$\begin{aligned}\overline{\Phi}_{\mathfrak{s}}^{(5)} = {}& \left.\overline{\Phi}_{\mathrm{SCPT},N_{\mathrm{SCPT}}=3,\mathfrak{s}}^{(5)}\right|_{\lambda\rightarrow\overline{\Phi}_{\mathfrak{s}}^{(2)}} \\ = {}& 480\left(\overline{\Phi}_{\mathfrak{s}}^{(2)}\right)^{3}\overline{G}_{\mathfrak{s}} .\end{aligned} \tag{287}$$

Moreover, for all our C-flow applications, we use the same cutoff function as in our 1PI-FRG study, namely at $N=1$:

$$C_{\mathfrak{s}}^{-1} = C^{-1} + R_{\mathfrak{s}} = m^{2} + R_{\mathfrak{s}} , \tag{288}$$

[40] The 2PI-FRG flow equations of the (0+0)-D $O(N)$ model can either be derived via standard derivation rules based on the identity $\mathcal{I} \equiv \frac{\partial G}{\partial G} = 1$ or by taking the (0+0)-D limit of their more general versions (written in terms of bosonic indices) using Eq. (282). We always follow the latter procedure in this study but solving the equations thus obtained in both situations leads in principle to identical results (see appendix F.6 of Ref. [46] for a more detailed discussion on this equivalence).

[41] Recall that, as discussed right below Eq. (270), it is pointless to investigate the mC-flow with $N_{\mathrm{SCPT}} \leq N_{\mathrm{max}}/2$ as it reduces to the tC-flow in this situation.

[42] According to the definition of $\overline{\Phi}_{\mathrm{sym},\mathfrak{s}}^{(2)}$ given by Eq. (263), we have $\overline{\Phi}_{\mathrm{sym},\mathfrak{s}}^{(2)} = \overline{\Phi}_{\mathfrak{s}}^{(2)}$ for the studied toy model at $N=1$.

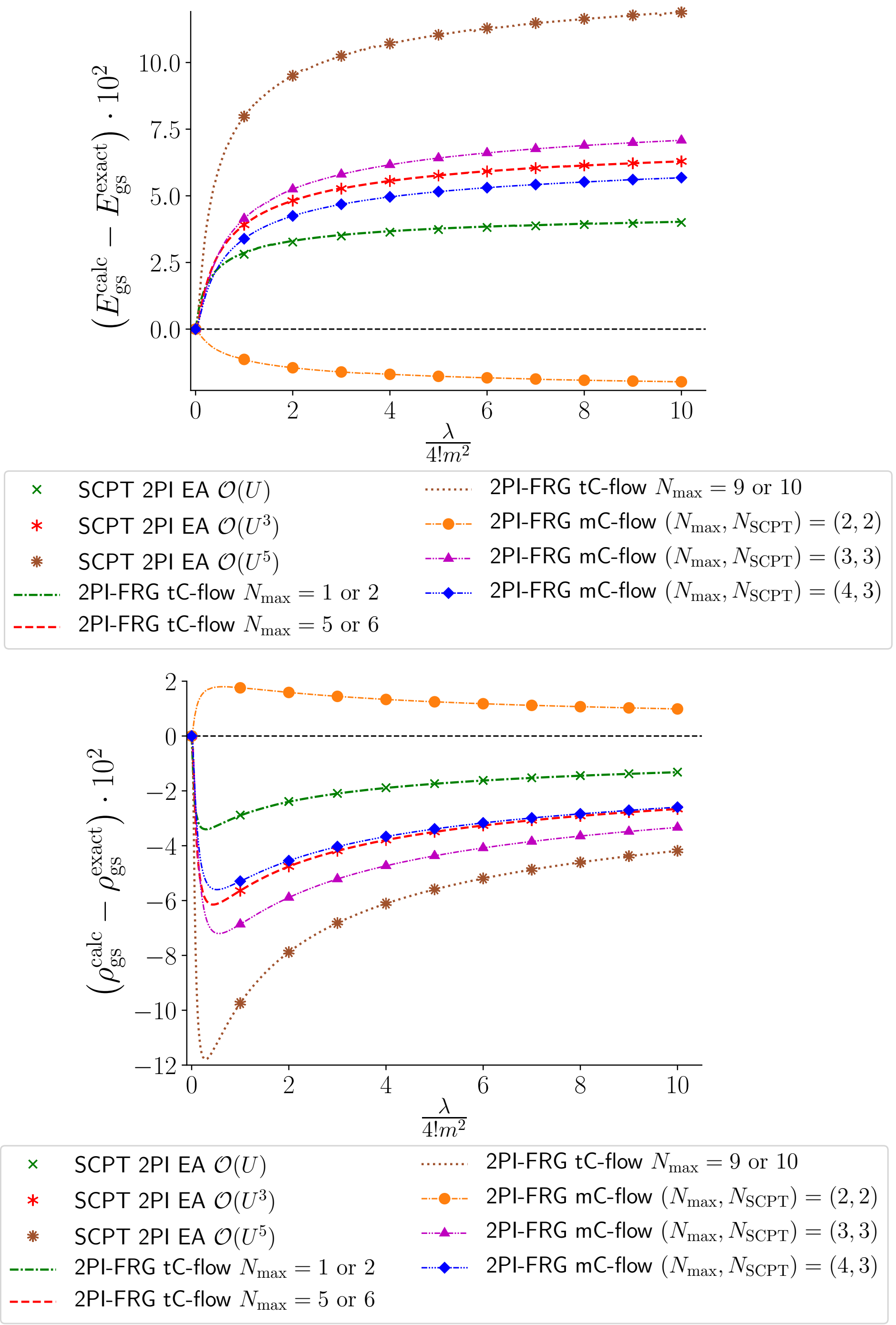

**Fig. 15.** Difference between the calculated gs energy $E_{\mathrm{gs}}^{\mathrm{calc}}$ or density $\rho_{\mathrm{gs}}^{\mathrm{calc}}$ and the corresponding exact solution $E_{\mathrm{gs}}^{\mathrm{exact}}$ or $\rho_{\mathrm{gs}}^{\mathrm{exact}}$ at $m^2 = +1$ and $N = 1$. The SCPT result labeled "SCPT 2PI EA $\mathcal{O}(U^n)$" is identical to that corresponding to the legend "SCPT original 2PI EA $\mathcal{O}\big(\hbar^{n+1}\big)$" in the figures of paper I.

with

$$R_{\mathfrak{s}} = \mathfrak{s}^{-1} - 1 \;, \tag{289}$$

which satisfies the required boundary conditions set by Eqs. (242) as the flow parameter $\mathfrak{s}$ still runs from $\mathfrak{s}_{\mathrm{i}} = 0$ to $\mathfrak{s}_{\mathrm{f}} = 1$ during the flow. Finally, the gs energy and density of our toy model are determined within the C-flow also from Eqs. (156) and (157) (that were previously exploited for the pU-flow), i.e.:

$$E_{\mathrm{gs}}^{\text{2PI-FRG;C-flow}} = \overline{\Gamma}^{(\mathrm{2PI})}_{\mathfrak{s}=\mathfrak{s}_{\mathrm{f}}} \underset{N=1}{=} \Delta\overline{\Gamma}^{(\mathrm{2PI})}_{\mathfrak{s}=\mathfrak{s}_{\mathrm{f}}} - \frac{1}{2}\ln\left(\frac{2\pi}{m^2}\right) , \tag{290}$$

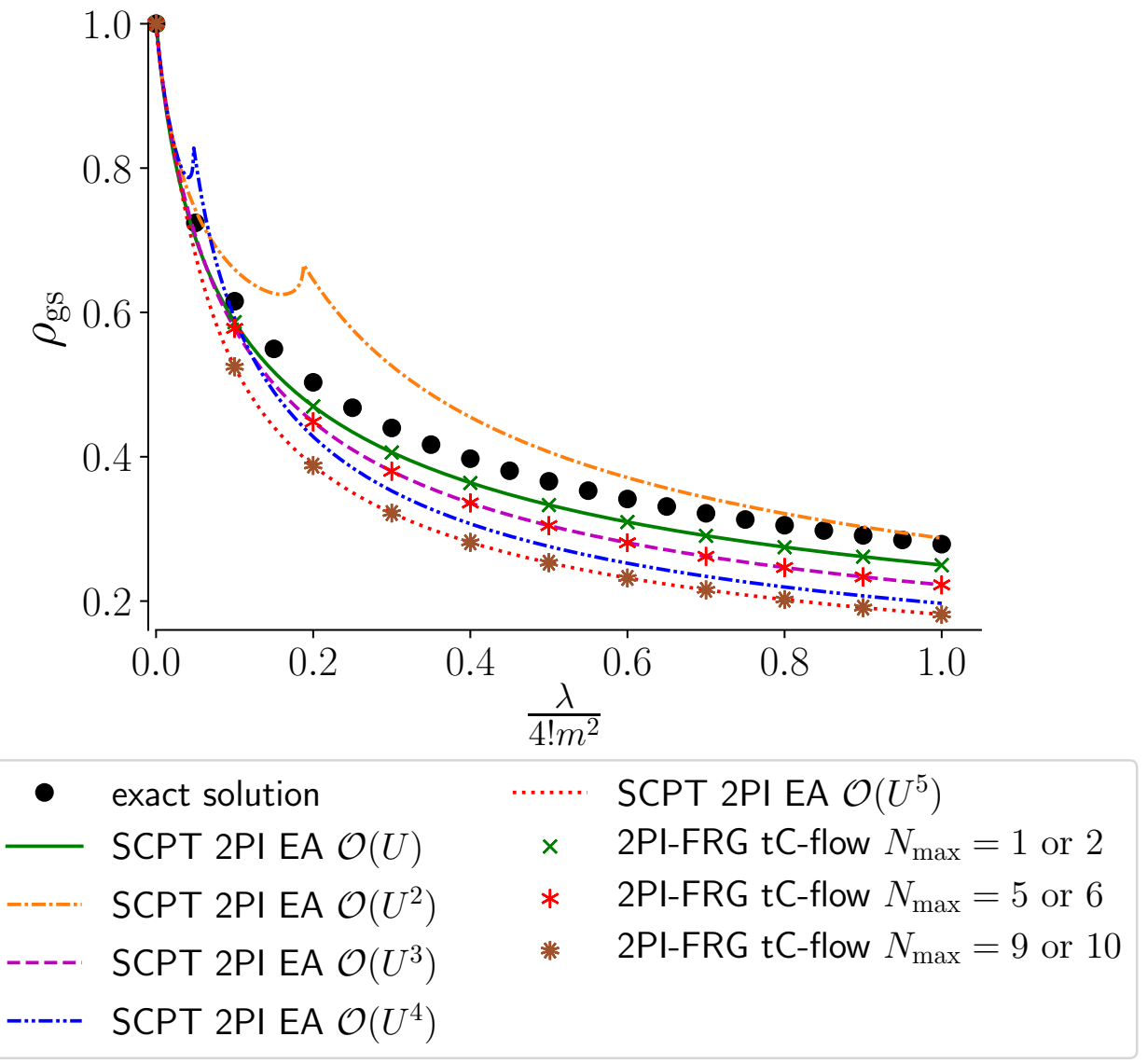

**Fig. 16.** Gs density $\rho_{\mathrm{gs}}$ calculated at $m^2 = +1$ and $N = 1$, and compared with the corresponding exact solution (black dots). The SCPT result labeled "SCPT 2PI EA $\mathcal{O}(U^n)$" is identical to that corresponding to the legend "SCPT original 2PI EA $\mathcal{O}(\hbar^{n+1})$" in the figures of paper I.

$$\rho_{\mathrm{gs}}^{\text{2PI-FRG;C-flow}} = \frac{1}{N}\sum_{a=1}^{N} \overline{G}_{\mathfrak{s}=\mathfrak{s}_f,aa} \underset{N=1}{=} \overline{G}_{\mathfrak{s}=\mathfrak{s}_f} . \tag{291}$$

The tC-flow and mC-flow results obtained in this way for the (0+0)-D $O(N)$ model at $N = 1$ are reported in Fig. 15. In this figure, we can see that, for both $E_{\mathrm{gs}}$ and $\rho_{\mathrm{gs}}$, the tC-flow curves are further and further away from the exact solution (except for $\lambda/4! \ll 1$) as the truncation order $N_{\mathrm{max}}$ increases. This behavior is consistent with the equivalence between the tC-flow and SCPT mentioned earlier and highlighted in Ref. [179]: such a worsening is thus a manifestation of the divergent character of the series underlying SCPT, which is at the heart of paper I.

We can also note that there are no tC-flow results for the truncation orders $N_{\mathrm{max}} = 3$ or 4 and $N_{\mathrm{max}} = 7$ or 8 in Fig. 15 as we face the same stiffness issues with the used numerical tools[12] as those encountered in section 2 for our 1PI-FRG applications in the broken-symmetry phase. However, all those tC-flow calculations (with or without stiffness issues) have been performed in the unbroken-symmetry phase, which suggests that the origins of these stiffness problems for the 1PI-FRG flow equations (only occurring in the broken-symmetry phase) on the one hand and for the 2PI-FRG tC-flow on the other hand are different. Despite such limitations, it is rather fruitful to further exploit the equivalence between the tC-flow implementation of the 2PI-FRG and SCPT. To that end, we recall that, as we have done in paper I, SCPT based on $\Gamma^{(2\mathrm{PI})}(G)$ is carried out by solving the gap equations extremizing $\Gamma^{(2\mathrm{PI})}(G)$ with respect to $G$ and then picking up the physical solution[43] $\overline{G}$. Remarkably, the initial conditions of the C-flow are such that the tC-flow results coincide with those of our physical solutions for $N_{\mathrm{max}} = 1$ or 2, $N_{\mathrm{max}} = 5$ or 6 and $N_{\mathrm{max}} = 9$ or 10 according to Fig. 15. Nevertheless, there is a change of physical solutions in the perturbative regime (i.e. for $\lambda/4! \ll 1$) for SCPT applied up to order $\mathcal{O}(\lambda^2) = \mathcal{O}(U^2)$ and $\mathcal{O}(\lambda^4) = \mathcal{O}(U^4)$, as can be seen in Fig. 16. These correspond respectively to the tC-flow approach with truncation orders $N_{\mathrm{max}} = 3$ or 4 and $N_{\mathrm{max}} = 7$ or 8, which are precisely the $N_{\mathrm{max}}$ values where the stiffness problem arises. This illustrates that the tC-flow is not suited to fully reproduce SCPT when there is a change of physical solutions involved in the latter framework for the chosen truncation of the 2PI EA. This limitation can be attributed to the fact that the initial conditions for the tC-flow are fixed once and for all (i.e. regardless of the values of coupling constants) from the perturbative expression (158) of the LW functional, which does not allow for reproducing the change of solutions observed in Fig. 16. This also implies that the stiffness problem arising in our tC-flow calculations is inherent to the C-flow formalism and not to the numerical tools exploited in this study[12].

Regarding the mC-flow, we can see in Fig. 15 that the corresponding ansatz underlying the truncation manages to cure the aforementioned stiffness problem for $N_{\mathrm{max}} = 3$ or 4, but not for all choices of $N_{\mathrm{SCPT}}$: the combination $(N_{\mathrm{max}}, N_{\mathrm{SCPT}}) = (3, 2)$ set by Eq. (285) still suffers from it for instance. Besides, it also introduces this issue at the truncation order $N_{\mathrm{max}} = 2$ which is not affected in the framework of the tC-flow: this problem arises e.g. at the truncation $(N_{\mathrm{max}}, N_{\mathrm{SCPT}}) = (2, 3)$ set by Eq. (284), hence its absence from Fig. 15. Furthermore, we can also see that mC-flow results might deteriorate as $N_{\mathrm{max}}$ and/or $N_{\mathrm{SCPT}}$ increase(s), as can be noticed by comparing the curves associated with $(N_{\mathrm{max}}, N_{\mathrm{SCPT}}) = (2, 2)$ and $(N_{\mathrm{max}}, N_{\mathrm{SCPT}}) = (3, 3)$ in Fig. 15. This is an important drawback as it shows that we lose accuracy while incorporating explicitly more information in our truncation. The ansatz underlying the mC-flow truncation is therefore not reliable.

We have only discussed applications of the C-flow version of the 2PI-FRG to the regime of the studied toy model with $m^2 > 0$ so far. The corresponding formalism is actually not suited to treat the regime with $m^2 < 0$: if one sets $m^2$ equal to a negative value to solve the equation system made of Eqs. (271) to (274), the results thus obtained would be unphysical (with e.g. a negative estimate of the gs density $\rho_{\mathrm{gs}}$). A more relevant way the tackle the unbroken-symmetry regime with a C-flow approach would be to add a linear source $J$ in the generating functional (106), i.e. to add a source term $\int_\alpha J_\alpha \widetilde{\psi}_\alpha$ alongside with $\frac{1}{2}\int_{\alpha,\alpha'} \widetilde{\psi}_\alpha K_{\alpha\alpha'} \widetilde{\psi}_{\alpha'}$ in definition (106). However, besides complicating the C-flow formalism presented in this appendix, we do not expect such an extension to lift the limitations of the C-flow approach identified from Fig. 15.

[43] The physical solution $\overline{G}$ leading to our SCPT results shown in Figs. 15 and 16 is defined here as the solution of the gap equation (for the propagator $G$) yielding the calculated complex gs density $\rho_{\mathrm{gs,comp}}^{\mathrm{calc}}$ that is closest to the corresponding exact solution $\rho_{\mathrm{gs}}^{\mathrm{exact}}$, i.e. that gives us the smallest norm $\left|\rho_{\mathrm{gs,comp}}^{\mathrm{calc}} - \rho_{\mathrm{gs}}^{\mathrm{exact}}\right|$. The term "norm" should be understood here as the norm of a complex number as $\rho_{\mathrm{gs,comp}}^{\mathrm{calc}}$ might have a non-zero imaginary part. The results displayed in Figs. 15 and 16 are then defined as $\rho_{\mathrm{gs}}^{\mathrm{calc}} = \mathcal{R}e\big(\rho_{\mathrm{gs,comp}}^{\mathrm{calc}}\big)$.

Hence, although we have only discussed here numerical applications of the C-flow to the unbroken-symmetry regime of our toy model at $N=1$,[44] we have highlighted here important limitations of this 2PI-FRG implementation, which are inherent to this approach rather than to the model under consideration. More precisely, neither the tC-flow nor the mC-flow are systematically improvable in a reliable fashion: i) the tC-flow worsens with increasing truncation orders and does not enable us to go reliably beyond the Hartree-Fock result of SCPT since it just reproduces SCPT without resummation; ii) the ansatz underlying the mC-flow does not seem reliable either as the corresponding results might also deteriorate as $N_{\text{max}}$ and/or $N_{\text{SCPT}}$ increase(s). As an outlook of the present study, one could further investigate the properties of the series (128) representing the 2PI EA in the 2PI-FRG framework to see whether a combination of the tC-flow with resummation theory might be relevant to overcome limitation i). We thus conclude that, unless one considers particularly weakly-coupled regimes where SCPT without resummation (and therefore the tC-flow) can improve with increasing truncation orders, the 2PI-FRG C-flow exploited here is not suited to design well-controlled systematically improvable approximation schemes.

### B.2 CU-flow

The CU-flow version of the 2PI-FRG [45] essentially amounts to combining the C-flow and the U-flow, and more specifically the tC-flow and the pU-flow, together. A noticeable fact is that, as the C-flow and (if the corresponding cutoff function is chosen accordingly) the U-flow, such an approach also carries out the Wilsonian momentum-shell integration. Hence, the CU-flow relies on the transformation $C^{-1} \rightarrow C^{-1}_{\mathfrak{s}} = C^{-1} + R^{(C)}_{\mathfrak{s}}$ (or equivalently $C^{-1} \rightarrow C^{-1}_{\mathfrak{s}} = R^{(C)}_{\mathfrak{s}} C^{-1}$) combined with $U \rightarrow U_{\mathfrak{s}} = U + R^{(U)}_{\mathfrak{s}}$ (or equivalently $U \rightarrow U_{\mathfrak{s}} = R^{(U)}_{\mathfrak{s}} U$), with:

$$\begin{cases} C_{\mathfrak{s}=\mathfrak{s}_{\text{i}},\gamma} = 0 \quad \forall\gamma \,. & (292\text{a}) \\ C_{\mathfrak{s}=\mathfrak{s}_{\text{f}}} = C \,. & (292\text{b}) \\ U_{\mathfrak{s}=\mathfrak{s}_{\text{i}},\gamma_1\gamma_2} = 0 \quad \forall\gamma_1,\gamma_2 \,. & (292\text{c}) \\ U_{\mathfrak{s}=\mathfrak{s}_{\text{f}}} = U \,. & (292\text{d}) \end{cases}$$

As can be seen from Eqs. (242) and (136), the boundary conditions for $C_{\mathfrak{s}}$ and $U_{\mathfrak{s}}$ are therefore exactly identical to those of the C-flow and the U-flow, respectively. Due to the initial condition (292a), the starting point of the CU-flow suffers from the same divergence problem as that of the C-flow, which is why we will consider the functional $\Delta\overline{\Gamma}^{(\text{2PI})}_{\mathfrak{s}}$ (defined by Eq. (248)) in the present framework as well. Furthermore, we can already expect from the master equation (126) expressing $\dot{\Gamma}^{(\text{2PI})}_{\mathfrak{s}}[G]$ that the differential equations underlying the CU-flow contain contributions from both the C-flow and the U-flow, besides the flow equation expressing $\dot{\overline{G}}_{\mathfrak{s}}$ which is given by Eq. (249) as in the C-flow. Actually, since the LW functional is an invariant of the C-flow (according to Eq. (243)), the CU-flow equations expressing $\dot{\overline{\Sigma}}_{\mathfrak{s}}$ and $\dot{\overline{\Phi}}^{(n)}_{\mathfrak{s}}$ (with $n \geq 2$) coincide in principle with those of the pU-flow, and notably with Eq. (141) for $\dot{\overline{\Phi}}^{(2)}_{\mathfrak{s}}$ and Eq. (142) for $\dot{\overline{\Phi}}^{(3)}_{\mathfrak{s}}$. There is however a subtlety that implies that Eq. (140) (which expresses $\dot{\overline{\Sigma}}_{\mathfrak{s}}$ for the pU-flow) is not valid in the present situation.[45] With the help of the chain rule of Eq. (139a), we can rewrite this equation in a form exploitable for the CU-flow, which gives us:

$$\begin{aligned} \dot{\overline{\Sigma}}_{\mathfrak{s},\gamma} = & -\frac{1}{3}\bigg[2\left(\mathcal{I}+\overline{\Pi}_{\mathfrak{s}}\overline{\Phi}^{(2)}_{\mathfrak{s}}\right)^{\text{inv}} \dot{U}_{\mathfrak{s}} \left(\mathcal{I}+\overline{\Pi}_{\mathfrak{s}}\overline{\Phi}^{(2)}_{\mathfrak{s}}\right)^{\text{inv}} \\ & + \dot{U}_{\mathfrak{s}}\bigg]_{\alpha\hat{\alpha}\hat{\alpha}'\alpha'} \overline{G}_{\mathfrak{s},\hat{\gamma}} \\ & + \frac{1}{6}\dot{U}_{\mathfrak{s},\hat{\gamma}_1\hat{\gamma}_2} \overline{W}^{(2)}_{\mathfrak{s},\hat{\gamma}_2\hat{\gamma}_3} \overline{\Phi}^{(3)}_{\mathfrak{s},\gamma\hat{\gamma}_3\hat{\gamma}_4} \overline{W}^{(2)}_{\mathfrak{s},\hat{\gamma}_4\hat{\gamma}_1} - \dot{\overline{G}}_{\mathfrak{s},\hat{\gamma}} \overline{\Phi}^{(2)}_{\mathfrak{s},\hat{\gamma}\gamma} \,. \end{aligned} \tag{293}$$

We then discuss the initial conditions for the CU-flow. On the one hand, as for the C-flow, imposing $C_{\mathfrak{s}=\mathfrak{s}_{\text{i}},\gamma} = 0$ $\forall\gamma$ (i.e. Eq. (292a)) induces that:

$$\overline{G}_{\mathfrak{s}=\mathfrak{s}_{\text{i}},\gamma} = 0 \quad \forall\gamma \,, \tag{294}$$

$$\Delta\overline{\Gamma}^{(\text{2PI})}_{\mathfrak{s}=\mathfrak{s}_{\text{i}}} = 0 \,. \tag{295}$$

On the other hand, as for the pU-flow, the condition $U_{\mathfrak{s}=\mathfrak{s}_{\text{i}},\gamma_1\gamma_2} = 0$ $\forall\gamma_1,\gamma_2$ (i.e. Eq. (292c)) implies that:

$$\overline{\Sigma}_{\mathfrak{s}=\mathfrak{s}_{\text{i}},\gamma} = 0 \quad \forall\gamma \,, \tag{296}$$

$$\overline{\Phi}^{(n)}_{\mathfrak{s}=\mathfrak{s}_{\text{i}},\gamma_1\cdots\gamma_n} = 0 \quad \forall\gamma_1,\cdots,\gamma_n, \ \forall n \geq 2 \,. \tag{297}$$

Moreover, the infinite tower of differential equations for the CU-flow is closed by enforcing the same truncation condition as in the tC-flow and the pU-flow, i.e.:

$$\dot{\overline{\Phi}}^{(n)}_{\mathfrak{s}} = \overline{\Phi}^{(n)}_{\mathfrak{s}=\mathfrak{s}_{\text{i}}} \quad \forall\mathfrak{s}, \ \forall n > N_{\text{max}} \,, \tag{298}$$

which is why we mentioned that the CU-flow is essentially a merger of the latter two approaches.

[44] See Ref. [46] for additional results from the 2PI-FRG C-flow at $N=1$ or 2 in the original or mixed representation of the (0+0)-D $O(N)$ model.

[45] We refer to appendix F.4.3 of Ref. [46] for a technical explanation on this point.

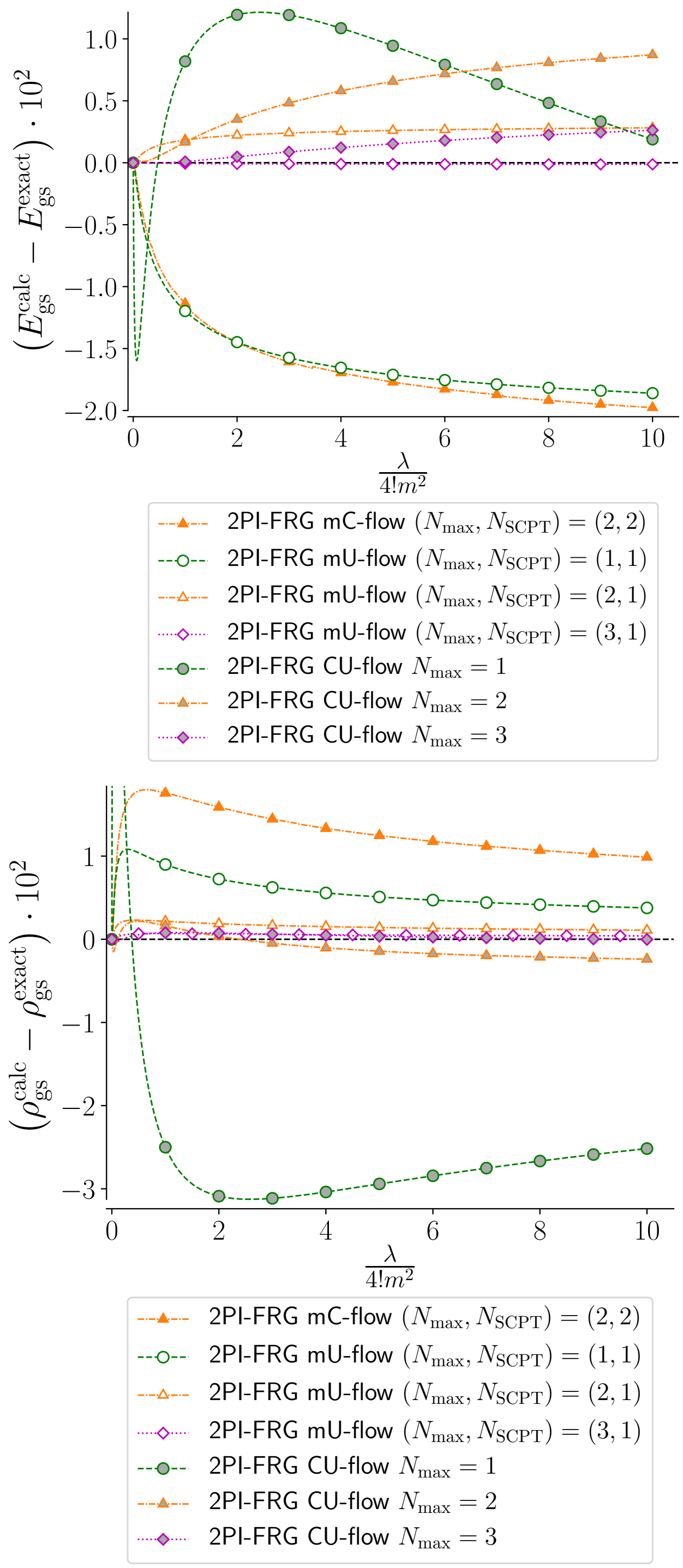

**Fig. 17.** Difference between the calculated gs energy $E_{\mathrm{gs}}^{\mathrm{calc}}$ or density $\rho_{\mathrm{gs}}^{\mathrm{calc}}$ and the corresponding exact solution $E_{\mathrm{gs}}^{\mathrm{exact}}$ or $\rho_{\mathrm{gs}}^{\mathrm{exact}}$ at $m^2 = +1$ and $N = 1$. The mC-flow curve represents the best C-flow result up to $N_{\mathrm{max}} = 3$ whereas the green, orange and purple mU-flow curves correspond to the best U-flow results obtained at $N_{\mathrm{max}} = 1, 2$ and 3, respectively.

As for the C-flow, we will only apply the CU-flow to the (0+0)-D $O(N)$ model at $N = 1$, which we will consider sufficient to make our point, showing notably that, in the unbroken-symmetry regime, the CU-flow version of the 2PI-FRG can achieve similar performances to those of the mU-flow at $N_{\mathrm{SCPT}} = 1$. We will thus exploit the following CU-flow equations to treat our toy model at $N = 1$:

$$\dot{\overline{G}}_{\mathfrak{s}} = -\overline{G}_{\mathfrak{s}}^2 \left(\dot{C}_{\mathfrak{s}}^{-1} - \dot{\overline{\Sigma}}_{\mathfrak{s}}\right) , \quad (299)$$

$$\begin{aligned}\Delta\dot{\Gamma}^{(2\mathrm{PI})}_{\mathfrak{s}} = & \frac{1}{2}\dot{C}_{\mathfrak{s}}^{-1}\left(\overline{G}_{\mathfrak{s}} - C_{\mathfrak{s}}\right) \\ & + \frac{\dot{U}_{\mathfrak{s}}}{24}\left(4\left(2\overline{G}_{\mathfrak{s}}^{-2} + \overline{\Phi}^{(2)}_{\mathfrak{s}}\right)^{-1} + \overline{G}_{\mathfrak{s}}^2\right) ,\end{aligned} \quad (300)$$

$$\begin{aligned}\dot{\overline{\Sigma}}_{\mathfrak{s}} = & \frac{\dot{U}_{\mathfrak{s}}}{3}\overline{G}_{\mathfrak{s}}\left(2 + \overline{G}_{\mathfrak{s}}^2\overline{\Phi}^{(2)}_{\mathfrak{s}}\right)^{-2}\left(\frac{1}{2}\overline{G}_{\mathfrak{s}}^3\overline{\Phi}^{(3)}_{\mathfrak{s}} - 4\right) \\ & - \frac{\dot{U}_{\mathfrak{s}}}{6}\overline{G}_{\mathfrak{s}} - \frac{1}{2}\dot{\overline{G}}_{\mathfrak{s}}\overline{\Phi}^{(2)}_{\mathfrak{s}} ,\end{aligned} \quad (301)$$

$$\begin{aligned}\dot{\overline{\Phi}}^{(2)}_{\mathfrak{s}} = & \frac{\dot{U}_{\mathfrak{s}}}{6}\left(2\left(2\overline{G}_{\mathfrak{s}}^{-2} + \overline{\Phi}^{(2)}_{\mathfrak{s}}\right)^{-3}\left(8\overline{G}_{\mathfrak{s}}^{-3} - \overline{\Phi}^{(3)}_{\mathfrak{s}}\right)^2\right. \\ & - 64\left(2\overline{G}_{\mathfrak{s}}^{-2} + \overline{\Phi}^{(2)}_{\mathfrak{s}}\right)^{-2}\overline{G}_{\mathfrak{s}}^{-4} \\ & \left. + \left(2\overline{G}_{\mathfrak{s}}^{-2} + \overline{\Phi}^{(2)}_{\mathfrak{s}}\right)^{-2}\left(16\overline{G}_{\mathfrak{s}}^{-4} - \overline{\Phi}^{(4)}_{\mathfrak{s}}\right) + 2\right) \\ & + \frac{1}{2}\dot{\overline{G}}_{\mathfrak{s}}\overline{\Phi}^{(3)}_{\mathfrak{s}} ,\end{aligned} \quad (302)$$

$$\begin{aligned}\dot{\overline{\Phi}}^{(3)}_{\mathfrak{s}} = & \frac{\dot{U}_{\mathfrak{s}}}{6}\left(2\overline{G}_{\mathfrak{s}}^{-2} + \overline{\Phi}^{(2)}_{\mathfrak{s}}\right)^{-2}\left(6\left(2\overline{G}_{\mathfrak{s}}^{-2} + \overline{\Phi}^{(2)}_{\mathfrak{s}}\right)^{-2}\right. \\ & \times\left(8\overline{G}_{\mathfrak{s}}^{-3} - \overline{\Phi}^{(3)}_{\mathfrak{s}}\right)^3 + 6\left(2\overline{G}_{\mathfrak{s}}^{-2} + \overline{\Phi}^{(2)}_{\mathfrak{s}}\right)^{-1} \\ & \times\left(8\overline{G}_{\mathfrak{s}}^{-3} - \overline{\Phi}^{(3)}_{\mathfrak{s}}\right)\left(16\overline{G}_{\mathfrak{s}}^{-4} - \overline{\Phi}^{(4)}_{\mathfrak{s}}\right) \\ & - 384\overline{G}_{\mathfrak{s}}^{-4}\left(\left(2\overline{G}_{\mathfrak{s}}^{-2} + \overline{\Phi}^{(2)}_{\mathfrak{s}}\right)^{-1}\left(8\overline{G}_{\mathfrak{s}}^{-3} - \overline{\Phi}^{(3)}_{\mathfrak{s}}\right)\right. \\ & \left.\left. - \overline{G}_{\mathfrak{s}}^{-1}\right) - \overline{\Phi}^{(5)}_{\mathfrak{s}}\right) + \frac{1}{2}\dot{\overline{G}}_{\mathfrak{s}}\overline{\Phi}^{(4)}_{\mathfrak{s}} ,\end{aligned} \quad (303)$$

where we used once again the shorthand notations $\overline{G}_{\mathfrak{s}} \equiv \overline{G}_{\mathfrak{s},11}$, $\overline{\Sigma}_{\mathfrak{s}} \equiv \overline{\Sigma}_{\mathfrak{s},11}$ and $\overline{\Phi}^{(n)}_{\mathfrak{s}} \equiv \overline{\Phi}^{(n)}_{\mathfrak{s},(1,1)\cdots(1,1)}$ $\forall n$. We will also use the same cutoff functions as in our previous C-flow and U-flow applications, i.e.:

$$C_{\mathfrak{s}}^{-1} = C^{-1} + R_{\mathfrak{s}}^{(C)} = m^2 + \mathfrak{s}^{-1} - 1 , \quad (304)$$

$$U_{\mathfrak{s}} = R_{\mathfrak{s}}^{(U)} U = \mathfrak{s}\lambda , \quad (305)$$

still at $N = 1$, and with $\mathfrak{s}$ also running from $\mathfrak{s}_{\mathrm{i}} = 0$ to $\mathfrak{s}_{\mathrm{f}} = 1$. The gs energy and density of the (0+0)-D $O(N)$ model will be extracted from the flowing objects of the CU-flow exactly as within the C-flow, namely:

$$E_{\mathrm{gs}}^{\text{2PI-FRG;CU-flow}} = \overline{\Gamma}^{(\text{2PI})}_{\mathfrak{s}=\mathfrak{s}_{\mathrm{f}}} \underset{N=1}{=} \Delta\overline{\Gamma}^{(\text{2PI})}_{\mathfrak{s}=\mathfrak{s}_{\mathrm{f}}} - \frac{1}{2}\ln\left(\frac{2\pi}{m^2}\right) , \tag{306}$$

$$\rho_{\mathrm{gs}}^{\text{2PI-FRG;CU-flow}} = \frac{1}{N}\sum_{a=1}^{N} \overline{G}_{\mathfrak{s}=\mathfrak{s}_{\mathrm{f}},aa} \underset{N=1}{=} \overline{G}_{\mathfrak{s}=\mathfrak{s}_{\mathrm{f}}} . \tag{307}$$

In Fig. 17 which shows the results thus obtained, the CU-flow estimates for $E_{\mathrm{gs}}$ and $\rho_{\mathrm{gs}}$ clearly outperform at $N_{\mathrm{max}} = 2$ the mC-flow curve which can be considered as our best C-flow result, as can be seen from Fig. 15. Furthermore, Fig. 17 shows that the CU-flow results are comparable to the best mU-flow ones (i.e. to the mU-flow results at $N_{\mathrm{SCPT}} = 1$) up to $N_{\mathrm{max}} = 3$ for both $E_{\mathrm{gs}}$ and $\rho_{\mathrm{gs}}$. Similarly to the previous analysis of our C-flow results in section B.1, we also expect the latter remarks to be general and notably that the CU-flow could also efficiently tackle $O(N)$ models at $N > 1$.[46] Note however that, for the same reasons as the C-flow, the CU-flow is not suited to treat the broken-symmetry regime of the studied toy model.

## C Treatment of the $O(N)$ symmetry

### C.1 1PI formalism

In this appendix, we will explain how the $O(N)$ symmetry is exploited to simplify the equations underlying the vertex expansion of the Wetterich equation within the unbroken-symmetry phase of the (0+0)-D $O(N)$ model. To that end, we start by considering the corresponding Wetterich equation, given by Eq. (28) that can be put in the form:

$$\dot{\Gamma}^{(\text{1PI})}_{k}\left(\vec{\phi}\right) = \frac{1}{2}\sum_{a_1,a_2=1}^{N} \dot{\boldsymbol{R}}_{k,a_1 a_2}\boldsymbol{G}_{k,a_2 a_1}\left(\vec{\phi}\right) , \tag{308}$$

where the propagator $\boldsymbol{G}_k(\vec{\phi})$ is defined by:

$$\boldsymbol{G}^{-1}_{k,a_1 a_2}\left(\vec{\phi}\right) \equiv \Gamma^{(\text{1PI})(2)}_{k,a_1 a_2}\left(\vec{\phi}\right) + \boldsymbol{R}_{k,a_1 a_2} . \tag{309}$$

We then briefly present the main ingredients of the vertex expansion of Eq. (308) in the unbroken-symmetry phase. In the latter phase, we have $\vec{\overline{\phi}}_k = \vec{0}\ \forall k$ and all 1PI vertices of odd order, i.e. $\overline{\Gamma}^{(\text{1PI})(n)}_k$ with $n$ odd, also vanish according to Eq. (44). According to this, we can simplify the Taylor expansion (30) of $\Gamma^{(\text{1PI})}_k(\vec{\phi})$ as follows:

$$\Gamma^{(\text{1PI})}_{k}\left(\vec{\phi}\right) = \overline{\Gamma}^{(\text{1PI})}_{k} + \sum_{\substack{n=2 \\ \{\text{n even}\}}}^{\infty} \frac{1}{n!} \sum_{a_1,\cdots,a_n=1}^{N} \overline{\Gamma}^{(\text{1PI})(n)}_{k,a_1\cdots a_n}\phi_{a_1}\cdots\phi_{a_n} , \tag{310}$$

[46] See Ref. [46] for additional results from the 2PI-FRG CU-flow in the framework of the (0+0)-D $O(N)$ model at $N = 2$.

where the curly braces below discrete sums indicate a constraint that must be satisfied by all terms of the sum in question and the last line of Eq. (310) thus only involves terms with even values of $n$. From Eq. (310), we can infer an expansion of the left-hand side of the Wetterich equation (308) in powers of the field, i.e.:

$$\dot{\Gamma}^{(\text{1PI})}_{k}\left(\vec{\phi}\right) = \dot{\overline{\Gamma}}^{(\text{1PI})}_{k} + \sum_{\substack{n=2 \\ \{\text{n even}\}}}^{\infty} \frac{1}{n!} \sum_{a_1,\cdots,a_n=1}^{N} \dot{\overline{\Gamma}}^{(\text{1PI})(n)}_{k,a_1\cdots a_n}\phi_{a_1}\cdots\phi_{a_n} . \tag{311}$$

Focusing then on the right-hand side of the Wetterich equation (308), one can first expand the propagator $\boldsymbol{G}_k(\vec{\phi})$ around $\vec{\phi} = \vec{\overline{\phi}}_k = \vec{0}\ \forall k$, still using Eq. (44) to simplify the relations thus derived. This leads to:

$\boldsymbol{G}_{k,a_1 a_2}\left(\vec{\phi}\right)$

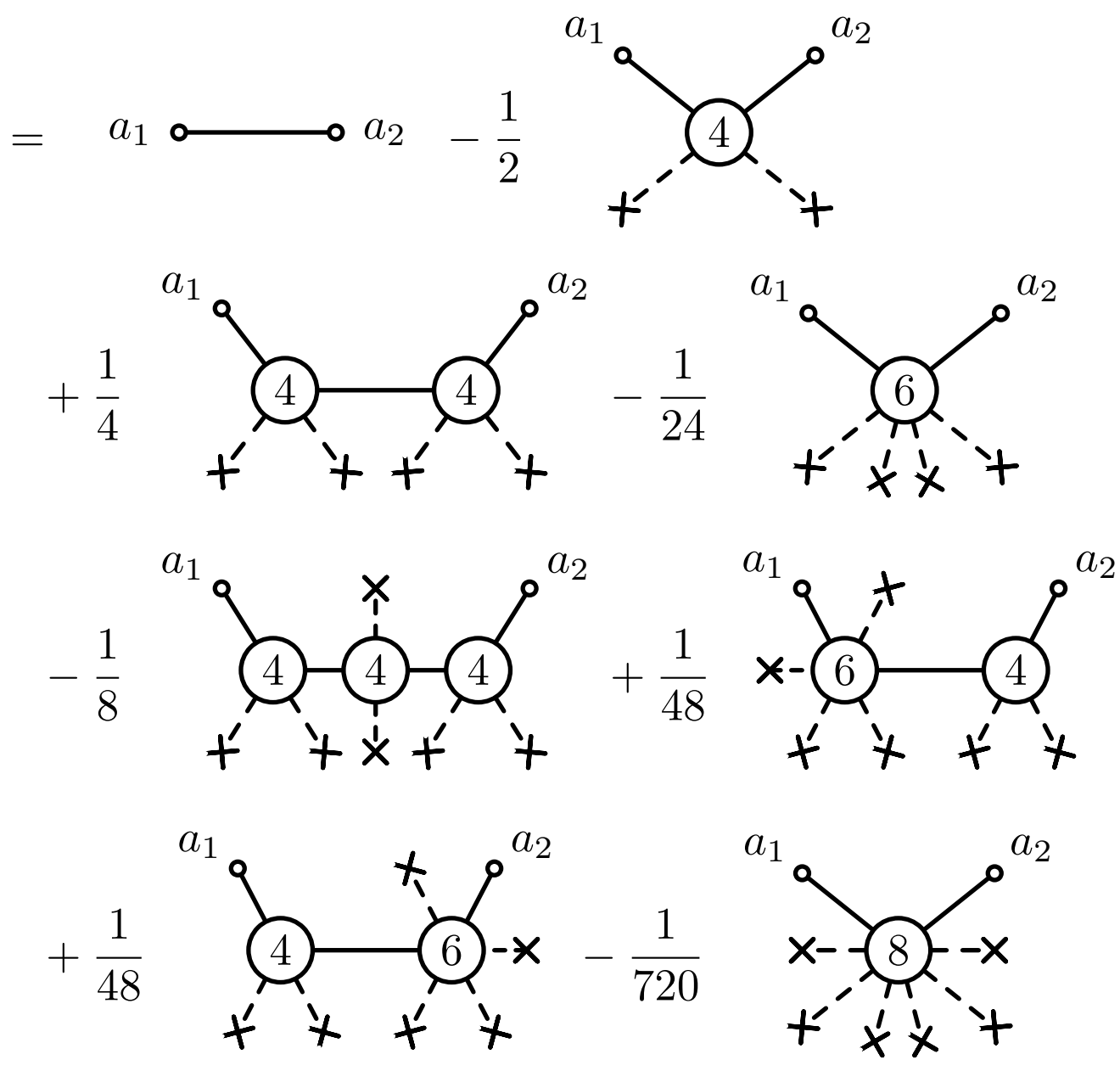

$$+\mathcal{O}\left(\left|\vec{\phi}\right|^8\right) , \tag{312}$$

where we have used the diagrammatic rules:

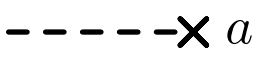
$$\rightarrow \quad \phi_a \,, \tag{313a}$$

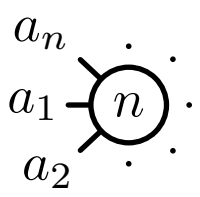
$$\rightarrow \quad \overline{\Gamma}^{(\mathrm{1PI})(n)}_{k,a_1\cdots a_n} \equiv \left.\frac{\partial^n \Gamma^{(\mathrm{1PI})}_k(\vec{\phi})}{\partial\phi_{a_1}\cdots\partial\phi_{a_n}}\right|_{\vec{\phi}=\vec{0}} \,, \tag{313b}$$

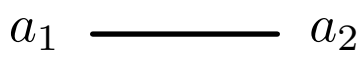
$$\rightarrow \quad \overline{\boldsymbol{G}}_{k,a_1a_2} \equiv \boldsymbol{G}_{k,a_1a_2}\left(\vec{\phi}=\vec{0}\right) \,, \tag{313c}$$

and the empty dots just indicate external points. As a next step, we insert this diagrammatic expression into the right-hand side of Eq. (308). As a result of sums over color indices in the latter relation, this basically amounts to joining the external points in Eq. (312) as follows:

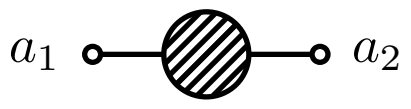
$$\longrightarrow \tag{314}$$
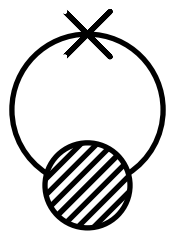

where the cross indicates an insertion of $\dot{\boldsymbol{R}}_k$ according to the rule:

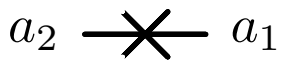
$$\rightarrow \quad \dot{\boldsymbol{R}}_{k,a_1a_2} = \dot{R}_k\delta_{a_1a_2} \,, \tag{315}$$

and the shaded blobs encompass any combinations of 1PI vertices and field insertions (313a) that can be found in Eq. (312). Hence, the right-hand side of Eq. (308) becomes in this way:

$$\frac{1}{2}\sum_{a_1,a_2=1}^{N} \dot{\boldsymbol{R}}_{k,a_1a_2}\boldsymbol{G}_{k,a_2a_1}\left(\vec{\phi}\right)$$

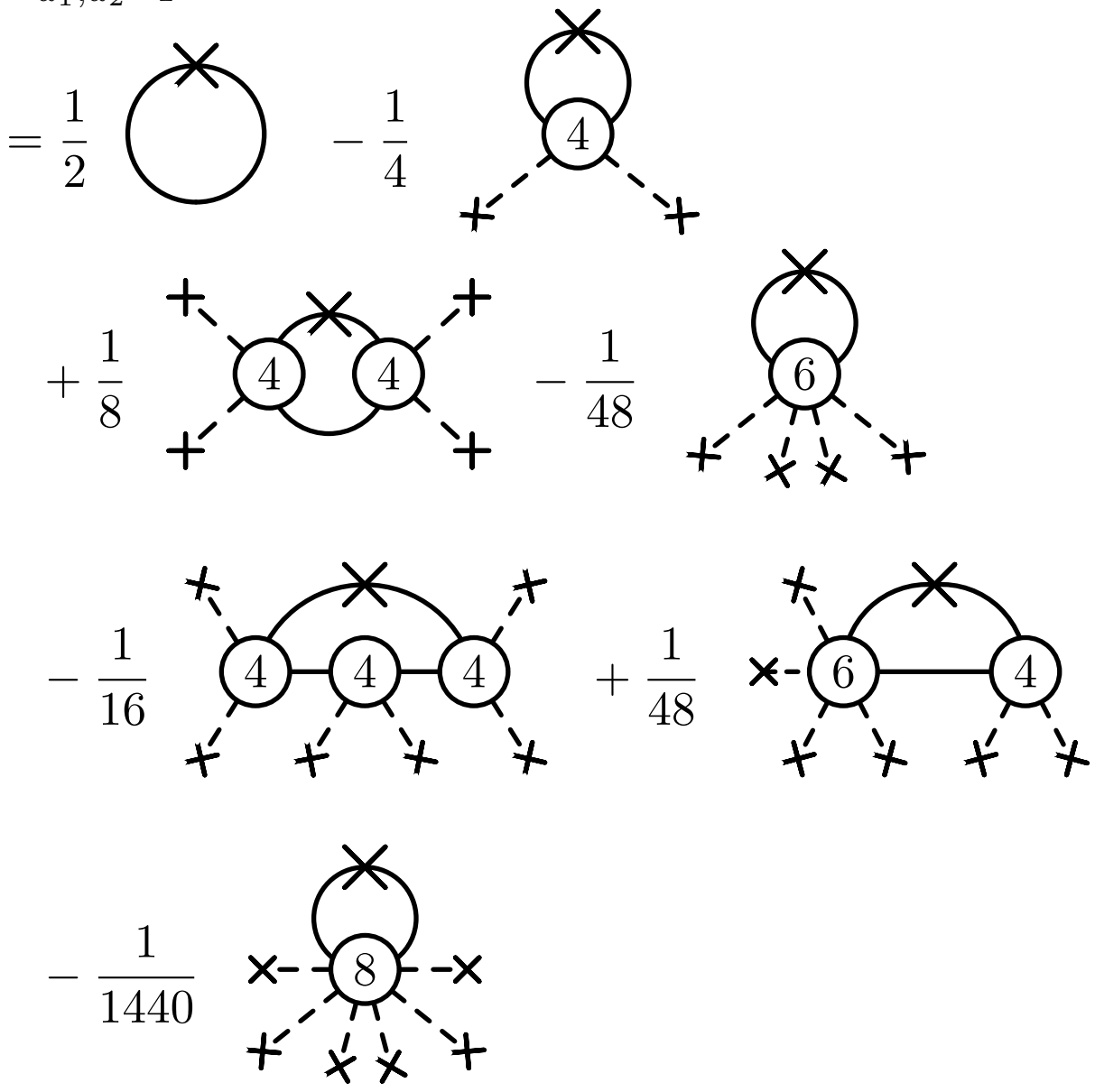

$$+\mathcal{O}\left(\left|\vec{\phi}\right|^8\right) . \tag{316}$$

According to Eq. (308), we can equate the right-hand sides of Eqs. (311) and (316) and then identify the terms with identical powers of the field, thus yielding:

$$\dot{\Gamma}^{(\mathrm{1PI})}_k = \frac{1}{2} \tag{317}$$
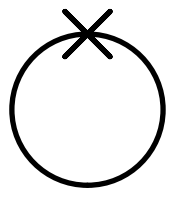

$$\sum_{a_1,a_2=1}^{N} \dot{\Gamma}^{(\mathrm{1PI})(2)}_{k,a_1a_2}\phi_{a_1}\phi_{a_2} = -\frac{1}{2} \tag{318}$$
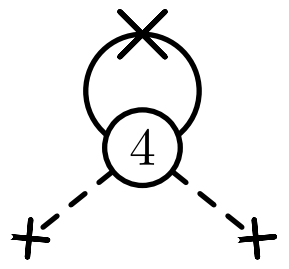

$$\sum_{a_1,a_2,a_3,a_4=1}^{N} \dot{\Gamma}^{(\mathrm{1PI})(4)}_{k,a_1a_2a_3a_4}\phi_{a_1}\phi_{a_2}\phi_{a_3}\phi_{a_4} \tag{319}$$
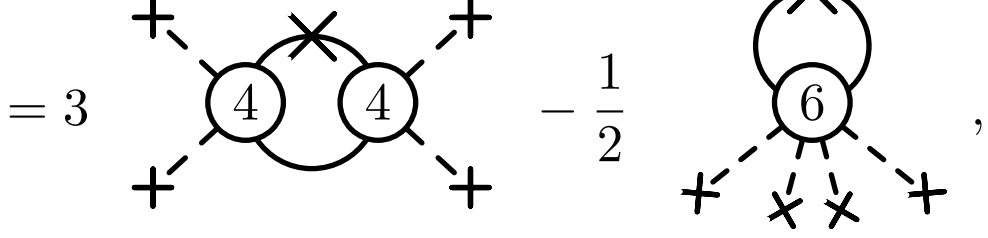

$$\sum_{a_1,a_2,a_3,a_4,a_5,a_6=1}^{N} \dot{\overline{\Gamma}}^{(\mathrm{1PI})(6)}_{k,a_1a_2a_3a_4a_5a_6}\phi_{a_1}\phi_{a_2}\phi_{a_3}\phi_{a_4}\phi_{a_5}\phi_{a_6}$$

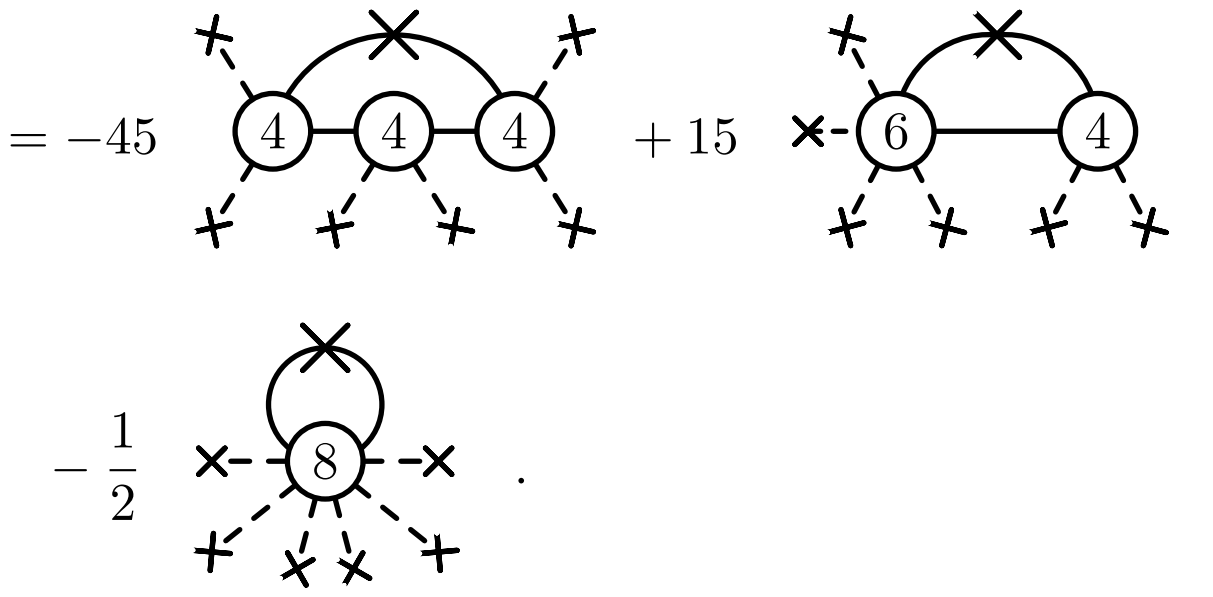

(320)

We are now arriving at the core of our derivations since we are about to explain how to simplify the sums over color indices involved in Eqs. (317) to (320) by using the definitions of symmetric parts of the cutoff function (given by Eq. (40)), of the 1PI vertices (given by Eqs. (41) to (43)) and of the propagator (given by Eq. (45)). We will explain in this way the appearance of the $N$-dependent factors in the right-hand sides of the resulting flow equations, expressed notably by Eqs. (34) to (37). As a first step, we can directly evaluate the sums over color indices in Eq. (317) (using Eqs. (40) and (45)) to obtain:

$$\dot{\overline{\Gamma}}^{(\mathrm{1PI})}_{k} = \frac{N}{2}\overline{G}_k \;, \tag{321}$$

since the sums over color indices yield a factor $N$ for any closed propagator loop in the present situation. In order to evaluate such sums in Eqs. (318) to (320), we must open up the 1PI vertices in favor of their symmetric part. For $\overline{\Gamma}^{(\mathrm{1PI})(4)}_{k}$, this amounts to writing:

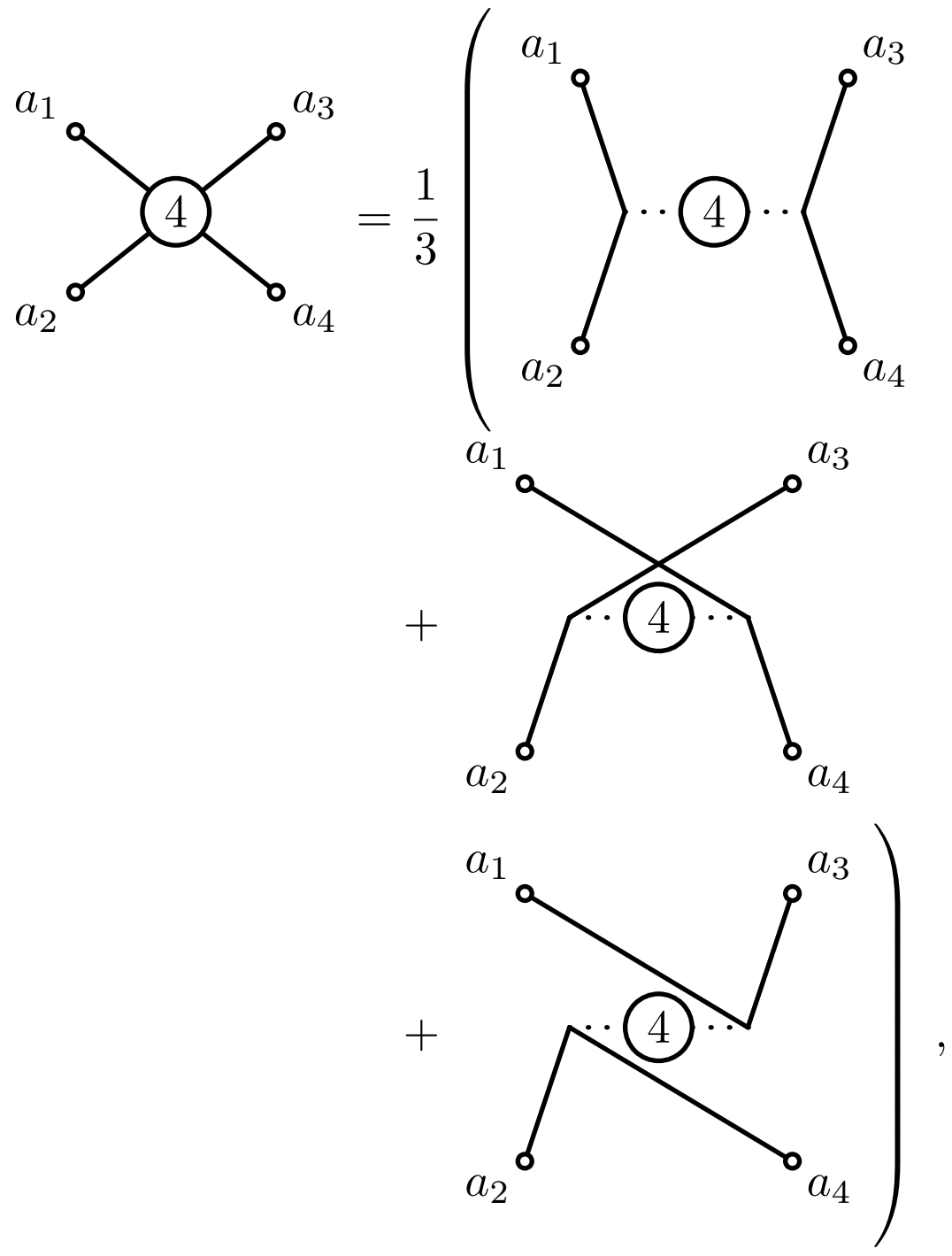

(322)

with the following rule:

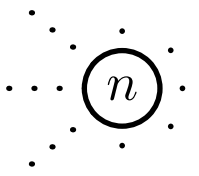

$$\rightarrow \quad \overline{\Gamma}^{(\mathrm{1PI})(n)}_{k} \;, \tag{323}$$

where there are $n/2$ dotted lines that leave the circle containing the integer $n$ (which is always even here). Rule (323) is to be distinguished from that numbered (313b) since the right-hand side of the latter (former) only involves the full (symmetric part of the) 1PI vertices. After inserting Eqs. (41) and (322) into result (318), we obtain:

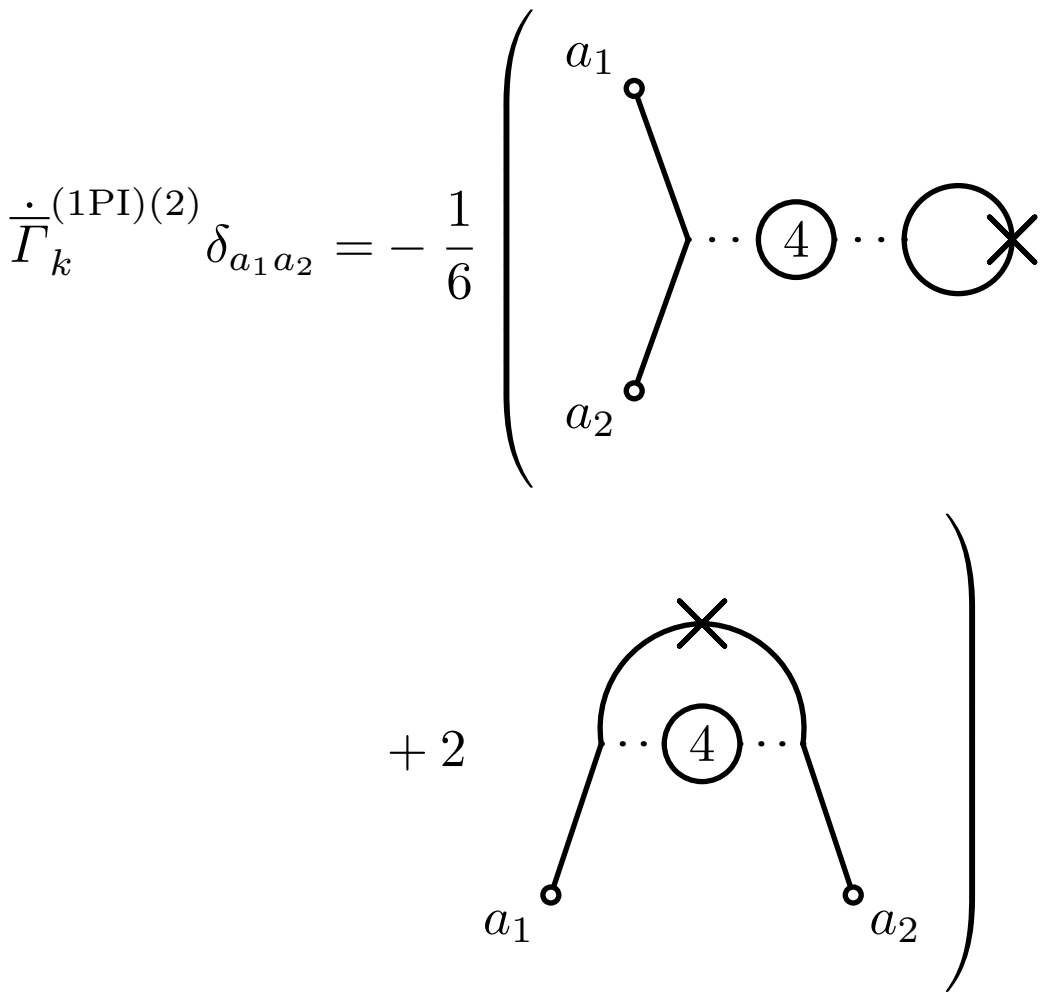

(324)

$$= -\frac{N+2}{6}\dot{R}_k\overline{G}^2_k\overline{\Gamma}^{(\mathrm{1PI})(4)}_{k}\delta_{a_1a_2} \;,$$

which, at $a_1 = a_2$, corresponds to our final expression for $\dot{\overline{\Gamma}}^{(\mathrm{1PI})(2)}_{k}$:

$$\dot{\overline{\Gamma}}^{(\mathrm{1PI})(2)}_{k} = -\frac{N+2}{6}\dot{R}_k\overline{G}^2_k\overline{\Gamma}^{(\mathrm{1PI})(4)}_{k} \;. \tag{325}$$

We have thus derived the flow equation (35). To derive the homologous flow equations for the vertices of higher orders, we must open up vertices of order $n > 4$. For Eq. (36) that expresses $\dot{\overline{\Gamma}}^{(\mathrm{1PI})(4)}_{k}$ in terms of $\overline{\Gamma}^{(\mathrm{1PI})(4)}_{k}$ and $\overline{\Gamma}^{(\mathrm{1PI})(6)}_{k}$, we also need to open up $\overline{\Gamma}^{(\mathrm{1PI})(6)}_{k}$ using this time the 15 permutations involved in Eq. (43). This results in:

$$\begin{aligned}&\frac{1}{3}\dot{\overline{\Gamma}}^{(\mathrm{1PI})(4)}_{k}\left(\delta_{a_1a_2}\delta_{a_3a_4}+\delta_{a_1a_3}\delta_{a_2a_4}+\delta_{a_1a_4}\delta_{a_2a_3}\right)\\ &= \left(\frac{N+8}{9}\dot{R}_k\overline{G}^3_k\left(\overline{\Gamma}^{(\mathrm{1PI})(4)}_{k}\right)^2 - \frac{N+4}{30}\dot{R}_k\overline{G}^2_k\overline{\Gamma}^{(\mathrm{1PI})(6)}_{k}\right)\\ &\quad\times\left(\delta_{a_1a_2}\delta_{a_3a_4}+\delta_{a_1a_3}\delta_{a_2a_4}+\delta_{a_1a_4}\delta_{a_2a_3}\right) \;,\end{aligned} \tag{326}$$

which directly yields Eq. (36). To then derive the flow equation (37) expressing $\dot{\overline{\Gamma}}^{(\mathrm{1PI})(6)}_{k}$, the 1PI vertex $\overline{\Gamma}^{(\mathrm{1PI})(8)}_{k}$ must also be opened up and so on.

### C.2 2PI formalism

As explained right below Eq. (125), the $O(N)$ symmetry can not be spontaneously broken down in the 2PI-FRG framework. For our toy model study, this means that no SSB can occur, whether we work in the unbroken- or broken-symmetry regime of the (0+0)-D $O(N)$ model. To explain clearly how such a conservation of the $O(N)$ symmetry impacts the 2PI-FRG formalism, it will be convenient to disentangle the resulting simplifications from those related to symmetry properties inherent to the 2PI EA framework. We have thus essentially two symmetry arguments that allow us to simplify the flow equations underlying all 2PI-FRG (C-flow, U-flow and CU-flow) approaches tested in this study:

- Symmetry argument inherent to the 2PI EA framework:
  The symmetry properties of the correlation functions $W^{(n)}$ given by Eqs. (112) are also exhibited by the propagator $\overline{G}_\mathfrak{s}$, the self-energy $\overline{\Sigma}_\mathfrak{s}$ and all other 2PI vertices $\overline{\Phi}^{(n)}_\mathfrak{s}$ (with $n \geq 2$). For the 2PI two-particle vertex $\overline{\Phi}^{(2)}_\mathfrak{s}$ of the (0+0)-D $O(N)$ model at $N = 2$ for instance, we have *a priori* $2^4 = 16$ components $\overline{\Phi}^{(2)}_{\mathfrak{s},(a_1,a_1')(a_2,a_2')}$ to consider for the flow. However, since $\overline{\Phi}^{(2)}_{\mathfrak{s},(a_1,a_1')(a_2,a_2')} = \overline{\Phi}^{(2)}_{\mathfrak{s},(a_1',a_1)(a_2,a_2')} = \overline{\Phi}^{(2)}_{\mathfrak{s},(a_1,a_1')(a_2',a_2)} = \overline{\Phi}^{(2)}_{\mathfrak{s},(a_1',a_1)(a_2',a_2)}$ according to Eq. (112a) and $\overline{\Phi}^{(2)}_{\mathfrak{s},(a_1,a_1')(a_2,a_2')} = \overline{\Phi}^{(2)}_{\mathfrak{s},(a_2,a_2')(a_1,a_1')}$ according to Eq. (112b), this set reduces to 6 flowing components, which are for instance: $\overline{\Phi}^{(2)}_{\mathfrak{s},(1,1)(1,1)}$, $\overline{\Phi}^{(2)}_{\mathfrak{s},(1,1)(1,2)}$, $\overline{\Phi}^{(2)}_{\mathfrak{s},(1,1)(2,2)}$, $\overline{\Phi}^{(2)}_{\mathfrak{s},(1,2)(1,2)}$, $\overline{\Phi}^{(2)}_{\mathfrak{s},(1,2)(2,2)}$ and $\overline{\Phi}^{(2)}_{\mathfrak{s},(2,2)(2,2)}$.
- Symmetry argument inherent to the $O(N)$ symmetry:
  Since the 2PI-FRG formalism was developed in a framework that can not exhibit any spontaneous breakdown of the $O(N)$ symmetry, all matrices reduce to scalars in color space throughout the entire flow. For the (0+0)-D $O(N)$ model, this translates into[47]:

$$C^{(-1)}_{\mathfrak{s},aa'} = C^{(-1)}_\mathfrak{s}\,\delta_{aa'} \quad \forall \mathfrak{s}\;, \tag{327}$$

$$\overline{G}_{\mathfrak{s},aa'} = \overline{G}_\mathfrak{s}\,\delta_{aa'} \quad \forall \mathfrak{s}\;, \tag{328}$$

$$\overline{\Sigma}_{\mathfrak{s},aa'} = \overline{\Sigma}_\mathfrak{s}\,\delta_{aa'} \quad \forall \mathfrak{s}\;. \tag{329}$$

After rewriting for example the C-flow equations (249) to (252) for the (0+0)-D $O(N)$ model, i.e.:

$$\dot{\overline{G}}_{\mathfrak{s},a_1a_1'} = -\sum_{a_2,a_2'=1}^{N} \overline{G}_{\mathfrak{s},a_1a_2}\left(\dot{C}^{-1}_\mathfrak{s} - \dot{\overline{\Sigma}}_\mathfrak{s}\right)_{a_2a_2'}\overline{G}_{\mathfrak{s},a_2'a_1'}\;, \tag{330}$$

$$\Delta\dot{\overline{\Gamma}}^{(2\mathrm{PI})}_\mathfrak{s} = \frac{1}{2}\sum_{a,a'=1}^{N}\dot{C}^{-1}_{\mathfrak{s},aa'}\left(\overline{G}_\mathfrak{s} - C_\mathfrak{s}\right)_{aa'}\;, \tag{331}$$

$$\dot{\overline{\Sigma}}_{\mathfrak{s},a_1a_1'} = -\frac{1}{2}\sum_{a_2,a_2'=1}^{N}\dot{\overline{G}}_{\mathfrak{s},a_2a_2'}\overline{\Phi}^{(2)}_{\mathfrak{s},(a_2,a_2')(a_1,a_1')}\;, \tag{332}$$

$$\dot{\overline{\Phi}}^{(n)}_{\mathfrak{s},(a_1,a_1')\cdots(a_n,a_n')} = \frac{1}{2}\sum_{a_{n+1},a_{n+1}'=1}^{N}\dot{\overline{G}}_{\mathfrak{s},a_{n+1}a_{n+1}'} \times \overline{\Phi}^{(n+1)}_{\mathfrak{s},(a_{n+1},a_{n+1}')(a_1,a_1')\cdots(a_n,a_n')} \quad \forall n \geq 2\;, \tag{333}$$

using Eqs. (327) to (329), we can see that the components of the 2PI vertices $\overline{\Phi}^{(n)}_\mathfrak{s}$ which have at least one bosonic index with distinct color indices (i.e. at least one index $\gamma = (a,a')$ with $a \neq a'$) are somehow cut out of the flow. In particular, this translates into the fact that they do not affect $\overline{\Gamma}^{(2\mathrm{PI})}_{\mathfrak{s}=\mathfrak{s}_\mathrm{f}}$ and $\overline{G}_{\mathfrak{s}=\mathfrak{s}_\mathrm{f}}$ which are the quantities of interest for us in the present study since the gs energy and density are directly determined from them (as can be seen e.g. from Eqs. (156) and (157)). Getting back to our example on $\overline{\Phi}^{(2)}_\mathfrak{s}$, the number of corresponding components of interest for the flow is thus further reduced from 6 to 3: $\overline{\Phi}^{(2)}_{\mathfrak{s},(1,1)(1,1)}$, $\overline{\Phi}^{(2)}_{\mathfrak{s},(1,1)(2,2)}$ and $\overline{\Phi}^{(2)}_{\mathfrak{s},(2,2)(2,2)}$. Finally, since the color space is isotropic in the absence of SSB,[48] we have $\overline{\Phi}^{(2)}_{\mathfrak{s},(2,2)(2,2)} = \overline{\Phi}^{(2)}_{\mathfrak{s},(1,1)(1,1)}$ $\forall \mathfrak{s}$, thus ending up with only 2 relevant components of $\overline{\Phi}^{(2)}_\mathfrak{s}$ for the flow. Therefore, with all the symmetry considerations outlined here, the coupled equations (330) to (333) with $n = 2$ reduce at $N = 2$ to the simpler set:

$$\dot{\overline{G}}_\mathfrak{s} = -\overline{G}^2_\mathfrak{s}\left(\dot{C}^{-1}_\mathfrak{s} - \dot{\overline{\Sigma}}_\mathfrak{s}\right)\;, \tag{334}$$

$$\Delta\dot{\overline{\Gamma}}^{(2\mathrm{PI})}_\mathfrak{s} = \dot{C}^{-1}_\mathfrak{s}\left(\overline{G}_\mathfrak{s} - C_\mathfrak{s}\right)\;, \tag{335}$$

$$\dot{\overline{\Sigma}}_\mathfrak{s} = -\frac{1}{2}\dot{\overline{G}}_\mathfrak{s}\left(\overline{\Phi}^{(2)}_{\mathfrak{s},(1,1)(1,1)} + \overline{\Phi}^{(2)}_{\mathfrak{s},(1,1)(2,2)}\right)\;, \tag{336}$$

$$\dot{\overline{\Phi}}^{(2)}_{\mathfrak{s},(1,1)(1,1)} = \frac{1}{2}\dot{\overline{G}}_\mathfrak{s}\left(\overline{\Phi}^{(3)}_{\mathfrak{s},(1,1)(1,1)(1,1)} + \overline{\Phi}^{(3)}_{\mathfrak{s},(1,1)(1,1)(2,2)}\right)\;, \tag{337}$$

[47] In particular, the cutoff function $R_\mathfrak{s}$ defining the propagator $C_\mathfrak{s}$ (via e.g. the relation $C^{-1}_\mathfrak{s} = C^{-1} + R_\mathfrak{s}$) within the C-flow or the CU-flow must therefore be chosen such that condition (327) is fulfilled.

[48] For $\overline{\Phi}^{(3)}_\mathfrak{s}$, this implies $\overline{\Phi}^{(3)}_{\mathfrak{s},(2,2)(2,2)(2,2)} = \overline{\Phi}^{(3)}_{\mathfrak{s},(1,1)(1,1)(1,1)}$ $\forall \mathfrak{s}$, but also $\overline{\Phi}^{(3)}_{\mathfrak{s},(1,1)(2,2)(2,2)} = \overline{\Phi}^{(3)}_{\mathfrak{s},(1,1)(1,1)(2,2)}$ $\forall \mathfrak{s}$.

$$\dot{\overline{\Phi}}^{(2)}_{\mathfrak{s},(1,1)(2,2)} = \frac{1}{2}\dot{\overline{G}}_{\mathfrak{s}}\Big(\overline{\Phi}^{(3)}_{\mathfrak{s},(1,1)(1,1)(2,2)} + \overline{\Phi}^{(3)}_{\mathfrak{s},(1,1)(2,2)(2,2)}\Big) , \quad (338)$$

with $\dot{C}^{-1}_{\mathfrak{s},aa'} = \dot{C}^{-1}_{\mathfrak{s}}\ \delta_{aa'}\ \forall\mathfrak{s}$, $\dot{\overline{G}}_{\mathfrak{s},aa'} = \dot{\overline{G}}_{\mathfrak{s}}\ \delta_{aa'}\ \forall\mathfrak{s}$ and $\dot{\overline{\Sigma}}_{\mathfrak{s},aa'} = \dot{\overline{\Sigma}}_{\mathfrak{s}}\ \delta_{aa'}\ \forall\mathfrak{s}$ as a result of Eqs. (327) to (329).

# D Additional flow equations

## D.1 Mixed 1PI functional renormalization group

We then outline the main steps of the vertex expansion procedure treating the Wetterich equation for the (0+0)-D $O(N)$ model in the framework of the mixed representation. The Wetterich equation to consider is already given by Eq. (62) in the form:

$$\dot{\Gamma}^{(\mathrm{1PI})}_{\mathrm{mix},k}\Big(\vec{\phi},\eta\Big) = \frac{1}{2}\mathcal{STr}\left[\dot{\mathcal{R}}_k\left(\Gamma^{(\mathrm{1PI})(2)}_{\mathrm{mix},k}\Big(\vec{\phi},\eta\Big)+\mathcal{R}_k\right)^{-1}\right] , \quad (339)$$

with

$$\mathcal{R}_k = \begin{pmatrix} \boldsymbol{R}^{(\phi)}_k & \vec{0} \\ \vec{0}^{\mathrm{T}} & R^{(\eta)}_k \end{pmatrix} = \begin{pmatrix} R_k\mathbb{I}_N & \vec{0} \\ \vec{0}^{\mathrm{T}} & R_k \end{pmatrix} = R_k\mathbb{I}_{N+1} . \quad (340)$$

As a first step, we expand the mixed 1PI EA around an extremum ($\overline{\Gamma}^{(\mathrm{1PI})(1\phi)}_{\mathrm{mix},k,a} = \overline{\Gamma}^{(\mathrm{1PI})(1\eta)}_{\mathrm{mix},k} = 0\ \forall a,k$) via Eq. (65) recalled below:

$$\Gamma^{(\mathrm{1PI})}_{\mathrm{mix},k}\Big(\vec{\phi},\eta\Big) = \overline{\Gamma}^{(\mathrm{1PI})}_{\mathrm{mix},k} + \sum_{n=2}^{\infty}\frac{1}{n!}\sum_{m=0}^{n}\binom{n}{m}\sum_{a_1,\cdots,a_m=1}^{N}\overline{\Gamma}^{(\mathrm{1PI})(m\phi,(n-m)\eta)}_{\mathrm{mix},k,a_1\cdots a_m}\Big(\vec{\phi}-\vec{\overline{\phi}}_k\Big)_{a_1}\cdots\Big(\vec{\phi}-\vec{\overline{\phi}}_k\Big)_{a_m}\left(\eta-\overline{\eta}_k\right)^{n-m} , \quad (341)$$

from which we deduce the following expression of the left-hand side of Eq. (339):

$$\begin{aligned}\dot{\Gamma}^{(\mathrm{1PI})}_{\mathrm{mix},k}\Big(\vec{\phi},\eta\Big) = {} & \dot{\overline{\Gamma}}^{(\mathrm{1PI})}_{\mathrm{mix},k} - \sum_{a_1=1}^{N}\left(\sum_{a_2=1}^{N}\dot{\overline{\phi}}_{k,a_2}\overline{\Gamma}^{(\mathrm{1PI})(2\phi)}_{\mathrm{mix},k,a_2a_1} + \dot{\overline{\eta}}_k\overline{\Gamma}^{(\mathrm{1PI})(1\phi,1\eta)}_{\mathrm{mix},k,a_1}\right)\Big(\vec{\phi}-\vec{\overline{\phi}}_k\Big)_{a_1} \\ & - \left(\sum_{a_1=1}^{N}\dot{\overline{\phi}}_{k,a_1}\overline{\Gamma}^{(\mathrm{1PI})(1\phi,1\eta)}_{\mathrm{mix},k,a_1} + \dot{\overline{\eta}}_k\overline{\Gamma}^{(\mathrm{1PI})(2\eta)}_{\mathrm{mix},k}\right)\left(\eta-\overline{\eta}_k\right) \\ & + \sum_{n=2}^{\infty}\frac{1}{n!}\sum_{m=0}^{n}\binom{n}{m}\sum_{a_1,\cdots,a_m=1}^{N}\left(\dot{\overline{\Gamma}}^{(\mathrm{1PI})(m\phi,(n-m)\eta)}_{\mathrm{mix},k,a_1\cdots a_m} - \sum_{a_{m+1}=1}^{N}\dot{\overline{\phi}}_{k,a_{m+1}}\overline{\Gamma}^{(\mathrm{1PI})((m+1)\phi,(n-m)\eta)}_{\mathrm{mix},k,a_{m+1}a_1\cdots a_m}\right. \\ & \left. - \dot{\overline{\eta}}_k\overline{\Gamma}^{(\mathrm{1PI})(m\phi,(n-m+1)\eta)}_{\mathrm{mix},k,a_1\cdots a_m}\right)\Big(\vec{\phi}-\vec{\overline{\phi}}_k\Big)_{a_1}\cdots\Big(\vec{\phi}-\vec{\overline{\phi}}_k\Big)_{a_m}\left(\eta-\overline{\eta}_k\right)^{n-m} .\end{aligned} \quad (342)$$

Then, the right-hand side of Eq. (339) is expanded by exploiting the matrices $\mathcal{P}_k$ and $\mathcal{F}_k$ already defined by Eqs. (69) to (72). Each component of $\mathcal{F}_k$ can also be expanded in the same way as in Eq. (341), thus leading to:

$$\begin{aligned}\mathcal{F}_{k,a_1a_2} \equiv {} & \Gamma^{(\mathrm{1PI})(2\phi)}_{\mathrm{mix},k,a_1a_2} - \overline{\Gamma}^{(\mathrm{1PI})(2\phi)}_{\mathrm{mix},k,a_1a_2} \\ = {} & \sum_{n=1}^{\infty}\frac{1}{n!}\sum_{m=0}^{n}\binom{n}{m}\sum_{a_3,\cdots,a_{m+2}=1}^{N}\overline{\Gamma}^{(\mathrm{1PI})((m+2)\phi,(n-m)\eta)}_{\mathrm{mix},k,a_1\cdots a_{m+2}} \\ & \times\Big(\vec{\phi}-\vec{\overline{\phi}}_k\Big)_{a_3}\cdots\Big(\vec{\phi}-\vec{\overline{\phi}}_k\Big)_{a_{m+2}}\left(\eta-\overline{\eta}_k\right)^{n-m} ,\end{aligned} \quad (343)$$

$$\begin{aligned}\mathcal{F}_{k,a_1\,N+1} = \mathcal{F}_{k,N+1\,a_1} \equiv {} & \Gamma^{(\mathrm{1PI})(1\phi,1\eta)}_{\mathrm{mix},k,a_1} - \overline{\Gamma}^{(\mathrm{1PI})(1\phi,1\eta)}_{\mathrm{mix},k,a_1} \\ = {} & \sum_{n=1}^{\infty}\frac{1}{n!}\sum_{m=0}^{n}\binom{n}{m}\sum_{a_2,\cdots,a_{m+1}=1}^{N}\overline{\Gamma}^{(\mathrm{1PI})((m+1)\phi,(n-m+1)\eta)}_{\mathrm{mix},k,a_1\cdots a_{m+1}} \\ & \times\Big(\vec{\phi}-\vec{\overline{\phi}}_k\Big)_{a_2}\cdots\Big(\vec{\phi}-\vec{\overline{\phi}}_k\Big)_{a_{m+1}}\left(\eta-\overline{\eta}_k\right)^{n-m} ,\end{aligned} \quad (344)$$

$$\begin{aligned}\mathcal{F}_{k,N+1\,N+1} \equiv {} & \Gamma^{(\mathrm{1PI})(2\eta)}_{\mathrm{mix},k} - \overline{\Gamma}^{(\mathrm{1PI})(2\eta)}_{\mathrm{mix},k} \\ = {} & \sum_{n=1}^{\infty}\frac{1}{n!}\sum_{m=0}^{n}\binom{n}{m}\sum_{a_1,\cdots,a_m=1}^{N}\overline{\Gamma}^{(\mathrm{1PI})(m\phi,(n-m+2)\eta)}_{\mathrm{mix},k,a_1\cdots a_m} \\ & \times\Big(\vec{\phi}-\vec{\overline{\phi}}_k\Big)_{a_1}\cdots\Big(\vec{\phi}-\vec{\overline{\phi}}_k\Big)_{a_m}\left(\eta-\overline{\eta}_k\right)^{n-m} .\end{aligned} \quad (345)$$

The expansion of the right-hand side of Eq. (339) is then carried out by matrix multiplications between $\mathcal{P}_k^{-1}$ and

$\mathcal{F}_k$ according to:

$$\begin{aligned}
&\frac{1}{2}\mathcal{STr}\left[\dot{\mathcal{R}}_k\left(\Gamma^{(1\mathrm{PI})(2)}_{\mathrm{mix},k}\left(\vec{\phi},\eta\right)+\mathcal{R}_k\right)^{-1}\right]\\
&=\frac{1}{2}\mathcal{STr}\left[\dot{\mathcal{R}}_k\left(\mathcal{P}_k+\mathcal{F}_k\right)^{-1}\right]\\
&=\frac{1}{2}\mathcal{STr}\left[\dot{\mathcal{R}}_k\mathcal{P}_k^{-1}\left(\mathbb{I}_{N+1}+\mathcal{P}_k^{-1}\mathcal{F}_k\right)^{-1}\right]\\
&=\frac{1}{2}\mathcal{STr}\left[\dot{\mathcal{R}}_k\mathcal{P}_k^{-1}\left(\mathbb{I}_{N+1}+\sum_{n=1}^{\infty}(-1)^n\left(\mathcal{P}_k^{-1}\mathcal{F}_k\right)^n\right)\right]\\
&=\frac{1}{2}\mathcal{STr}\left[\dot{\mathcal{R}}_k\mathcal{P}_k^{-1}\right]\\
&\quad+\frac{1}{2}\sum_{n=1}^{\infty}(-1)^n\mathcal{STr}\left[\dot{\mathcal{R}}_k\mathcal{P}_k^{-1}\left(\mathcal{P}_k^{-1}\mathcal{F}_k\right)^n\right]\ .
\end{aligned} \tag{346}$$

In the unbroken-symmetry phase, the $O(N)$ symmetry can be used to simplify the products between $\mathcal{P}_k^{-1}$ and $\mathcal{F}_k$, especially since it enables us to replace the 1PI vertices in favor of their symmetric parts (according to Eq. (73) in particular) and imposes that the 1PI vertices $\overline{\Gamma}^{(1\mathrm{PI})(n\phi,m\eta)}_{\mathrm{mix},k}$ with $n$ odd all vanish (according to Eq. (75)). Those simplifications do not apply in the framework of the broken-symmetry regime and the corresponding equations are thus particularly cumbersome. Hence, we will focus on the unbroken-symmetry phase from now on until the end of appendix D.1, although the general recipe of the vertex expansion outlined here is essentially the same for all phases. Using the aforementioned simplifications, definition (72) of $\mathcal{P}_k$ reduces to:

$$\mathcal{P}_k=\begin{pmatrix}\left(R_k+\overline{\Gamma}^{(1\mathrm{PI})(2\phi)}_{\mathrm{mix},k}\right)\mathbb{I}_N & \vec{0}\\ \vec{0}^{\mathrm{T}} & R_k+\overline{\Gamma}^{(1\mathrm{PI})(2\eta)}_{\mathrm{mix},k}\end{pmatrix}\ , \tag{347}$$

and the inversion of $\mathcal{P}_k$ thus becomes trivial such that $\mathcal{P}_k^{-1}$ is now a diagonal matrix, i.e.:

$$\begin{aligned}
\mathcal{P}_k^{-1}&=\begin{pmatrix}\left(R_k+\overline{\Gamma}^{(2\phi)}_{\mathrm{mix},k}\right)^{-1}\mathbb{I}_N & \vec{0}\\ \vec{0}^{\mathrm{T}} & \left(R_k+\overline{\Gamma}^{(2\eta)}_{\mathrm{mix},k}\right)^{-1}\end{pmatrix}\\
&\equiv\begin{pmatrix}\overline{G}^{(\phi)}_k\mathbb{I}_N & \vec{0}\\ \vec{0}^{\mathrm{T}} & \overline{G}^{(\eta)}_k\end{pmatrix}\ .
\end{aligned} \tag{348}$$

Moreover, as $\vec{\overline{\phi}}_k=\vec{0}\ \forall k$ in the unbroken-symmetry regime, the expressions of the components of $\mathcal{F}_k$ given by Eqs. (343) to (345) can be simplified as:

$$\begin{aligned}
\mathcal{F}_{k,a_1a_2}=&\sum_{n=1}^{\infty}\frac{1}{n!}\sum_{\substack{m=0\\ \{\text{m even}\}}}^{n}\binom{n}{m}\\
&\times\sum_{a_3,\cdots,a_{m+2}=1}^{N}\overline{\Gamma}^{(1\mathrm{PI})((m+2)\phi,(n-m)\eta)}_{\mathrm{mix},k,a_1\cdots a_{m+2}}\\
&\times\phi_{a_3}\cdots\phi_{a_{m+2}}\left(\eta-\overline{\eta}_k\right)^{n-m}\ ,
\end{aligned} \tag{349}$$

$$\begin{aligned}
\mathcal{F}_{k,a_1\,N+1}=&\,\mathcal{F}_{k,N+1\,a_1}\\
=&\sum_{n=1}^{\infty}\frac{1}{n!}\sum_{\substack{m=0\\ \{\text{m odd}\}}}^{n}\binom{n}{m}\\
&\times\sum_{a_2,\cdots,a_{m+1}=1}^{N}\overline{\Gamma}^{(1\mathrm{PI})((m+1)\phi,(n-m+1)\eta)}_{\mathrm{mix},k,a_1\cdots a_{m+1}}\\
&\times\phi_{a_2}\cdots\phi_{a_{m+1}}\left(\eta-\overline{\eta}_k\right)^{n-m}\ ,
\end{aligned} \tag{350}$$

$$\begin{aligned}
\mathcal{F}_{k,N+1\,N+1}=&\sum_{n=1}^{\infty}\frac{1}{n!}\sum_{\substack{m=0\\ \{\text{m even}\}}}^{n}\binom{n}{m}\\
&\times\sum_{a_1,\cdots,a_m=1}^{N}\overline{\Gamma}^{(1\mathrm{PI})(m\phi,(n-m+2)\eta)}_{\mathrm{mix},k,a_1\cdots a_m}\\
&\times\phi_{a_1}\cdots\phi_{a_m}\left(\eta-\overline{\eta}_k\right)^{n-m}\ ,
\end{aligned} \tag{351}$$

on the one hand and, on the other hand, the left-hand side of Eq. (339), given by Eq. (342), reduces to:

$$\begin{aligned}
\dot{\Gamma}^{(1\mathrm{PI})}_{\mathrm{mix},k}\left(\vec{\phi},\eta\right)=&\,\dot{\overline{\Gamma}}^{(1\mathrm{PI})}_{\mathrm{mix},k}-\dot{\overline{\eta}}_k\overline{\Gamma}^{(1\mathrm{PI})(2\eta)}_{\mathrm{mix},k}\left(\eta-\overline{\eta}_k\right)\\
&+\sum_{n=2}^{\infty}\frac{1}{n!}\sum_{\substack{m=0\\ \{\text{m even}\}}}^{n}\binom{n}{m}\\
&\times\sum_{a_1,\cdots,a_m=1}^{N}\left(\dot{\overline{\Gamma}}^{(1\mathrm{PI})(m\phi,(n-m)\eta)}_{\mathrm{mix},k,a_1\cdots a_m}\right.\\
&\left.-\dot{\overline{\eta}}_k\overline{\Gamma}^{(1\mathrm{PI})(m\phi,(n-m+1)\eta)}_{\mathrm{mix},k,a_1\cdots a_m}\right)\phi_{a_1}\cdots\phi_{a_m}\\
&\times\left(\eta-\overline{\eta}_k\right)^{n-m}\ ,
\end{aligned} \tag{352}$$

where the restrictions "m even" and "m odd" for the sums are a direct consequence of Eq. (75). Finally, we carry out the matrix products $(\mathcal{P}_k^{-1}\mathcal{F}_k)^n$ involved in Eq. (346) (where the expansion is performed up to a finite truncation order $n=N_{\mathrm{max}}$ in practice) with $\mathcal{P}_k^{-1}$ given by Eq. (348) and $\mathcal{F}_k$ specified by Eqs. (349) to (351). In this way, we have expanded the right-hand side of the Wetterich equation given by Eq. (339). By identifying the terms of the relation thus derived with those of Eq. (352) involving the same powers of the fields $\vec{\phi}$ and $\eta-\overline{\eta}_k$, we obtain the following tower of differential equations for $N=2$:

$$\dot{\overline{\Gamma}}^{(\mathrm{1PI})}_{\mathrm{mix},k} = \left(\overline{G}^{(\phi)}_{k} - \overline{G}^{(\phi)(0)}_{k}\right) + \frac{1}{2}\left(\overline{G}^{(\eta)}_{k} - \overline{G}^{(\eta)(0)}_{k}\right) , \tag{353}$$

$$\dot{\overline{\eta}}_{k} = \frac{\dot{R}_k}{2\overline{\Gamma}^{(\mathrm{1PI})(2\eta)}_{\mathrm{mix},k}}\left(\overline{\Gamma}^{(\mathrm{1PI})(3\eta)}_{\mathrm{mix},k}\left(\overline{G}^{(\eta)}_{k}\right)^2 + 2\overline{\Gamma}^{(\mathrm{1PI})(2\phi,1\eta)}_{\mathrm{mix},k}\left(\overline{G}^{(\phi)}_{k}\right)^2\right) , \tag{354}$$

$$\begin{split}\dot{\overline{\Gamma}}^{(\mathrm{1PI})(2\phi)}_{\mathrm{mix},k} = {} & \dot{\overline{\eta}}_{k}\overline{\Gamma}^{(\mathrm{1PI})(2\phi,1\eta)}_{\mathrm{mix},k} - \frac{1}{6}\dot{R}_k\Bigg(3\overline{\Gamma}^{(\mathrm{1PI})(2\phi,2\eta)}_{\mathrm{mix},k}\left(\overline{G}^{(\eta)}_{k}\right)^2 + 4\overline{\Gamma}^{(\mathrm{1PI})(4\phi)}_{\mathrm{mix},k}\left(\overline{G}^{(\phi)}_{k}\right)^2 \\ & - 6\left(\overline{\Gamma}^{(\mathrm{1PI})(2\phi,1\eta)}_{\mathrm{mix},k}\right)^2\overline{G}^{(\eta)}_{k}\overline{G}^{(\phi)}_{k}\left(\overline{G}^{(\eta)}_{k}+\overline{G}^{(\phi)}_{k}\right)\Bigg) ,\end{split} \tag{355}$$

$$\begin{split}\dot{\overline{\Gamma}}^{(\mathrm{1PI})(2\eta)}_{\mathrm{mix},k} = {} & \dot{\overline{\eta}}_{k}\overline{\Gamma}^{(\mathrm{1PI})(3\eta)}_{\mathrm{mix},k} + \dot{R}_k\Bigg(-\frac{1}{2}\overline{\Gamma}^{(\mathrm{1PI})(4\eta)}_{\mathrm{mix},k}\left(\overline{G}^{(\eta)}_{k}\right)^2 + \left(\overline{\Gamma}^{(\mathrm{1PI})(3\eta)}_{\mathrm{mix},k}\right)^2\left(\overline{G}^{(\eta)}_{k}\right)^3 - \overline{\Gamma}^{(\mathrm{1PI})(2\phi,2\eta)}_{\mathrm{mix},k}\left(\overline{G}^{(\phi)}_{k}\right)^2 \\ & + 2\left(\overline{\Gamma}^{(\mathrm{1PI})(2\phi,1\eta)}_{\mathrm{mix},k}\right)^2\left(\overline{G}^{(\phi)}_{k}\right)^3\Bigg) ,\end{split} \tag{356}$$

$$\begin{split}\dot{\overline{\Gamma}}^{(\mathrm{1PI})(2\phi,1\eta)}_{\mathrm{mix},k} = {} & \dot{\overline{\eta}}_{k}\overline{\Gamma}^{(\mathrm{1PI})(2\phi,2\eta)}_{\mathrm{mix},k} + \dot{R}_k\overline{\Gamma}^{(\mathrm{1PI})(3\eta)}_{\mathrm{mix},k}\left(\overline{G}^{(\eta)}_{k}\right)^2\left(\overline{\Gamma}^{(\mathrm{1PI})(2\phi,2\eta)}_{\mathrm{mix},k}\overline{G}^{(\eta)}_{k} - \left(\overline{\Gamma}^{(\mathrm{1PI})(2\phi,1\eta)}_{\mathrm{mix},k}\right)^2\overline{G}^{(\phi)}_{k}\left(2\overline{G}^{(\eta)}_{k}+\overline{G}^{(\phi)}_{k}\right)\right) \\ & + \frac{1}{3}\dot{R}_k\overline{\Gamma}^{(\mathrm{1PI})(2\phi,1\eta)}_{\mathrm{mix},k}\overline{G}^{(\phi)}_{k}\Bigg(6\overline{\Gamma}^{(\mathrm{1PI})(2\phi,2\eta)}_{\mathrm{mix},k}\overline{G}^{(\eta)}_{k}\left(\overline{G}^{(\eta)}_{k}+\overline{G}^{(\phi)}_{k}\right) + \overline{G}^{(\phi)}_{k}\Bigg(4\overline{\Gamma}^{(\mathrm{1PI})(4\phi)}_{\mathrm{mix},k}\overline{G}^{(\phi)}_{k} \\ & - 3\left(\overline{\Gamma}^{(\mathrm{1PI})(2\phi,1\eta)}_{\mathrm{mix},k}\right)^2\overline{G}^{(\eta)}_{k}\left(\overline{G}^{(\eta)}_{k}+2\overline{G}^{(\phi)}_{k}\right)\Bigg)\Bigg) \\ & + \mathcal{F}^{(N=2)}_{1}\left(\overline{\Gamma}^{(\mathrm{1PI})(n\phi,m\eta)}_{\mathrm{mix},k};\ n+m\leq 5\right) ,\end{split} \tag{357}$$

$$\begin{split}\dot{\overline{\Gamma}}^{(\mathrm{1PI})(3\eta)}_{\mathrm{mix},k} = {} & \dot{\overline{\eta}}_{k}\overline{\Gamma}^{(\mathrm{1PI})(4\eta)}_{\mathrm{mix},k} \\ & + \dot{R}_k\Bigg(3\overline{\Gamma}^{(\mathrm{1PI})(3\eta)}_{\mathrm{mix},k}\overline{\Gamma}^{(\mathrm{1PI})(4\eta)}_{\mathrm{mix},k}\left(\overline{G}^{(\eta)}_{k}\right)^3 - 3\left(\overline{\Gamma}^{(\mathrm{1PI})(3\eta)}_{\mathrm{mix},k}\right)^3\left(\overline{G}^{(\eta)}_{k}\right)^4 + 6\overline{\Gamma}^{(\mathrm{1PI})(2\phi,1\eta)}_{\mathrm{mix},k}\left(\overline{G}^{(\phi)}_{k}\right)^3\Bigg(\overline{\Gamma}^{(\mathrm{1PI})(2\phi,2\eta)}_{\mathrm{mix},k} \\ & - \left(\overline{\Gamma}^{(\mathrm{1PI})(2\phi,1\eta)}_{\mathrm{mix},k}\right)^2\overline{G}^{(\phi)}_{k}\Bigg)\Bigg) \\ & + \mathcal{F}^{(N=2)}_{2}\left(\overline{\Gamma}^{(\mathrm{1PI})(n\phi,m\eta)}_{\mathrm{mix},k};\ n+m\leq 5\right) ,\end{split} \tag{358}$$

$$\begin{split}\dot{\overline{\Gamma}}^{(\mathrm{1PI})(4\phi)}_{\mathrm{mix},k} = {} & \frac{1}{3}\dot{R}_k\Bigg(9\left(\overline{\Gamma}^{(\mathrm{1PI})(2\phi,2\eta)}_{\mathrm{mix},k}\right)^2\left(\overline{G}^{(\eta)}_{k}\right)^3 - 18\left(\overline{\Gamma}^{(\mathrm{1PI})(2\phi,1\eta)}_{\mathrm{mix},k}\right)^2\overline{\Gamma}^{(\mathrm{1PI})(2\phi,2\eta)}_{\mathrm{mix},k}\left(\overline{G}^{(\eta)}_{k}\right)^2\overline{G}^{(\phi)}_{k}\left(2\overline{G}^{(\eta)}_{k}+\overline{G}^{(\phi)}_{k}\right) \\ & + 2\left(\overline{G}^{(\phi)}_{k}\right)^2\Bigg(5\left(\overline{\Gamma}^{(\mathrm{1PI})(4\phi)}_{\mathrm{mix},k}\right)^2\overline{G}^{(\phi)}_{k} + 18\left(\overline{\Gamma}^{(\mathrm{1PI})(2\phi,1\eta)}_{\mathrm{mix},k}\right)^4\left(\overline{G}^{(\eta)}_{k}\right)^2\left(\overline{G}^{(\eta)}_{k}+\overline{G}^{(\phi)}_{k}\right) \\ & - 9\left(\overline{\Gamma}^{(\mathrm{1PI})(2\phi,1\eta)}_{\mathrm{mix},k}\right)^2\overline{\Gamma}^{(\mathrm{1PI})(4\phi)}_{\mathrm{mix},k}\overline{G}^{(\eta)}_{k}\left(\overline{G}^{(\eta)}_{k}+2\overline{G}^{(\phi)}_{k}\right)\Bigg)\Bigg) \\ & + \mathcal{F}^{(N=2)}_{3}\left(\overline{\Gamma}^{(\mathrm{1PI})(n\phi,m\eta)}_{\mathrm{mix},k};\ n+m\leq 6\right) ,\end{split} \tag{359}$$

$$\begin{split}\dot{\overline{\Gamma}}^{(\mathrm{1PI})(4\eta)}_{\mathrm{mix},k} = {} & 3\dot{R}_k\Bigg(\left(\overline{\Gamma}^{(\mathrm{1PI})(4\eta)}_{\mathrm{mix},k}\right)^2\left(\overline{G}^{(\eta)}_{k}\right)^3 - 6\left(\overline{\Gamma}^{(\mathrm{1PI})(3\eta)}_{\mathrm{mix},k}\right)^2\overline{\Gamma}^{(\mathrm{1PI})(4\eta)}_{\mathrm{mix},k}\left(\overline{G}^{(\eta)}_{k}\right)^4 + 4\left(\overline{\Gamma}^{(\mathrm{1PI})(3\eta)}_{\mathrm{mix},k}\right)^4\left(\overline{G}^{(\eta)}_{k}\right)^5 \\ & + 2\left(\overline{G}^{(\phi)}_{k}\right)^3\left(\left(\overline{\Gamma}^{(\mathrm{1PI})(2\phi,2\eta)}_{\mathrm{mix},k}\right)^2 - 6\left(\overline{\Gamma}^{(\mathrm{1PI})(2\phi,1\eta)}_{\mathrm{mix},k}\right)^2\overline{\Gamma}^{(\mathrm{1PI})(2\phi,2\eta)}_{\mathrm{mix},k}\overline{G}^{(\phi)}_{k} + 4\left(\overline{\Gamma}^{(\mathrm{1PI})(2\phi,1\eta)}_{\mathrm{mix},k}\right)^4\left(\overline{G}^{(\phi)}_{k}\right)^2\right)\Bigg) \\ & + \mathcal{F}^{(N=2)}_{4}\left(\overline{\Gamma}^{(\mathrm{1PI})(n\phi,m\eta)}_{\mathrm{mix},k};\ n+m\leq 6\right) ,\end{split} \tag{360}$$

$$\begin{aligned}\dot{\overline{\mathit{\Gamma}}}^{(\mathrm{1PI})(2\phi,2\eta)}_{\mathrm{mix},k} = & \frac{1}{3}\dot{R}_k\Bigg(12\overline{\mathit{\Gamma}}^{(\mathrm{1PI})(3\eta)}_{\mathrm{mix},k}\overline{\mathit{\Gamma}}^{(\mathrm{1PI})(2\phi,1\eta)}_{\mathrm{mix},k}\left(\overline{G}^{(\eta)}_k\right)^2\overline{G}^{(\phi)}_k\Bigg(\left(\overline{\mathit{\Gamma}}^{(\mathrm{1PI})(2\phi,1\eta)}_{\mathrm{mix},k}\right)^2\overline{G}^{(\phi)}_k\left(\overline{G}^{(\eta)}_k+\overline{G}^{(\phi)}_k\right)\\
& -\overline{\mathit{\Gamma}}^{(\mathrm{1PI})(2\phi,2\eta)}_{\mathrm{mix},k}\left(2\overline{G}^{(\eta)}_k+\overline{G}^{(\phi)}_k\right)\Bigg)+3\overline{\mathit{\Gamma}}^{(\mathrm{1PI})(4\eta)}_{\mathrm{mix},k}\left(\overline{G}^{(\eta)}_k\right)^2\Bigg(\overline{\mathit{\Gamma}}^{(\mathrm{1PI})(2\phi,2\eta)}_{\mathrm{mix},k}\overline{G}^{(\eta)}_k\\
& -\left(\overline{\mathit{\Gamma}}^{(\mathrm{1PI})(2\phi,1\eta)}_{\mathrm{mix},k}\right)^2\overline{G}^{(\phi)}_k\left(2\overline{G}^{(\eta)}_k+\overline{G}^{(\phi)}_k\right)\Bigg)+3\left(\overline{\mathit{\Gamma}}^{(\mathrm{1PI})(3\eta)}_{\mathrm{mix},k}\right)^2\left(\overline{G}^{(\eta)}_k\right)^3\Bigg(-3\overline{\mathit{\Gamma}}^{(\mathrm{1PI})(2\phi,2\eta)}_{\mathrm{mix},k}\overline{G}^{(\eta)}_k\\
& +2\left(\overline{\mathit{\Gamma}}^{(\mathrm{1PI})(2\phi,1\eta)}_{\mathrm{mix},k}\right)^2\overline{G}^{(\phi)}_k\left(3\overline{G}^{(\eta)}_k+\overline{G}^{(\phi)}_k\right)\Bigg)+\overline{G}^{(\phi)}_k\Bigg(6\left(\overline{\mathit{\Gamma}}^{(\mathrm{1PI})(2\phi,2\eta)}_{\mathrm{mix},k}\right)^2\overline{G}^{(\eta)}_k\left(\overline{G}^{(\eta)}_k+\overline{G}^{(\phi)}_k\right)\\
& +\overline{\mathit{\Gamma}}^{(\mathrm{1PI})(2\phi,2\eta)}_{\mathrm{mix},k}\overline{G}^{(\phi)}_k\left(4\overline{\mathit{\Gamma}}^{(\mathrm{1PI})(4\phi)}_{\mathrm{mix},k}\overline{G}^{(\phi)}_k-15\left(\overline{\mathit{\Gamma}}^{(\mathrm{1PI})(2\phi,1\eta)}_{\mathrm{mix},k}\right)^2\overline{G}^{(\eta)}_k\left(\overline{G}^{(\eta)}_k+2\overline{G}^{(\phi)}_k\right)\right)\\
& +6\left(\overline{\mathit{\Gamma}}^{(\mathrm{1PI})(2\phi,1\eta)}_{\mathrm{mix},k}\right)^2\left(\overline{G}^{(\phi)}_k\right)^2\left(-2\overline{\mathit{\Gamma}}^{(\mathrm{1PI})(4\phi)}_{\mathrm{mix},k}\overline{G}^{(\phi)}_k+\left(\overline{\mathit{\Gamma}}^{(\mathrm{1PI})(2\phi,1\eta)}_{\mathrm{mix},k}\right)^2\overline{G}^{(\eta)}_k\left(\overline{G}^{(\eta)}_k+3\overline{G}^{(\phi)}_k\right)\right)\Bigg)\Bigg)\\
& +\mathcal{F}^{(N=2)}_5\left(\overline{\mathit{\Gamma}}^{(\mathrm{1PI})(n\phi,m\eta)}_{\mathrm{mix},k};\ n+m\leq 6\right)\;,\end{aligned} \tag{361}$$

where the propagators $\overline{G}^{(\phi)(0)}_k$ and $\overline{G}^{(\eta)(0)}_k$ are defined by[49]:

$$\begin{aligned}\left(\overline{\boldsymbol{G}}^{(\phi)(0)}_k\right)^{-1}_{a_1 a_2} &= \left(\overline{G}^{(\phi)(0)}_k\right)^{-1}\delta_{a_1 a_2}\\
&= \left(\overline{\mathit{\Gamma}}^{(2\phi)}_{\mathrm{mix},k=k_\mathrm{i}}+R_k\right)\delta_{a_1 a_2}\quad\forall a_1,a_2\;,\end{aligned} \tag{362}$$

$$\left(\overline{G}^{(\eta)(0)}_k\right)^{-1} = \overline{\mathit{\Gamma}}^{(2\eta)}_{\mathrm{mix},k=k_\mathrm{i}}+R_k\;. \tag{363}$$

The functions $\mathcal{F}^{(N)}_n$, involved in Eqs. (357) to (361), depend on 1PI vertices of order 5 (and 6 for $n\geq 3$) so that they do not contribute to the above equations for $N_\mathrm{max}\leq 4$, according to the initial conditions set by Eq. (85).

## D.2 2PI functional renormalization group U-flow

### D.2.1 Plain U-flow

In this appendix, we present additional pU-flow equations for the (0+0)-D $O(N)$ model, showing notably what the resolution of the Bethe-Salpeter equation amounts to in this framework. Owing to the full conservation of the $O(N)$ symmetry within the 2PI-FRG scheme, the propagator $\overline{G}_\mathfrak{s}$ and the self-energy $\overline{\Sigma}_\mathfrak{s}$ satisfy respectively Eqs. (328) and (329) in the present situation. This implies that the corresponding pair propagator, defined previously by Eq. (127), reduces to:

$$\begin{aligned}\overline{\mathit{\Pi}}_{\mathfrak{s},(a_1,a_1')(a_2,a_2')} &= \overline{G}_{\mathfrak{s},a_1 a_2'}\overline{G}_{\mathfrak{s},a_1' a_2}+\overline{G}_{\mathfrak{s},a_1 a_2}\overline{G}_{\mathfrak{s},a_1' a_2'}\\
&= \overline{G}^2_\mathfrak{s}\left(\delta_{a_1 a_2'}\delta_{a_1' a_2}+\delta_{a_1 a_2}\delta_{a_1' a_2'}\right)\;,\end{aligned} \tag{364}$$

and, according to Eqs. (147) and (149) expressing respectively the bosonic identity matrix and the derivative $\overline{W}^{(2)}_\mathfrak{s}$, this also leads to:

$$\begin{aligned}\overline{W}^{(2)}_{\mathfrak{s},(a_1,a_1')(a_2,a_2')} &= \left(\overline{\mathit{\Pi}}^{\mathrm{inv}}_\mathfrak{s}+\overline{\mathit{\Phi}}^{(2)}_\mathfrak{s}\right)^{\mathrm{inv}}_{(a_1,a_1')(a_2,a_2')}\\
&= \frac{1}{2}\sum^N_{a_3,a_3'=1}\overline{\mathit{\Pi}}_{\mathfrak{s},(a_1,a_1')(a_3,a_3')}\left(\mathcal{I}+\overline{\mathit{\Pi}}_\mathfrak{s}\overline{\mathit{\Phi}}^{(2)}_\mathfrak{s}\right)^{\mathrm{inv}}_{(a_3,a_3')(a_2,a_2')}\\
&= \overline{G}^2_\mathfrak{s}\left(\mathcal{I}+\overline{\mathit{\Pi}}_\mathfrak{s}\overline{\mathit{\Phi}}^{(2)}_\mathfrak{s}\right)^{\mathrm{inv}}_{(a_1,a_1')(a_2,a_2')}\;,\end{aligned} \tag{365}$$

with

$$\begin{aligned}&\left(\mathcal{I}+\overline{\mathit{\Pi}}_\mathfrak{s}\overline{\mathit{\Phi}}^{(2)}_\mathfrak{s}\right)_{(a_1,a_1')(a_2,a_2')}\\
&= \mathcal{I}_{(a_1,a_1')(a_2,a_2')}+\frac{1}{2}\sum^N_{a_3,a_3'=1}\overline{\mathit{\Pi}}_{\mathfrak{s},(a_1,a_1')(a_3,a_3')}\overline{\mathit{\Phi}}^{(2)}_{\mathfrak{s},(a_3,a_3')(a_2,a_2')}\\
&= \delta_{a_1 a_2}\delta_{a_1' a_2'}+\delta_{a_1 a_2'}\delta_{a_1' a_2}+\overline{G}^2_\mathfrak{s}\overline{\mathit{\Phi}}^{(2)}_{\mathfrak{s},(a_1,a_1')(a_2,a_2')}\;.\end{aligned} \tag{366}$$

Furthermore, as mentioned in section 3.2, we have exploited the following analytical expression of $U_\mathfrak{s}$ for all our U-flow calculations:

$$\begin{aligned}U_{\mathfrak{s},a_1 a_2 a_3 a_4} &= R_\mathfrak{s} U_{a_1 a_2 a_3 a_4}\\
&= \mathfrak{s}U_{a_1 a_2 a_3 a_4}\\
&= \frac{\mathfrak{s}\lambda}{3}\left(\delta_{a_1 a_2}\delta_{a_3 a_4}+\delta_{a_1 a_3}\delta_{a_2 a_4}+\delta_{a_1 a_4}\delta_{a_2 a_3}\right)\;,\end{aligned} \tag{367}$$

where the last line was obtained using expression (135) of the two-body interaction $U$. The Kronecker deltas introduced via Eqs. (364) and (367) as well as Eqs. (328) and (329) can be exploited to evaluate sums over color indices involved in the generic pU-flow equations (137), (138) and (140) to (142). For the flow equations expressing $\dot{\overline{\mathit{\Gamma}}}^{(\mathrm{2PI})}_\mathfrak{s}$ and $\dot{\overline{\Sigma}}_\mathfrak{s}$ (i.e. Eqs. (138) and (140)) for instance, this gives us:

[49] Note that $\overline{G}^{(\phi)(0)}_k$ and $\overline{G}^{(\eta)(0)}_k$ are introduced in the flow equations for the same reasons as $\overline{G}^{(0)}_k$ in the framework of the original 1PI-FRG. We thus refer to footnote 9 for more details on this point.

$$\dot{\overline{\Gamma}}_{\mathfrak{s}}^{(\mathrm{2PI})} = \frac{\lambda}{72}\overline{G}_{\mathfrak{s}}^{2}\left(\sum_{a_1,a_2=1}^{N}\left(\mathcal{I}+\overline{\Pi}_{\mathfrak{s}}\overline{\Phi}_{\mathfrak{s}}^{(2)}\right)^{\mathrm{inv}}_{(a_1,a_1)(a_2,a_2)} + 2\sum_{a_1,a_1'=1}^{N}\left(\mathcal{I}+\overline{\Pi}_{\mathfrak{s}}\overline{\Phi}_{\mathfrak{s}}^{(2)}\right)^{\mathrm{inv}}_{(a_1,a_1')(a_1,a_1')} + N\left(2+N\right)\right) , \tag{368}$$

$$\begin{split}\dot{\overline{\Sigma}}_{\mathfrak{s}} = & -\frac{\lambda}{72}\overline{G}_{\mathfrak{s}}\sum_{a_1,a_1',a_2=1}^{N}\left(\mathcal{I}+\overline{\Pi}_{\mathfrak{s}}\overline{\Phi}_{\mathfrak{s}}^{(2)}\right)^{\mathrm{inv}}_{(1,1)(a_1,a_1')} \\ & \times\left(\sum_{a_3,a_4=1}^{N}\left(\mathcal{I}+\overline{\Pi}_{\mathfrak{s}}\overline{\Phi}_{\mathfrak{s}}^{(2)}\right)^{\mathrm{inv}}_{(a_1,a_2)(a_3,a_3)}\left(\mathcal{I}+\overline{\Pi}_{\mathfrak{s}}\overline{\Phi}_{\mathfrak{s}}^{(2)}\right)^{\mathrm{inv}}_{(a_4,a_4)(a_2,a_1')}\right. \\ & \left.+2\sum_{a_3,a_3'=1}^{N}\left(\mathcal{I}+\overline{\Pi}_{\mathfrak{s}}\overline{\Phi}_{\mathfrak{s}}^{(2)}\right)^{\mathrm{inv}}_{(a_1,a_2)(a_3,a_3')}\left(\mathcal{I}+\overline{\Pi}_{\mathfrak{s}}\overline{\Phi}_{\mathfrak{s}}^{(2)}\right)^{\mathrm{inv}}_{(a_3,a_3')(a_2,a_1')}\right) \\ & -\frac{\lambda}{36}\overline{G}_{\mathfrak{s}}\left(N+2\right)\sum_{a_1=1}^{N}\left(\mathcal{I}+\overline{\Pi}_{\mathfrak{s}}\overline{\Phi}_{\mathfrak{s}}^{(2)}\right)^{\mathrm{inv}}_{(1,1)(a_1,a_1)} \\ & +\frac{\lambda}{576}\overline{G}_{\mathfrak{s}}^{4}\sum_{a_1,a_1',a_2,a_3,a_4,a_4',a_5,a_5'=1}^{N}\left(\mathcal{I}+\overline{\Pi}_{\mathfrak{s}}\overline{\Phi}_{\mathfrak{s}}^{(2)}\right)^{\mathrm{inv}}_{(1,1)(a_1,a_1')}\left(\mathcal{I}+\overline{\Pi}_{\mathfrak{s}}\overline{\Phi}_{\mathfrak{s}}^{(2)}\right)^{\mathrm{inv}}_{(a_2,a_2)(a_4,a_4')} \\ & \times\overline{\Phi}^{(3)}_{\mathfrak{s},(a_1,a_1')(a_4,a_4')(a_5,a_5')}\left(\mathcal{I}+\overline{\Pi}_{\mathfrak{s}}\overline{\Phi}_{\mathfrak{s}}^{(2)}\right)^{\mathrm{inv}}_{(a_5,a_5')(a_3,a_3)} \\ & +\frac{\lambda}{288}\overline{G}_{\mathfrak{s}}^{4}\sum_{a_1,a_1',a_2,a_2',a_3,a_3',a_4,a_4'=1}^{N}\left(\mathcal{I}+\overline{\Pi}_{\mathfrak{s}}\overline{\Phi}_{\mathfrak{s}}^{(2)}\right)^{\mathrm{inv}}_{(1,1)(a_1,a_1')}\left(\mathcal{I}+\overline{\Pi}_{\mathfrak{s}}\overline{\Phi}_{\mathfrak{s}}^{(2)}\right)^{\mathrm{inv}}_{(a_2,a_2')(a_3,a_3')} \\ & \times\overline{\Phi}^{(3)}_{\mathfrak{s},(a_1,a_1')(a_3,a_3')(a_4,a_4')}\left(\mathcal{I}+\overline{\Pi}_{\mathfrak{s}}\overline{\Phi}_{\mathfrak{s}}^{(2)}\right)^{\mathrm{inv}}_{(a_4,a_4')(a_2,a_2')} ,\end{split} \tag{369}$$

where the color indices set equal to 1 in the right-hand side of Eq. (369) just result from our convention $\overline{\Sigma}_{\mathfrak{s}} \equiv \overline{\Sigma}_{\mathfrak{s},11}$ (which is arbitrary in the sense that $\overline{\Sigma}_{\mathfrak{s},11} = \overline{\Sigma}_{\mathfrak{s},aa}$ $\forall a$ according to Eq. (329)). Rewriting the pU-flow equations explicitly for our toy model as in Eqs. (368) and (369) can become quickly cumbersome, especially for the flow equations (141) and (142) expressing respectively $\dot{\overline{\Phi}}_{\mathfrak{s}}^{(2)}$ and $\dot{\overline{\Phi}}_{\mathfrak{s}}^{(3)}$. One could also just replace $U_{\mathfrak{s}}$, $\overline{\Pi}_{\mathfrak{s}}$, $\frac{\delta\overline{\Pi}_{\mathfrak{s}}}{\delta\overline{G}_{\mathfrak{s},\gamma}}$ and $\frac{\delta^2\overline{\Pi}_{\mathfrak{s}}}{\delta\overline{G}_{\mathfrak{s},\gamma_1}\delta\overline{G}_{\mathfrak{s},\gamma_2}}$ by their expressions in terms of $\overline{G}_{\mathfrak{s}}$ and Kronecker deltas in the generic pU-flow equations given in section 3.2, and then solve numerically the (quite lengthy) differential equations thus obtained. Another possibility is to exploit the (0+0)-D character of the studied model with the equality $\Gamma^{(\mathrm{2PI})}(G) = \Gamma^{(\mathrm{2PPI})}(\rho)$. More specifically, this amounts to rederiving the flow equations underlying a 2PI-FRG approach in the 2PPI-FRG framework, yielding in this way less cumbersome flow equations (as compared to Eqs. (368), (369) and their counterparts expressing $\dot{\overline{\Phi}}_{\mathfrak{s}}^{(n)}$ with $n \geq 2$). This is achieved by treating the exact flow equation for $\Gamma^{(\mathrm{2PPI})}(\rho)$ (i.e. the (0+0)-D version of Eq. (198)) with a vertex expansion: for the 2PI-FRG pU-flow, such a procedure is equivalent to implementing the KS-FRG according to our findings of section 4.[50]

We can also see e.g. from Eqs. (368) and (369) that, for the (0+0)-D $O(N)$ model, the resolution of the Bethe-Salpeter equation within the pU-flow simply amounts to inverting the bosonic matrix $\mathcal{I}+\overline{\Pi}_{\mathfrak{s}}\overline{\Phi}_{\mathfrak{s}}^{(2)}$, i.e. to solve:

$$\mathcal{I}_{(a_1,a_1')(a_2,a_2')} = \frac{1}{2}\sum_{a_3,a_3'=1}^{N}\left(\mathcal{I}+\overline{\Pi}_{\mathfrak{s}}\overline{\Phi}_{\mathfrak{s}}^{(2)}\right)_{(a_1,a_1')(a_3,a_3')} \times\left(\mathcal{I}+\overline{\Pi}_{\mathfrak{s}}\overline{\Phi}_{\mathfrak{s}}^{(2)}\right)^{\mathrm{inv}}_{(a_3,a_3')(a_2,a_2')} , \tag{370}$$

which is nothing other than a set of $N^4$ coupled algebraic equations. We will discuss in more detail the resolution of such a system of equations at $N=2$ in the upcoming section on the mU-flow.

### D.2.2 Modified U-flow

As a next step, we address further the mU-flow equations. As opposed to the pU-flow (with Eqs. (137), (138)

[50] See also appendix F.8.2 of Ref. [46] for an application of this idea for the 2PI-FRG mU-flow, where we develop in this way a mU-flow version of the 2PPI-FRG only valid in the (0+0)-D framework.

and (140) to (142)), generic differential equations underpinning the mU-flow were not given in section 3. We remedy this below for the mU-flow with Hartree-Fock starting point, i.e. with $N_{\text{SCPT}} = 1$:

$$\dot{\overline{\boldsymbol{G}}}_{\mathfrak{s},\alpha_1\alpha'_1} = \int_{\alpha_2,\alpha'_2} \overline{\boldsymbol{G}}_{\mathfrak{s},\alpha_1\alpha_2}\dot{\overline{\boldsymbol{\Sigma}}}_{\mathfrak{s},\alpha_2\alpha'_2}\overline{\boldsymbol{G}}_{\mathfrak{s},\alpha'_2\alpha'_1} \;, \tag{371}$$

$$\dot{\overline{\boldsymbol{\Gamma}}}^{(\text{2PI})}_{\mathfrak{s}} = \frac{1}{6}\dot{U}_{\mathfrak{s},\hat{\gamma}_1\hat{\gamma}_2}\left(\overline{W}^{(2)}_{\mathfrak{s}} - \overline{\Pi}_{\mathfrak{s}}\right)_{\hat{\gamma}_2\hat{\gamma}_1} \;, \tag{372}$$

$$\begin{aligned}\dot{\overline{\boldsymbol{\Sigma}}}_{\mathfrak{s},\gamma} = & -\frac{1}{3}\left(\mathcal{I}+\overline{\Pi}_{\mathfrak{s}}\overline{\boldsymbol{\Phi}}^{(2)}_{\mathfrak{s}}\right)^{\text{inv}}_{\gamma\hat{\gamma}_1} \\ & \times\Bigg(\left[2\left(\mathcal{I}+\overline{\Pi}_{\mathfrak{s}}\left(\overline{\boldsymbol{\Phi}}^{(2)}_{\mathfrak{s}}-U+U_{\mathfrak{s}}\right)\right)^{\text{inv}}\dot{U}_{\mathfrak{s}}\left(\mathcal{I}+\overline{\Pi}_{\mathfrak{s}}\left(\overline{\boldsymbol{\Phi}}^{(2)}_{\mathfrak{s}}-U+U_{\mathfrak{s}}\right)\right)^{\text{inv}}+\dot{U}_{\mathfrak{s}}\right]_{\hat{\alpha}_1\hat{\alpha}_2\hat{\alpha}'_2\hat{\alpha}'_1}\overline{\boldsymbol{G}}_{\mathfrak{s},\hat{\gamma}_2} \\ & -\frac{1}{2}\dot{U}_{\mathfrak{s},\hat{\gamma}_2\hat{\gamma}_3}\overline{W}^{(2)}_{\mathfrak{s},\hat{\gamma}_3\hat{\gamma}_4}\overline{\boldsymbol{\Phi}}^{(3)}_{\mathfrak{s},\hat{\gamma}_1\hat{\gamma}_4\hat{\gamma}_5}\overline{W}^{(2)}_{\mathfrak{s},\hat{\gamma}_5\hat{\gamma}_2} - 3\dot{U}_{\mathfrak{s},\hat{\gamma}_1\hat{\gamma}_2}\overline{\boldsymbol{G}}_{\mathfrak{s},\hat{\gamma}_2}\Bigg) \;,\end{aligned} \tag{373}$$

$$\begin{aligned}\dot{\overline{\boldsymbol{\Phi}}}^{(2)}_{\mathfrak{s},\gamma_1\gamma_2} = & \frac{1}{3}\dot{U}_{\mathfrak{s},\hat{\gamma}_1\hat{\gamma}_2}\Bigg[\overline{W}^{(2)}_{\mathfrak{s},\hat{\gamma}_2\hat{\gamma}_3}\left(\overline{\Pi}^{\text{inv}}_{\mathfrak{s},\hat{\gamma}_3\hat{\gamma}_4}\frac{\delta\overline{\Pi}_{\mathfrak{s},\hat{\gamma}_4\hat{\gamma}_5}}{\delta\overline{\boldsymbol{G}}_{\mathfrak{s},\gamma_1}}\overline{\Pi}^{\text{inv}}_{\mathfrak{s},\hat{\gamma}_5\hat{\gamma}_6}-\overline{\boldsymbol{\Phi}}^{(3)}_{\mathfrak{s},\gamma_1\hat{\gamma}_3\hat{\gamma}_6}\right)\overline{W}^{(2)}_{\mathfrak{s},\hat{\gamma}_6\hat{\gamma}_7} \\ & \times\left(\overline{\Pi}^{\text{inv}}_{\mathfrak{s},\hat{\gamma}_7\hat{\gamma}_8}\frac{\delta\overline{\Pi}_{\mathfrak{s},\hat{\gamma}_8\hat{\gamma}_9}}{\delta\overline{\boldsymbol{G}}_{\mathfrak{s},\gamma_2}}\overline{\Pi}^{\text{inv}}_{\mathfrak{s},\hat{\gamma}_9\hat{\gamma}_{10}}-\overline{\boldsymbol{\Phi}}^{(3)}_{\mathfrak{s},\gamma_2\hat{\gamma}_7\hat{\gamma}_{10}}\right)\overline{W}^{(2)}_{\mathfrak{s},\hat{\gamma}_{10}\hat{\gamma}_1} \\ & -\overline{W}^{(2)}_{\mathfrak{s},\hat{\gamma}_2\hat{\gamma}_3}\overline{\Pi}^{\text{inv}}_{\mathfrak{s},\hat{\gamma}_3\hat{\gamma}_4}\frac{\delta\overline{\Pi}_{\mathfrak{s},\hat{\gamma}_4\hat{\gamma}_5}}{\delta\overline{\boldsymbol{G}}_{\mathfrak{s},\gamma_1}}\overline{\Pi}^{\text{inv}}_{\mathfrak{s},\hat{\gamma}_5\hat{\gamma}_6}\frac{\delta\overline{\Pi}_{\mathfrak{s},\hat{\gamma}_6\hat{\gamma}_7}}{\delta\overline{\boldsymbol{G}}_{\mathfrak{s},\gamma_2}}\overline{\Pi}^{\text{inv}}_{\mathfrak{s},\hat{\gamma}_7\hat{\gamma}_8}\overline{W}^{(2)}_{\mathfrak{s},\hat{\gamma}_8\hat{\gamma}_1} \\ & +\frac{1}{2}\overline{W}^{(2)}_{\mathfrak{s},\hat{\gamma}_2\hat{\gamma}_3}\left(\overline{\Pi}^{\text{inv}}_{\mathfrak{s},\hat{\gamma}_3\hat{\gamma}_4}\frac{\delta^2\overline{\Pi}_{\mathfrak{s},\hat{\gamma}_4\hat{\gamma}_5}}{\delta\overline{\boldsymbol{G}}_{\mathfrak{s},\gamma_1}\delta\overline{\boldsymbol{G}}_{\mathfrak{s},\gamma_2}}\overline{\Pi}^{\text{inv}}_{\mathfrak{s},\hat{\gamma}_5\hat{\gamma}_6}-\overline{\boldsymbol{\Phi}}^{(4)}_{\mathfrak{s},\gamma_1\gamma_2\hat{\gamma}_3\hat{\gamma}_6}\right)\overline{W}^{(2)}_{\mathfrak{s},\hat{\gamma}_6\hat{\gamma}_1}+\frac{1}{4}\frac{\delta^2\overline{\Pi}_{\mathfrak{s},\hat{\gamma}_2\hat{\gamma}_1}}{\delta\overline{\boldsymbol{G}}_{\mathfrak{s},\gamma_1}\delta\overline{\boldsymbol{G}}_{\mathfrak{s},\gamma_2}}\Bigg]-\dot{U}_{\mathfrak{s},\gamma_1\gamma_2} \\ & +\dot{\overline{\boldsymbol{G}}}_{\mathfrak{s},\hat{\gamma}}\overline{\boldsymbol{\Phi}}^{(3)}_{\mathfrak{s},\hat{\gamma}\gamma_1\gamma_2} \;,\end{aligned} \tag{374}$$

$$\begin{aligned}\dot{\overline{\boldsymbol{\Phi}}}^{(3)}_{\mathfrak{s},\gamma_1\gamma_2\gamma_3} = & \frac{1}{3}\dot{U}_{\mathfrak{s},\hat{\gamma}_1\hat{\gamma}_2}\Bigg[3\overline{W}^{(2)}_{\mathfrak{s},\hat{\gamma}_2\hat{\gamma}_3}\left(\overline{\Pi}^{\text{inv}}_{\mathfrak{s},\hat{\gamma}_3\hat{\gamma}_4}\frac{\delta\overline{\Pi}_{\mathfrak{s},\hat{\gamma}_4\hat{\gamma}_5}}{\delta\overline{\boldsymbol{G}}_{\mathfrak{s},\gamma_1}}\overline{\Pi}^{\text{inv}}_{\mathfrak{s},\hat{\gamma}_5\hat{\gamma}_6}-\overline{\boldsymbol{\Phi}}^{(3)}_{\mathfrak{s},\gamma_1\hat{\gamma}_3\hat{\gamma}_6}\right)\overline{W}^{(2)}_{\mathfrak{s},\hat{\gamma}_6\hat{\gamma}_7}\Bigg(\overline{\Pi}^{\text{inv}}_{\mathfrak{s},\hat{\gamma}_7\hat{\gamma}_8}\frac{\delta\overline{\Pi}_{\mathfrak{s},\hat{\gamma}_8\hat{\gamma}_9}}{\delta\overline{\boldsymbol{G}}_{\mathfrak{s},\gamma_2}}\overline{\Pi}^{\text{inv}}_{\mathfrak{s},\hat{\gamma}_9\hat{\gamma}_{10}} \\ & -\overline{\boldsymbol{\Phi}}^{(3)}_{\mathfrak{s},\gamma_2\hat{\gamma}_7\hat{\gamma}_{10}}\Bigg)\overline{W}^{(2)}_{\mathfrak{s},\hat{\gamma}_{10}\hat{\gamma}_{11}}\left(\overline{\Pi}^{\text{inv}}_{\mathfrak{s},\hat{\gamma}_{11}\hat{\gamma}_{12}}\frac{\delta\overline{\Pi}_{\mathfrak{s},\hat{\gamma}_{12}\hat{\gamma}_{13}}}{\delta\overline{\boldsymbol{G}}_{\mathfrak{s},\gamma_3}}\overline{\Pi}^{\text{inv}}_{\mathfrak{s},\hat{\gamma}_{13}\hat{\gamma}_{14}}-\overline{\boldsymbol{\Phi}}^{(3)}_{\mathfrak{s},\gamma_3\hat{\gamma}_{11}\hat{\gamma}_{14}}\right)\overline{W}^{(2)}_{\mathfrak{s},\hat{\gamma}_{14}\hat{\gamma}_1} \\ & +\Bigg(\overline{W}^{(2)}_{\mathfrak{s},\hat{\gamma}_2\hat{\gamma}_3}\left(\overline{\Pi}^{\text{inv}}_{\mathfrak{s},\hat{\gamma}_3\hat{\gamma}_4}\frac{\delta\overline{\Pi}_{\mathfrak{s},\hat{\gamma}_4\hat{\gamma}_5}}{\delta\overline{\boldsymbol{G}}_{\mathfrak{s},\gamma_1}}\overline{\Pi}^{\text{inv}}_{\mathfrak{s},\hat{\gamma}_5\hat{\gamma}_6}-\overline{\boldsymbol{\Phi}}^{(3)}_{\mathfrak{s},\gamma_1\hat{\gamma}_3\hat{\gamma}_6}\right)\overline{W}^{(2)}_{\mathfrak{s},\hat{\gamma}_6\hat{\gamma}_7} \\ & \times\left(\overline{\Pi}^{\text{inv}}_{\mathfrak{s},\hat{\gamma}_7\hat{\gamma}_8}\frac{\delta^2\overline{\Pi}_{\mathfrak{s},\hat{\gamma}_8\hat{\gamma}_9}}{\delta\overline{\boldsymbol{G}}_{\mathfrak{s},\gamma_2}\delta\overline{\boldsymbol{G}}_{\mathfrak{s},\gamma_3}}\overline{\Pi}^{\text{inv}}_{\mathfrak{s},\hat{\gamma}_9\hat{\gamma}_{10}}-\overline{\boldsymbol{\Phi}}^{(4)}_{\mathfrak{s},\gamma_2\gamma_3\hat{\gamma}_7\hat{\gamma}_{10}}\right)\overline{W}^{(2)}_{\mathfrak{s},\hat{\gamma}_{10}\hat{\gamma}_1}+(\gamma_2,\gamma_1,\gamma_3)+(\gamma_3,\gamma_1,\gamma_2)\Bigg) \\ & -2\Bigg(\overline{W}^{(2)}_{\mathfrak{s},\hat{\gamma}_2\hat{\gamma}_3}\left(\overline{\Pi}^{\text{inv}}_{\mathfrak{s},\hat{\gamma}_3\hat{\gamma}_4}\frac{\delta\overline{\Pi}_{\mathfrak{s},\hat{\gamma}_4\hat{\gamma}_5}}{\delta\overline{\boldsymbol{G}}_{\mathfrak{s},\gamma_1}}\overline{\Pi}^{\text{inv}}_{\mathfrak{s},\hat{\gamma}_5\hat{\gamma}_6}-\overline{\boldsymbol{\Phi}}^{(3)}_{\mathfrak{s},\gamma_1\hat{\gamma}_3\hat{\gamma}_6}\right)\overline{W}^{(2)}_{\mathfrak{s},\hat{\gamma}_6\hat{\gamma}_7} \\ & \times\overline{\Pi}^{\text{inv}}_{\mathfrak{s},\hat{\gamma}_7\hat{\gamma}_8}\frac{\delta\overline{\Pi}_{\mathfrak{s},\hat{\gamma}_8\hat{\gamma}_9}}{\delta\overline{\boldsymbol{G}}_{\mathfrak{s},\gamma_2}}\overline{\Pi}^{\text{inv}}_{\mathfrak{s},\hat{\gamma}_9\hat{\gamma}_{10}}\frac{\delta\overline{\Pi}_{\mathfrak{s},\hat{\gamma}_{10}\hat{\gamma}_{11}}}{\delta\overline{\boldsymbol{G}}_{\mathfrak{s},\gamma_3}}\overline{\Pi}^{\text{inv}}_{\mathfrak{s},\hat{\gamma}_{11}\hat{\gamma}_{12}}\overline{W}^{(2)}_{\mathfrak{s},\hat{\gamma}_{12}\hat{\gamma}_1}+(\gamma_2,\gamma_1,\gamma_3)+(\gamma_3,\gamma_1,\gamma_2)\Bigg) \\ & +3\overline{W}^{(2)}_{\mathfrak{s},\hat{\gamma}_2\hat{\gamma}_3}\overline{\Pi}^{\text{inv}}_{\mathfrak{s},\hat{\gamma}_3\hat{\gamma}_4}\frac{\delta\overline{\Pi}_{\mathfrak{s},\hat{\gamma}_4\hat{\gamma}_5}}{\delta\overline{\boldsymbol{G}}_{\mathfrak{s},\gamma_1}}\overline{\Pi}^{\text{inv}}_{\mathfrak{s},\hat{\gamma}_5\hat{\gamma}_6}\frac{\delta\overline{\Pi}_{\mathfrak{s},\hat{\gamma}_6\hat{\gamma}_7}}{\delta\overline{\boldsymbol{G}}_{\mathfrak{s},\gamma_2}}\overline{\Pi}^{\text{inv}}_{\mathfrak{s},\hat{\gamma}_7\hat{\gamma}_8}\frac{\delta\overline{\Pi}_{\mathfrak{s},\hat{\gamma}_8\hat{\gamma}_9}}{\delta\overline{\boldsymbol{G}}_{\mathfrak{s},\gamma_3}}\overline{\Pi}^{\text{inv}}_{\mathfrak{s},\hat{\gamma}_9\hat{\gamma}_{10}}\overline{W}^{(2)}_{\mathfrak{s},\hat{\gamma}_{10}\hat{\gamma}_1} \\ & -\Bigg(\overline{W}^{(2)}_{\mathfrak{s},\hat{\gamma}_2\hat{\gamma}_3}\overline{\Pi}^{\text{inv}}_{\mathfrak{s},\hat{\gamma}_3\hat{\gamma}_4}\frac{\delta\overline{\Pi}_{\mathfrak{s},\hat{\gamma}_4\hat{\gamma}_5}}{\delta\overline{\boldsymbol{G}}_{\mathfrak{s},\gamma_1}}\overline{\Pi}^{\text{inv}}_{\mathfrak{s},\hat{\gamma}_5\hat{\gamma}_6}\frac{\delta^2\overline{\Pi}_{\mathfrak{s},\hat{\gamma}_6\hat{\gamma}_7}}{\delta\overline{\boldsymbol{G}}_{\mathfrak{s},\gamma_2}\delta\overline{\boldsymbol{G}}_{\mathfrak{s},\gamma_3}}\overline{\Pi}^{\text{inv}}_{\mathfrak{s},\hat{\gamma}_7\hat{\gamma}_8}\overline{W}^{(2)}_{\mathfrak{s},\hat{\gamma}_8\hat{\gamma}_1}+(\gamma_2,\gamma_1,\gamma_3)+(\gamma_3,\gamma_1,\gamma_2)\Bigg) \\ & -\frac{1}{2}\overline{W}^{(2)}_{\mathfrak{s},\hat{\gamma}_2\hat{\gamma}_3}\overline{\boldsymbol{\Phi}}^{(5)}_{\mathfrak{s},\gamma_1\gamma_2\gamma_3\hat{\gamma}_3\hat{\gamma}_4}\overline{W}^{(2)}_{\mathfrak{s},\hat{\gamma}_4\hat{\gamma}_1}\Bigg]+\dot{\overline{\boldsymbol{G}}}_{\mathfrak{s},\hat{\gamma}}\overline{\boldsymbol{\Phi}}^{(4)}_{\mathfrak{s},\hat{\gamma}\gamma_1\gamma_2\gamma_3} \;,\end{aligned} \tag{375}$$

where we have used again notation (144) and exploited a slight modification of definition (143), i.e.:

$$\frac{\delta^n \overline{\Pi}_{\mathfrak{s}}}{\delta\overline{\boldsymbol{G}}_{\mathfrak{s},\gamma_1}\cdots\delta\overline{\boldsymbol{G}}_{\mathfrak{s},\gamma_n}} \equiv \left.\frac{\delta^n \Pi[G]}{\delta G_{\gamma_1}\cdots\delta G_{\gamma_n}}\right|_{G=\overline{\boldsymbol{G}}_{\mathfrak{s}}} . \quad (376)$$

In short, Eq. (371) is obtained by differentiating Dyson equation in the form of Eq. (164) whereas Eqs. (372) to (375) are derived in the same way as their pU-flow counterparts (i.e. Eqs. (138) and (140) to (142)) but using in addition the following substitutions in order to replace the original flowing quantities of the pU-flow by their bold counterparts[51]:

$$\begin{cases} \Gamma^{(2\mathrm{PI})}_{\mathfrak{s}}[G] \rightarrow \boldsymbol{\Gamma}^{(2\mathrm{PI})}_{\mathfrak{s}}[G] + \frac{1}{2}\Delta U_{\mathfrak{s},\hat{\gamma}_1\hat{\gamma}_2} G_{\hat{\gamma}_1} G_{\hat{\gamma}_2} \,, & (377\text{a}) \\ \Sigma_{\mathfrak{s},\gamma}[G] \rightarrow \boldsymbol{\Sigma}_{\mathfrak{s},\gamma}[G] - \Delta U_{\mathfrak{s},\gamma\hat{\gamma}} G_{\hat{\gamma}} \,, & (377\text{b}) \\ \Phi^{(2)}_{\mathfrak{s},\gamma_1\gamma_2}[G] \rightarrow \boldsymbol{\Phi}^{(2)}_{\mathfrak{s},\gamma_1\gamma_2}[G] + \Delta U_{\mathfrak{s},\gamma_1\gamma_2} \,, & (377\text{c}) \\ \Phi^{(n)}_{\mathfrak{s},\gamma_1\cdots\gamma_n}[G] \rightarrow \boldsymbol{\Phi}^{(n)}_{\mathfrak{s},\gamma_1\cdots\gamma_n}[G] \quad \forall n \geq 3 \,, & (377\text{d}) \end{cases}$$

where $\Delta U_{\mathfrak{s}} \equiv U_{\mathfrak{s}} - U$. Note that these substitutions can all be inferred from definition (159) at $N_{\mathrm{SCPT}} = 1$. Moreover, we stress that, according to transformation (377c) (combined with Eq. (149)), the Bethe-Salpeter equation takes the form:

$$\overline{W}^{(2)}_{\mathfrak{s}} = \left(\overline{\Pi}^{\mathrm{inv}}_{\mathfrak{s}} + \overline{\boldsymbol{\Phi}}^{(2)}_{\mathfrak{s}} + \Delta U_{\mathfrak{s}}\right)^{\mathrm{inv}} , \quad (378)$$

in the framework of the mU-flow at $N_{\mathrm{SCPT}} = 1$.

We then continue with a discussion similar to that of section D.2.1. To that end, we still focus on the mU-flow with $N_{\mathrm{SCPT}} = 1$ and give the corresponding flow equations in the framework of the (0+0)-D $O(N)$ model for all $N$. The underlying idea here is once again to illustrate how the conservation of the $O(N)$ symmetry is used to simplify generic flow equations (given e.g. by Eqs. (371) to (375) in the studied case) and explain how the Bethe-Salpeter equation is solved throughout the flow. For the toy model under consideration, the bold quantities underlying the mU-flow at $N_{\mathrm{SCPT}} = 1$ rely on the following modified LW functional:

$$\begin{aligned} \boldsymbol{\Phi}_{\mathfrak{s}}(G) \equiv & \; \Phi_{\mathfrak{s}}(G) + \Phi_{\mathrm{SCPT},N_{\mathrm{SCPT}}=1}(U,G) \\ & - \Phi_{\mathrm{SCPT},N_{\mathrm{SCPT}}=1}(U_{\mathfrak{s}},G) \\ = & \; \Phi_{\mathfrak{s}}(G) + \lambda\left(1-\mathfrak{s}\right)\left(\frac{1}{24}\left(\sum_{a_1=1}^{N} G_{a_1 a_1}\right)^2\right. \quad (379) \\ & \left. + \frac{1}{12}\sum_{a_1,a_2=1}^{N} G^2_{a_1 a_2}\right) , \end{aligned}$$

with

$$\Phi_{\mathrm{SCPT},N_{\mathrm{SCPT}}=1}(U,G) = \frac{1}{24}$$ 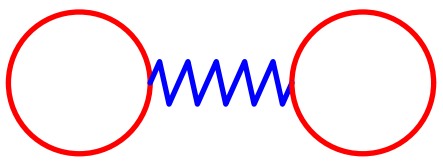

$$+ \frac{1}{12}$$ 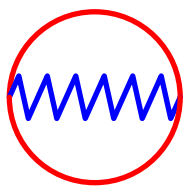 (380)

$$\begin{aligned} & = \frac{\lambda}{24}\left(\sum_{a_1=1}^{N} G_{a_1 a_1}\right)^2 \\ & + \frac{\lambda}{12}\sum_{a_1,a_2=1}^{N} G^2_{a_1 a_2} . \end{aligned}$$

Moreover, the conservation of the $O(N)$ symmetry implies:

$$\overline{\boldsymbol{G}}_{\mathfrak{s},aa'} = \overline{\boldsymbol{G}}_{\mathfrak{s}}\, \delta_{aa'} \quad \forall \mathfrak{s} \,, \quad (381)$$

$$\overline{\boldsymbol{\Sigma}}_{\mathfrak{s},aa'} = \overline{\boldsymbol{\Sigma}}_{\mathfrak{s}}\, \delta_{aa'} \quad \forall \mathfrak{s} \,, \quad (382)$$

$$\overline{\Pi}_{\mathfrak{s},(a_1,a'_1)(a_2,a'_2)} = \overline{\boldsymbol{G}}^2_{\mathfrak{s}}\left(\delta_{a_1 a'_2}\delta_{a'_1 a_2} + \delta_{a_1 a_2}\delta_{a'_1 a'_2}\right) , \quad (383)$$

and the expression of $\overline{W}^{(2)}_{\mathfrak{s}}$ to consider now can be directly inferred from Eq. (378) (alongside with Eqs. (379), (380) and (383)):

$$\begin{aligned} & \overline{W}^{(2)}_{\mathfrak{s},(a_1,a'_1)(a_2,a'_2)} \\ & = \Big(\overline{\Pi}^{\mathrm{inv}}_{\mathfrak{s}} + \overline{\boldsymbol{\Phi}}^{(2)}_{\mathfrak{s}} + \Phi^{(2)}_{\mathrm{SCPT},N_{\mathrm{SCPT}}=1}\big(U_{\mathfrak{s}}, G=\overline{\boldsymbol{G}}_{\mathfrak{s}}\big) \\ & \quad - \Phi^{(2)}_{\mathrm{SCPT},N_{\mathrm{SCPT}}=1}\big(U, G=\overline{\boldsymbol{G}}_{\mathfrak{s}}\big)\Big)^{\mathrm{inv}}_{(a_1,a'_1)(a_2,a'_2)} \quad (384) \\ & = \frac{1}{2}\sum_{a_3,a'_3=1}^{N} \overline{\Pi}_{\mathfrak{s},(a_1,a'_1)(a_3,a'_3)} \overline{\Upsilon}^{\mathrm{inv}}_{\mathfrak{s},(a_3,a'_3)(a_2,a'_2)} \\ & = \overline{\boldsymbol{G}}^2_{\mathfrak{s}} \overline{\Upsilon}^{\mathrm{inv}}_{\mathfrak{s},(a_1,a'_1)(a_2,a'_2)} \,, \end{aligned}$$

with

$$\begin{aligned} & \overline{\Upsilon}_{\mathfrak{s},(a_1,a'_1)(a_2,a'_2)} \\ & \equiv \Big(\mathcal{I} + \overline{\Pi}_{\mathfrak{s}}\Big(\overline{\boldsymbol{\Phi}}^{(2)}_{\mathfrak{s}} + \Phi^{(2)}_{\mathrm{SCPT},N_{\mathrm{SCPT}}=1}\big(U_{\mathfrak{s}}, G=\overline{\boldsymbol{G}}_{\mathfrak{s}}\big) \\ & \quad - \Phi^{(2)}_{\mathrm{SCPT},N_{\mathrm{SCPT}}=1}\big(U, G=\overline{\boldsymbol{G}}_{\mathfrak{s}}\big)\Big)\Big)_{(a_1,a'_1)(a_2,a'_2)} \\ & = \delta_{a_1 a_2}\delta_{a'_1 a'_2} + \delta_{a_1 a'_2}\delta_{a'_1 a_2} + \overline{\boldsymbol{G}}^2_{\mathfrak{s}}\overline{\boldsymbol{\Phi}}^{(2)}_{\mathfrak{s},(a_1,a'_1)(a_2,a'_2)} \\ & \quad - \frac{\lambda}{3}(1-\mathfrak{s})\overline{\boldsymbol{G}}^2_{\mathfrak{s}}\left(\delta_{a_1 a'_1}\delta_{a_2 a'_2} + \delta_{a_1 a_2}\delta_{a'_1 a'_2} + \delta_{a_1 a'_2}\delta_{a'_1 a_2}\right) , \end{aligned} \quad (385)$$

where the last equality simply follows from definition (281) of the bosonic identity matrix and:

[51] We refer to appendix F.4.2 of Ref. [46] for more details on these derivations.

$$\left(\Phi^{(2)}_{\mathrm{SCPT},N_{\mathrm{SCPT}}=1}\left(U_{\mathfrak{s}},G=\overline{\boldsymbol{G}}_{\mathfrak{s}}\right)-\Phi^{(2)}_{\mathrm{SCPT},N_{\mathrm{SCPT}}=1}\left(U,G=\overline{\boldsymbol{G}}_{\mathfrak{s}}\right)\right)_{(a_1,a_1')(a_2,a_2')}=-\frac{\lambda}{3}(1-\mathfrak{s})\left(\delta_{a_1a_1'}\delta_{a_2a_2'}+\delta_{a_1a_2}\delta_{a_1'a_2'}+\delta_{a_1a_2'}\delta_{a_1'a_2}\right)\;, \tag{386}$$

as a result of Eq. (380). We also stress once again that the upper bars label functionals evaluated at $G=\overline{\boldsymbol{G}}_{\mathfrak{s}}$ instead of $G=\overline{G}_{\mathfrak{s}}$ in the framework of the mU-flow. With relations (379) to (385), we can rewrite for example the mU-flow equations (372) and (373) for the (0+0)-D $O(N)$ model and obtain:

$$\dot{\overline{\boldsymbol{\Gamma}}}^{(\mathrm{2PI})}_{\mathfrak{s}}=\frac{\lambda}{72}\overline{\boldsymbol{G}}_{\mathfrak{s}}^2\left(\sum_{a_1,a_2=1}^{N}\overline{\Upsilon}^{\mathrm{inv}}_{\mathfrak{s},(a_1,a_1)(a_2,a_2)}+2\sum_{a_1,a_1'=1}^{N}\overline{\Upsilon}^{\mathrm{inv}}_{\mathfrak{s},(a_1,a_1')(a_1,a_1')}-2N\left(N+2\right)\right)\;, \tag{387}$$

$$\begin{aligned}\dot{\overline{\boldsymbol{\Sigma}}}_{\mathfrak{s}}=&-\frac{\lambda}{72}\overline{\boldsymbol{G}}_{\mathfrak{s}}\sum_{a_1,a_1',a_2=1}^{N}\left(\mathcal{I}+\overline{\Pi}_{\mathfrak{s}}\overline{\boldsymbol{\Phi}}^{(2)}_{\mathfrak{s}}\right)^{\mathrm{inv}}_{(1,1)(a_1,a_1')}\\&\times\left(\sum_{a_3,a_4=1}^{N}\overline{\Upsilon}^{\mathrm{inv}}_{\mathfrak{s},(a_1,a_2)(a_3,a_3)}\overline{\Upsilon}^{\mathrm{inv}}_{\mathfrak{s},(a_4,a_4)(a_2,a_1')}+2\sum_{a_3,a_3'=1}^{N}\overline{\Upsilon}^{\mathrm{inv}}_{\mathfrak{s},(a_1,a_2)(a_3,a_3')}\overline{\Upsilon}^{\mathrm{inv}}_{\mathfrak{s},(a_3,a_3')(a_2,a_1')}\right)\\&+\frac{\lambda}{18}\overline{\boldsymbol{G}}_{\mathfrak{s}}\left(N+2\right)\sum_{a_1=1}^{N}\left(\mathcal{I}+\overline{\Pi}_{\mathfrak{s}}\overline{\boldsymbol{\Phi}}^{(2)}_{\mathfrak{s}}\right)^{\mathrm{inv}}_{(1,1)(a_1,a_1)}\\&+\frac{\lambda}{576}\overline{\boldsymbol{G}}_{\mathfrak{s}}^4\sum_{a_1,a_1',a_2,a_3,a_4,a_4',a_5,a_5'=1}^{N}\left(\mathcal{I}+\overline{\Pi}_{\mathfrak{s}}\overline{\boldsymbol{\Phi}}^{(2)}_{\mathfrak{s}}\right)^{\mathrm{inv}}_{(1,1)(a_1,a_1')}\overline{\Upsilon}^{\mathrm{inv}}_{\mathfrak{s},(a_2,a_2)(a_4,a_4')}\overline{\boldsymbol{\Phi}}^{(3)}_{\mathfrak{s},(a_1,a_1')(a_4,a_4')(a_5,a_5')}\overline{\Upsilon}^{\mathrm{inv}}_{\mathfrak{s},(a_5,a_5')(a_3,a_3)}\\&+\frac{\lambda}{288}\overline{\boldsymbol{G}}_{\mathfrak{s}}^4\sum_{a_1,a_1',a_2,a_2',a_3,a_3',a_4,a_4'=1}^{N}\left(\mathcal{I}+\overline{\Pi}_{\mathfrak{s}}\overline{\boldsymbol{\Phi}}^{(2)}_{\mathfrak{s}}\right)^{\mathrm{inv}}_{(1,1)(a_1,a_1')}\overline{\Upsilon}^{\mathrm{inv}}_{\mathfrak{s},(a_2,a_2')(a_3,a_3')}\overline{\boldsymbol{\Phi}}^{(3)}_{\mathfrak{s},(a_1,a_1')(a_3,a_3')(a_4,a_4')}\overline{\Upsilon}^{\mathrm{inv}}_{\mathfrak{s},(a_4,a_4')(a_2,a_2')}\;,\end{aligned} \tag{388}$$

where the color indices equal to 1 in the right-hand side of Eq. (388) follow from the convention $\overline{\boldsymbol{\Sigma}}_{\mathfrak{s}}\equiv\overline{\boldsymbol{\Sigma}}_{\mathfrak{s},11}=\overline{\boldsymbol{\Sigma}}_{\mathfrak{s},aa}$ $\forall a$, similarly to Eq. (369). We stress also that our previous comments, put forward right below Eq. (369) and concerning the derivation of the pU-flow equations for our (0+0)-D toy model, also apply in the mU-flow framework. In particular, one can exploit the 2PPI-FRG framework in (0+0)-D so as to implement the 2PI-FRG mU-flow by solving flow equations simpler than Eqs. (387), (388) and their counterparts expressing $\dot{\overline{\boldsymbol{\Phi}}}^{(n)}_{\mathfrak{s}}$ with $n\geq 2$[50].

Let us then focus more deeply on the quantities involved in Eqs. (387) and (388), which will enable us to further clarify how the Bethe-Salpeter equation is solved within our (0+0)-D applications of the mU-flow at $N_{\mathrm{SCPT}}=1$. The bosonic matrix $\mathcal{I}+\overline{\Pi}_{\mathfrak{s}}\overline{\boldsymbol{\Phi}}^{(2)}_{\mathfrak{s}}$ satisfies the equality:

$$\left(\mathcal{I}+\overline{\Pi}_{\mathfrak{s}}\overline{\boldsymbol{\Phi}}^{(2)}_{\mathfrak{s}}\right)_{(a_1,a_1')(a_2,a_2')}=\delta_{a_1a_2}\delta_{a_1'a_2'}+\delta_{a_1a_2'}\delta_{a_1'a_2}+\overline{\boldsymbol{G}}_{\mathfrak{s}}^2\overline{\boldsymbol{\Phi}}^{(2)}_{\mathfrak{s},(a_1,a_1')(a_2,a_2')}\;, \tag{389}$$

which can be obtained by following derivation (366) with $\overline{G}_{\mathfrak{s}}$ and $\overline{\Phi}^{(2)}_{\mathfrak{s}}$ respectively replaced by $\overline{\boldsymbol{G}}_{\mathfrak{s}}$ and $\overline{\boldsymbol{\Phi}}^{(2)}_{\mathfrak{s}}$. As can be seen from Eqs. (387) and (388), solving the Bethe-Salpeter equation now translates into inverting the bosonic matrices $\mathcal{I}+\overline{\Pi}_{\mathfrak{s}}\overline{\boldsymbol{\Phi}}^{(2)}_{\mathfrak{s}}$ and $\overline{\Upsilon}_{\mathfrak{s}}$. Actually, by comparing Eqs. (385) and (389), we can notice that $\overline{\Upsilon}_{\mathfrak{s}}$ reduces to $\mathcal{I}+\overline{\Pi}_{\mathfrak{s}}\overline{\boldsymbol{\Phi}}^{(2)}_{\mathfrak{s}}$ by setting $\mathfrak{s}=1$ in the very last line of expression (385). Hence, the components of $\left(\mathcal{I}+\overline{\Pi}_{\mathfrak{s}}\overline{\boldsymbol{\Phi}}^{(2)}_{\mathfrak{s}}\right)^{\mathrm{inv}}$ can be directly deduced from those of $\overline{\Upsilon}^{\mathrm{inv}}_{\mathfrak{s}}$ as well and we can content ourselves to invert $\overline{\Upsilon}_{\mathfrak{s}}$, i.e. to solve:

$$\mathcal{I}_{(a_1,a_1')(a_2,a_2')}=\frac{1}{2}\sum_{a_3,a_3'=1}^{N}\overline{\Upsilon}_{\mathfrak{s},(a_1,a_1')(a_3,a_3')}\overline{\Upsilon}^{\mathrm{inv}}_{\mathfrak{s},(a_3,a_3')(a_2,a_2')}\;. \tag{390}$$

At $N=1$, the resolution of Eq. (390) simply yields:

$$\overline{\Upsilon}^{\mathrm{inv}}_{\mathfrak{s}}=\frac{4}{\overline{\Upsilon}_{\mathfrak{s}}}\;, \tag{391}$$

with $\overline{\Upsilon}_{\mathfrak{s}}\equiv\overline{\Upsilon}_{\mathfrak{s},(1,1)(1,1)}$ whereas, at $N=2$, $\overline{\Upsilon}^{\mathrm{inv}}_{\mathfrak{s}}$ is determined by solving $2^4=16$ coupled algebraic equations deduced from Eq. (390) combined with Eq. (385). Among the 16 components of $\overline{\Upsilon}_{\mathfrak{s}}^{\mathrm{inv}}$ at $N=2$, only 8 differ from zero and are given by:

$$\begin{aligned}\overline{\Upsilon}^{\mathrm{inv}}_{\mathfrak{s},(1,1)(1,1)} &= \overline{\Upsilon}^{\mathrm{inv}}_{\mathfrak{s},(2,2)(2,2)} \\ &= \frac{9\left(2+\mathfrak{s}\lambda\overline{\boldsymbol{G}}_{\mathfrak{s}}^{2}\right)}{9+9\mathfrak{s}\lambda\overline{\boldsymbol{G}}_{\mathfrak{s}}^{2}+2\mathfrak{s}^{2}\lambda^{2}\overline{\boldsymbol{G}}_{\mathfrak{s}}^{4}}\;,\end{aligned} \tag{392}$$

$$\begin{aligned}\overline{\Upsilon}^{\mathrm{inv}}_{\mathfrak{s},(1,1)(2,2)} &= \overline{\Upsilon}^{\mathrm{inv}}_{\mathfrak{s},(2,2)(1,1)} \\ &= -\frac{3\mathfrak{s}\lambda\overline{\boldsymbol{G}}_{\mathfrak{s}}^{2}}{9+9\mathfrak{s}\lambda\overline{\boldsymbol{G}}_{\mathfrak{s}}^{2}+2\mathfrak{s}^{2}\lambda^{2}\overline{\boldsymbol{G}}_{\mathfrak{s}}^{4}}\;,\end{aligned} \tag{393}$$

$$\begin{aligned}\overline{\Upsilon}^{\mathrm{inv}}_{\mathfrak{s},(1,2)(1,2)} = \overline{\Upsilon}^{\mathrm{inv}}_{\mathfrak{s},(1,2)(2,1)} = \overline{\Upsilon}^{\mathrm{inv}}_{\mathfrak{s},(2,1)(1,2)} = \overline{\Upsilon}^{\mathrm{inv}}_{\mathfrak{s},(2,1)(2,1)} \\ = \frac{3}{3+\mathfrak{s}\lambda\overline{\boldsymbol{G}}_{\mathfrak{s}}^{2}}\;.\end{aligned} \tag{394}$$

Following our previous explanation between Eqs. (389) and (390), we directly infer from this the 8 non-vanishing components of $\left(\mathcal{I}+\overline{\Pi}_{\mathfrak{s}}\overline{\boldsymbol{\Phi}}_{\mathfrak{s}}^{(2)}\right)^{\mathrm{inv}}$ (and therefore of $\left(\mathcal{I}+\overline{\Pi}_{\mathfrak{s}}\overline{\Phi}_{\mathfrak{s}}^{(2)}\right)^{\mathrm{inv}}$ involved in the pU-flow equations such as Eq. (369)):

$$\begin{aligned}\left(\mathcal{I}+\overline{\Pi}_{\mathfrak{s}}\overline{\boldsymbol{\Phi}}_{\mathfrak{s}}^{(2)}\right)^{\mathrm{inv}}_{(1,1)(1,1)} &= \left(\mathcal{I}+\overline{\Pi}_{\mathfrak{s}}\overline{\boldsymbol{\Phi}}_{\mathfrak{s}}^{(2)}\right)^{\mathrm{inv}}_{(2,2)(2,2)} \\ &= \frac{9\left(2+\lambda\overline{\boldsymbol{G}}_{\mathfrak{s}}^{2}\right)}{9+9\lambda\overline{\boldsymbol{G}}_{\mathfrak{s}}^{2}+2\lambda^{2}\overline{\boldsymbol{G}}_{\mathfrak{s}}^{4}}\;,\end{aligned} \tag{395}$$

$$\begin{aligned}\left(\mathcal{I}+\overline{\Pi}_{\mathfrak{s}}\overline{\boldsymbol{\Phi}}_{\mathfrak{s}}^{(2)}\right)^{\mathrm{inv}}_{(1,1)(2,2)} &= \left(\mathcal{I}+\overline{\Pi}_{\mathfrak{s}}\overline{\boldsymbol{\Phi}}_{\mathfrak{s}}^{(2)}\right)^{\mathrm{inv}}_{(2,2)(1,1)} \\ &= -\frac{3\lambda\overline{\boldsymbol{G}}_{\mathfrak{s}}^{2}}{9+9\lambda\overline{\boldsymbol{G}}_{\mathfrak{s}}^{2}+2\lambda^{2}\overline{\boldsymbol{G}}_{\mathfrak{s}}^{4}}\;,\end{aligned} \tag{396}$$

$$\begin{aligned}\left(\mathcal{I}+\overline{\Pi}_{\mathfrak{s}}\overline{\boldsymbol{\Phi}}_{\mathfrak{s}}^{(2)}\right)^{\mathrm{inv}}_{(1,2)(1,2)} &= \left(\mathcal{I}+\overline{\Pi}_{\mathfrak{s}}\overline{\boldsymbol{\Phi}}_{\mathfrak{s}}^{(2)}\right)^{\mathrm{inv}}_{(1,2)(2,1)} \\ &= \left(\mathcal{I}+\overline{\Pi}_{\mathfrak{s}}\overline{\boldsymbol{\Phi}}_{\mathfrak{s}}^{(2)}\right)^{\mathrm{inv}}_{(2,1)(1,2)} \\ &= \left(\mathcal{I}+\overline{\Pi}_{\mathfrak{s}}\overline{\boldsymbol{\Phi}}_{\mathfrak{s}}^{(2)}\right)^{\mathrm{inv}}_{(2,1)(2,1)} \\ &= \frac{3}{3+\lambda\overline{\boldsymbol{G}}_{\mathfrak{s}}^{2}}\;.\end{aligned} \tag{397}$$

We have also treated the mU-flow with $N_{\mathrm{SCPT}}$ up to 3 at $N=1$. To that end, we exploited the following expression of the modified LW functional:

$$\begin{aligned}\boldsymbol{\Phi}_{\mathfrak{s}}(G) &= \Phi_{\mathfrak{s}}(G) + \Phi_{\mathrm{SCPT},N_{\mathrm{SCPT}}=3}(U,G) \\ &\quad - \Phi_{\mathrm{SCPT},N_{\mathrm{SCPT}}=3}(U_{\mathfrak{s}},G) \\ &= \Phi_{\mathfrak{s}}(G) + \frac{1}{8}\lambda G^{2}\left(1-\mathfrak{s}\right) - \frac{1}{48}\lambda^{2}G^{4}\left(1-\mathfrak{s}^{2}\right) \\ &\quad + \frac{1}{48}\lambda^{3}G^{6}\left(1-\mathfrak{s}^{3}\right)\;,\end{aligned} \tag{398}$$

which can be deduced from Eq. (275) and from our choice (367) for the cutoff function, i.e. $U_{\mathfrak{s}}=\mathfrak{s}\lambda$ at $N=1$. With Eq. (398), we derive the differential equations underlying the mU-flow up to $N_{\mathrm{SCPT}}=3$ for the (0+0)-D $O(N)$ model at $N=1$, namely:

– At $N_{\mathrm{SCPT}}=1$:

$$\dot{\boldsymbol{G}}_{\mathfrak{s}} = \overline{\boldsymbol{G}}_{\mathfrak{s}}^{2}\dot{\boldsymbol{\Sigma}}_{\mathfrak{s}}\;, \tag{399}$$

$$\dot{\boldsymbol{\Gamma}}_{\mathfrak{s}}^{(\mathrm{2PI})} = \frac{\lambda}{24}\left(4\left(2\overline{\boldsymbol{G}}_{\mathfrak{s}}^{-2}+\overline{\boldsymbol{\Phi}}_{\mathfrak{s}}^{(2)}-\lambda\left(1-\mathfrak{s}\right)\right)^{-1}+\overline{\boldsymbol{G}}_{\mathfrak{s}}^{2}\right)-\frac{1}{8}\lambda\overline{\boldsymbol{G}}_{\mathfrak{s}}^{2}\;, \tag{400}$$

$$\dot{\boldsymbol{\Sigma}}_{\mathfrak{s}} = -\frac{\lambda}{3}\left(2+\overline{\boldsymbol{G}}_{\mathfrak{s}}^{2}\overline{\boldsymbol{\Phi}}_{\mathfrak{s}}^{(2)}\right)^{-1}\left(\left(2\overline{\boldsymbol{G}}_{\mathfrak{s}}^{-2}+\overline{\boldsymbol{\Phi}}_{\mathfrak{s}}^{(2)}-\lambda\left(1-\mathfrak{s}\right)\right)^{-2}\left(8\overline{\boldsymbol{G}}_{\mathfrak{s}}^{-3}-\overline{\boldsymbol{\Phi}}_{\mathfrak{s}}^{(3)}\right)-2\overline{\boldsymbol{G}}_{\mathfrak{s}}\right)\;, \tag{401}$$

$$\begin{aligned}\dot{\boldsymbol{\Phi}}_{\mathfrak{s}}^{(2)} = \frac{\lambda}{6}&\left(2\left(2\overline{\boldsymbol{G}}_{\mathfrak{s}}^{-2}+\overline{\boldsymbol{\Phi}}_{\mathfrak{s}}^{(2)}-\lambda\left(1-\mathfrak{s}\right)\right)^{-3}\left(8\overline{\boldsymbol{G}}_{\mathfrak{s}}^{-3}-\overline{\boldsymbol{\Phi}}_{\mathfrak{s}}^{(3)}\right)^{2}-64\overline{\boldsymbol{G}}_{\mathfrak{s}}^{-4}\left(2\overline{\boldsymbol{G}}_{\mathfrak{s}}^{-2}+\overline{\boldsymbol{\Phi}}_{\mathfrak{s}}^{(2)}-\lambda\left(1-\mathfrak{s}\right)\right)^{-2}\right. \\ &\left.+\left(2\overline{\boldsymbol{G}}_{\mathfrak{s}}^{-2}+\overline{\boldsymbol{\Phi}}_{\mathfrak{s}}^{(2)}-\lambda\left(1-\mathfrak{s}\right)\right)^{-2}\left(16\overline{\boldsymbol{G}}_{\mathfrak{s}}^{-4}-\overline{\boldsymbol{\Phi}}_{\mathfrak{s}}^{(4)}\right)+2\right)-\lambda+\frac{1}{2}\dot{\boldsymbol{G}}_{\mathfrak{s}}\overline{\boldsymbol{\Phi}}_{\mathfrak{s}}^{(3)}\;,\end{aligned} \tag{402}$$

$$\begin{aligned}\dot{\overline{\boldsymbol{\Phi}}}_{\mathfrak{s}}^{(3)} = {} & \frac{\lambda}{6}\left(2\overline{\boldsymbol{G}}_{\mathfrak{s}}^{-2}+\overline{\boldsymbol{\Phi}}_{\mathfrak{s}}^{(2)}-\lambda\left(1-\mathfrak{s}\right)\right)^{-2}\left(6\left(2\overline{\boldsymbol{G}}_{\mathfrak{s}}^{-2}+\overline{\boldsymbol{\Phi}}_{\mathfrak{s}}^{(2)}-\lambda\left(1-\mathfrak{s}\right)\right)^{-2}\left(8\overline{\boldsymbol{G}}_{\mathfrak{s}}^{-3}-\overline{\boldsymbol{\Phi}}_{\mathfrak{s}}^{(3)}\right)^{3}\right. \\ & \left.+6\left(2\overline{\boldsymbol{G}}_{\mathfrak{s}}^{-2}+\overline{\boldsymbol{\Phi}}_{\mathfrak{s}}^{(2)}-\lambda\left(1-\mathfrak{s}\right)\right)^{-1}\left(8\overline{\boldsymbol{G}}_{\mathfrak{s}}^{-3}-\overline{\boldsymbol{\Phi}}_{\mathfrak{s}}^{(3)}\right)\left(-48\overline{\boldsymbol{G}}_{\mathfrak{s}}^{-4}-\overline{\boldsymbol{\Phi}}_{\mathfrak{s}}^{(4)}\right)+384\overline{\boldsymbol{G}}_{\mathfrak{s}}^{-5}-\overline{\boldsymbol{\Phi}}_{\mathfrak{s}}^{(5)}\right)+\frac{1}{2}\dot{\overline{\boldsymbol{G}}}_{\mathfrak{s}}\overline{\boldsymbol{\Phi}}_{\mathfrak{s}}^{(4)}\;,\end{aligned} \tag{403}$$

– At $N_{\mathrm{SCPT}}=2$:

$$\dot{\overline{\boldsymbol{G}}}_{\mathfrak{s}}=\overline{\boldsymbol{G}}_{\mathfrak{s}}^{2}\dot{\overline{\boldsymbol{\Sigma}}}_{\mathfrak{s}}\;, \tag{404}$$

$$\dot{\overline{\boldsymbol{\Gamma}}}_{\mathfrak{s}}^{(\mathrm{2PI})}=\frac{\lambda}{24}\left(4\left(2\overline{\boldsymbol{G}}_{\mathfrak{s}}^{-2}+\overline{\boldsymbol{\Phi}}_{\mathfrak{s}}^{(2)}-\lambda\left(1-\mathfrak{s}\right)+\lambda^{2}\overline{\boldsymbol{G}}_{\mathfrak{s}}^{2}\left(1-\mathfrak{s}^{2}\right)\right)^{-1}+\overline{\boldsymbol{G}}_{\mathfrak{s}}^{2}\right)-\frac{1}{8}\lambda\overline{\boldsymbol{G}}_{\mathfrak{s}}^{2}+\frac{1}{24}\mathfrak{s}\lambda^{2}\overline{\boldsymbol{G}}_{\mathfrak{s}}^{4}\;, \tag{405}$$

$$\begin{aligned}\dot{\overline{\boldsymbol{\Sigma}}}_{\mathfrak{s}}= {} & -\frac{\lambda}{3}\left(2+\overline{\boldsymbol{G}}_{\mathfrak{s}}^{2}\overline{\boldsymbol{\Phi}}_{\mathfrak{s}}^{(2)}\right)^{-1}\left(\left(2\overline{\boldsymbol{G}}_{\mathfrak{s}}^{-2}+\overline{\boldsymbol{\Phi}}_{\mathfrak{s}}^{(2)}-\lambda\left(1-\mathfrak{s}\right)+\lambda^{2}\overline{\boldsymbol{G}}_{\mathfrak{s}}^{2}\left(1-\mathfrak{s}^{2}\right)\right)^{-2}\left(8\overline{\boldsymbol{G}}_{\mathfrak{s}}^{-3}-\overline{\boldsymbol{\Phi}}_{\mathfrak{s}}^{(3)}-4\lambda^{2}\overline{\boldsymbol{G}}_{\mathfrak{s}}\left(1-\mathfrak{s}^{2}\right)\right)\right. \\ & \left.-2\overline{\boldsymbol{G}}_{\mathfrak{s}}+2\mathfrak{s}\lambda\overline{\boldsymbol{G}}_{\mathfrak{s}}^{3}\right)\;,\end{aligned} \tag{406}$$

$$\begin{aligned}\dot{\overline{\boldsymbol{\Phi}}}_{\mathfrak{s}}^{(2)}= {} & \frac{\lambda}{6}\left(2\left(2\overline{\boldsymbol{G}}_{\mathfrak{s}}^{-2}+\overline{\boldsymbol{\Phi}}_{\mathfrak{s}}^{(2)}-\lambda\left(1-\mathfrak{s}\right)+\lambda^{2}\overline{\boldsymbol{G}}_{\mathfrak{s}}^{2}\left(1-\mathfrak{s}^{2}\right)\right)^{-3}\left(8\overline{\boldsymbol{G}}_{\mathfrak{s}}^{-3}-\overline{\boldsymbol{\Phi}}_{\mathfrak{s}}^{(3)}-4\lambda^{2}\overline{\boldsymbol{G}}_{\mathfrak{s}}\left(1-\mathfrak{s}^{2}\right)\right)^{2}\right. \\ & -64\overline{\boldsymbol{G}}_{\mathfrak{s}}^{-4}\left(2\overline{\boldsymbol{G}}_{\mathfrak{s}}^{-2}+\overline{\boldsymbol{\Phi}}_{\mathfrak{s}}^{(2)}-\lambda\left(1-\mathfrak{s}\right)+\lambda^{2}\overline{\boldsymbol{G}}_{\mathfrak{s}}^{2}\left(1-\mathfrak{s}^{2}\right)\right)^{-2} \\ & \left.+\left(2\overline{\boldsymbol{G}}_{\mathfrak{s}}^{-2}+\overline{\boldsymbol{\Phi}}_{\mathfrak{s}}^{(2)}-\lambda\left(1-\mathfrak{s}\right)+\lambda^{2}\overline{\boldsymbol{G}}_{\mathfrak{s}}^{2}\left(1-\mathfrak{s}^{2}\right)\right)^{-2}\left(16\overline{\boldsymbol{G}}_{\mathfrak{s}}^{-4}-\overline{\boldsymbol{\Phi}}_{\mathfrak{s}}^{(4)}-8\lambda^{2}\left(1-\mathfrak{s}^{2}\right)\right)+2\right) \\ & -\lambda+2\mathfrak{s}\lambda^{2}\overline{\boldsymbol{G}}_{\mathfrak{s}}^{2}+\frac{1}{2}\dot{\overline{\boldsymbol{G}}}_{\mathfrak{s}}\overline{\boldsymbol{\Phi}}_{\mathfrak{s}}^{(3)}\;,\end{aligned} \tag{407}$$

$$\begin{aligned}\dot{\overline{\boldsymbol{\Phi}}}_{\mathfrak{s}}^{(3)}= {} & \frac{\lambda}{6}\left(2\overline{\boldsymbol{G}}_{\mathfrak{s}}^{-2}+\overline{\boldsymbol{\Phi}}_{\mathfrak{s}}^{(2)}-\lambda\left(1-\mathfrak{s}\right)+\lambda^{2}\overline{\boldsymbol{G}}_{\mathfrak{s}}^{2}\left(1-\mathfrak{s}^{2}\right)\right)^{-2} \\ & \times\left(6\left(2\overline{\boldsymbol{G}}_{\mathfrak{s}}^{-2}+\overline{\boldsymbol{\Phi}}_{\mathfrak{s}}^{(2)}-\lambda\left(1-\mathfrak{s}\right)+\lambda^{2}\overline{\boldsymbol{G}}_{\mathfrak{s}}^{2}\left(1-\mathfrak{s}^{2}\right)\right)^{-2}\left(8\overline{\boldsymbol{G}}_{\mathfrak{s}}^{-3}-\overline{\boldsymbol{\Phi}}_{\mathfrak{s}}^{(3)}-4\lambda^{2}\overline{\boldsymbol{G}}_{\mathfrak{s}}\left(1-\mathfrak{s}^{2}\right)\right)^{3}\right. \\ & +6\left(2\overline{\boldsymbol{G}}_{\mathfrak{s}}^{-2}+\overline{\boldsymbol{\Phi}}_{\mathfrak{s}}^{(2)}-\lambda\left(1-\mathfrak{s}\right)+\lambda^{2}\overline{\boldsymbol{G}}_{\mathfrak{s}}^{2}\left(1-\mathfrak{s}^{2}\right)\right)^{-1}\left(8\overline{\boldsymbol{G}}_{\mathfrak{s}}^{-3}-\overline{\boldsymbol{\Phi}}_{\mathfrak{s}}^{(3)}-4\lambda^{2}\overline{\boldsymbol{G}}_{\mathfrak{s}}\left(1-\mathfrak{s}^{2}\right)\right) \\ & \left.\times\left(-48\overline{\boldsymbol{G}}_{\mathfrak{s}}^{-4}-\overline{\boldsymbol{\Phi}}_{\mathfrak{s}}^{(4)}-8\lambda^{2}\left(1-\mathfrak{s}^{2}\right)\right)+384\overline{\boldsymbol{G}}_{\mathfrak{s}}^{-5}-\overline{\boldsymbol{\Phi}}_{\mathfrak{s}}^{(5)}\right)+8\mathfrak{s}\lambda^{2}\overline{\boldsymbol{G}}_{\mathfrak{s}}+\frac{1}{2}\dot{\overline{\boldsymbol{G}}}_{\mathfrak{s}}\overline{\boldsymbol{\Phi}}_{\mathfrak{s}}^{(4)}\;,\end{aligned} \tag{408}$$

– At $N_{\mathrm{SCPT}}=3$:

$$\dot{\overline{\boldsymbol{G}}}_{\mathfrak{s}}=\overline{\boldsymbol{G}}_{\mathfrak{s}}^{2}\dot{\overline{\boldsymbol{\Sigma}}}_{\mathfrak{s}}\;, \tag{409}$$

$$\begin{aligned}\dot{\overline{\boldsymbol{\Gamma}}}_{\mathfrak{s}}^{(\mathrm{2PI})}= {} & \frac{\lambda}{24}\left(4\left(2\overline{\boldsymbol{G}}_{\mathfrak{s}}^{-2}+\overline{\boldsymbol{\Phi}}_{\mathfrak{s}}^{(2)}-\lambda\left(1-\mathfrak{s}\right)+\lambda^{2}\overline{\boldsymbol{G}}_{\mathfrak{s}}^{2}\left(1-\mathfrak{s}^{2}\right)-\frac{5}{2}\lambda^{3}\overline{\boldsymbol{G}}_{\mathfrak{s}}^{4}\left(1-\mathfrak{s}^{3}\right)\right)^{-1}+\overline{\boldsymbol{G}}_{\mathfrak{s}}^{2}\right)-\frac{1}{8}\lambda\overline{\boldsymbol{G}}_{\mathfrak{s}}^{2}+\frac{1}{24}\mathfrak{s}\lambda^{2}\overline{\boldsymbol{G}}_{\mathfrak{s}}^{4} \\ & -\frac{1}{16}\mathfrak{s}^{2}\lambda^{3}\overline{\boldsymbol{G}}_{\mathfrak{s}}^{6}\;,\end{aligned} \tag{410}$$

$$\begin{aligned}\dot{\overline{\boldsymbol{\Sigma}}}_{\mathfrak{s}}= {} & -\frac{\lambda}{3}\left(2+\overline{\boldsymbol{G}}_{\mathfrak{s}}^{2}\overline{\boldsymbol{\Phi}}_{\mathfrak{s}}^{(2)}\right)^{-1}\left(\left(2\overline{\boldsymbol{G}}_{\mathfrak{s}}^{-2}+\overline{\boldsymbol{\Phi}}_{\mathfrak{s}}^{(2)}-\lambda\left(1-\mathfrak{s}\right)+\lambda^{2}\overline{\boldsymbol{G}}_{\mathfrak{s}}^{2}\left(1-\mathfrak{s}^{2}\right)-\frac{5}{2}\lambda^{3}\overline{\boldsymbol{G}}_{\mathfrak{s}}^{4}\left(1-\mathfrak{s}^{3}\right)\right)^{-2}\right. \\ & \left.\times\left(8\overline{\boldsymbol{G}}_{\mathfrak{s}}^{-3}-\overline{\boldsymbol{\Phi}}_{\mathfrak{s}}^{(3)}-4\lambda^{2}\overline{\boldsymbol{G}}_{\mathfrak{s}}\left(1-\mathfrak{s}^{2}\right)+20\lambda^{3}\overline{\boldsymbol{G}}_{\mathfrak{s}}^{3}\left(1-\mathfrak{s}^{3}\right)\right)-2\overline{\boldsymbol{G}}_{\mathfrak{s}}+2\mathfrak{s}\lambda\overline{\boldsymbol{G}}_{\mathfrak{s}}^{3}-\frac{9}{2}\mathfrak{s}^{2}\lambda^{2}\overline{\boldsymbol{G}}_{\mathfrak{s}}^{5}\right)\;,\end{aligned} \tag{411}$$

$$\begin{aligned}\dot{\overline{\boldsymbol{\Phi}}}_{\mathfrak{s}}^{(2)} =& \frac{\lambda}{6}\Bigg(2\left(2\overline{\boldsymbol{G}}_{\mathfrak{s}}^{-2}+\overline{\boldsymbol{\Phi}}_{\mathfrak{s}}^{(2)}-\lambda\left(1-\mathfrak{s}\right)+\lambda^{2}\overline{\boldsymbol{G}}_{\mathfrak{s}}^{2}\left(1-\mathfrak{s}^{2}\right)-\frac{5}{2}\lambda^{3}\overline{\boldsymbol{G}}_{\mathfrak{s}}^{4}\left(1-\mathfrak{s}^{3}\right)\right)^{-3}\\ &\times\left(8\overline{\boldsymbol{G}}_{\mathfrak{s}}^{-3}-\overline{\boldsymbol{\Phi}}_{\mathfrak{s}}^{(3)}-4\lambda^{2}\overline{\boldsymbol{G}}_{\mathfrak{s}}\left(1-\mathfrak{s}^{2}\right)+20\lambda^{3}\overline{\boldsymbol{G}}_{\mathfrak{s}}^{3}\left(1-\mathfrak{s}^{3}\right)\right)^{2}\\ &-64\overline{\boldsymbol{G}}_{\mathfrak{s}}^{-4}\left(2\overline{\boldsymbol{G}}_{\mathfrak{s}}^{-2}+\overline{\boldsymbol{\Phi}}_{\mathfrak{s}}^{(2)}-\lambda\left(1-\mathfrak{s}\right)+\lambda^{2}\overline{\boldsymbol{G}}_{\mathfrak{s}}^{2}\left(1-\mathfrak{s}^{2}\right)-\frac{5}{2}\lambda^{3}\overline{\boldsymbol{G}}_{\mathfrak{s}}^{4}\left(1-\mathfrak{s}^{3}\right)\right)^{-2}\\ &+\left(2\overline{\boldsymbol{G}}_{\mathfrak{s}}^{-2}+\overline{\boldsymbol{\Phi}}_{\mathfrak{s}}^{(2)}-\lambda\left(1-\mathfrak{s}\right)+\lambda^{2}\overline{\boldsymbol{G}}_{\mathfrak{s}}^{2}\left(1-\mathfrak{s}^{2}\right)-\frac{5}{2}\lambda^{3}\overline{\boldsymbol{G}}_{\mathfrak{s}}^{4}\left(1-\mathfrak{s}^{3}\right)\right)^{-2}\\ &\times\left(16\overline{\boldsymbol{G}}_{\mathfrak{s}}^{-4}-\overline{\boldsymbol{\Phi}}_{\mathfrak{s}}^{(4)}-8\lambda^{2}\left(1-\mathfrak{s}^{2}\right)+120\lambda^{3}\overline{\boldsymbol{G}}_{\mathfrak{s}}^{2}\left(1-\mathfrak{s}^{3}\right)\right)+2\Bigg)-\lambda+2\mathfrak{s}\lambda^{2}\overline{\boldsymbol{G}}_{\mathfrak{s}}^{2}-\frac{15}{2}\mathfrak{s}^{2}\lambda^{3}\overline{\boldsymbol{G}}_{\mathfrak{s}}^{4}+\frac{1}{2}\dot{\overline{\boldsymbol{G}}}_{\mathfrak{s}}\overline{\boldsymbol{\Phi}}_{\mathfrak{s}}^{(3)}\,,\end{aligned} \tag{412}$$

$$\begin{aligned}\dot{\overline{\boldsymbol{\Phi}}}_{\mathfrak{s}}^{(3)} =& \frac{\lambda}{6}\left(2\overline{\boldsymbol{G}}_{\mathfrak{s}}^{-2}+\overline{\boldsymbol{\Phi}}_{\mathfrak{s}}^{(2)}-\lambda\left(1-\mathfrak{s}\right)+\lambda^{2}\overline{\boldsymbol{G}}_{\mathfrak{s}}^{2}\left(1-\mathfrak{s}^{2}\right)-\frac{5}{2}\lambda^{3}\overline{\boldsymbol{G}}_{\mathfrak{s}}^{4}\left(1-\mathfrak{s}^{3}\right)\right)^{-2}\\ &\times\Bigg(6\left(2\overline{\boldsymbol{G}}_{\mathfrak{s}}^{-2}+\overline{\boldsymbol{\Phi}}_{\mathfrak{s}}^{(2)}-\lambda\left(1-\mathfrak{s}\right)+\lambda^{2}\overline{\boldsymbol{G}}_{\mathfrak{s}}^{2}\left(1-\mathfrak{s}^{2}\right)-\frac{5}{2}\lambda^{3}\overline{\boldsymbol{G}}_{\mathfrak{s}}^{4}\left(1-\mathfrak{s}^{3}\right)\right)^{-2}\\ &\times\left(8\overline{\boldsymbol{G}}_{\mathfrak{s}}^{-3}-\overline{\boldsymbol{\Phi}}_{\mathfrak{s}}^{(3)}-4\lambda^{2}\overline{\boldsymbol{G}}_{\mathfrak{s}}\left(1-\mathfrak{s}^{2}\right)+20\lambda^{3}\overline{\boldsymbol{G}}_{\mathfrak{s}}^{3}\left(1-\mathfrak{s}^{3}\right)\right)^{3}\\ &+6\left(2\overline{\boldsymbol{G}}_{\mathfrak{s}}^{-2}+\overline{\boldsymbol{\Phi}}_{\mathfrak{s}}^{(2)}-\lambda\left(1-\mathfrak{s}\right)+\lambda^{2}\overline{\boldsymbol{G}}_{\mathfrak{s}}^{2}\left(1-\mathfrak{s}^{2}\right)-\frac{5}{2}\lambda^{3}\overline{\boldsymbol{G}}_{\mathfrak{s}}^{4}\left(1-\mathfrak{s}^{3}\right)\right)^{-1}\\ &\times\left(8\overline{\boldsymbol{G}}_{\mathfrak{s}}^{-3}-\overline{\boldsymbol{\Phi}}_{\mathfrak{s}}^{(3)}-4\lambda^{2}\overline{\boldsymbol{G}}_{\mathfrak{s}}\left(1-\mathfrak{s}^{2}\right)+20\lambda^{3}\overline{\boldsymbol{G}}_{\mathfrak{s}}^{3}\left(1-\mathfrak{s}^{3}\right)\right)\\ &\times\left(-48\overline{\boldsymbol{G}}_{\mathfrak{s}}^{-4}-\overline{\boldsymbol{\Phi}}_{\mathfrak{s}}^{(4)}-8\lambda^{2}\left(1-\mathfrak{s}^{2}\right)+120\lambda^{3}\overline{\boldsymbol{G}}_{\mathfrak{s}}^{2}\left(1-\mathfrak{s}^{3}\right)\right)+384\overline{\boldsymbol{G}}_{\mathfrak{s}}^{-5}-\overline{\boldsymbol{\Phi}}_{\mathfrak{s}}^{(5)}+480\lambda^{3}\overline{\boldsymbol{G}}_{\mathfrak{s}}\left(1-\mathfrak{s}^{3}\right)\Bigg)\\ &+8\mathfrak{s}\lambda^{2}\overline{\boldsymbol{G}}_{\mathfrak{s}}-60\mathfrak{s}^{2}\lambda^{3}\overline{\boldsymbol{G}}_{\mathfrak{s}}^{3}+\frac{1}{2}\dot{\overline{\boldsymbol{G}}}_{\mathfrak{s}}\overline{\boldsymbol{\Phi}}_{\mathfrak{s}}^{(4)}\,,\end{aligned} \tag{413}$$

where we have used our usual shorthand notations at $N=1$, i.e. $\overline{\boldsymbol{G}}_{\mathfrak{s}}\equiv\overline{\boldsymbol{G}}_{\mathfrak{s},11}$, $\overline{\boldsymbol{\Sigma}}_{\mathfrak{s}}\equiv\overline{\boldsymbol{\Sigma}}_{\mathfrak{s},11}$ and $\overline{\boldsymbol{\Phi}}_{\mathfrak{s}}^{(n)}\equiv\overline{\boldsymbol{\Phi}}_{\mathfrak{s},(1,1)\cdots(1,1)}^{(n)}$ $\forall n$.

### D.3 Kohn-Sham functional renormalization group

The generic differential equations that we solved to implement the KS-FRG were not given in section 4 that contains our 2PPI-FRG analysis. These flow equations, which are the counterparts of Eqs. (202) to (205) for the KS-FRG, belong to an infinite hierarchy of coupled differential equations for the vertices $\overline{\gamma}_{\mathfrak{s}}^{(n)}$ and are obtained from a vertex expansion of the master equation (223), as was explained right below Eq. (224). The first equations of this hierarchy read[31]:

$$\dot{\overline{\gamma}}_{\mathfrak{s}}=\int_{\alpha}\dot{V}_{\mathfrak{s},\alpha}\overline{\rho}_{\mathfrak{s},\alpha}+\int_{\alpha_1,\alpha_2}\overline{\rho}_{\mathfrak{s},\alpha_1}\overline{\Gamma}^{(2)}_{\mathrm{KS},\mathfrak{s},\alpha_1\alpha_2}\dot{\overline{\rho}}_{\mathfrak{s},\alpha_2}+\frac{1}{2}\int_{\alpha_1,\alpha_2}\dot{U}_{\mathfrak{s},\alpha_1\alpha_2}\left(\overline{\boldsymbol{G}}_{\mathfrak{s},\alpha_2\alpha_1}+\overline{\rho}_{\mathfrak{s},\alpha_2}\overline{\rho}_{\mathfrak{s},\alpha_1}\right)\,, \tag{414}$$

$$\dot{\overline{\rho}}_{\mathfrak{s},\alpha_1}=\int_{\alpha_2}\overline{\boldsymbol{G}}_{\mathfrak{s},\alpha_1\alpha_2}\left(-\dot{V}_{\mathfrak{s},\alpha_2}+\frac{1}{2}\int_{\alpha_3,\cdots,\alpha_6}\dot{U}_{\mathfrak{s},\alpha_3\alpha_4}\overline{\boldsymbol{G}}_{\mathfrak{s},\alpha_3\alpha_5}\left(\overline{\Gamma}^{(3)}_{\mathrm{KS},\mathfrak{s}}+\overline{\gamma}^{(3)}_{\mathfrak{s}}\right)_{\alpha_2\alpha_5\alpha_6}\overline{\boldsymbol{G}}_{\mathfrak{s},\alpha_6\alpha_4}-\int_{\alpha_3}\dot{U}_{\mathfrak{s},\alpha_2\alpha_3}\overline{\rho}_{\mathfrak{s},\alpha_3}\right)\,, \tag{415}$$

$$\begin{aligned}\dot{\overline{\gamma}}^{(2)}_{\mathfrak{s},\alpha_1\alpha_2}=&\int_{\alpha_3}\dot{\overline{\rho}}_{\mathfrak{s},\alpha_3}\overline{\gamma}^{(3)}_{\mathfrak{s},\alpha_3\alpha_1\alpha_2}+\dot{U}_{\mathfrak{s},\alpha_1\alpha_2}\\ &+\int_{\alpha_3,\cdots,\alpha_8}\dot{U}_{\mathfrak{s},\alpha_3\alpha_4}\overline{\boldsymbol{G}}_{\mathfrak{s},\alpha_3\alpha_5}\left(\overline{\Gamma}^{(3)}_{\mathrm{KS},\mathfrak{s}}+\overline{\gamma}^{(3)}_{\mathfrak{s}}\right)_{\alpha_1\alpha_5\alpha_6}\overline{\boldsymbol{G}}_{\mathfrak{s},\alpha_6\alpha_7}\left(\overline{\Gamma}^{(3)}_{\mathrm{KS},\mathfrak{s}}+\overline{\gamma}^{(3)}_{\mathfrak{s}}\right)_{\alpha_2\alpha_7\alpha_8}\overline{\boldsymbol{G}}_{\mathfrak{s},\alpha_8\alpha_4}\\ &-\frac{1}{2}\int_{\alpha_3,\cdots,\alpha_6}\dot{U}_{\mathfrak{s},\alpha_3\alpha_4}\overline{\boldsymbol{G}}_{\mathfrak{s},\alpha_3\alpha_5}\left(\overline{\Gamma}^{(4)}_{\mathrm{KS},\mathfrak{s}}+\overline{\gamma}^{(4)}_{\mathfrak{s}}\right)_{\alpha_1\alpha_2\alpha_5\alpha_6}\overline{\boldsymbol{G}}_{\mathfrak{s},\alpha_6\alpha_4}\,,\end{aligned} \tag{416}$$

$$\begin{aligned}\dot{\overline{\gamma}}^{(3)}_{\mathfrak{s},\alpha_1\alpha_2\alpha_3} = & \int_{\alpha_4} \dot{\overline{\rho}}_{\mathfrak{s},\alpha_4} \overline{\gamma}^{(4)}_{\mathfrak{s},\alpha_4\alpha_1\alpha_2\alpha_3} \\ & - \Bigg( \int_{\alpha_4,\cdots,\alpha_{11}} \dot{U}_{\mathfrak{s},\alpha_4\alpha_5} \overline{\boldsymbol{G}}_{\mathfrak{s},\alpha_4\alpha_6} \left(\overline{\Gamma}^{(3)}_{\mathrm{KS},\mathfrak{s}} + \overline{\gamma}^{(3)}_{\mathfrak{s}}\right)_{\alpha_1\alpha_6\alpha_7} \overline{\boldsymbol{G}}_{\mathfrak{s},\alpha_7\alpha_8} \left(\overline{\Gamma}^{(3)}_{\mathrm{KS},\mathfrak{s}} + \overline{\gamma}^{(3)}_{\mathfrak{s}}\right)_{\alpha_2\alpha_8\alpha_9} \overline{\boldsymbol{G}}_{\mathfrak{s},\alpha_9\alpha_{10}} \\ & \times \left(\overline{\Gamma}^{(3)}_{\mathrm{KS},\mathfrak{s}} + \overline{\gamma}^{(3)}_{\mathfrak{s}}\right)_{\alpha_3\alpha_{10}\alpha_{11}} \overline{\boldsymbol{G}}_{\mathfrak{s},\alpha_{11}\alpha_5} + (\alpha_2,\alpha_1,\alpha_3) + (\alpha_1,\alpha_3,\alpha_2) \Bigg) \\ & + \Bigg( \int_{\alpha_4,\cdots,\alpha_9} \dot{U}_{\mathfrak{s},\alpha_4\alpha_5} \overline{\boldsymbol{G}}_{\mathfrak{s},\alpha_4\alpha_6} \left(\overline{\Gamma}^{(4)}_{\mathrm{KS},\mathfrak{s}} + \overline{\gamma}^{(4)}_{\mathfrak{s}}\right)_{\alpha_1\alpha_2\alpha_6\alpha_7} \overline{\boldsymbol{G}}_{\mathfrak{s},\alpha_7\alpha_8} \left(\overline{\Gamma}^{(3)}_{\mathrm{KS},\mathfrak{s}} + \overline{\gamma}^{(3)}_{\mathfrak{s}}\right)_{\alpha_3\alpha_8\alpha_9} \overline{\boldsymbol{G}}_{\mathfrak{s},\alpha_9\alpha_5} \\ & + (\alpha_1,\alpha_3,\alpha_2) + (\alpha_2,\alpha_3,\alpha_1) \Bigg) - \frac{1}{2} \int_{\alpha_4,\cdots,\alpha_7} \dot{U}_{\mathfrak{s},\alpha_4\alpha_5} \overline{\boldsymbol{G}}_{\mathfrak{s},\alpha_4\alpha_6} \left(\overline{\Gamma}^{(5)}_{\mathrm{KS},\mathfrak{s}} + \overline{\gamma}^{(5)}_{\mathfrak{s}}\right)_{\alpha_1\alpha_2\alpha_3\alpha_6\alpha_7} \overline{\boldsymbol{G}}_{\mathfrak{s},\alpha_7\alpha_5} \ ,\end{aligned} \tag{417}$$

where the propagator $\overline{\boldsymbol{G}}_{\mathfrak{s}}$ satisfies:

$$\overline{\boldsymbol{G}}_{\mathfrak{s}}^{-1} \equiv \overline{\Gamma}^{(2\mathrm{PPI})(2)}_{\mathfrak{s}} = \overline{\Gamma}^{(2)}_{\mathrm{KS},\mathfrak{s}} + \overline{\gamma}^{(2)}_{\mathfrak{s}} \ . \tag{418}$$

Finally, the differential equations that we have solved to obtain the KS-FRG results presented in section 4 for the (0+0)-D $O(N)$ model can be directly inferred from Eqs. (414) to (417) combined with Eqs. (194), (196) and (197). Another important point to address is the resolution of the Kohn-Sham equation and the determination of the derivatives $\overline{\Gamma}^{(n)}_{\mathrm{KS}}$ that should be carried out alongside the treatment of these differential equations to implement the KS-FRG. For the studied toy model, we can once again benefit from the (0+0)-D framework and exploit the connection between 2PI and 2PPI EAs to deduce an explicit expression of $\Gamma_{\mathrm{KS}}(\rho)$ in terms of $\rho$. In particular, expression (118) of the free 2PI EA $\Gamma^{(2\mathrm{PI})}_0$ reduces for the (0+0)-D $O(N)$ model to:

$$\Gamma^{(2\mathrm{PI})}_0(\boldsymbol{G}) = -\frac{1}{2}\mathrm{Tr}_a\big[\ln(2\pi\boldsymbol{G})\big] + \frac{1}{2}\mathrm{Tr}_a\big(C^{-1}\boldsymbol{G}\big) - \frac{N}{2} \ . \tag{419}$$

After comparing the classical actions (133) and (195) at vanishing two-body interaction $U$, one deduces that the mean-field part $\Gamma_{\mathrm{KS},\mathfrak{s}}(\rho)$ of our toy model can also be expressed after performing the substitutions $\boldsymbol{G}_{a_1a_2} \rightarrow \rho_{a_1}\delta_{a_1a_2}$ and $C^{-1}_{a_1a_2} \rightarrow 2V_{\mathrm{KS},\mathfrak{s},a_1}\delta_{a_1a_2}$ in Eq. (419). This yields:

$$\Gamma_{\mathrm{KS},\mathfrak{s}}(\rho) = -\frac{1}{2}\sum_{a=1}^{N}\ln(2\pi\rho_a) + \sum_{a=1}^{N} V_{\mathrm{KS},\mathfrak{s},a}\rho_a - \frac{N}{2} \ . \tag{420}$$

As a result, the derivatives $\overline{\Gamma}^{(n)}_{\mathrm{KS},\mathfrak{s}}$ can be directly obtained by differentiating the latter equation, i.e.:

$$\overline{\Gamma}^{(2\mathrm{PPI})(n)}_{\mathrm{KS},\mathfrak{s},a_1\cdots a_n} = -\frac{1}{2}\frac{\partial^n}{\partial\rho_{a_1}\cdots\partial\rho_{a_n}}\sum_{a_{n+1}=1}^{N}\ln(\rho_{a_{n+1}})\bigg|_{\rho=\overline{\rho}_{\mathfrak{s}}} \quad \forall n \geq 2 \ . \tag{421}$$

However, it is in principle not possible to find an explicit expression for $\Gamma_{\mathrm{KS},\mathfrak{s}}[\rho]$ in terms of $\rho$ (such as that of Eq. (420)) at finite dimensions for the same reason that we can not determine such an analytical expression for $\Gamma^{(2\mathrm{PPI})}[\rho]$ or its non-interacting version $\Gamma^{(2\mathrm{PPI})}_0[\rho]$, namely that Legendre transforms underlying 2PPI EAs can not be carried out explicitly[34]. Therefore, for the implementation of the KS-FRG at finite dimensions, we rely on the same lengthy procedure to determine the derivatives $\overline{\Gamma}^{(n)}_{\mathrm{KS},\mathfrak{s}}$ as that required to determine the initial conditions of the standard 2PPI-FRG (see Eqs. (212) and (213)). In this situation, the derivatives $\overline{\Gamma}^{(n)}_{\mathrm{KS},\mathfrak{s}}$ are thus determined from the relations:

$$\overline{\Gamma}^{(2)}_{\mathrm{KS},\mathfrak{s},\alpha_1\alpha_2} = \left(W^{(2)}_{\mathrm{KS},\mathfrak{s}}[K=0]\right)^{-1}_{\alpha_1\alpha_2} \ , \tag{422}$$

$$\begin{aligned}\overline{\Gamma}^{(n)}_{\mathrm{KS},\mathfrak{s},\alpha_1\cdots\alpha_n} = & \int_{\alpha_{2n-2}} \left(W^{(2)}_{\mathrm{KS},\mathfrak{s}}[K]\right)^{-1}_{\alpha_n\alpha_{2n-2}} \frac{\delta}{\delta K_{\alpha_{2n-2}}} \cdots \\ & \times \int_{\alpha_{n+1}} \left(W^{(2)}_{\mathrm{KS},\mathfrak{s}}[K]\right)^{-1}_{\alpha_3\alpha_{n+1}} \\ & \times \frac{\delta}{\delta K_{\alpha_{n+1}}} \left(W^{(2)}_{\mathrm{KS},\mathfrak{s}}[K]\right)^{-1}_{\alpha_1\alpha_2}\bigg|_{K=0} \\ & \forall n \geq 3 \ ,\end{aligned} \tag{423}$$

where

$$\begin{aligned}Z_{\mathrm{KS},\mathfrak{s}}[K] = & \ e^{W_{\mathrm{KS},\mathfrak{s}}[K]} \\ = & \int \mathcal{D}\widetilde{\psi}^\dagger\mathcal{D}\widetilde{\psi} \ e^{-\int_\alpha \widetilde{\psi}^\dagger_\alpha\left(\hat{O}_{\mathrm{kin},\alpha}+V_{\mathrm{KS},\mathfrak{s},\alpha}-\mu\right)\widetilde{\psi}_\alpha + \int_\alpha K_\alpha \widetilde{\psi}^\dagger_\alpha\widetilde{\psi}_\alpha} \ .\end{aligned} \tag{424}$$

Finally, we explain how the Kohn-Sham equation is solved at each step of the KS-FRG flow in the framework of the (0+0)-D $O(N)$ model. This amounts to determining the Kohn-Sham potential $V_{\mathrm{KS},\mathfrak{s}}$ that satisfies the equality:

$$\frac{\partial\Gamma_{\mathrm{KS},\mathfrak{s}}(\rho)}{\partial\rho_a}\bigg|_{\rho=\overline{\rho}_{\mathfrak{s}}} = 0 \quad \forall a,\mathfrak{s} \ , \tag{425}$$

with $\Gamma_{\mathrm{KS},\mathfrak{s}}(\rho)$ given by Eq. (420). The solution of this procedure is simply:

$$V_{\mathrm{KS},\mathfrak{s},a} = \frac{1}{2}\overline{\rho}^{-1}_{\mathfrak{s},a} \quad \forall a,\mathfrak{s} \ . \tag{426}$$

After plugging result (426) into Eq. (420) at vanishing source $K$, one directly obtains:

$$\overline{\Gamma}_{\mathrm{KS},\mathfrak{s}} = -\frac{1}{2}\sum_{a=1}^{N} \ln\left(2\pi\overline{\rho}_{\mathfrak{s},a}\right) , \tag{427}$$

which is consistent with relation (231).